\documentclass[pdflatex,sn-mathphys-ay]{sn-jnl}

\usepackage[T1]{fontenc}
\usepackage[utf8]{inputenc} % only needed for pdfLaTeX
\usepackage{lmodern}

\makeatletter

\renewcommand\normalsize{%
   \@setfontsize\normalsize{12bp}{14.4bp}%
   \abovedisplayskip 12\p@ \@plus2\p@ \@minus1\p@
   \abovedisplayshortskip \z@ \@plus3\p@
   \belowdisplayshortskip 3\p@ \@plus3\p@ \@minus3\p@
   \belowdisplayskip \abovedisplayskip
   \let\@listi\@listI
}
\normalsize

\renewcommand\small{%
   \@setfontsize\small{11bp}{13.2bp}%
}

\renewcommand\footnotesize{%
   \@setfontsize\footnotesize{10bp}{12bp}%
}

\renewcommand\scriptsize{%
   \@setfontsize\scriptsize{9bp}{10.5bp}%
}

\renewcommand\tiny{%
   \@setfontsize\tiny{8bp}{9.5bp}%
}

\renewcommand\large{%
   \@setfontsize\large{14bp}{16.8bp}%
}

\renewcommand\Large{%
   \@setfontsize\Large{16bp}{19.2bp}%
}

\renewcommand\LARGE{%
   \@setfontsize\LARGE{18bp}{21.6bp}%
}

\renewcommand\huge{%
   \@setfontsize\huge{20bp}{24bp}%
}

\renewcommand\Huge{%
   \@setfontsize\Huge{24bp}{28.8bp}%
}

\def\Artcatfont{%
  \reset@font\fontsize{10bp}{12bp}\selectfont
}

\def\Titlefont{%
  \reset@font\fontsize{20bp}{24bp}\selectfont\titraggedcenter
}

\def\SubTitlefont{%
  \reset@font\fontsize{16bp}{19bp}\selectfont\titraggedcenter
}

\def\Authorfont{%
  \reset@font\fontsize{13bp}{15.5bp}\selectfont\boldmath\titraggedcenter
}

\def\addressfont{%
  \reset@font\fontsize{12bp}{14.4bp}\selectfont\titraggedcenter
}

\def\abstractheadfont{%
  \reset@font\fontsize{12bp}{14.4bp}\bfseries\selectfont\titraggedcenter
}

\def\abstractsubheadfont{%
  \reset@font\fontsize{12bp}{14.4bp}\bfseries\selectfont
}

\def\abstractfont{%
  \reset@font\fontsize{12bp}{14.4bp}\selectfont
  \leftskip=24pt
  \rightskip=24pt
  \parfillskip=0pt plus 1fil
}

\def\keywordfont{%
  \reset@font\fontsize{11bp}{13.2bp}\selectfont
  \leftskip=24pt
  \rightskip=24pt plus0.5fill
}

\def\historyfont{%
  \reset@font\fontsize{11bp}{13.2bp}\selectfont
  \leftskip=24pt
  \rightskip=24pt plus0.5fill
}

\def\mottofont{%
  \reset@font\fontfamily{\rmdefault}%
  \fontsize{11bp}{13.2bp}\fontshape{it}\selectfont\raggedright
}

\def\opheaderfont{%
  \reset@font\fontsize{11bp}{13bp}\selectfont
}

\def\headerfont{%
  \reset@font\fontsize{11bp}{13bp}\selectfont
}

\def\footerfont{%
  \reset@font\fontsize{11bp}{13bp}\selectfont
}

\def\sectionfont{%
  \reset@font\fontfamily{\rmdefault}%
  \fontsize{16bp}{19bp}\bfseries\selectfont\raggedright\boldmath
}

\def\subsectionfont{%
  \reset@font\fontfamily{\rmdefault}%
  \fontsize{14bp}{16.8bp}\bfseries\selectfont\raggedright\boldmath
}

\def\subsubsectionfont{%
  \reset@font%
  \fontsize{13bp}{15.5bp}\bfseries\selectfont\raggedright\boldmath
}

\def\paragraphfont{%
  \reset@font%
  \fontsize{12bp}{14.4bp}\bfseries\itshape\selectfont\raggedright
}

\def\bmheadfont{%
  \reset@font\fontfamily{\rmdefault}%
  \fontsize{12bp}{14.4bp}\bfseries\selectfont\raggedright\boldmath
}

\def\figurecaptionfont{%
  \reset@font\fontfamily{\rmdefault}%
  \fontsize{11bp}{13.2bp}\selectfont
}

\def\tablecaptionfont{%
  \reset@font\fontsize{11bp}{13.2bp}\selectfont
}

\def\tablebodyfont{%
  \reset@font\fontsize{11bp}{13.2bp}\selectfont
}

\def\tablecolheadfont{%
  \reset@font\fontsize{11bp}{13.2bp}\selectfont\bfseries\boldmath
}

\def\tablefootnotefont{%
  \reset@font\fontsize{10bp}{12bp}\selectfont
}

\def\quotefont{%
  \reset@font\fontfamily{\rmdefault}%
  \fontsize{11bp}{13.2bp}\selectfont
}

\def\authbiotextfont{%
  \reset@font\fontsize{10.5bp}{12.5bp}\selectfont
}

\def\eqnheadfont{%
  \reset@font\fontfamily{\rmdefault}%
  \fontsize{16bp}{19bp}\bfseries\selectfont
}

\makeatother

\usepackage{graphicx}%
\usepackage{multirow}%
\usepackage{amsmath,amssymb,amsfonts}%
\usepackage{bm}
\usepackage{amsthm}%
\usepackage{mathrsfs}%
\usepackage[title]{appendix}%
\usepackage{xcolor}%
\usepackage{textcomp}%
\usepackage{manyfoot}%
\usepackage{booktabs}%
\usepackage{algorithm}%
\usepackage{algorithmicx}%
\usepackage{algpseudocode}%
\usepackage{listings}%
\usepackage{enumerate}
\usepackage{subfig}
\newcommand{\ri}{\mathrm{i}}

\newcommand{\diag}{\mathrm{diag}}

\usepackage{todonotes}
\usepackage{comment}
\usepackage{soul}

\usepackage{tikz}
\usetikzlibrary{shapes.geometric, arrows}
\tikzstyle{specif} = [rectangle, rounded corners, 
minimum width=3cm, 
minimum height=1.24cm,
draw=black,
thick,
fill=blue!10]
\tikzstyle{process} = [rectangle,
minimum width=3cm, 
minimum height=1.24cm,
draw=black,
thick,
fill=red!10]
\tikzstyle{arrow} = [thick,->,>=stealth]

\theoremstyle{thmstyletwo}%

\theoremstyle{thmstylethree}%

\begin{document}

\title[Monolithic kinetic algorithm for heterogeneous porous media systems using a continuous one-domain approach]{Monolithic kinetic algorithm for heterogeneous porous media systems using a continuous one-domain approach}

% Monolithic kinetic algorithm for heterogeneous porous media systems using a continuous one-domain approach
% Lattice Boltzmann model for heterogeneous porous media systems using a macroscopic one-domain approach
% Lattice Boltzmann model for volume-averaged fluid dynamics in heterogeneous non-Darcian porous media

\author*[1]{\fnm{Nikita O.} \sur{Gusev}}\email{ngusev@ethz.ch} 
\author*[1]{\fnm{Ilya V.} \sur{Karlin}}\email{ikarlin@ethz.ch} 
\affil[1]{\orgdiv{Department of Mechanical and Process Engineering}, \orgname{ETH Z{\"u}rich}, \orgaddress{\city{8092 Zurich}, \country{Switzerland}}}

\abstract{We propose a lattice Boltzmann model (LBM) on standard lattices for simulating multi-dimensional, weakly compressible, isothermal flows within and around isotropic heterogeneous porous media. The model incorporates Darcy--Forchheimer drag and a Brinkman-like effective viscous stress tensor. In the hydrodynamic limit, it recovers a generalized volume-averaged formulation valid in both free-fluid and porous-medium regions. By relying on a single kinetic equation and a monolithic LBM algorithm, the formulation provides a one-domain solver for free-fluid/porous-medium interactions. Unlike previous LBM formulations for porous media, the proposed model recovers the correct porosity scaling of both the pressure and convective terms, while preserving the isotropy, and hence the Galilean invariance, of the viscous stress tensor. Linear and nonlinear drag, variable-porosity corrections, and additional body forces are incorporated through a consistent generalized forcing scheme. The model allows the speed of sound to be specified independently thereby improving computational efficiency. In addition, it includes a freely tunable effective bulk viscosity that can be used to enhance numerical stability. Model performance was evaluated using 2D benchmark flow problems. The ability of the proposed LBM model to simulate transport between free-fluid and heterogeneous porous regions within a one-domain framework enables a broad range of applications, particularly in early-stage, device-scale design studies of engineered porous structures with spatially varying porosity.}

\keywords{Lattice Boltzmann method, heterogeneous porous media, volume-averaging, Darcy-Brinkman-Forchheimer model, non-Darcian flow, Free-fluid/porous-medium system.}

\maketitle
\clearpage
\newpage
\section*{Article Highlights}
\begin{itemize}
    \setlength{\itemsep}{1em}
    \item \textbf{Generalized Darcy-Forchheimer-Brinkman macroscopic model} for weakly compressible, isothermal flow in heterogeneous isotropic porous media.
    \item \textbf{Continuous one-domain formulation} applicable throughout the entire computational domain, enabling a unified treatment of heterogeneous porous media and pure-fluid channel regions through a smooth transition zone.
    \item \textbf{Monolithic solution algorithm} based on a single generalized kinetic equation and realized using a lattice Boltzmann scheme on standard lattices.
    \item \textbf{Improved computational efficiency} achieved through the independent specification of the speed of sound and the effective kinematic bulk viscosity.
    \item \textbf{Adapted wet-node boundary-condition schemes} that extend common lattice Boltzmann boundary treatments to the proposed porous-media model.
    \item \textbf{Consistent hydrodynamic limit} recovering an isotropic viscous stress with the correct porosity scaling, validated through dispersion--dissipation analysis and two-dimensional benchmark problems.
    \item \textbf{Numerical demonstrations} including confined flow past a permeable cylinder and lid-driven cavity flow with a porous obstacle, illustrating the ability of the proposed model to capture transport between free-fluid and porous-medium regions within a one-domain framework.
\end{itemize}
\clearpage
\newpage
\tableofcontents
\vspace{0.4cm}
{\color{gray}\hrule}
\vspace{0.4cm}

\section{Introduction}\label{sec:intro}
\subsection{Motivation}
Flow in porous media is a classical problem in fluid mechanics, motivated both by its fundamental significance and by its wide range of applications across engineering, earth and environmental sciences, energy systems, and related fields (\cite{Feder_2022_Porous_Media,Wood_He_Apte_2020,Nield_Bejan_2017}). Porous media may be either naturally occurring, such as geological formations, or engineered, such as porous materials and structured components used in industrial devices. Their practical importance arises from distinctive properties including high specific surface area, tunable permeability, and selective transport capability, which make porous structures especially useful for controlling momentum, heat, and mass transfer. Understanding the mechanisms that govern flow through such media is therefore essential for the design, optimisation, and prediction of many natural and engineered systems.

Classical examples are found throughout chemical and process engineering, where engineered porous structures are used in filtration and separation systems (\cite{Rabiee_2019,Valverde_Griffiths_2024}), flow distributors (\cite{Banhart_2001_metal_foams,Kumar_2024_flowdist}), heat exchangers (\cite{Khoso_2026_HEX,Rashidi_2019}), energy storage and conversion devices (\cite{Zhang_2021_FC,Mathias_2010}), and chemical reactors (\cite{Michaud_2023,Zhiqiang_2025}). More recently, porous medium combustion has also emerged as a promising technology for achieving efficient and environmentally favourable fuel combustion (\cite{Banerjee_2021_PMC_review,Gharehghani_2021_PMCT}). At the same time, many important applications involve naturally occurring porous domains, including reservoir engineering for oil and gas recovery from petroliferous formations (\cite{Ahmed_2019_reservoir_rock,Ahmed_2019_Fluid_Flow}), geothermal resource exploration (\cite{Flovenz_2012_Geothermal_Energy,Cumming_2016_Geophysics}), and hydrogeology and geoenvironmental engineering (\cite{Ladd_Szymczak_2021,Mohamed_2018_GeoEngineering}).

Realistic porous structures, whether natural or engineered, are often heterogeneous, with porosity and permeability varying spatially. In addition, the broad range of characteristic length and velocity scales encountered in practical systems gives rise to a variety of flow regimes in which viscous and inertial effects may both be important (\cite{Wood_2007_Inertial,Feder_2022_Porous_Media,Wood_He_Apte_2020,Jin_Kuznetsov_2024}). These features make porous-media flows not only practically important, but also physically rich and challenging to model.

In recent years, increasing attention has been given to coupled flows in domains that contain both pure-fluid regions and permeable media. Such configurations differ from the classical problem of flow through an extended porous medium because the permeable regions are finite and are directly adjacent to regions of unobstructed fluid. Flows over permeable surfaces, as well as flows through and around permeable barriers, arise in many engineering applications, including fuel cells (\cite{Divisek_2010,Jarauta_Compressible_Porous_2020}), water purification systems (\cite{Shannon_2008}), and fog-water harvesting devices (\cite{Labbe_2019,Ghosh_2023}). A related class of problems is found in canopy flows, where fluid moves through and above arrays of obstacles protruding from a surface, such as forests, aquatic vegetation, coral reefs, urban city blocks, or wind farms (\cite{Foggi_Rota_Monti_Olivieri_Rosti_2024,He_Liu_Shen_2022}). Because the characteristic size of an individual canopy element is typically much smaller than the overall extent of the canopy, such systems are often represented as porous or permeable structures across a range of physical scales (\cite{Zampogna_etal_2016,Battiato_Rubol_2014,Papke_2013,Battiato_2012}).

A particularly interesting subset of these coupled-flow problems concerns flow past permeable bluff bodies, which appears in several natural phenomena and engineering applications (\cite{Ciuti_2021_porous_sphere,Seol_Kim_Kim_2024}). Unlike impermeable obstacles, permeable bodies allow part of the incoming fluid to pass through their internal structure. This permeability alters the interaction between the flow and the body and can substantially modify the downstream wake dynamics. These effects are especially relevant for passive flow-control strategies, where porous materials may be used either as coatings on bluff bodies or as the main structural material of bodies composed of patterned obstructions. Such approaches have been shown to promote drag reduction (\cite{Klausmann_Ruck_2017,Du_2022a,Du_2022b,Xu_2022}), mitigate vortex-induced vibrations, and suppress aerodynamic noise (\cite{Sueki_2009,Sato_Hattori_2021,Yuan_etal_2021,Geyer_2020}).

\subsection{Porous media modeling}
A deeper understanding of porous-media flow dynamics is essential for advancing the design and optimisation of the systems and technologies discussed above. However, experimental investigation of flow inside porous structures remains challenging, particularly when the internal geometry is complex or optically inaccessible (\cite{Seol_Kim_Kim_2024}). Mathematical modelling therefore provides an indispensable tool for studying such systems. Since the resulting models rarely admit analytical solutions in practical configurations, numerical simulations are typically required.

Modelling approaches for porous-media flows can generally be divided into two broad categories: microscale and macroscale modelling. Microscale modelling resolves the geometry of the porous matrix explicitly and describes the fluid motion around and through the pore space by solving the governing transport equations that are valid at the pore scale (\cite{Icardi_2014_pore_scale,Chen_2022_Pore_scale,Feriadi_Arbie_Fauzi_Fariduzzaman_2024}). While this approach provides detailed insight into the local flow physics, it is computationally demanding because realistic porous media often possess highly complex geometries and involve a wide range of characteristic length scales. As a result, direct numerical simulations of an entire macroscopic device or natural system are usually prohibitively expensive. In practice, pore-scale simulations are therefore often restricted to small representative portions of the domain, which may limit the generality of the results and make them sensitive to the chosen microscopic configuration.

Macroscale modelling offers an alternative route for reducing the computational cost of such multiscale problems. Instead of resolving the pore geometry directly, the aim is to derive continuum equations that describe the averaged behaviour of the fluid-solid system while retaining the essential geometrical information and physical effects of the porous matrix. This is achieved by exploiting the separation of scales between the pore-scale dynamics and the macroscopic length scale of the system. The governing transport equations, which are originally valid only within the fluid phase, are upscaled to obtain effective equations that are valid throughout the macroscopic domain.

Two principal techniques are commonly used for this purpose: homogenisation and the method of volume averaging. Homogenisation relies on asymptotic multiscale expansions and performs the upscaling by considering the limit in which the microscopic length scale tends to zero (\cite{Hornung_1997}). The method of volume averaging, by contrast, uses pore-scale information together with statistical smoothing and spatial averaging procedures to derive continuum equations for macroscopic fields (\cite{Whitaker_VA_method_1998}). In both approaches, the resulting equations represent the influence of the porous matrix on the fluid through effective macroscopic quantities and closure relations.

Both homogenisation and volume averaging have played an important role in the rigorous theoretical development of closures that connect pore-scale structure with macroscopic transport behaviour. They often recover, justify, or extend classical relations that were originally obtained heuristically or experimentally, including Darcy’s law (\cite{Darcy_1856}) and its extensions such as the Brinkman law (\cite{Brinkman_1949a,Brinkman_1949b}) and the Forchheimer law (\cite{Forchheimer_1901}).

When coupled free-fluid and porous-medium flows are modelled at the macroscopic level, additional difficulties arise from the variation of material and transport properties across the interface between the clear-fluid and porous regions. Two main strategies are commonly used to describe such coupled systems: two-domain and one-domain formulations (\cite{Battiato_2012}). In a two-domain formulation, different governing equations are solved in the two regions. The porous medium is typically described by Darcy’s law or one of its extensions, whereas the clear-fluid region is described by the Navier--Stokes equations or by appropriate simplified forms. The two sets of equations are then coupled through additional interfacial conditions imposed at the boundary separating the two domains.

A wide variety of such interfacial conditions has been proposed. These include conditions allowing a discontinuity of the interfacial velocity (\cite{Beavers_Joseph_1967,Taylor_1971,Richardson_1971,Saffman_1971}), continuity of pressure (\cite{Ene_Sanchez_1975,Levy_Sanchez_1975}), continuity of tangential velocity combined with a discontinuity of shear stress (\cite{Ochoa_Tapia_Whitaker_1995a,Ochoa_Tapia_Whitaker_1995b,Ochoa_Tapia_Whitaker_1998}), continuity of both velocity and stress (\cite{Neale_Nader_1974,Vafai_Kim_1990}), and discontinuity of both tangential velocity and shear stress (\cite{Cieszko_Kubik_1999}). These different formulations reflect different assumptions about momentum transfer across the interface and about the effective representation of the interfacial region. Detailed comparisons of the differences and common features of these conditions can be found in \cite{Alazmi_Vafai_2001,BARS_WORSTER_2006}.

In contrast, the one-domain approach applies a single set of conservation equations throughout the entire computational domain. The interface is not treated as an explicit internal boundary; instead, it is represented implicitly through spatial variations of the physical properties. This avoids the need to prescribe separate interfacial boundary conditions. Within this framework, the transition between the clear-fluid and porous regions may be described either as a continuous transition zone, where the physical variables undergo possibly strong but continuous variations (\cite{Verma_Tomar_2023,Chen_Wang_2014,Arico_Helmig_Puleo_Schneider_2024}), or as a sharp interface, where the physical variables may be discontinuous (\cite{Basu_Khalili_1999,Chikh_Boumedien_Bouhadef_Lauriat_1998,Silva_Lemos_2003}). These two representations correspond to different levels of description of the interfacial region (\cite{Chandesris_Jamet_2007}).

As the local properties of the medium change from those of a porous material to those of a clear fluid, the one-domain equations recover the appropriate governing forms in each region. At the same time, the conservation laws implicitly enforce the relevant hydrodynamic matching conditions, such as continuity of velocity and stress, across the interface. In the case of a sharp interface, however, the abrupt variation of material properties can introduce numerical difficulties and may require additional treatment. For example, harmonic averaging has been used to regularise singularities and define effective interfacial properties (\cite{Chikh_Boumedien_Bouhadef_Lauriat_1998}). Further discussion of the differences, similarities, and asymptotic equivalence between one-domain and two-domain formulations can be found in \cite{Goyeau_Lhuillier_2003,Chandesris_Jamet_2006,Jamet_Chandesris_Goyeau_2008,Hirata_Goyeau_2009}. More recently, hybrid formulations have also been introduced, in which a complex interface is embedded within a finite transition zone (\cite{Ruan_Rybak_2025,Ruan_Rybak_2026}).

\subsection{Lattice Boltzmann models}
The lattice Boltzmann method (LBM) can be interpreted as a bottom-up strategy for constructing numerical approximations to a prescribed set of macroscopic transport equations (\cite{Kruger_LBM_2016,succi_LBM_2018}). In contrast to conventional top-down methods, which discretise the hydrodynamic partial differential equations directly, the LBM originates from the discretisation of a Boltzmann-type kinetic equation in phase space. The macroscopic equations recovered in the hydrodynamic limit are therefore determined by the particular kinetic ingredients of the method, including the choice of equilibrium distribution, collision model, forcing scheme, and discrete velocity set.

In the context of porous-media flows, LBM has been used extensively at the pore scale as a direct numerical simulation method. In such models, the fluid motion is resolved directly within realistic porous geometries, which makes LBM particularly attractive because of its ability to handle complex boundaries and multiscale flow physics. Pore-scale LBM simulations have been used, for example, to determine drag-force and permeability correlations in porous media (\cite{Fattahi_Waluga_etal_2016}). More recently, there has also been growing interest in hybrid approaches that couple LBM with pore-network models, with the aim of combining the accuracy of pore-scale LBM with the computational efficiency of pore-network representations (\cite{Zhao_Liu_Qin_Fei_2023}).

At the macroscopic level, LBM models for porous media can be broadly divided into two main classes: “gray” schemes and force-based models (\cite{Ginzburg_Silva_Talon_2015}). Gray schemes, first introduced by \cite{Dardis_McCloskey_1998}, modify the kinetic population dynamics by partially reverting the post-collision populations through a partial-bounce-back mechanism. These schemes provide an efficient way of representing porous resistance, but they are generally limited to Darcy or Darcy-Brinkman-type models. Force-based models, first introduced by \cite{Spaid_Phelan_1997}, instead represent the effect of the porous matrix through a momentum sink, typically written as a drag force. Since the drag contribution is introduced through a forcing scheme, this class of models can, in principle, incorporate a wider range of porous-media closures.

One of the most widely used force-based formulations is the volume-averaged Navier-Stokes LBM with a Darcy-Forchheimer-Brinkman closure proposed by \cite{Guo_Porous_2002}. This model was the first to include porosity directly in the equilibrium distribution while accounting for both linear Darcy drag and nonlinear Forchheimer drag through a forcing term. However, the recovered momentum equation contains inconsistent porosity scaling in the pressure-gradient and viscous-stress terms, which limits the formulation to domains with homogeneous porosity. A different formulation was later proposed by \cite{Zhang_2014}, which corrected the porosity scaling of the pressure gradient. Nevertheless, for variable-porosity media, the recovered Brinkman viscous stress remained inconsistent because it was expressed in terms of gradients of the intrinsic velocity rather than gradients of the superficial, or Darcy, velocity.

Related issues also arise in more recent kinetic formulations. For example, \cite{Sawant_Karlin_2025} presented a model for chemically reacting multicomponent flow in homogenised porous media based on the Brinkman penalisation method. While this formulation recovers the correct porosity scaling of the pressure gradient, the viscous stress remains inconsistent in variable-porosity domains, again restricting the model to homogeneous porous media. To the best of our knowledge, no existing LBM formulation recovers a Brinkman viscous stress tensor with fully consistent porosity scaling, expressed in terms of gradients of the mean velocity in the porous mass, also called the Darcy or superficial velocity, in accordance with the original proposal of \cite{Brinkman_1949a,Brinkman_1949b}. Such a formulation is necessary for applying macroscopic LBM models consistently to flows in heterogeneous porous media with spatially varying porosity.

LBM has also been applied to coupled free-fluid and porous-medium flows. Several studies have developed models for coupled free-fluid and homogeneous porous-medium configurations using a discontinuous one-domain approach (\cite{Rong_Guo_Lu_Shi_2011,Pepona_Favier_2016,Wang_Wang_Guo_Mi_2015}), while others have used a two-domain formulation with interfacial coupling through stress-jump conditions (\cite{Bai_Yu_Winoto_Low_2009}). Another interesting direction is the hybrid upscaled LBM model of \cite{Li_Brown_2017}, in which a macroscopic simulation on a coarse grid is coupled with pore-scale microscopic simulations on fine grids. In this approach, the effective permeability is computed from microscopic local problems and then supplied to the upscaled macroscopic model. For a broader overview of LBM approaches for transport in porous media, the reader is referred to the review by \cite{He_Liu_Li_Tao_2019}.

\subsection{Present study}
The present study develops a generalized Darcy-Forchheimer-Brinkman macroscopic model for weakly compressible, isothermal flow in heterogeneous isotropic porous media. The target macroscopic formulation, summarized in Section \ref{sec:target_macro}, is written as a continuous one-domain model that is applicable throughout the entire computational domain. This enables a unified treatment of heterogeneous porous media and pure-fluid channel regions by representing the transition between them through a smooth variation of the material properties, rather than by imposing explicit interfacial coupling conditions.

To solve the resulting macroscopic model, we propose a monolithic numerical algorithm based on a single generalized kinetic equation, realized through a lattice Boltzmann scheme on standard lattices. The kinetic solver, presented in Section \ref{sec:kinetic_solver}, is constructed so that the speed of sound and the effective kinematic bulk viscosity can be specified independently, improving computational efficiency and flexibility. We also adapt wet-node boundary-condition schemes to extend common lattice Boltzmann boundary treatments to the proposed porous-media formulation.

A central feature of the proposed method is its consistent hydrodynamic limit. In particular, the model recovers an isotropic viscous stress tensor with the correct porosity scaling, expressed in terms of the mean velocity in the porous mass, thereby making the formulation suitable for variable-porosity domains. This consistency is assessed through dispersion and dissipation analysis in Section \ref{numerical_results:spectral}, and validated using two-dimensional benchmark problems in Section \ref{sec:Benchmark_validation}, including both homogeneous and heterogeneous porosity cases. Finally, Section \ref{sec:Numerical_studies} presents numerical demonstrations involving confined flow past a permeable cylinder and lid-driven cavity flow with a porous obstacle. These examples illustrate the ability of the proposed one-domain framework to capture transport between free-fluid and porous-medium regions within a single unified formulation.

\section{Macroscopic model}\label{sec:target_macro}
We consider an isothermal two-phase system comprised of a moving fluid phase and a static inhomogeneous porous matrix, i.e., a stationary solid phase, with no phase change, impermeable solid walls, and no-slip condition at the fluid-solid interface.

Following the method of volume averaging (see e.g., \cite{Whitaker_1967,Whitaker_Advances_1969,Whitaker_1986,Whitaker_VA_method_1998}), we associate each point $\bm{x}$ in the space domain with a representative elementary volume (REV) $V_{\mathrm{REV}}$, which can be used as an averaging volume for a conserved hydrodynamic field, provided that it is large enough to meaningfully describe the behavior of the average values of this field. This condition can be expressed as:
\begin{equation}
    L_{d} \ll L_{\mathrm{REV}} \ll L_{c}, \label{VA_length_scale_assumption}
\end{equation}
with $L_{d}$ being the microstructural distance over which the considered field varies significantly, $L_{\mathrm{REV}}\propto\sqrt[3]{V_{\mathrm{REV}}}$ is the characteristic length for the averaging volume, and $L_{c}$ is the characteristic macroscopic length of the process. Details on the method of volume averaging, including the definitions, local averaging procedure, and the averaging theorems, although standard, are included in Appendix \ref{Append:volume_averaging} for the sake of completeness.

As derived in Appendix \ref{Append:derivation_va_equations}, the dynamics of the fluid phase can be described by the following volume-averaged continuity and momentum equations in weakly-compressible form,
\begin{gather}
    \partial_t(\epsilon\rho) + \bm{\nabla}\cdot(\rho\bm{v}) = 0, \label{continuity_target} \\
    \partial_t(\rho\bm{v}) + \bm{\nabla}\cdot(\rho\bm{u}\otimes\bm{v}) = - \epsilon\bm{\nabla}p + \bm{\nabla}\cdot\bm{\sigma}_{\mathrm{eff}} - \bm{R} + \epsilon\rho\bm{b} ,\label{momentum_target}
\end{gather}
where $\rho$ is the intrinsic fluid density; $\bm{v}$ is the superficial fluid velocity; $\bm{u}=\bm{v}/\epsilon$ is the intrinsic fluid velocity; $p$ is the intrinsic fluid pressure; $\bm{b}$ is the acceleration due to some external body force; $\epsilon$ is the porosity, defined as the local volume fraction of the fluid phase $\epsilon(\bm{x})=V_{f}(\bm{x})/V_{\mathrm{REV}}$; $\bm{R}$ is interpreted as the total resistance to fluid flow due to fluid-solid interaction at the interface; and $\bm{\sigma}_{\mathrm{eff}}$ is the effective viscous stress tensor given by
\begin{equation}
    \bm{\sigma}_{\mathrm{eff}} = \rho\nu_{\mathrm{eff}}\left[\bm{\nabla}\bm{v} + (\bm{\nabla}\bm{v})^{\mathrm{T}} + \left(\frac{\eta_{\mathrm{eff}}}{\nu_{\mathrm{eff}}}-\frac{2}{D}\right)(\bm{\nabla}\cdot\bm{v})\mathbf{I}\right]. \label{visc_stress_target}
\end{equation}
with $\nu_{\mathrm{eff}}$ and $\eta_{\mathrm{eff}}$ denoting the effective (or sometimes referred to  as “apparent”) shear and bulk kinematic viscosity, respectively, and $D$ the number of spatial dimensions.

Note that these macroscopic effective viscosities are not necessarily equal to the actual fluid viscosities, but are model parameters used to account for various unresolved dissipative effects. As a consequence of the volume-averaging procedure, the viscous stress tensor $\bm{\sigma}_{\mathrm{eff}}$ in \eqref{visc_stress_target} is expressed in terms of gradients of the superficial velocity $\bm{v}$. If one were to use the fluid viscosities in this closure, $\bm{\sigma}_{\mathrm{eff}}$ alone may fail to represent microscopic viscous effects associated with unresolved pore-scale motion, as well as dispersive fluxes induced by velocity fluctuations. The effective-viscosity model adopted here, following the proposals of (\cite{Hassanizadeh_Gray_1980,Ni_Beckermann_1991,Ishii_Hibiki_2005}), introduces generalized effective parameters: within the porous medium they can account for its tortuosity and the associated effects, like pockets of trapped fluid and recirculation zones, while in the pure-fluid region ($\epsilon=1$) they can account for dispersive fluxes arising from turbulence. The use of effective viscosities can be considered a generalization of the model proposed by \cite{Brinkman_1949a,Brinkman_1949b}, which has seen numerous experimental (\cite{Givler_Altobelli_1994,Rinehart_2021}) and theoretical (\cite{Hornung_1997,Valdes_Ochoa_Tapia_2007}) justification.

The term $\bm{R}$ in \eqref{momentum_target} accounts for dissipative interfacial forces due to viscous and form drag. As argued in \cite{Whitaker_Advances_1969}, in general, $\bm{R}$ depends on the flow regime, fluid properties and porous medium structure. Multiple suitable constitutive relations and correlations with different ranges of validity may be used in general simulations of non-Darcian flow behavior. Here we consider the Darcy-Forchheimer model (see e.g., \cite{Nield_Bejan_2017,Whitaker_Forchheimer_1996,Joseph_1982_Nonlinear,Irmay_1958}), which in general can be expressed as:
\begin{equation}
    \bm{R} = \rho\left(\nu_\mathrm{f} \mathbf{D} + \lvert\bm{v}\rvert \mathbf{F} \right) \bm{v} \label{closure_general}
\end{equation}
where $\mathbf{D}$ and $\mathbf{F}$ are the Darcy and Forchheimer drag tensors, respectively, and $\nu_\mathrm{f}$ is the molecular (fluid) kinematic viscosity. Using the effective viscous stress tensor \eqref{visc_stress_target} together with the porous-media drag closure \eqref{closure_general} is commonly referred to in the porous-media literature as the Darcy-Forchheimer-Brinkman model. Note that the dependence on the velocity magnitude $\lvert\bm{v}\rvert$ in \eqref{closure_general} makes the closure non-Galilean-invariant. Strictly speaking, the form \eqref{closure_general} is therefore valid only in the inertial frame attached to the porous matrix, i.e., the reference frame in which the porous medium is stationary.

In porous media with isotropic permeability we can define:
\begin{align}
    \nu_\mathrm{f}\mathbf{D} &= \mu_\mathrm{D}\mathbf{I} \label{Darcy_isotropic} \\
    \mathbf{F} &= \mu_\mathrm{F}\mathbf{I} \label{Forchheimer_isotropic} 
\end{align}
with the coefficients $\mu_\mathrm{D}$ and $\mu_\mathrm{F}$ defined by the fluid and porous medium properties:
\begin{align}
    \mu_\mathrm{D} &= \frac{\epsilon\nu_\mathrm{f}}{\kappa}, \label{mu_D_Ergun_perm} \\
    \mu_\mathrm{F} &= \frac{\epsilon F_{\epsilon}}{\sqrt{\kappa}}. \label{mu_F_Ergun_perm}
\end{align}
where $\kappa$ is interpreted as the medium permeability and $F_{\epsilon}/\kappa$ as the inertial permeability.

In packed beds, the commonly used correlation by \cite{Ergun_1952} gives
\begin{align}
    \kappa &= \frac{d_\mathrm{p}^2}{150}\frac{\epsilon^{3}}{(1-\epsilon)^2}, \label{kappa_Ergun} \\
    F_{\epsilon} &= \frac{1.75}{\sqrt{150\epsilon^3}}, \label{F_eps_Ergun}
\end{align}
where $d_\mathrm{p}$ is the equivalent spherical diameter of the packing.

As discussed previously, the governing equations (\ref{continuity_target}-\ref{visc_stress_target}) are valid throughout the entire computational domain, thereby enabling a unified treatment of heterogeneous porous media and pure-fluid channel regions. The correlations (\ref{mu_D_Ergun_perm}-\ref{F_eps_Ergun}) are consistent with this formulation and are applicable to heterogeneous porous media. Moreover, since $\bm{R}\rightarrow\bm{0}$ as $\epsilon\rightarrow1$, the model naturally reduces to the pure-fluid limit and therefore remains valid in clear-fluid regions.

In the present formulation, no special treatment is introduced for possible discontinuities at the interface between porous and pure-fluid regions. Instead, we adopt a continuous transition zone in which the porosity, and consequently the drag coefficients $(\mu_\mathrm{D},\mu_\mathrm{F})$, may vary strongly but continuously. This representation is physically meaningful because the porosity field $\epsilon(\bm{x})$, obtained through a local volume-averaging procedure over a finite REV (see Appendix \ref{Append:Local_averaging_procedure}), is smooth provided that an appropriate separation of scales, expressed by \eqref{VA_length_scale_assumption}, is satisfied.

This remains true even when the underlying geometry corresponds to a non-smooth heterogeneous porous medium or to a homogeneous porous region adjacent to a pure-fluid region, including cases with sharp geometric features. The reason is that local volume averaging acts as a convolution, or spatial filtering, of the underlying discontinuous phase-indicator field.

\section{Kinetic solver}\label{sec:kinetic_solver}
\subsection{Discrete velocity kinetic model}\label{sec:kinetic_model}
The fluid is modeled by tracking the evolution of a volume-averaged single-particle distribution function, defined as $f_{i}=f_{i}(\bm{x},\bm{c}_{i},t)$ for a particle with a discrete velocity $\bm{c}_{i}$ at position $\bm{x}$ and time $t$, referred to as the populations. We consider the standard discrete velocity sets $D(n)Q({3^n})$, $n\in\{1,2,3\}$, with $D=n$ being the number of dimensions, and $Q=3^n$ the number of discrete velocities, $c_{i\alpha}\in\{-1,0,1\}$, $i=0,1,\dots,Q-1$. We then introduce a discrete-velocity kinetic equation:
\begin{equation}
    \partial_{t} f_{i} + \bm{c}_{i}\cdot\bm{\nabla} f_{i} = \Omega_{i} + \mathcal{K}_{i}. \label{kinetic-eq}
\end{equation}
The left-hand side of equation \eqref{kinetic-eq} describes the advection of the particle populations, while the right-hand side accounts for particle interactions. The term $\Omega_{i}$ accounts for the interparticle interaction, while $\mathcal{K}_{i}$ is a generalized forcing term. Using the lattice Bhatnagar-Gross-Krook (LBGK) approximation, $\Omega_{i}$ is given by:
\begin{equation}
    \Omega_{i} = -\frac{1}{\tau}(f_{i}-f_{i}^{\mathrm{eq}}), \label{omega_BGK}
\end{equation}
where $\tau$ is the bare relaxation time to the local equilibrium $f_{i}^{\mathrm{eq}}$. For $\mathcal{K}_{i}$ we use the Relax-to-Force method (RtFM) recently introduced by (\cite{Hosseini_Karlin_Shallow_Water_2025,Karlin_Hosseini_PracticalModels_2025,Hosseini_Feinberg_Karlin_2026}):
\begin{equation}
    \mathcal{K}_{i}= \frac{1}{\lambda}(f_{i}^{*}-f_{i}^{\mathrm{eq}}),  \label{F_RtFM}
\end{equation}
where $f_{i}^{*}$ are the shifted-equilibrium populations to be specified below and $\lambda>0$ is a time-parameter. 

To recover the target hydrodynamics, we interpret $f_{i}$, $f_{i}^{\mathrm{eq}}$ and $f_{i}^{*}$ as volume-averaged distribution functions at the REV scale. This averaging is of conceptual significance, and while it does not alter the form of the kinetic evolution equation, it defines the physical meaning of the moments of the distribution functions, i.e., of the locally conserved fields, by incorporating porosity scaling.

The locally conserved fields are the volume-averaged density $\epsilon\rho$ and the intrinsic fluid velocity $\bm{u}=\bm{v}/\epsilon$. These are defined via the zeroth- and the first-order moments of the populations $f_{i}$, respectively,
\begin{align}
    \Pi_0 &= \sum_{i=0}^{Q-1}{f_{i}} = \epsilon\rho, \label{density_moment} \\
    \bm{\Pi}_1 &= \sum_{i=0}^{Q-1}{\bm{c}_{i}f_{i}} = \epsilon\rho\bm{u} = \rho\bm{v}. \label{momentum_moment}
\end{align}
These local conservation relations can be interpreted as follows: Since the fluid occupies only a fraction of the total volume, characterized by the porosity $\epsilon$, and the distribution functions are volume-averaged quantities, the zeroth moment  \eqref{density_moment} yields the volume-averaged density. Accordingly, the first moment \eqref{momentum_moment} represents the volume-averaged momentum, defined as the product of the volume-averaged density and the intrinsic fluid velocity. For convenience, we keep the majority of algorithm equations in terms of the intrinsic fluid density $\rho$ and superficial flow velocity $\bm{v}$. 

We define the discrete equilibrium distribution as a function of locally conserved fields, $f_{i}^{\mathrm{eq}} = f_{i}^{\mathrm{eq}}(\epsilon\rho,\bm{v}/\epsilon)$, which naturally includes the porosity scaling. Similarly, the shifted-equilibrium populations are defined as a local equilibrium parameterized by the local values of volume-averaged density and shifted intrinsic velocity $\bm{v}_{\lambda}^{*}/\epsilon$, i.e., $f_{i}^{*} =f_{i}^{\mathrm{eq}}(\epsilon\rho,\bm{v}_{\lambda}^{*}/\epsilon)$, where $\bm{v}_{\lambda}^{*}$ is defined as:
\begin{equation}
    \bm{v}_{\lambda}^{*} = \bm{v} + \lambda\bm{a}, \label{shifted_v_def}
\end{equation}
with $\bm{a}$ representing the total superficial acceleration due to the presence of a porous medium and any other external body forces (e.g., gravity).
{Finally, the parameter $\lambda$ has a dimension of time and shall be specified later.}

\subsection{Local discrete equilibrium}\label{sec:equilibrium}
To describe the correct macroscopic hydrodynamics, the equilibrium distribution function must be defined such that the mass and momentum are conserved, and hence it needs to correctly account for porosity scaling in convective terms in the hydrodynamic limit. Consequently, the required moments of the volume-averaged discrete equilibrium distribution are found as:
\begin{align}
    \Pi_{0}^{\mathrm{eq}} &= \sum_{i=0}^{Q-1}{f_{i}^{\mathrm{eq}}} = \epsilon\rho, \label{Pi_0_eq} \\
    \Pi_{\alpha}^{\mathrm{eq}} &= \sum_{i=0}^{Q-1}{c_{i\alpha}f_{i}^{\mathrm{eq}}} = \rho v_{\alpha}, \label{Pi_1_eq} \\
    \Pi_{\alpha\beta}^{\mathrm{eq}} &= \sum_{i=0}^{Q-1}{c_{i\alpha}c_{i\beta}f_{i}^{\mathrm{eq}}} = \frac{\rho v_{\alpha}v_{\beta}}{\epsilon} + \epsilon\rho RT. \label{Pi_2_eq}
\end{align}
The reader is cautioned that in this part of the paper repeated indices here do not imply a summation.

Discrete equilibria for volume-averaged Navier-Stokes formulations that reproduce the target moments (\ref{Pi_0_eq}-\ref{Pi_2_eq}) have previously been proposed in the context of multiphase-flow modeling (\cite{Wang_Wang_2005,Xiong_Madadi_Lorenzini_2014,Zhang_2014}). These formulations, however, were based on the so-called second-order equilibrium functional, which does not generally lead to a consistent hydrodynamic limit (\cite{Extended_LBM_Saadat_2021}). More recently, \cite{Sawant_Karlin_2025} employed a product-form equilibrium.

As shown in Appendix \ref{Append:CE_analysis}, the second-order equilibrium can be interpreted as a truncated version of the product-form equilibrium. This truncation introduces larger errors, as will be demonstrated numerically in Section \ref{numerical_results:spectral}, while offering no tangible computational advantage. We therefore employ the product-form discrete equilibrium to define both the equilibrium and shifted-equilibrium populations. Following the product-form formalism of \cite{Karlin_Factorization_2010}, we introduce a set of functions of two variables, $\xi_{\alpha}$ and $\mathcal{P}_{\alpha\alpha}$,
\begin{equation}
    \Psi_{i\alpha}(\xi_{\alpha},\mathcal{P}_{\alpha\alpha}) = 1 - c_{i\alpha}^{2} + \frac{(3c_{i\alpha}^{2}-2)\mathcal{P}_{\alpha\alpha} + c_{i\alpha}\xi_{\alpha}}{2}, \label{1d_scaled_eq}
\end{equation}
and  construct the populations as products of these functions,
\begin{equation}
    f_i^{\mathrm{prod}} = \epsilon\rho\prod_{\alpha}{\Psi_{i\alpha}(\xi_{\alpha},\mathcal{P}_{\alpha\alpha})}.\label{prod_form}
\end{equation}
In general, we set $\xi_{\alpha}$ and $\mathcal{P}_{\alpha\alpha}$ as first- and the second-order moments of the populations, respectively, scaled by zeroth-order moment, i.e., $\epsilon\rho$. Specifically, for the equilibrium populations we set,
\begin{align}
    \xi_{\alpha}^{\mathrm{eq}} &= \frac{v_{\alpha}}{\epsilon}, \label{xi_alpha_eq} \\
    \mathcal{P}_{\alpha\alpha}^{\mathrm{eq}} &= (\xi_{\alpha}^{\mathrm{eq}})^2 + RT, \label{P_alpha_alpha_eq}
\end{align}
and for the shifted-equilibrium populations we set,
\begin{align}
    \xi_{\alpha}^{*} &= \frac{v_{\alpha}}{\epsilon} + \lambda\left(\frac{a_\alpha}{\epsilon}\right), \label{xi_alpha_*} \\
    \mathcal{P}_{\alpha\alpha}^{*} &= (\xi_{\alpha}^{*})^2 + RT + \left(\frac{\lambda}{\epsilon\rho}\right)(\Phi_{\alpha\alpha}^{\mathrm{ext}}+\Phi^{\eta}), \label{P_aa_*}
\end{align}
where the functions $\Phi_{\alpha\alpha}^{\mathrm{ext}}$ and $\Phi^{\eta}$ are correction terms given by
\begin{align}
    \Phi_{\alpha\alpha}^{\mathrm{ext}} &= \partial_{\alpha} \left[\rho v_\alpha (1-3RT)-\frac{\rho v_{\alpha}^{3}}{\epsilon^{2}}\right], \label{ex_corr} \\
    \Phi^{\eta} &= \rho RT\left(\frac{2}{D}-\frac{\eta_\mathrm{eff}}{\nu_{\mathrm{eff}}}\right)\left(\bm{\nabla}\cdot\bm{v}\right). \label{eta_corr}
\end{align}
These exact functional forms of the correction terms are found from the multiscale analysis (see Appendix \ref{Append:CE_analysis}) of the lattice Boltzmann equation \eqref{dicrete_scheme} introduced in the next section.

The correction term $\Phi_{\alpha\alpha}^{\mathrm{ext}}$, as first introduced in \cite{Extended_LBM_Saadat_2021}, is intended to restore the isotropy, and hence the Galilean invariance, of the effective viscous stress tensor in the hydrodynamic limit, by addressing the anomaly in the diagonal elements of the third-order moment tensor $\Pi_{\alpha\alpha\alpha}^{\mathrm{prod}}$ (see Appendix \ref{Append:eq_moments}). This modification is often referred to as the extended equilibrium (or extended LBM). By lifting the anisotropic restrictions, it extends the model’s applicability to higher flow velocities and to temperatures different from the lattice reference temperature ($RT\neq RT_\mathrm{L} = 1/3$), therefore improving computational efficiency.

The second correction term $\Phi^{\eta}$ allows for achieving a variable bulk viscosity in the hydrodynamic limit, similarly to what was done in \cite{Renard_2021, Sawant_Karlin_2025, Hosseini_Karlin_Shallow_Water_2025}. This makes the kinematic effective bulk viscosity an independent free parameter $\eta_\mathrm{eff} \geq 0$, that can be set according to correlations or stability considerations.

\subsection{Lattice Boltzmann realization}\label{sec:lbe}
The second-order accurate in time discretization of the kinetic equation \eqref{kinetic-eq} (see details in Appendix \ref{Append:time_discretization}) results in the following lattice Boltzmann equation (LBE)
\begin{equation}
    \bar{f}_{i}(\bm{x}+\bm{c}_{i}\delta t, t+\delta t) = \bar{f}_{i} + 2\beta(f_{i}^{\mathrm{eq}}-\bar{f}_{i}) + S_i^{\lambda}\delta t, \label{dicrete_scheme}
\end{equation}
where the right-hand side is evaluated at $(\bm{x},t)$, $\delta t$ is the time step, $S_i^{\lambda}$ is the source term (i.e., rate of momentum injection due to a force),
\begin{equation}
    S_i^{\lambda} = (1-\beta)\mathcal{K}_{i} = \left(\frac{1-\beta}{\lambda}\right)(f_{i}^{*}-f_{i}^{\mathrm{eq}}), \label{momentum_rate}
\end{equation}
and $\beta\in[0,1]$ is a dimensionless relaxation parameter defined as
\begin{equation}
    \beta = \frac{\delta t}{2\tau + \delta t}. \label{beta_def}
\end{equation}
We have also introduced a transformation ${f}_{i}\rightarrow\bar{f}_{i}$ as
\begin{equation}
    \bar{f}_{i} = \left(1+\frac{\delta t}{2\tau}\right)f_i + \frac{\delta t}{2}\left(\frac{1}{\lambda}-\frac{1}{\tau}\right)f_{i}^{\mathrm{eq}} - \frac{\delta t}{2\lambda}f_{i}^{*}, \label{f_bar_transform}
\end{equation}
and so the pertinent macroscopic moments of the transformed populations $\bar{f}_{i}$ (see Appendix \ref{Append:transform_fields}) are
\begin{align}
    \bar{\Pi}_0 &= \sum_{i=0}^{Q-1}{\bar{f}_{i}} = \epsilon\rho, \label{transformed_density_moment} \\
    \bar{\bm{\Pi}}_1 &= \sum_{i=0}^{Q-1}{\bm{c}_{i}\bar{f}_{i}} = \rho\bm{v} - \frac{\delta t}{2}\rho\bm{a}. \label{transformed_momentum_moment}
\end{align}
The multiscale analysis, conducted by expanding the LBE \eqref{dicrete_scheme} to second order in the time step $\delta t$ and taking moments of the resulting expansion (see Appendix \ref{Append:CE_analysis}), reveals that, at the Navier-Stokes level, the recovered macroscopic hydrodynamic equations are the target continuity equation \eqref{continuity_target} and a momentum equation of the form
\begin{equation}
    \partial_t(\rho \bm{v}) + \bm{\nabla}\cdot(\rho \bm{u} \otimes \bm{v}) = -\bm{\nabla}(\epsilon p) + \bm{\nabla}\cdot\bm{\sigma}_{\mathrm{LB}} + \rho \bm{a}, \label{momentum_LBM}
\end{equation}
with the intrinsic fluid pressure given by an ideal gas equation of state, $p=\rho RT$, and the viscous stress given by
\begin{equation}
    \bm{\sigma}_{\mathrm{LB}} = \rho\nu_{\mathrm{eff}}\left[\epsilon \left(\bm{\nabla}\bm{u} + (\bm{\nabla}\bm{u})^{\mathrm{T}}\right) + \left(\frac{\eta_{\mathrm{eff}}}{\nu_{\mathrm{eff}}}-\frac{2}{D}\right)(\bm{\nabla}\cdot\bm{v})\mathbf{I}\right], \label{visc_stress_LBM}
\end{equation}
where the effective kinematic shear viscosity is defined by,
\begin{equation}
    \nu_{\mathrm{eff}} = \delta t \left(\frac{1}{2\beta}-\frac{1}{2}\right)RT.\label{nu_eff_def}
\end{equation}
Note that, without the correction term $\Phi^{\eta}$ as defined by \eqref{eta_corr}, the effective bulk viscosity would be fixed by the effective shear viscosity as
\begin{equation}
    \eta_{0} = \frac{2\nu_{\mathrm{eff}}}{D}. \label{eta_0_no_corr}
\end{equation}
By comparing the recovered viscous stress tensor $\bm{\sigma}_{\mathrm{LB}}$ in \eqref{visc_stress_LBM} with the target effective viscous stress tensor $\bm{\sigma}_{\mathrm{eff}}$ in \eqref{visc_stress_target}, we observe that the bulk contribution is identical in both expressions, whereas the shear contribution in \eqref{visc_stress_LBM} is expressed in terms of gradients of the intrinsic velocity $\bm{u}$ rather than the superficial velocity $\bm{v}$. Also, a comparison of recovered momentum equation \eqref{momentum_LBM} with the target momentum equation \eqref{momentum_target} reveals a mismatch in the porosity scaling of the pressure gradient term.

Consequently, the acceleration $\bm{a}$ must not only account for external body forces and the porous-medium resistance, but also include the necessary corrections to the pressure and viscous terms so that the formulation remains valid for heterogeneous domains with spatially varying porosity. Taking into account (\ref{momentum_target}, \ref{visc_stress_target}, \ref{momentum_LBM}, \ref{visc_stress_LBM}) together with the isotropic porous-media closure relations (\ref{closure_general}, \ref{Darcy_isotropic}, \ref{Forchheimer_isotropic}), the target momentum equation is recovered by setting
\begin{equation}
    \bm{a} = \epsilon \bm{b} - \mu_\mathrm{D} \bm{v} -\mu_\mathrm{F}|\bm{v}|\bm{v} + RT\bm{\nabla}\epsilon + \frac{1}{\rho}\bm{\Delta}_\nu(\rho\bm{u}),  \label{accel}
\end{equation}
with $\bm{\Delta}_\nu(\bm{m})$ being the viscous stress correction accounting for spatially varying porosity of the medium, given by
\begin{equation}
    \bm{\Delta}_\nu(\bm{m}) = \bm{\nabla}\cdot\left[ \nu_\mathrm{eff}\left(\bm{m}\otimes\bm{\nabla}\epsilon +\bm{\nabla}\epsilon\otimes\bm{m}\right)\right]. \label{porosity_correction}
\end{equation}
Some comments are in order regarding the recovered momentum equation. The pressure-gradient term in \eqref{momentum_LBM} depends on the local variation of porosity, which is nonphysical and must therefore be corrected. This issue is removed by deriving and introducing, through the forcing scheme, an additional correction term equal to $p\bm{\nabla}\epsilon$ in the momentum balance. Similar corrections have been used by \cite{Zhang_2014} for porous-media flows, by \cite{Wang_Wang_2005} for two-fluid Euler-Euler models, and by \cite{Song_Wang_Li_2013} and \cite{Xiong_Madadi_Lorenzini_2014} in the context of particle-fluid flows.

A second inconsistency concerns the form of the viscous stress. While a viscous stress written in terms of gradients of the intrinsic fluid velocity is consistent with two-fluid formulations (\cite{Zachariah_etal_2026}), it is not appropriate for the porous-media model considered here. As discussed above, for the present constitutive relations and closure models, the Brinkman viscous stress should instead be expressed in terms of gradients of the superficial average velocity. The additional correction term $\bm{\Delta}_\nu(\rho\bm{u})$ in the momentum equation is therefore introduced to remove this inconsistency, which is common in existing LBM formulations for porous media. To the best of our knowledge, this is the first implementation of such a correction in the context of porous-media flows.

Note that the transport coefficients $\nu_\mathrm{f}$ and $\eta_\mathrm{eff}$ are free model parameters, and only $\nu_{\mathrm{eff}}$ depends on the choice of $\beta$ and $RT$, as defined by \eqref{nu_eff_def}. It is also important to note that the hydrodynamic limit does not depend on the relaxation parameter $\lambda$, which in that regard is a free parameter. Following (\cite{Hosseini_Karlin_Shallow_Water_2025, Karlin_Hosseini_PracticalModels_2025, Hosseini_Feinberg_Karlin_2026}), we set $\lambda = \delta t$, which is consistent with the requirement $\lambda\gg\tau$, maintained in the over-relaxation regime $\beta\rightarrow1$.

At every time step, one needs to find the locally conserved fields, entering the equilibrium and the shifted equilibrium populations. The density at any $(\bm{x},t)$ is readily computed using \eqref{transformed_density_moment}. However, the acceleration $\bm{a}$ as given by \eqref{accel}, is a nonlinear function of $\bm{v}$, and so the definition for velocity \eqref{transformed_momentum_moment} is implicit. The proposed three-step algorithm to compute the superficial velocity $\bm{v}$ is as follows. First, evaluate
\begin{equation}
    \bm{v}^{(0)} = \frac{\bar{\bm{\Pi}}_1}{\rho} + \frac{\delta t}{2}\left(\epsilon\bm{b}+RT\bm{\nabla}\epsilon\right),\label{v_(0)}
\end{equation}
then find
\begin{equation}
    \bm{v}^{(1)} = \frac{2\bm{v}^{(0)}}{c_1 + \sqrt{c_1^2 + 4c_2|\bm{v}^{(0)}|}}, \label{v_(1)}
\end{equation}
with
\begin{equation}
    c_1 = 1 + \frac{\delta t}{2}\mu_\mathrm{D}, \quad c_2 = \frac{\delta t}{2}\mu_\mathrm{F} \label{velocity_coefficients}
\end{equation}
Finally, compute the velocity from
\begin{equation}
    \bm{v} = \bm{v}^{(1)}+\frac{\delta t}{2\rho}\bm{\Delta}_\nu\left(\frac{\rho\bm{v}^{(1)}}{\epsilon}\right). \label{v_final}
\end{equation}
The proposed algorithm is not a fully implicit solution of the original coupled system, with the nontrivial force component $\bm{\Delta}_\nu$ being treated explicitly in a lagged manner. Consequently, for the full coupled equation the scheme is formally first-order accurate in time, because the viscous correction is evaluated using the intermediate predicted velocity $\bm{v}^{(1)}$ rather than the final velocity $\bm{v}$. This lagged strategy is well suited to steady-state computations and to transient problems with slowly varying dynamics. However, the correction becomes stiff, then the error introduced by the explicit update in \eqref{v_final} may become significant, since the viscous contribution can no longer be regarded as a small additive correction. In such cases, an alternative algorithm like Picard (fixed-point) iteration scheme with the intended second-order accuracy (see Appendix \ref{Append:Picard_iter}), is a more appropriate choice.

Note, however, that steps (\ref{v_(0)}-\ref{v_(1)}) still solve the local nonlinear drag relation (Darcy-Forchheimer) implicitly and in closed form (see Appendix \ref{Append:transform_fields}). Therefore, in homogeneous porous media with $\epsilon=\text{const.}$, the update reduces to $\bm{v}=\bm{v}^{(1)}$, yielding a second-order accurate-in-time solution.

\subsection{Boundary conditions}\label{sec:boundary_conditions}
\begin{figure}[h]
\centering
\includegraphics[width=1.0\textwidth]{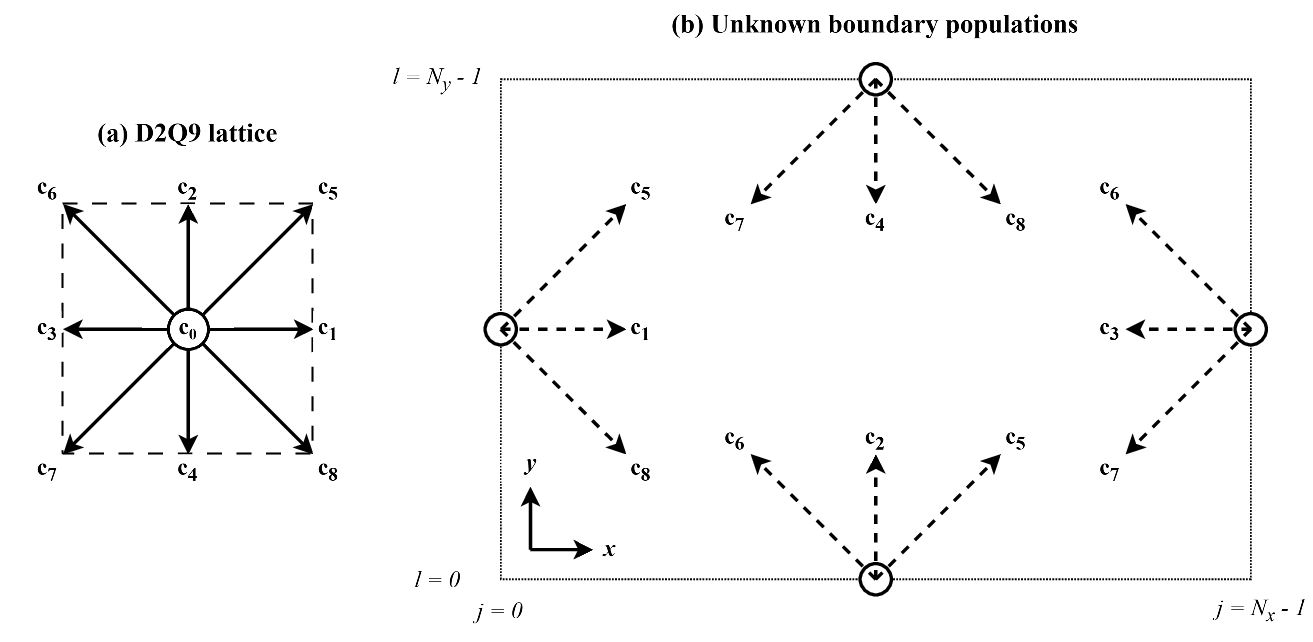}
\caption{Schematic of (a) D2Q9 lattice vectors and (b) unknown incoming populations on the boundary nodes of a 2D LBM computational domain. Empty circles represent computational nodes, and dotted edges indicate the effective boundaries. Dashed arrows are the discrete velocity vectors $\bm{c}_i$ of incoming populations $i\in\mathcal{P}_-$. $N_x$ and $N_y$ are the numbers of grid nodes in $x$ and $y$ directions, respectively, with $j\in\{0,\dots, N_x-1\}$ and $l\in\{0,\dots, N_y-1\}$ being the corresponding  grid point indices.}\label{bc_schematic}
\end{figure}

This section describes the implementation of boundary conditions for the proposed porous-media model. We focus on velocity Dirichlet conditions applied to straight, on-lattice (wet-node), axis-aligned boundaries. For clarity, we consider a two-dimensional computational domain with a D2Q9 lattice, as illustrated in Figure \ref{bc_schematic}(a).

As discussed in detail elsewhere (see, e.g., \cite{Kruger_LBM_2016}), the problem of boundary conditions in LBM is concerned with the post-advection populations at the boundary nodes. After streaming, some populations on the boundary are incoming from outside the computational domain and are therefore unknown. Nevertheless, these missing populations must encode the prescribed boundary information. The lattice vectors corresponding to the unknown incoming populations in a 2D domain are shown in Figure \ref{bc_schematic}(b) as dashed arrows. The goal of a boundary-condition algorithm is to reconstruct the unknown or all populations at boundary nodes in a manner consistent with the model dynamics and yielding the desired macroscopic behavior at the domain boundary.

Because we consider velocity boundaries, the superficial fluid velocity at the boundary, $\bm{v}_\mathrm{b}(t)=\bm{v}(\bm{x}_\mathrm{b},t)$, is prescribed by definition. However, the intrinsic fluid density on the boundary nodes $\rho_\mathrm{b}(t)=\rho(\bm{x}_\mathrm{b},t)$ cannot be obtained from \eqref{transformed_density_moment} due to the missing populations, and must instead be evaluated differently than on the bulk fluid nodes. The procedure for finding the density is described in the next section. We then show how several common boundary-condition strategies, including the equilibrium scheme, the non-equilibrium extrapolation method, and the Tamm--Mott-Smith approximation, can be adapted to the present porous-media model. While the underlying ideas behind the said boundary conditions remain unchanged, the practical implementation must account for the redefinition of the locally conserved fields and for the presence of the forcing term.

\subsubsection{Density on boundaries}
On straight on-lattice boundaries which are aligned with lattice directions, we can use the flow continuity condition as a constraint to find the value of the intrinsic fluid density from the known populations and prescribed fluid velocity, as discussed in \cite{Zou_He_1997_BC, Latt_etal_2008_DirichletBC}. To that end, we split the density at a boundary node into three components,
\begin{equation}
    \rho_\mathrm{b} = \rho_\mathrm{b}^{-} + \rho_\mathrm{b}^{0} + \rho_\mathrm{b}^{+}, \label{boundary_density_split}
\end{equation}
that are defined by
\begin{equation}
    \epsilon_\mathrm{b}\rho_\mathrm{b}^{-}(t) = \sum_{i\in\mathcal{P}_{-}}{\bar{f}_i(\bm{x}_\mathrm{b},t)},\quad \epsilon_\mathrm{b}\rho_\mathrm{b}^{0}(t) = \sum_{i\in\mathcal{P}_{0}}{\bar{f}_i(\bm{x}_\mathrm{b},t)},\quad \epsilon_\mathrm{b}\rho_\mathrm{b}^{+}(t) = \sum_{i\in\mathcal{P}_{+}}{\bar{f}_i(\bm{x}_\mathrm{b},t)}. \label{bc_rho_components_def}
\end{equation}
Here we have split the populations into three pairwise disjoint sets $i\in\{0,\dots,Q-1\}=(\mathcal{P}_{-}\cup\mathcal{P}_{0}\cup\mathcal{P}_{+})=\mathcal{P}$. These are defined as follows:
\begin{itemize}
    \item $\mathcal{P}_{-}$ is a set of the unknown incoming populations at the boundary node, corresponding to discrete velocities that have a component normal to the boundary and pointing inside the computational domain, see Figure \ref{bc_schematic}.
    \item $\mathcal{P}_{0}$ is a set of known populations with lattice velocities that are either tangential to the boundary or zero, i.e., $\bm{c}_i\cdot\bm{n}_\mathrm{b}$=0, where $\bm{n}_\mathrm{b}$ is the boundary normal unit vector.
    \item $\mathcal{P}_{+}$ is a set of known outgoing populations, whose lattice velocities are opposite to the unknown ones, i.e., those with lattice velocities having a component that is boundary normal and pointing outside of the computational domain.
\end{itemize}
Let $\phi_{\mathrm{b},\perp}=\bm{\phi}_{\mathrm{b}}\cdot\bm{n}_\mathrm{b}$ denote the normal component of any vector $\bm{\phi}_{\mathrm{b}}$ at the boundary node, taken along the outward-pointing unit normal (i.e., pointing out of the computational domain). Then, from relation \eqref{transformed_momentum_moment}, the outgoing momentum flux normal to the boundary is defined by
\begin{equation}
    \rho_\mathrm{b} v_{\mathrm{b},\perp} - \frac{\delta t}{2} \rho_\mathrm{b} a_{\mathrm{b},\perp} = \sum_{i\notin\mathcal{P}_{0}} {(\bm{c}_{i}\cdot\bm{n}_\mathrm{b})\bar{f}_{i}} = \epsilon_\mathrm{b}\rho_\mathrm{b}^{+} - \epsilon_\mathrm{b}\rho_\mathrm{b}^{-}. \label{boundary_norm_mom}
\end{equation}
Then, by combining equations \eqref{boundary_density_split} and \eqref{boundary_norm_mom}, we can express $\rho_\mathrm{b}$ independently of the unknown component $\rho_\mathrm{b}^{-}$,
\begin{equation}
    \rho_\mathrm{b} = \frac{\epsilon_\mathrm{b}\left(\rho_\mathrm{b}^{0}+2\rho_\mathrm{b}^{+}\right)}{\epsilon_\mathrm{b} + v_{\mathrm{b},\perp} - (\delta t/2)a_{\mathrm{b},\perp}}, \label{bc_density_impermWall}
\end{equation}
The normal boundary acceleration $a_{\mathrm{b},\perp}$ in \eqref{bc_density_impermWall} is obtained from \eqref{accel} and \eqref{porosity_correction}. In general, \eqref{bc_density_impermWall} defines an implicit relation and therefore requires an iterative solution. In special cases, for example when the porosity is locally constant up to second order, or when both the velocity and its gradient vanish, $a_{\mathrm{b},\perp}$ reduces to the contribution from the external body force alone, and \eqref{bc_density_impermWall} becomes explicit. In the more common case where $(\bm{\nabla}\epsilon)_\mathrm{b}\neq\bm{0}$ and $\bm{v}_\mathrm{b}\neq\bm{0}$, but the density gradient at the boundary is small, the viscous correction term in \eqref{porosity_correction} can be accurately approximated by taking the density outside the divergence operator, yielding
\begin{equation}
    \begin{split}
        a_{\mathrm{b},\perp} \simeq \epsilon_\mathrm{b} b_{\mathrm{b},\perp} &- \mu_\mathrm{D} v_{\mathrm{b},\perp} -\mu_\mathrm{F}\lvert\bm{v}_\mathrm{b}\rvert v_{\mathrm{b},\perp} + RT(\nabla\epsilon)_{\mathrm{b}, \perp} + \bm{\nabla}\cdot\left( \nu_\mathrm{eff}\left[\frac{\bm{v}_\mathrm{b}}{\epsilon_\mathrm{b}}(\nabla\epsilon)_{\mathrm{b}, \perp} +(\bm{\nabla}\epsilon)_{\mathrm{b}}\frac{v_{\mathrm{b},\perp}}{\epsilon_\mathrm{b}}\right]\right),  \label{accel_approx_boundary}
    \end{split}
\end{equation}
As an example, suppose the flow domain in Figure \ref{bc_schematic}(b) is space periodic in $x$, with solid walls at $l=0$ and $l=N_y-1$, which do not move in the $y$-direction, and the first and second derivatives of porosity at the boundary are vanishing. Using (\ref{bc_rho_components_def}, \ref{bc_density_impermWall}, \ref{accel_approx_boundary}), we obtain the following expression for the intrinsic fluid density at the wall boundary nodes
\begin{equation}
    \rho_\mathrm{b} = \frac{\bar{f}_0+\bar{f}_1+\bar{f}_3 + 2\sum_{i\in\mathcal{P}_{+}}\bar{f}_i}{\epsilon_\mathrm{b}\left[1-\delta t\left(\frac{l}{N_y-1}-\frac{1}{2}\right)b_y \right]},\quad l\in\{0,N_y-1\},\quad \mathcal{P}_{+} =
    \begin{cases}
        \{4,7,8\}, & l=0 \\
        \{2,5,6\}, & l=N_y-1.
    \end{cases}
\end{equation}
The method \eqref{bc_density_impermWall} is mass conserving, as it stems from the continuity condition, and it works for all planar boundaries in both 2D and 3D domains.
The major limitation of this method, apart from potentially requiring an iterative evaluation, is that it can only be applied to boundaries aligned with one of the coordinate axes. On boundaries defined by geometrical features where multiple straight surfaces intersect (2D: corners, 3D: edges and corners), the locally available information on a boundary node is almost always insufficient for the evaluation of the density. Two common remedies include: (i) extrapolating the density from neighboring bulk fluid nodes (\cite{Latt_etal_2008_DirichletBC}); or (ii) performing a bounce-back operation solely to compute the density at the boundary nodes (\cite{Chikatamarla_Karlin_2013_TMSBC}, \cite{Sawant_Dorschner_Karlin_2022_Reactive}). In the latter approach, for solid boundaries, the unknown incoming post-advection (pa) populations $i\in\mathcal{P}_{-}$ are set equal to the post-collision (pc) outgoing populations $i\in\mathcal{P}_{+}$ with mirrored lattice velocities, hence the boundary-node density is simply evaluated as
\begin{equation} \label{bc_density_BounceBack}
    \epsilon_\mathrm{b}\rho_\mathrm{bb}(\bm{x}_\mathrm{b},t) = \sum_{i\in\mathcal{P}_{+}}{\bar{f}_i^\mathrm{pc}}(\bm{x}_\mathrm{b},t) + \sum_{i\notin\mathcal{P}_{-}}{\bar{f}_i^\mathrm{pa}(\bm{x}_\mathrm{b},t)}.
\end{equation}
We also note that while \eqref{bc_density_impermWall} is used for imposing velocity boundary conditions, a similar approach could in principle be used to find a velocity component normal to the boundary to be used for imposing pressure/density at the boundary (see e.g., \cite{Kruger_LBM_2016}). 
 
\subsubsection{Equilibrium scheme}
The equilibrium scheme is the simplest approach for prescribing boundary conditions (\cite{Latt_etal_2008_DirichletBC, He_1997_NonSlipBC, Mohamad_2009_EqBC}). It enforces the equilibrium distribution for the post-advection populations at boundary nodes,
\begin{equation}
f_i(\bm{x}_\mathrm{b},t) = f_{i}^{\mathrm{eq}}(\epsilon_\mathrm{b}, \rho_\mathrm{b}, \bm{v}_\mathrm{b}). \label{eq_bc_f}
\end{equation}
Due to the population transform \eqref{f_bar_transform}, one cannot directly set the transformed populations $\bar{f}_i(\bm{x}_\mathrm{b},t)$ equal to the equilibrium ones. However, by using \eqref{f_bar_transform} and noting that both $f_{i}^{\mathrm{eq}}$ and $f_i^{*}$ are directly available from the prescribed macroscopic boundary data, \eqref{eq_bc_f} can be written in terms of $\bar{f}_i$ as
\begin{equation}
\bar{f}_i(\bm{x}_\mathrm{b},t) = \left(1 + \frac{\delta t}{2\lambda}\right)f_{i}^{\mathrm{eq}}(\epsilon_\mathrm{b}, \rho_\mathrm{b}, \bm{v}_\mathrm{b}) - \frac{\delta t}{2\lambda}f_{i}^{\mathrm{eq}}(\epsilon_\mathrm{b}, \rho_\mathrm{b},\bm{v}_{\lambda,\mathrm{b}}^{*}), \label{eq_bc_f_bar}
\end{equation}
where $\bm{v}_{\mathrm{b},\lambda}^{*} = \bm{v}_\mathrm{b} + \lambda\bm{a}_\mathrm{b}$ is defined in \eqref{shifted_v_def}. Note that \eqref{eq_bc_f_bar} is applied to all transformed boundary populations, $i\in\mathcal{P}$, rather than only to the unknown incoming ones.

In general, the equilibrium scheme is not an accurate boundary treatment because it neglects the non-equilibrium contribution of the boundary populations, $f_{i}^{\mathrm{neq}} = f_i - f_{i}^{\mathrm{eq}}$. The following sections describe two common techniques for approximating this non-equilibrium part.

\subsubsection{Non-equilibrium extrapolation method}
One way to approximate the non-equilibrium part is to extrapolate it from the adjacent fluid region, where $f_i^{\mathrm{neq}}$ is known, as was originally proposed by \cite{Guo_NEEM_BC_2, Guo_NEEM_BC_1}. The approach assumes that the non-equilibrium contribution of each post-advection population at the boundary node equals that at the nearest interior fluid node, i.e.,
\begin{equation}
f_i(\bm{x}_\mathrm{b},t) - f_{i}^{\mathrm{eq}}(\epsilon_\mathrm{b}, \rho_\mathrm{b}, \bm{v}_\mathrm{b}) =
f_i(\bm{x}_\mathrm{f},t)-f_{i}^{\mathrm{eq}}(\epsilon_\mathrm{f}, \rho_\mathrm{f}, \bm{v}_\mathrm{f}), \label{NEEM_bc_f}
\end{equation}
where $\bm{x}_\mathrm{f}$ denotes the fluid node adjacent to the boundary node $\bm{x}_\mathrm{b}$ along the boundary normal. Condition \eqref{NEEM_bc_f} with the transform \eqref{f_bar_transform} translates into the following operation applied to all of the transformed post-advection boundary populations $i\in\mathcal{P}$,
\begin{equation}
    \begin{split}
        \bar{f}_i(\bm{x}_\mathrm{b},t) = \zeta_\beta\bar{f}_i(\bm{x}_\mathrm{f},t) &+ \left(1+\frac{\delta t}{2\lambda}\right)\left[f_i^\mathrm{eq}(\epsilon_\mathrm{b}, \rho_\mathrm{b}, \bm{v}_\mathrm{b}) - \zeta_\beta f_i^\mathrm{eq}(\epsilon_\mathrm{f}, \rho_\mathrm{f}, \bm{v}_\mathrm{f})\right] \\
        &+ \frac{\delta t}{2\lambda}\left[\zeta_\beta f_i^\mathrm{eq}(\epsilon_\mathrm{f}, \rho_\mathrm{f}, \bm{v}_{\lambda,\mathrm{f}}^{*})-f_i^\mathrm{eq}(\epsilon_\mathrm{b}, \rho_\mathrm{b}, \bm{v}_{\lambda,\mathrm{b}}^{*})\right], \label{NEEM_bc_f_bar}
    \end{split}
\end{equation}
with
\begin{equation}
    \zeta_\beta = \frac{1-\beta(\bm{x}_\mathrm{f},t)}{1-\beta(\bm{x}_\mathrm{b},t)}. \label{zeta_beta_NEEM}
\end{equation}
Note that the relaxation parameter $\beta$ in \eqref{zeta_beta_NEEM}, as defined by \eqref{beta_def}, can be a local quantity, and therefore $\zeta_\beta$ is not necessarily equal to one. Through \eqref{nu_eff_def}, $\beta$ determines the effective shear viscosity $\nu_\mathrm{eff}$, which may generally vary in space and time.

\subsubsection{Tamm--Mott-Smith scheme}
The Tamm--Mott-Smith (TMS) boundary condition, as introduced in \cite{Chikatamarla_Karlin_2013_TMSBC} and further expanded in \cite{Sawant_Dorschner_Karlin_2022_Reactive}, specifies the non-equilibrium parts of the post-advection boundary populations as
\begin{equation}
    f_i^\mathrm{neq}(\bm{x}_\mathrm{b},t) = \begin{cases}
        f_i^\mathrm{eq}(\epsilon_\mathrm{b}, \rho_\mathrm{b}, \bm{v}_\mathrm{b}) -f_i^\mathrm{eq}(\epsilon_\mathrm{b}, \rho_\mathrm{loc}, \bm{v}_\mathrm{loc}), & i\in\mathcal{P_{-}} \\
        f_i(\bm{x}_\mathrm{b},t) -f_i^\mathrm{eq}(\epsilon_\mathrm{b}, \rho_\mathrm{loc}, \bm{v}_\mathrm{loc}), & i\notin\mathcal{P_{-}}
    \end{cases} \label{f_i_neq_TMS}
\end{equation}
where the local density $\rho_\mathrm{loc}$ and local velocity $\bm{v}_\mathrm{loc}$ at the boundary nodes are found by setting the unknown incoming populations to the target equilibrium, i.e., $f_i(\bm{x}_\mathrm{b},t)=f_i^\mathrm{eq}(\epsilon_\mathrm{b}, \rho_\mathrm{b}, \bm{v}_\mathrm{b})\,\forall\,i\in\mathcal{P}_{-}$, and are thus defined by
\begin{gather}
    \epsilon_\mathrm{b}\rho_\mathrm{loc} = \epsilon_\mathrm{b}\rho_\mathrm{b} + \sum_{i\notin\mathcal{P}_{-}}{f_i^\mathrm{neq}(\bm{x}_\mathrm{b},t)},\\
    \rho_\mathrm{loc}\bm{v}_\mathrm{loc} = \rho_\mathrm{b}\bm{v}_\mathrm{b} + \sum_{i\notin\mathcal{P}_{-}}{\bm{c}_i f_i^\mathrm{neq}(\bm{x}_\mathrm{b},t)},
\end{gather}
By construction, non-equilibrium contributions $f_i^\mathrm{neq}$ in \eqref{f_i_neq_TMS} satisfy the mass and momentum conservation: $\sum_{i\in\mathcal{P}}f_i^\mathrm{neq}=0$, $\sum_{i\in\mathcal{P}}\bm{c}_if_i^\mathrm{neq}=0$, and hence the TMS boundary condition enforces the target macroscopic values at the boundary nodes. Conceptually, the method approximates the non-equilibrium part of the unknown incoming populations using the local equilibrium state, evaluated using the locally reconstructed flow variables $(\rho_\mathrm{loc}, \bm{v}_\mathrm{loc})$ rather than the imposed target values $(\rho_\mathrm{b}, \bm{v}_\mathrm{b})$, thus representing a non-equilibrium state relative to the target equilibrium. The naming is inspired by the classical Tamm--Mott-Smith approximation of a shock-wave solution of the Boltzmann equation using a linear combination of two Maxwellian distributions (\cite{Mott_Smith_1951, Cercignani_Boltzmann_1988}).

Let us introduce a shorthand notation for the target equilibrium $f_i^\mathrm{tgt}$, the shifted equilibrium $f_i^*$, and the local equilibrium $f_i^\mathrm{loc}$ at the boundary nodes, specified uniquely as
\begin{gather}
f_i^\mathrm{tgt} = f_i^\mathrm{eq}(\epsilon_\mathrm{b}, \rho_\mathrm{b}, \bm{v}_\mathrm{b}),\quad f_i^* = f_i^\mathrm{eq}(\epsilon_\mathrm{b}, \rho_\mathrm{b}, \bm{v}_{\lambda,\mathrm{b}}^{*}),\quad f_i^\mathrm{loc} = f_i^\mathrm{eq}(\epsilon_\mathrm{b}, \rho_\mathrm{loc}, \bm{v}_\mathrm{loc}).
\end{gather}
The procedure to specify all of the transformed post-advection boundary populations $\bar{f}_i(\bm{x}_\mathrm{b},t)$ using the TMS scheme is as follows:
\begin{enumerate}
\item Compute the local density $\rho_\mathrm{loc}$ and velocity $\bm{v}_\mathrm{loc}$ at the boundary node using
\begin{gather}
\epsilon_\mathrm{b}\rho_\mathrm{loc} = \epsilon_\mathrm{b}\rho_\mathrm{b} + (1-\beta_\mathrm{b})\sum_{i\notin\mathcal{P}_{-}}{\left[\bar{f}_i^\mathrm{pa}-\left(1+\frac{\delta t}{2\lambda}\right)f_i^\mathrm{tgt} + \frac{\delta t}{2\lambda}f_i^*\right]},\\
\rho_\mathrm{loc}\bm{v}_\mathrm{loc} = \rho_\mathrm{b}\bm{v}_\mathrm{b} + (1-\beta_\mathrm{b})\sum_{i\notin\mathcal{P}_{-}}{\bm{c}_i\left[\bar{f}_i^\mathrm{pa}-\left(1+\frac{\delta t}{2\lambda}\right)f_i^\mathrm{tgt} + \frac{\delta t}{2\lambda}f_i^*\right]},
\end{gather}
where $\bar{f}_i^\mathrm{pa}$, $i\notin\mathcal{P}_{-}$, are the known post-advection transformed populations.
\item Using these local parameters, evaluate the corresponding local equilibrium populations $f_i^\mathrm{loc}$.
\item Finally, update all transformed post-advection boundary populations according to
\begin{align}
    \bar{f}_i^\mathrm{TMS} &= \frac{f_i^\mathrm{tgt}-f_i^\mathrm{loc}}{1-\beta_\mathrm{b}} + \left(1+\frac{\delta t}{2\lambda}\right)f_i^\mathrm{tgt} - \frac{\delta t}{2\lambda}f_i^*, & \forall\ i\in\mathcal{P_{-}}, \\
    \bar{f}_i^\mathrm{TMS} &= \frac{f_i^\mathrm{tgt}-f_i^\mathrm{loc}}{1-\beta_\mathrm{b}} + \bar{f}_i^\mathrm{pa}, & \forall\ i\notin\mathcal{P_{-}}.
\end{align}
\end{enumerate}

\subsection{Implementation summary} \label{sec:impl_summary}
The populations $\bar{f}_i$ evolve according to the collide-and-stream LBE algorithm given by (\ref{dicrete_scheme}-\ref{momentum_rate}), with $\lambda=\delta t$. The equilibrium populations $f_i^\mathrm{eq}$ are functions of the porosity $\epsilon$, intrinsic density $\rho$, and superficial velocity $\bm{v}$, as defined in (\ref{1d_scaled_eq}-\ref{P_alpha_alpha_eq}). The shifted-equilibrium populations $f_i^*$ are defined through (\ref{1d_scaled_eq}-\ref{prod_form},\ref{xi_alpha_*}-\ref{P_aa_*}), and depend additionally on the acceleration $\bm{a}$ (\ref{accel},\ref{porosity_correction_expanded}), and correction terms ${\Phi}_{\alpha\alpha}^{\mathrm{ext}}$ \eqref{ex_corr} and $\Phi^{\eta}$ \eqref{eta_corr}. The intrinsic density $\rho$ is obtained from \eqref{density_moment}, while the superficial velocity $\bm{v}$ is evaluated from (\ref{v_(0)}-\ref{v_final},\ref{porosity_correction_expanded}). The relaxation parameter $\beta$ is related to the effective kinematic viscosity $\nu_\mathrm{eff}$ through \eqref{nu_eff_def}, whereas the effective bulk viscosity $\eta_\mathrm{eff}$ enters the shifted equilibrium directly through $\Phi^{\eta}$ in \eqref{eta_corr}.

\begin{figure}
\centering
\begin{tikzpicture}[node distance=1cm]
\node (params) [specif, align=center] {Parameters:\\$RT,\,\nu_\mathrm{f},\,\bm{b}$};
\node (porosity) [specif, align=center, left=of params, xshift=-2cm] {Porosity profile:\\ $\epsilon(\bm{x}),\,\bm{\nabla}\epsilon(\bm{x}),\,\mathbf{H}_\epsilon(\bm{x})$};
\node (initConds) [specif, align=center, left=of porosity] {Initial conditions:\\$\rho_0(\bm{x}),\,\bm{v}_0(\bm{x}),\,t=0$};
\node (dragParams) [process, align=center, below=of params] {Compute: $\mu_\mathrm{D}$ \eqref{mu_D_Ergun_perm}, $\mu_\mathrm{F}$ \eqref{mu_F_Ergun_perm},\\$\eta_\mathrm{eff}(\epsilon)$, $\nu_\mathrm{eff}(\epsilon)$, $\nu_\mathrm{eff}'(\epsilon)$};
\node (initAcc) [process, align=center, below=of porosity] {Compute: $\bm{a}_0$ (\ref{accel},\ref{porosity_correction_expanded}),\\ $\boldsymbol{\Phi}_{0}^{\mathrm{ext}}$ \eqref{ex_corr}, $\Phi_0^{\eta}$ \eqref{eta_corr}};
\node (beta) [process, align=center, below=of dragParams] {Compute: $\beta$ \eqref{nu_eff_def}};
\node (initPop) [process, align=center, left=of beta] {Initialize:\\ $\bar{f}_{0,i}=\bar{f}_{0,i}^\mathrm{eq}$ \eqref{eq_bc_f_bar}};
\node (initfeq) [process, align=center, left=of initPop] {Compute:\\${f}_{0,i}^\mathrm{eq}$, ${f}_{0,i}^*$ (\ref{1d_scaled_eq}-\ref{P_aa_*})};
\node (lbe) [process, align=center, below=of initPop] {Collide \& Stream: $\bar{f}_{i}$ (\ref{dicrete_scheme}-\ref{momentum_rate})};
\node (dt) [process, align=center, left=of lbe] {$t=t+\delta t$};
\node (bc_rho) [process, align=center, below=of lbe] {Compute:\\$\rho_\mathrm{b}$ (e.g., \ref{bc_density_BounceBack})};
\node (bc_v) [specif, align=center, left=of bc_rho] {Boundary velocity: $\bm{v}_\mathrm{b}$};
\node (moments) [process, align=center, right=of bc_rho, xshift=1cm] {Compute: $\rho_\mathrm{f}$ \eqref{density_moment},\\ $\bm{v}_\mathrm{f}$ (\ref{v_(0)}-\ref{v_final},\ref{porosity_correction_expanded})};
\node (accel) [process, align=center, below=of bc_rho] {Compute: $\bm{a}$ (\ref{accel},\ref{porosity_correction_expanded})};
\node (corr) [process, align=center, below=of moments] {Compute:\\$\boldsymbol{\Phi}^{\mathrm{ext}}$ \eqref{ex_corr}, $\Phi^{\eta}$ \eqref{eta_corr}};
\node (eq) [process, align=center, below=of bc_v] {Compute:\\$f_i^\mathrm{eq}$ (\ref{1d_scaled_eq}-\ref{P_alpha_alpha_eq})};
\node (fstar) [process, align=center, below=of accel] {Compute:\\$f_i^*$ (\ref{1d_scaled_eq}-\ref{prod_form},\ref{xi_alpha_*}-\ref{P_aa_*})};
\node (bc_f) [process, align=center, below=of eq] {Compute:\\$\bar{f}_{\mathrm{b},i}$ (e.g., \ref{NEEM_bc_f_bar}-\ref{zeta_beta_NEEM})};
\draw [arrow] (initConds) |- (initAcc);
\draw [arrow] (porosity) -- (initAcc);
\draw [arrow] (initConds) |- (initfeq);
\draw [arrow] ([yshift=-0.5cm]porosity.south) -| (dragParams);
\draw [arrow] (params) -- (dragParams);
\draw [arrow] (dragParams) -- (initAcc);
\draw [arrow] (dragParams) -- (beta);
\draw [arrow] (initAcc.south) -- ++(0,-0.5cm) node[red] {} -| (initfeq.north);
\draw [arrow] (initfeq) -- (initPop);
\draw [arrow] (initPop) -- (lbe);
\draw [arrow] (beta.south) -- ++(0,-0.5cm) node[red] {} -| (lbe.north);
\draw [arrow] (initfeq.south) -- ++(0,-0.5cm) node[red] {} -| (lbe.north);
\draw [arrow] (lbe) -- (bc_rho);
\draw [arrow] (lbe.south) -- ++(0,-0.5cm) node[red] {} -| (moments.north);
\draw [arrow] (bc_v) -- (bc_rho);
\draw [arrow] (bc_rho.south) -- ++(0,-0.5cm) node[red] {} -| (accel.north);
\draw [arrow] (bc_rho.south) -- ++(0,-0.5cm) node[red] {} -| (corr.north);
\draw [arrow] (bc_rho.south) -- ++(0,-0.5cm) node[red] {} -| (eq.north);
\draw [arrow] (bc_v.south) -- ++(0,-0.5cm) node[red] {} -| (accel.north);
\draw [arrow] (bc_v.south) -- ++(0,-0.5cm) node[red] {} -| (corr.north);
\draw [arrow] (bc_v.south) -- ++(0,-0.5cm) node[red] {} -| (eq.north);
\draw [arrow] (moments.south) -- ++(0,-0.5cm) node[red] {} -| (accel.north);
\draw [arrow] (moments.south) -- ++(0,-0.5cm) node[red] {} -| (corr.north);
\draw [arrow] (moments.south) -- ++(0,-0.5cm) node[red] {} -| (eq.north);
\draw [arrow] (corr) |- (fstar);
\draw [arrow] (accel) -- (fstar);
\draw [arrow] (fstar) -- (bc_f);
\draw [arrow] (eq) -- (bc_f);
\coordinate (conn_1) at ([xshift=0.5cm]moments.east);
\coordinate (conn_2) at ([yshift=-0.5cm]fstar.south);
\coordinate (conn_3) at ([yshift=-0.5cm]bc_f.south);
\coordinate (conn_4) at ([xshift=-1.1cm]eq.west);
\draw [arrow] (lbe.east) -| (conn_1) |- (conn_2) -- (conn_3) -| (conn_4) |- (dt.west);
\draw [arrow] (fstar.south) -- (conn_2);
\draw [arrow] (bc_f.south) -- (conn_3);
\draw [arrow] (eq.west) -- (conn_4);
\draw [arrow] (dt) -- (lbe);
\end{tikzpicture}
\smallskip
\caption{Flowchart showing the overall structure of the proposed algorithm for simulating flows in heterogeneous porous media. Blue rounded boxes denote input quantities, whereas red rectangular boxes denote computed quantities. The subscripts $\{0,\mathrm{b},\mathrm{f}\}$ refer to initial values, boundary-node values, and interior fluid-node values, respectively. The boundary-density relation based on the bounce-back scheme, \eqref{bc_density_BounceBack}, and the boundary-population reconstruction based on non-equilibrium extrapolation, (\ref{NEEM_bc_f_bar}-\ref{zeta_beta_NEEM}), are shown only as examples; any of the boundary-condition techniques described in Section~\ref{sec:boundary_conditions} may be used instead.} \label{Fig:FlowChart}
\end{figure}
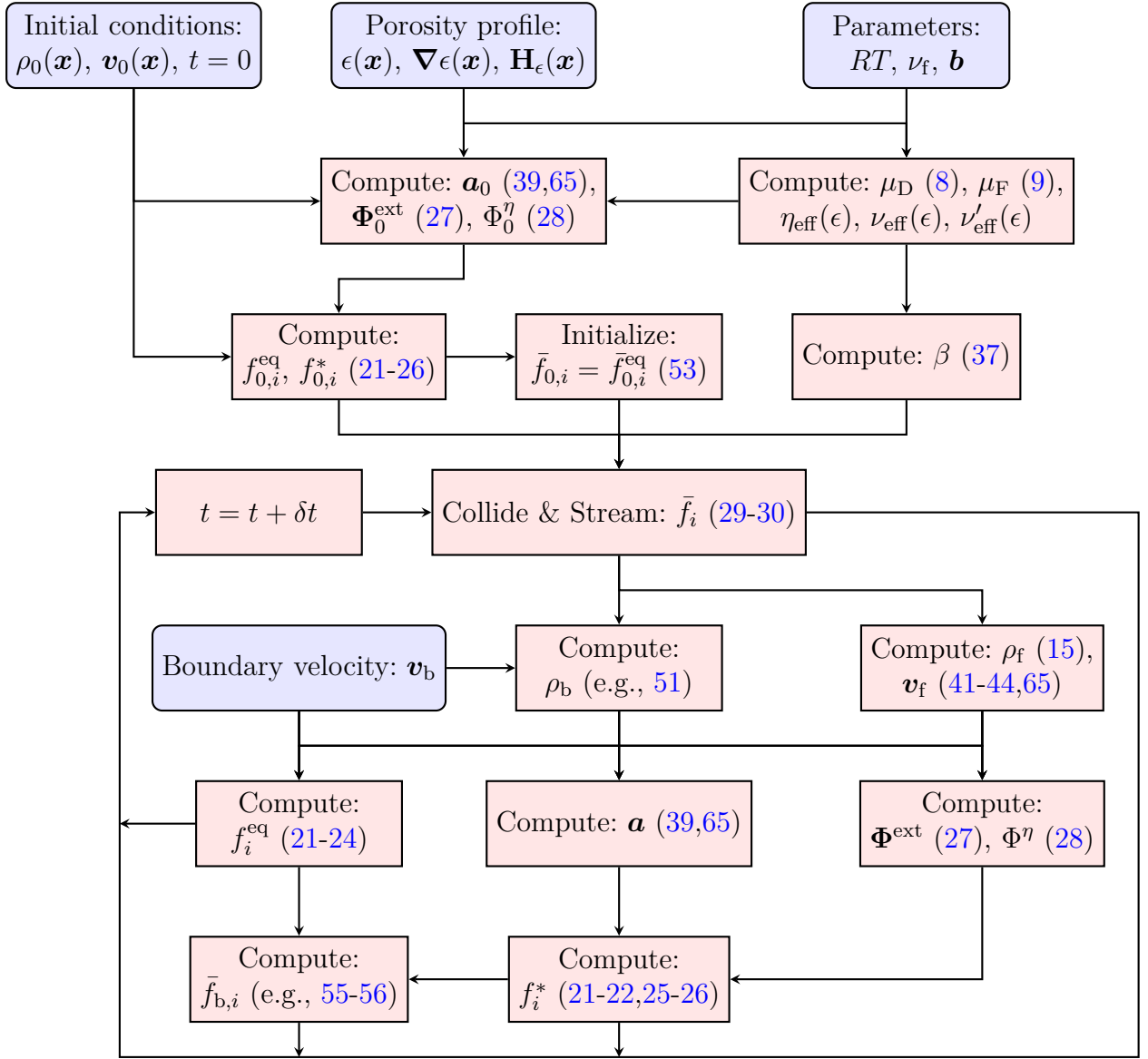

The boundary treatment follows the same structure. The boundary density can be evaluated either from the post-advection boundary populations alone using mass continuity \eqref{bc_density_impermWall}, or from both post-collision and post-advection populations using the bounce-back treatment \eqref{bc_density_BounceBack}. The reconstruction of boundary populations, for example by non-equilibrium extrapolation, then requires both $f_i^\mathrm{eq}$ and $f_i^*$ as inputs, as described in (\ref{NEEM_bc_f_bar}-\ref{zeta_beta_NEEM}). The overall structure of the proposed algorithm is summarized in Figure \ref{Fig:FlowChart}.

A central numerical ingredient of the algorithm is the evaluation of the viscous-stress correction $\bm{\Delta}_{\nu}(\bm{m})$, which is required for the computation of both the acceleration $\bm{a}$ in \eqref{accel} and the velocity $\bm{v}$ in \eqref{v_final}. Whenever available, this correction is most conveniently evaluated using analytical derivatives of $\epsilon(\bm{x})$ and $\nu_\mathrm{eff}(\bm{x})$. Analytical derivatives offer several advantages: they improve accuracy, reduce numerical noise, preserve isotropy more effectively, and provide cleaner behaviour near boundaries. In practice, they also improve stability and efficiency, since coarser grids can remain reliable even in the presence of sharp porosity transitions. This is particularly useful for porous structures with complex geometric features, such as curved interfaces, where numerical differentiation of $\epsilon(\bm{x})$ may otherwise appear under-resolved even when the porosity field itself is smooth.

This use of smooth analytical porosity fields is not merely a numerical fitting device. As argued in Section \ref{sec:target_macro}, a smooth porosity field $\epsilon(\bm{x})$ is the natural outcome of volume averaging a porous microstructure over a finite control volume. Smooth analytical functions can therefore be used to construct physically meaningful porous domains that mimic the result of averaging a geometrically complex, and possibly sharp, microstructure over a finite volume.

The viscous-stress correction term \eqref{porosity_correction} can be expanded as
\begin{equation}\label{porosity_correction_expanded} \begin{split} \bm{\Delta}_{\nu}(\bm{m}) &= \nu_\mathrm{eff}(\epsilon) \left\{\left[\mathbf{H}_\epsilon+\mathrm{tr}{(\mathbf{H}_\epsilon)\mathbf{I}}\right]\bm{m}+(\bm{\nabla}\cdot\bm{m})\bm{\nabla}\epsilon+(\bm{\nabla}\bm{m})\bm{\nabla}\epsilon\right\} \\ &+ \nu_\mathrm{eff}'(\epsilon)\left[\lvert\bm{\nabla\epsilon}\rvert^2\bm{m} + (\bm{m}\cdot\bm{\nabla}\epsilon)\bm{\nabla}\epsilon\right] \end{split} \end{equation}
where $\mathbf{H}_\epsilon=\bm{\nabla}\bm{\nabla}\epsilon$ is the Hessian matrix of the porosity field. Evaluation of \eqref{porosity_correction_expanded} requires the porosity field to be at least twice continuously differentiable, that is, $\epsilon(\bm{x})\in C^2$, although smoother choices are generally preferable. If variable effective viscosities are considered, the effective shear-viscosity field should satisfy at least $\nu_\mathrm{eff}(\bm{x})\in C^1$. In practice, it is both convenient and physically meaningful to prescribe $\nu_\mathrm{eff}$ through a correlation with porosity (\cite{Rinehart_2021}). In that case, if $\epsilon(\bm{x})\in C^2$, then a linear scalar relation $\nu_\mathrm{eff}(\epsilon)$ is sufficient.

Examples of smooth porous domains constructed using bump functions are included in the numerical validation and demonstration cases. These include the heterogeneous Poiseuille flow in Section \ref{numerical_results:Poiseuille:Heterogeneous}, the confined flow past a permeable cylinder in Section \ref{numerical_results:Poiseuille:Cylinder}, and the lid-driven cavity with a porous obstacle in Section \ref{sec:results:cavity}.

The correction terms ${\Phi}_{\alpha\alpha}^{\mathrm{ext}}$ and $\Phi^{\eta}$ also involve spatial derivatives (\ref{ex_corr}-\ref{eta_corr}). In addition, evaluation of $\bm{\Delta}_{\nu}(\bm{m})$ through \eqref{porosity_correction_expanded} requires spatial derivatives of $\bm{m}(x,y)$. In the numerical applications presented in this study, these derivatives are computed using second-order central differences at interior points and second-order one-sided forward or backward differences at the boundaries.

\section{Consistency validation}\label{numerical_results:spectral} % Dispersion and dissipation analysis
To validate the hydrodynamic consistency of the proposed model, we compare (i) the propagation speed and (ii) the dissipation rates of small-amplitude plane-wave disturbances obtained from numerical simulations with analytical predictions from a linear temporal spectral analysis of the target hydrodynamic limit (\ref{continuity_target}, \ref{momentum_target}, \ref{visc_stress_target}). The analytical results are derived via a normal-mode method (see Appendix \ref{Append:Spectral_Analysis}).

We consider isothermal flow through homogeneous, isotropic, stationary porous media, so that $\nu_\mathrm{eff}$, $\eta_\mathrm{eff}$, $\epsilon$, $\mu_\mathrm{D}$, $\mu_\mathrm{F}$, and $T$ are spatially and temporally constant. The two-dimensional computational domain is a square of size $\bm{L}=(L,L)$ and periodic boundary conditions in $\bm{x}=(x,y)$. A homogeneous steady base flow (advection) is imposed. Maintaining this base flow requires a constant uniform body force that exactly counteracts the porous-media drag.

Monochromatic plane-wave disturbances are introduced with a wave-vector $\bm{k}$ such that the base flow is either parallel or perpendicular to $\bm{k}$. To ensure linear behavior, we use a small disturbance amplitude $a_0=1\times 10^{-5}$. Unless stated otherwise, we set $L=200$ and $\epsilon=0.5$ in all simulations, and initialize all populations at equilibrium using \eqref{eq_bc_f_bar}.

To measure the dispersion and dissipation properties from numerical simulations, we need to determine the complex Fourier coefficient for a mode $\bm{k}=(k_x,k_y)$ using the 2D discrete Fourier transform (DFT) of a relevant flow field at that mode. For an arbitrary scalar flow field $\phi(x,y,t)$ sampled on a uniform square grid with $N_x=N_y=L$ grid nodes in each direction, the complex Fourier coefficient at mode $\bm{k}=2\pi(n_x,n_y)/L$, $\{n_x,n_y\}=0,\dots,L-1$, at time $t$ can be expressed as
\begin{equation}
    \mathcal{F}_{(n_x,n_y)}[\phi(x,y,t)] = \frac{1}{L^2}\sum_{j=0}^{L-1}\sum_{l=0}^{L-1}\phi_{j,l}(t)\exp{\left[-\ri \frac{2\pi}{L}(n_xj + n_yl)\right]} \label{2D-DFT}
\end{equation}
where $j$ and $l$ are the grid point indices in $x$ and $y$ directions, respectively. In practice, the DFT in \eqref{2D-DFT} is evaluated by a fast Fourier transform (FFT) algorithm.

\subsection{Propagation of normal modes}
To measure the propagation speed of normal eigenmodes in numerical simulations, i.e., the acoustic phase speed $c_\mathrm{ac}$, we excite a small-amplitude acoustic wave with a single normal mode, let it propagate, and extract $c_\mathrm{ac}$ from the phase. To that end, we initialize the simulation as
\begin{gather}
    \rho(x,y,0) = \rho_0 + a_0\cos{\left(\frac{2\pi x}{L}\right)}, \nonumber\\
    v_x(x,y,0) = A_v\cos{\left(\frac{2\pi x}{L}+ \vartheta_v\right)},\quad v_y(x,y,0) = \epsilon\mathrm{Ma}_0\sqrt{RT},
\end{gather}
with 
\begin{gather}
    A_v = \frac{\epsilon a_0}{\rho_0}\sqrt{RT}, \\
    \vartheta_v = \arg{\left(- \ri\chi + \omega^\mathrm{analyt}\right)}
\end{gather}
where
\begin{gather}
    \omega^\mathrm{analyt} = \sqrt{RT\left(\frac{2\pi}{L}\right)^2 -\chi^2}, \label{omega_analyt} \\
    \chi = \frac{1}{2}\left[(\nu_\mathrm{eff}+\eta_\mathrm{eff})\left(\frac{2\pi}{L}\right)^2 + \mu_\mathrm{D} + \mu_\mathrm{F} \epsilon\mathrm{Ma}_0\sqrt{RT}\right], \label{chi_perp}
\end{gather}
and $\mathrm{Ma}_0$ is the advection Mach number. Details of the derivation of the functional form of $\omega^\mathrm{analyt}$ \eqref{omega_analyt} and $\chi$ \eqref{chi_perp} are presented in Appendix \ref{Append:Spectral_Analysis:General}.

We set the base density as $\rho_0=1$ and impose the following uniform acceleration in \eqref{accel}
\begin{equation}
    b_x = 0,\quad b_y = \mu_\mathrm{D}\mathrm{Ma}_0\sqrt{RT} + \mu_\mathrm{F}\epsilon\mathrm{Ma}_0^2RT.
\end{equation}
This acceleration counteracts the porous-media drag to achieve steady advection.

We let the simulation run, and at each time step $t_n=n\delta t$ we compute $\hat{\rho}(t_n)=\mathcal{F}_{(1,0)}[\rho(t_n)-\rho_0]$ as defined by \eqref{2D-DFT} using FFT, and extract the phase (angle) $\vartheta_\rho(t_n)=\arg{(\hat{\rho}(t_n))}$. We then unwrap $\vartheta_\rho$ in time (to remove $2\pi$ jumps) and fit it to the following linear relation
\begin{equation}
    \vartheta_\rho(t_n) \propto - \omega^{\mathrm{sim}} t_n,
\end{equation}
The frequencies $\omega^{\{\mathrm{analyt/sim}\}}$, found either analytically or from simulations, are related to the acoustic phase speed as
\begin{equation}
    c_{\mathrm{ac}}^{\{\mathrm{analyt/sim}\}} = \omega^{\{\mathrm{analyt/sim}\}}\left(\frac{L}{2\pi}\right) \label{c_ac_analyt_sim}
\end{equation}
We then define the effective speed of sound as the long-wavelength limit of the acoustic phase speed $c_\mathrm{s,eff}=\lim_{k\rightarrow 0}{(\omega/k)}$, $k=2\pi/L$, as in the hydrodynamic regime the acoustic dispersion is linear to leading order, $\omega=c_\mathrm{s,eff}k+\mathcal{O}(k^2)$, while dissipative and dispersive corrections enter at higher order and are model- and scale-dependent. In the present volume-averaged porous-media formulation, $c_\mathrm{s,eff}$ should be interpreted as the propagation speed of small pressure--density perturbations supported by the homogenized fluid-phase equations (\ref{continuity_target}, \ref{momentum_target}). It describes neither the elastic-wave propagation in the solid matrix nor a poroelastic (two-phase) coupling, since the porous medium is treated here as stationary and rigid and enters only through porosity and drag terms. Capturing matrix elasticity and coupled poroelastic modes would require a two-phase/poroelastic framework, which is beyond the scope of the present model.

To obtain $c_\mathrm{s,eff}$ to leading order in the long-wavelength limit, we prescribe the porous-media drag parameters $\mu_\mathrm{D}$ and $\mu_\mathrm{F}$ with an explicit $k$-scaling (equivalently an $L$-scaling). We consider two parameter choices. In the first case, we enforce a viscous-drag balance (VDB) by setting
\begin{equation}
\mu_\mathrm{D}^{\mathrm{VDB}} = \nu_\mathrm{f}\left(\frac{2\pi}{L}\right)^2, \quad \mu_\mathrm{F}^{\mathrm{VDB}} = \frac{\nu_\mathrm{f}}{\epsilon\sqrt{RT}}\left(\frac{2\pi}{L}\right)^2. \label{VDB_params}
\end{equation}
From \eqref{chi_perp}, the choice \eqref{VDB_params} makes the viscous and drag-induced damping contributions comparable in magnitude and ensures that both scale with $L$ in the same way. Using the analysis in Appendix \ref{Append:Spectral_Analysis:VDB}, the resulting effective speed of sound is
\begin{equation}
c_\mathrm{s,eff}^\mathrm{VDB} = \sqrt{RT}. \label{cs_approx_VDB}
\end{equation}
In the second case, we determine $\mu_\mathrm{D}$ and $\mu_\mathrm{F}$ from the closure relations \eqref{mu_D_Ergun_perm} and \eqref{mu_F_Ergun_perm}, with the permeability specified via the Darcy number $\mathrm{Da}=\kappa/L^2$ and with $F_\epsilon$ given by the correlation \eqref{F_eps_Ergun}. In the simulations we set $\mathrm{Da}=1\times 10^{-4}$. As derived in Appendix \ref{Append:Spectral_Analysis:constDa}, the effective speed of sound in this case is
\begin{equation}
c_\mathrm{s,eff}^\mathrm{closure} = \sqrt{RT\left(1-\frac{\epsilon^4 F_\epsilon^2 \mathrm{Ma}_0^2}{16\pi^2 \mathrm{Da}}\right)}. \label{cs_approx_closure}
\end{equation}
The expression \eqref{cs_approx_closure} indicates that, under the choices (\ref{F_eps_Ergun}, \ref{mu_D_Ergun_perm}, \ref{mu_F_Ergun_perm}) and for a fixed $\mathrm{Da}$, the effective speed of sound is influenced by the inertial (Forchheimer) drag. This follows from the $L$-scalings (i.e., $\mu_\mathrm{F}\propto L^{-1}$ and $\mu_\mathrm{D}\propto L^{-2}$) that are different than with the choice \eqref{VDB_params}. The dependence of $c_\mathrm{eff}$ on $\mathrm{Ma}_0$ arises because the closure \eqref{closure_general} is not Galilean invariant.

\begin{figure}[h]
\centering
\includegraphics[width=0.85\textwidth]{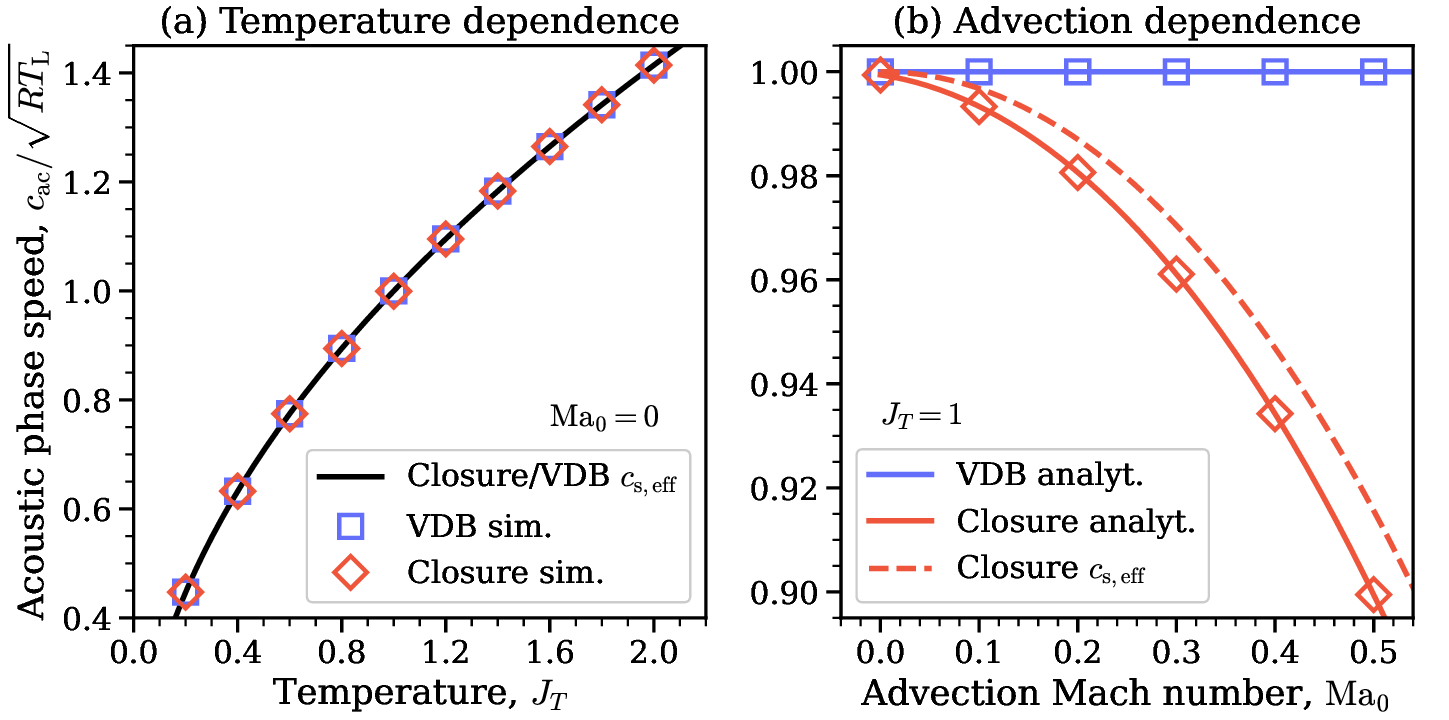}
\caption{Dependence of acoustic phase speed on (a) temperature and (b) advection speed. Markers denote values measured from numerical simulations: blue squares correspond to the VDB parameter set \eqref{VDB_params}, while red diamonds correspond to the closure relations (\ref{F_eps_Ergun}, \ref{mu_D_Ergun_perm}, \ref{mu_F_Ergun_perm}) with $\mathrm{Da}=\kappa/L^2=1\times 10^{-4}$. In panel (a), the black solid line shows the effective speed of sound predicted by (\ref{cs_approx_VDB}, \ref{cs_approx_closure}). In panel (b), the blue and red solid lines show analytical solutions from (\ref{omega_analyt}, \ref{chi_perp}, \ref{c_ac_analyt_sim}) using the VDB and closure parameterizations, respectively. The red dashed line indicates the effective speed of sound for the closure parameterization given by \eqref{cs_approx_closure}.
}\label{cs_eff_plots}
\end{figure}

For convenience, we introduce the temperature ratio $J_{T}$, effective shear viscosity ratio $J_{\nu_{\mathrm{eff}}}$ and the effective bulk viscosity ratio $J_{\eta_{\mathrm{eff}}}$, defined as, respectively,
\begin{equation}
    J_{T} = \frac{T}{T_\mathrm{L}}, \quad J_{\nu_{\mathrm{eff}}} = \frac{\nu_{\mathrm{eff}}}{\nu_\mathrm{f}}, \quad J_{\eta_{\mathrm{eff}}} = \frac{\eta_{\mathrm{eff}}}{\nu_\mathrm{f}}. \label{ratios_def}
\end{equation}
For both parameterizations, simulations were performed by sweeping $J_T\in[0.2,2.0]$ at $\mathrm{Ma}_0=0$, and $\mathrm{Ma}_0\in[0.1,0.5]$ at $J_T=1$. The fluid viscosity was set to $\nu_\mathrm{f}=0.01$, and the effective viscosity ratios were set to one, $J_{\nu_\mathrm{eff}}=J_{\eta_\mathrm{eff}}=1$. The results are shown in Figure \ref{cs_eff_plots}. In the absence of advection ($\mathrm{Ma}_0=0$), the acoustic phase speed $c_\mathrm{ac}$ extracted from the numerical simulations agrees excellently with the effective speed of sound $c_{\mathrm{s,eff}}$ given by (\ref{cs_approx_VDB}, \ref{cs_approx_closure}), as seen in Figure \ref{cs_eff_plots}(a). Note that, in this case, both parameterizations yield the same temperature dependence.

For varying advection speed, the results in Figure \ref{cs_eff_plots}(b) likewise show excellent agreement with the analytical solutions obtained from (\ref{omega_analyt}, \ref{chi_perp}, \ref{c_ac_analyt_sim}) for both the VDB and closure choices, confirming that the dispersion is recovered consistently in the hydrodynamic limit. Although the underlying equation system is not Galilean invariant, and so the normal-mode propagation speed depends on the advection velocity, the close match between the numerically measured phase speeds and the analytical predictions indicates that the proposed LBM model does not introduce additional anisotropic artifacts.

Finally, Figure \ref{cs_eff_plots}(b) shows that, for the closure parameterization, the effective speed of sound given by \eqref{cs_approx_closure} increasingly deviates from the acoustic phase speed as $\mathrm{Ma}_0$ grows. This is primarily because the chosen domain size is not sufficiently large to realize the long-wavelength limit. With the closure (\ref{F_eps_Ergun}, \ref{mu_D_Ergun_perm}, \ref{mu_F_Ergun_perm}), the drag terms are much larger in magnitude than for the VDB choice \eqref{VDB_params}. Consequently, the Darcy drag (set by $\mu_\mathrm{D}$) remains non-negligible in the present setup, leading to a noticeable separation between $c_\mathrm{s,eff}^\mathrm{closure}$ and $c_\mathrm{ac}$ as $\mathrm{Ma}_0$ increases.

\subsection{Dissipation of shear \& acoustic modes}\label{dissip_shear_normal_modes}
To compare the performance of the product-form discrete equilibrium to that of the second-order equilibrium (see \eqref{so_eq_v2} in Appendix \ref{Append:eq_moments}), like the one proposed by \cite{Zhang_2014}, and to validate the effectiveness of the proposed correction terms (\ref{ex_corr}, \ref{eta_corr}), we examine the shear and acoustic (normal) dissipation rates. To reduce the influence of the reference-frame dependence inherent to the non-Galilean-invariant closure \eqref{closure_general}, and to suppress possible spurious contributions from higher-order (non-hydrodynamic) moments, we employ the VDB parameterization \eqref{VDB_params} in the following simulations.

To measure the dissipation rates of transverse and longitudinal plane-wave disturbances (i.e., the shear and acoustic decay rates) in numerical simulations, we impose a small-amplitude sinusoidal velocity perturbation of the form
\begin{equation}
\tilde{v}_\alpha(x,y) = a_{\alpha}\sin{\left[\frac{2\pi}{L}(n_xx+n_yy)\right]} \label{general_v_perturb}
\end{equation}
on top of a constant base flow (advection), allow it to propagate, and then extract the decay rates as described below. We define the absolute values of the Fourier coefficients of the vorticity $\left(\bm{\nabla}\times\bm{v}\right)_z$ and the divergence $(\bm{\nabla}\cdot\bm{v})$ of the superficial fluid velocity field as
\begin{equation}
\hat{\mathcal{V}}_{(n_x,n_y)}(t) = \left\lvert\mathcal{F}_{(n_x,n_y)}[\partial_x v_y - \partial_y v_x]\right\rvert,\quad \hat{\mathcal{D}}_{(n_x,n_y)}(t) = \left\lvert\mathcal{F}_{(n_x,n_y)}[\partial_x v_x + \partial_y v_y]\right\rvert, \label{vort_div_Fourier}
\end{equation}
where the mode indices $(n_x,n_y)$ match those of the imposed perturbation \eqref{general_v_perturb}.

To measure the shear (transverse) decay rate $r_\mathrm{sh}^\mathrm{sim}$, we run the simulation and, at each time step $t_n=n\delta t$, compute $\hat{\mathcal{V}}_{(n_x,n_y)}(t_n)$, as defined in \eqref{vort_div_Fourier}, using a second-order central difference scheme followed by FFT. We then estimate $r_\mathrm{sh}^\mathrm{sim}$ by fitting the data to the log-linear relation
\begin{equation}
\ln{\left[\hat{\mathcal{V}}_{(n_x,n_y)}(t_n)\right]} \sim - r_\mathrm{sh}^\mathrm{sim} t_n,
\end{equation}
using a least-squares fit.

To measure the acoustic (longitudinal) decay rate $r_\mathrm{ac}^\mathrm{sim}$, we analogously compute $\hat{\mathcal{D}}_{(n_x,n_y)}(t_n)$ in \eqref{vort_div_Fourier} at each time step via a second-order central difference and FFT. As the target hydrodynamic limit admits two propagating normal eigenmodes traveling in opposite directions, $\hat{\mathcal{D}}_{(n_x,n_y)}(t_n)$ is oscillatory. We therefore extract the decay rate from its peak envelope, by detecting the local maxima (peaks) at times $t_k$ and fitting the peak values to the log-linear relation
\begin{equation}
\ln{\left[\hat{\mathcal{D}}_{(n_x,n_y)}(t_k)\right]} \propto - r_\mathrm{ac}^\mathrm{sim} t_k.
\end{equation}
An alternative approach to measure the shear and acoustic dissipation rates, commonly used in literature (see e.g., \cite{Extended_LBM_Saadat_2021, Hosseini_Karlin_Shallow_Water_2025}), is to track the time evolution of a point-wise quantity, e.g., the maximum of a velocity component along the perturbation direction (relative to the base flow), and fit its exponential decay. In contrast, here we extract dissipation rates from the amplitude of the excited Fourier mode at the prescribed wavenumber $(n_x,n_y)$, using the vorticity and divergence Fourier coefficients \eqref{vort_div_Fourier}. This mode-selective, domain-global measure is consistent with linear eigenmode theory and is less sensitive to advection-induced phase drift, grid-translation effects, and local noise or waveform distortions that can bias point-wise extrema. Using vorticity and divergence further separates shear and acoustic dynamics, reducing mode cross-contamination, and the method extends straightforwardly to oblique perturbations and higher dimensions without requiring a direction-dependent sampling procedure.

We consider four test setups, all with a uniform initial density field $\rho(x,y,0) = \rho_0 = 1$, but each defined by a distinct perturbation of the form \eqref{general_v_perturb} applied to the initial velocity field. These are as follows:
\begin{enumerate}[(I)]
  \item An axis-aligned shear wave perpendicular to the base flow advection. The velocity initial conditions are
  \begin{equation}
      v_x(x,y,0) = a_0\sin{\left(\frac{2\pi}{L}y\right)},\quad
      v_y(x,y,0) = \epsilon\mathrm{Ma}_0\sqrt{RT}, \label{v_init_setup_I}
  \end{equation}
  the following uniform acceleration is imposed
  \begin{equation}
    b_x(x,y) = 0,\quad b_y(x,y) = \mu_\mathrm{D}\mathrm{Ma}_0\sqrt{RT} + \mu_\mathrm{F}\epsilon\mathrm{Ma}_0^2RT,
  \end{equation}
  and the analytical shear decay rate is
  \begin{equation}
      r_\mathrm{sh}^\mathrm{analyt} = \frac{4\pi^2\nu_\mathrm{f}}{L^2}\left(J_{\nu_\mathrm{eff}} + 1 + \mathrm{Ma}_0\right).
  \end{equation}
  \item An axis-aligned acoustic wave parallel to the base flow advection. The velocity initial conditions are
  \begin{equation}
      v_x(x,y,0) = \epsilon\mathrm{Ma}_0\sqrt{RT} + a_0\sin{\left(\frac{2\pi}{L}x\right)},\quad
      v_y(x,y,0) = 0, \label{v_init_setup_II}
  \end{equation}
  the following uniform acceleration is imposed
  \begin{equation}
    b_x(x,y) = \mu_\mathrm{D}\mathrm{Ma}_0\sqrt{RT} + \mu_\mathrm{F}\epsilon\mathrm{Ma}_0^2RT,\quad b_y(x,y) = 0,
  \end{equation}
  and the analytical acoustic decay rate is
  \begin{equation}
      r_\mathrm{ac}^\mathrm{analyt} = \frac{2\pi^2\nu_\mathrm{f}}{L^2}\left(J_{\nu_\mathrm{eff}} + J_{\eta_\mathrm{eff}} + 1 + 2\mathrm{Ma}_0\right).
  \end{equation}
  \item A diagonal shear wave perpendicular to the base flow advection. The velocity initial conditions are
  \begin{gather}
      v_x(x,y,0) = \frac{1}{\sqrt{2}}\left[\epsilon\mathrm{Ma}_0\sqrt{RT}+a_0\sin{\left(\frac{2\pi}{L}(x+y)\right)}\right], \nonumber \\
      v_y(x,y,0) = \frac{1}{\sqrt{2}}\left[\epsilon\mathrm{Ma}_0\sqrt{RT}-a_0\sin{\left(\frac{2\pi}{L}(x+y)\right)}\right], \label{v_init_setup_III}
  \end{gather}
  the following uniform acceleration is imposed
  \begin{equation}
    b_x(x,y) = b_y(x,y) = \frac{\mathrm{Ma}_0}{\sqrt{2}}\left(\mu_\mathrm{D}\sqrt{RT} + \mu_\mathrm{F}\epsilon\mathrm{Ma}_0RT\right),
  \end{equation}
  and the analytical shear decay rate is
  \begin{equation}
      r_\mathrm{sh}^\mathrm{analyt} = \frac{8\pi^2\nu_\mathrm{f}}{L^2}\left(J_{\nu_\mathrm{eff}} + 1 + \mathrm{Ma}_0\right).
  \end{equation}
  \item An axis-aligned acoustic wave perpendicular to the base flow advection. The velocity initial conditions are
  \begin{equation}
      v_x(x,y,0) =  a_0\sin{\left(\frac{2\pi}{L}x\right)},\quad
      v_y(x,y,0) = \epsilon\mathrm{Ma}_0\sqrt{RT}, \label{v_init_setup_IV}
  \end{equation}
  the following uniform acceleration is imposed
  \begin{equation}
   b_x(x,y) = 0,\quad b_y(x,y) = \mu_\mathrm{D}\mathrm{Ma}_0\sqrt{RT} + \mu_\mathrm{F}\epsilon\mathrm{Ma}_0^2RT,
  \end{equation}
  and the analytical acoustic decay rate is
  \begin{equation}
      r_\mathrm{ac}^\mathrm{analyt} = \frac{2\pi^2\nu_\mathrm{f}}{L^2}\left(J_{\nu_\mathrm{eff}} + J_{\eta_\mathrm{eff}} + 1 + \mathrm{Ma}_0\right). \label{r_ac_analyt_case_IV}
  \end{equation}
\end{enumerate}
In demonstration of the preservation of isotropy and Galilean invariance in the resulting effective viscous stress tensor, as well as temperature independence, of the proposed model in the hydrodynamic limit, it is instructive to examine the correction term \eqref{ex_corr}, introduced to remedy the diagonal anomaly of the product-form equilibrium \eqref{prod_form}. We note that this correction, and hence the underlying error it compensates, is non-zero whenever at least one diagonal derivative of the velocity field is non-zero, i.e., $\diag(\bm{\nabla}\bm{v})\neq\bm{0}$. Moreover, the magnitude of this error increases as the temperature $T$ departs from the lattice reference temperature $T_\mathrm{L}$, and as the corresponding velocity component $v_\alpha$ increases while varying more strongly along the $\alpha$-direction.

\begin{figure}[h!]
\centering
\includegraphics[width=1.0\textwidth]{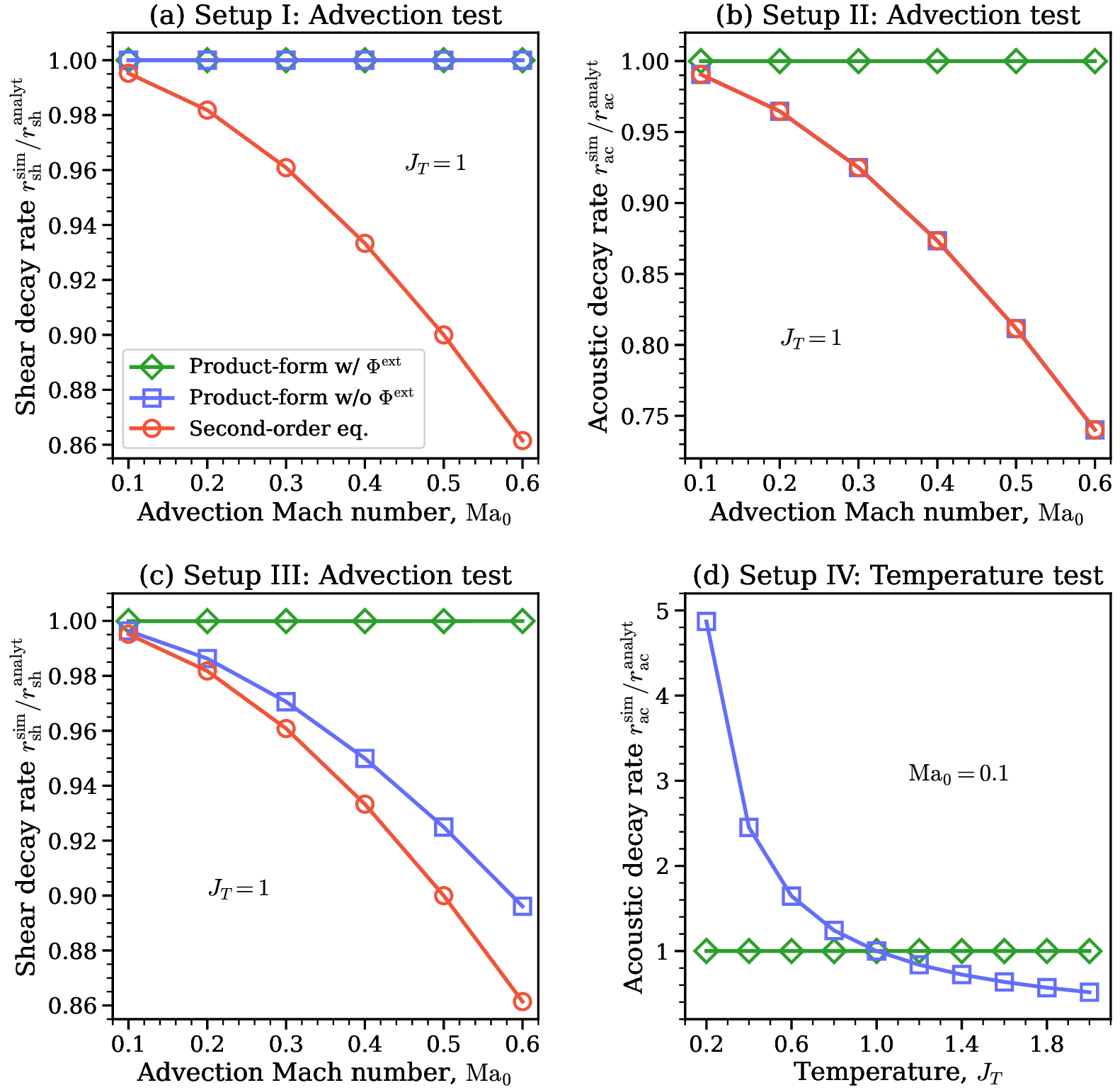}
\caption{Decay rate measurements from numerical simulations. (a-c) Advection tests for setups (I-III) at $T=T_\mathrm{L}$ over $\mathrm{Ma}_0\in[0.1,0.6]$. (d) Temperature test for setup (IV) at $\mathrm{Ma}_0=0.1$ over $J_T\in[0.2,2.0]$. The exact solution corresponds to $r^\mathrm{sim}_{\{sh/ac\}}/r^\mathrm{analyt}_{\{sh/ac\}}=1$. Green diamonds denote the product-form equilibrium \eqref{prod_form} with the correction \eqref{ex_corr}; blue squares denote \eqref{prod_form} without \eqref{ex_corr}; and red circles denote the second-order equilibrium \eqref{so_eq_v2}.}\label{sh_ac_decay_rates}
\end{figure}

Setups (I-III) are used to assess Galilean invariance by performing simulations at $T=T_\mathrm{L}$ over a range $\mathrm{Ma}_0\in[0.1,0.6]$, comparing the product-form equilibrium \eqref{prod_form} with and without the correction \eqref{ex_corr} against the second-order equilibrium \eqref{so_eq_v2}. The setup (IV) is used for the temperature independence testing over $J_T\in[0.2,2.0]$ at fixed $\mathrm{Ma}_0=0.1$. In all cases, the fluid kinematic viscosity is set to $\nu_\mathrm{f}=0.01$, and the effective kinematic viscosity ratio is set to unity, i.e., $J_{\nu_\mathrm{eff}}=1$. For simulations using \eqref{prod_form}, the bulk-viscosity correction is disabled, $\Phi^\eta=0$, which effectively enforces $J_{\eta_\mathrm{eff}}=1$ via \eqref{eta_0_no_corr}, same as with \eqref{so_eq_v2}. The resulting numerical measurements of the decay rates are presented in Figure \ref{sh_ac_decay_rates}.

With the velocity perturbation \eqref{v_init_setup_I} in setup (I), the diagonal anomaly in the third-order moment of the product-form equilibrium \eqref{prod_form} remains dormant and does not induce spurious effects, since both diagonal velocity derivatives vanish. Consequently, the correction term \eqref{ex_corr} has no impact in this case. By contrast, for the second-order equilibrium \eqref{so_eq_v2}, the non-vanishing anomaly in the off-diagonal components of the third-order moment introduces anisotropy in the effective viscous stress tensor, which effectively reduces the apparent numerical value of $\nu_\mathrm{eff}$. This behavior is evident in Figure \ref{sh_ac_decay_rates}(a): the shear decay rate obtained with the second-order equilibrium exhibits a strong reference-frame dependence, decreasing with increasing $\mathrm{Ma}_0$, whereas the product-form equilibrium, both with and without the correction, remains in excellent agreement with the analytical predictions.

The results for setup (II) in Figure \ref{sh_ac_decay_rates}(b) show that the product-form equilibrium \eqref{prod_form} without the correction \eqref{ex_corr} becomes effectively equivalent to the second-order equilibrium \eqref{so_eq_v2}: both exhibit the same Galilean-invariance violation in the acoustic decay rates. This behavior is expected, since the velocity perturbation \eqref{v_init_setup_II} excites only the diagonal components of the viscous stress, and the associated diagonal anomaly is the same for \eqref{prod_form} and \eqref{so_eq_v2}.

For setup (III), Figure \ref{sh_ac_decay_rates}(c) shows that both \eqref{prod_form} without \eqref{ex_corr} and \eqref{so_eq_v2} produce advection-dependent shear decay rates, albeit with different error levels. Here, the perturbation \eqref{v_init_setup_III} affects both diagonal and off-diagonal components of the effective viscous stress tensor, so the two equilibria incur distinct anisotropic errors. By contrast, \eqref{prod_form} augmented with the correction \eqref{ex_corr} recovers the imposed analytical decay rates, i.e., it remains Galilean invariant.

The setup (IV) then employs the same velocity perturbation \eqref{v_init_setup_IV} as setup (II), which affects only the diagonal components of the viscous stress, but adds the base-flow advection to the other (unperturbed) velocity component. As a result, the diagonal anomaly is activated, yet it remains independent of the advection velocity. Departing from the lattice reference temperature,  $J_T\neq 1$, then amplifies the resulting error, which scales linearly with the diagonal velocity components of the viscous stress tensor. This setup isolates the temperature-induced contribution as advection-related effects are negligible.

The results of the temperature-test in Figure \ref{sh_ac_decay_rates}(d) show that, for \eqref{prod_form} without \eqref{ex_corr}, the diagonal-anomaly effects are negligible only at the lattice reference temperature. In contrast, incorporating the correction \eqref{ex_corr} yields accurate acoustic decay rates across a wide range of temperatures.

Finally, we demonstrate that the present model can set the effective bulk viscosity independently through the correction term \eqref{eta_corr}. To this end, we performed tests with setup (IV) using \eqref{prod_form} supplemented by \eqref{ex_corr}, fixing $J_T=1$ and $\mathrm{Ma}_0=0.1$, and varying the effective shear and bulk viscosity ratios. The analytical expression \eqref{r_ac_analyt_case_IV} shows that the acoustic decay rate depends linearly on both $J_{\nu_\mathrm{eff}}$ and $J_{\eta_\mathrm{eff}}$, with identical weights (i.e., the same slope with respect to either parameter). Therefore, in the hydrodynamic limit and at a fixed advection speed, plotting $2r_\mathrm{ac}/(\nu_\mathrm{f} k^2)$ (with $k=2\pi/L$) versus $J_{\nu_\mathrm{eff}}$ at constant $J_{\eta_\mathrm{eff}}$, or versus $J_{\eta_\mathrm{eff}}$ at constant $J_{\nu_\mathrm{eff}}$, should yield a straight line with unit slope.

\begin{figure}[h]
\centering
\includegraphics[width=1.0\textwidth]{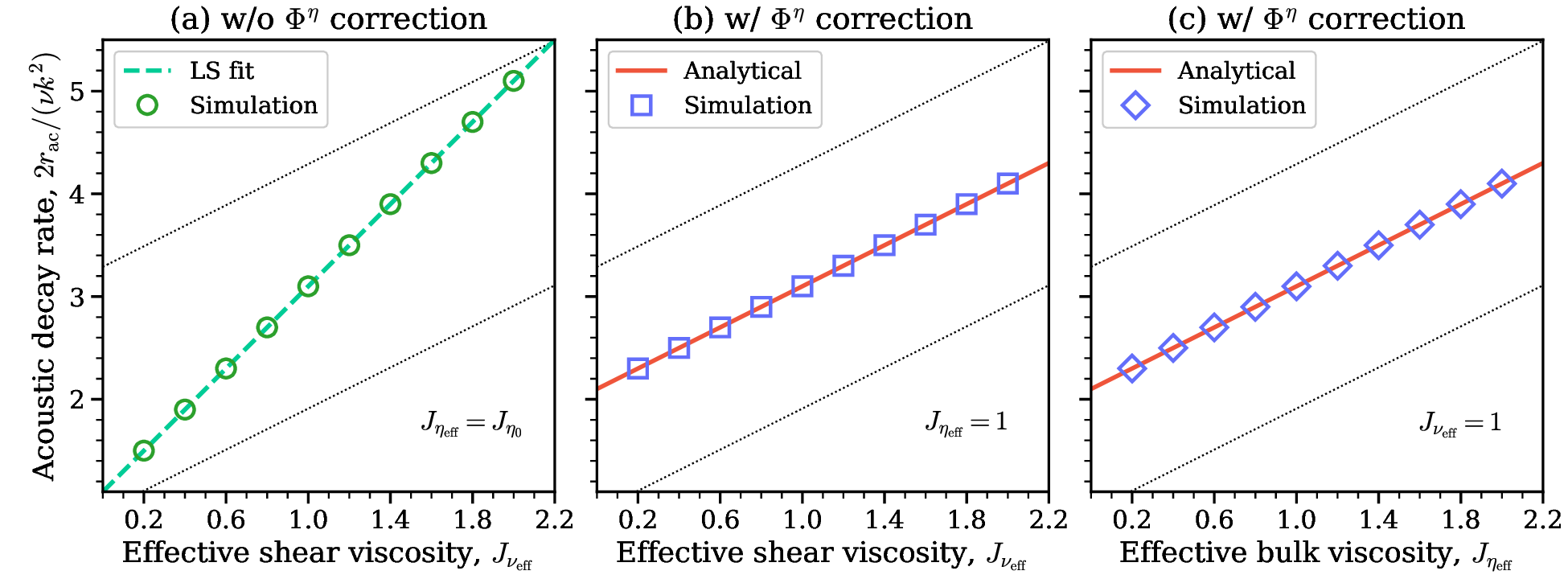}
\caption{Acoustic decay-rate measurements from numerical simulations. (a,b) Effective shear-viscosity tests over $J_{\nu_\mathrm{eff}}\in[0.2,2.0]$, without the correction \eqref{eta_corr} (effectively setting $J_{\eta_\mathrm{eff}}=J_{\eta_0}$), and with the correction \eqref{eta_corr} (fixing $J_{\eta_\mathrm{eff}}=1$), respectively. (c) Effective bulk-viscosity test over $J_{\eta_\mathrm{eff}}\in[0.2,2.0]$ including the correction \eqref{eta_corr} (fixing $J_{\nu_\mathrm{eff}}=1$). Markers indicate simulation results. The light-green dashed line in (a) is a least-squares fit through the simulation data. The red lines in (b,c) show the analytical predictions \eqref{r_ac_analyt_case_IV}. Dash-dotted lines denote unit-slope isolines, included to facilitate visual comparison across the test cases.}\label{Phi_eta_plots}
\end{figure}

The first set of simulations was performed without correction \eqref{eta_corr}, varying $J_{\nu_\mathrm{eff}}\in[0.2,2.0]$. The results are shown in Figure \ref{Phi_eta_plots}(a). A linear fit of the data yields a slope of $\approx 2$. This is expected because, in the hydrodynamic limit, $\eta_\mathrm{eff}$ is not independent of $\nu_\mathrm{eff}$ when \eqref{eta_corr} is omitted. As shown by the multiscale analysis in Appendix \ref{Append:CE_analysis}, the two are linked by \eqref{eta_0_no_corr}, implying $J_{\eta_\mathrm{eff}}=J_{\eta_0}=J_{\nu_\mathrm{eff}}$. This coupling is a well-known artifact of isothermal models employing the LBGK approximation \eqref{omega_BGK}.

The remaining two test cases used the correction \eqref{eta_corr}. In these simulations, we fixed either $J_{\nu_\mathrm{eff}}$ or $J_{\eta_\mathrm{eff}}$ and varied the other parameter. The results in Figure \ref{Phi_eta_plots}(b,c) show excellent agreement with the analytical acoustic decay rate \eqref{r_ac_analyt_case_IV}, and the data exhibit a unit slope in both cases. This confirms that \eqref{eta_corr} enables independent specification of the effective bulk viscosity $\eta_\mathrm{eff}$ in the hydrodynamic limit.

In weakly compressible models, the ability to tune the bulk viscosity is particularly useful: it can improve numerical stability and help diagnose or mitigate artifacts in underresolved (coarse-grid) simulations, even for flows that are nominally incompressible and shear-dominated.

\section{Benchmark validation}\label{sec:Benchmark_validation}
To further validate the proposed lattice Boltzmann model, we consider two standard two-dimensional benchmark problems: planar Couette flow under a gravitational field between two parallel moving plates, and generalized Poiseuille flow driven by a constant body force. A homogeneous porosity profile is used for the Couette-flow problem, whereas both homogeneous and heterogeneous porosity profiles are considered for the Poiseuille-flow problem. In each case, the simulation results are compared against reference numerical solutions.

\subsection{Couette flow}\label{numerical_results:Couette}
Here we consider a planar Couette flow under a gravitational field. The configuration consists of a two-dimensional solenoidal flow between two parallel horizontal walls that form a channel of width $H$, filled with a homogeneous porous medium of porosity $\epsilon$. The flow is driven by the tangential motion of one or both walls. To induce a density gradient, we apply a constant and uniform body acceleration (gravity) in the $-y$ direction, $\bm{b}=(0,-g)$. Under the assumption of a fully developed steady flow along the channel, the governing hydrodynamic equations (\ref{continuity_target}, \ref{momentum_target}, \ref{visc_stress_target}) reduce to the following two ordinary differential equations (ODEs),
\begin{gather}
    \frac{d\rho}{dy}+\frac{\rho g}{RT} = 0, \label{Couette_rho_ODE} \\
    \frac{d^2v}{dy^2} - \frac{g}{RT}\left(\frac{dv}{dy}\right) - \frac{\mu_\mathrm{D}v+\mu_\mathrm{F}\lvert v \rvert v}{\nu_\mathrm{eff}} = 0,\label{Couette_v_ODE}
\end{gather}
where $v$ refers to the streamwise superficial flow velocity (in the $x$-direction).

Integrating equation \eqref{Couette_rho_ODE} results in an exponential density profile as a function of $y$, and by defining the average density,
\begin{equation}
    \rho_0 = \frac{1}{H}\int^H_0{\rho(y)}\,dy
\end{equation}
we can express the density profile as
\begin{equation} \label{Couette_rho_analyt}
    \rho(y) =
    \begin{cases}
        \frac{\rho_0 gH e^{-gy/RT}}{RT\left(1-e^{-gH/RT}\right)}, & g\neq0 \\
        \rho_0, & g=0
    \end{cases}
\end{equation}
We impose no-slip boundary conditions at the walls,
\begin{equation}
    v(y=0)=v_\mathrm{bot},\quad v(y=H)=v_\mathrm{top} \label{Couette_BC}
\end{equation}
where $v_\mathrm{bot}$ and $v_\mathrm{top}$ denote the tangential velocities of the bottom and top walls, respectively. Together with \eqref{Couette_v_ODE}, these conditions define a two-point boundary-value problem (BVP).

The ODE \eqref{Couette_v_ODE} includes the Forchheimer drag term, and so it is nonlinear and, in general, does not admit an elementary closed-form solution satisfying both boundary conditions. To obtain a reference velocity profile solution for comparison with the LBM results, we solve the BVP (\ref{Couette_v_ODE}, \ref{Couette_BC}) numerically using a fourth-order collocation-based method with residual control and adaptive mesh refinement, as implemented in SciPy’s (\cite{2020_SciPy}) BVP solver. The computation is initialized on a uniform mesh of 1000 points in the $y$ direction.

In the absence of the Forchheimer drag term, i.e. for $\mu_\mathrm{F}=0$, the BVP (\ref{Couette_v_ODE}, \ref{Couette_BC}) admits the following closed-form Darcy-limit analytical solution for the steady superficial velocity profile:
\begin{equation}
    v_\mathrm{D}(y) = \frac{\left(v_\mathrm{top}-v_\mathrm{bot} e^{r_-H}\right)e^{r_+ y} + \left(v_\mathrm{bot} e^{r_+H}-v_\mathrm{top}\right)e^{r_- y}}{e^{r_+ H}-e^{r_-H}},\label{Couette_v_Darcy_analyt}
\end{equation}
where $r_\pm$ are the characteristic roots of the ODE \eqref{Couette_v_ODE},
\begin{equation}
    r_{\pm} = \frac{(g/RT)\pm \sqrt{(g/RT)^2 + 4(\mu_\mathrm{D}/\nu_\mathrm{eff})}}{2}
\end{equation}
The analytical Darcy-limit solution \eqref{Couette_v_Darcy_analyt} is used as the initial guess for the iterative numerical solution of the full BVP (\ref{Couette_v_ODE}, \ref{Couette_BC}) including the Forchheimer drag contribution.

Two wall-motion configurations are considered in the simulations. In the first, the bottom wall moves with speed $v_\mathrm{bot}=-V_0,$ and the top wall moves with speed $v_\mathrm{top}=V_0$. In the second configuration, the bottom wall is stationary, $v_\mathrm{bot}=0$, while the top wall moves tangentially with speed $v_\mathrm{top}=2V_0$.

The flow is characterized by the following three non-dimensional parameters: the Reynolds number $\mathrm{Re}$, the gravity-compressibility number $\Gamma$, and the Darcy number $\mathrm{Da}$, defined as, respectively,
\begin{equation}\label{Couette_non_dim_numbers}
    \mathrm{Re} = \frac{2V_0 H}{\nu_\mathrm{f}},\quad \Gamma = \frac{gH}{RT},\quad \mathrm{Da}=\frac{\kappa}{H^2}
\end{equation}
In all simulations, the porosity is set to $\epsilon=0.1$ and the channel height to $H=200$, while the temperature ratio and the effective-viscosity ratios \eqref{ratios_def} are all fixed at unity. The kinematic shear viscosity was prescribed as
\begin{equation} \label{Couette_nu}
    \nu_\mathrm{f} = \min{\left(0.1,\ \frac{\epsilon H \sqrt{RT}}{2\mathrm{Re}}\right)},
\end{equation}
and the drag coefficients $\mu_\mathrm{D}$ and $\mu_\mathrm{F}$ are determined from the closure relations \eqref{mu_D_Ergun_perm} and \eqref{mu_F_Ergun_perm}, respectively, with the permeability specified through the $\mathrm{Da}$ number and with $F_\epsilon$ evaluated using the correlation \eqref{F_eps_Ergun}.

The computational domain is taken as a square lattice of $N\times N$ grid points, where $N=H+1$. The domain is periodic in $x$, with solid walls at the bottom ($y=0$) and top ($y=H$). The non-slip boundary conditions \eqref{Couette_BC} are imposed by first determining the density from the mass continuity \eqref{bc_density_impermWall}, and then evaluating the boundary populations using the non-equilibrium extrapolation scheme \eqref{NEEM_bc_f_bar}. Since the domain is periodic in $x$, no special corner treatment is required when evaluating the density.

The populations are initialized at equilibrium using \eqref{eq_bc_f_bar}. The initial density field is prescribed from the analytical solution \eqref{Couette_rho_analyt} with $\rho_0=1$, while the initial velocity is set to zero throughout the domain. Each simulation was run until a steady state was reached.

\begin{figure}[h]
\centering
\includegraphics[width=1.0\textwidth]{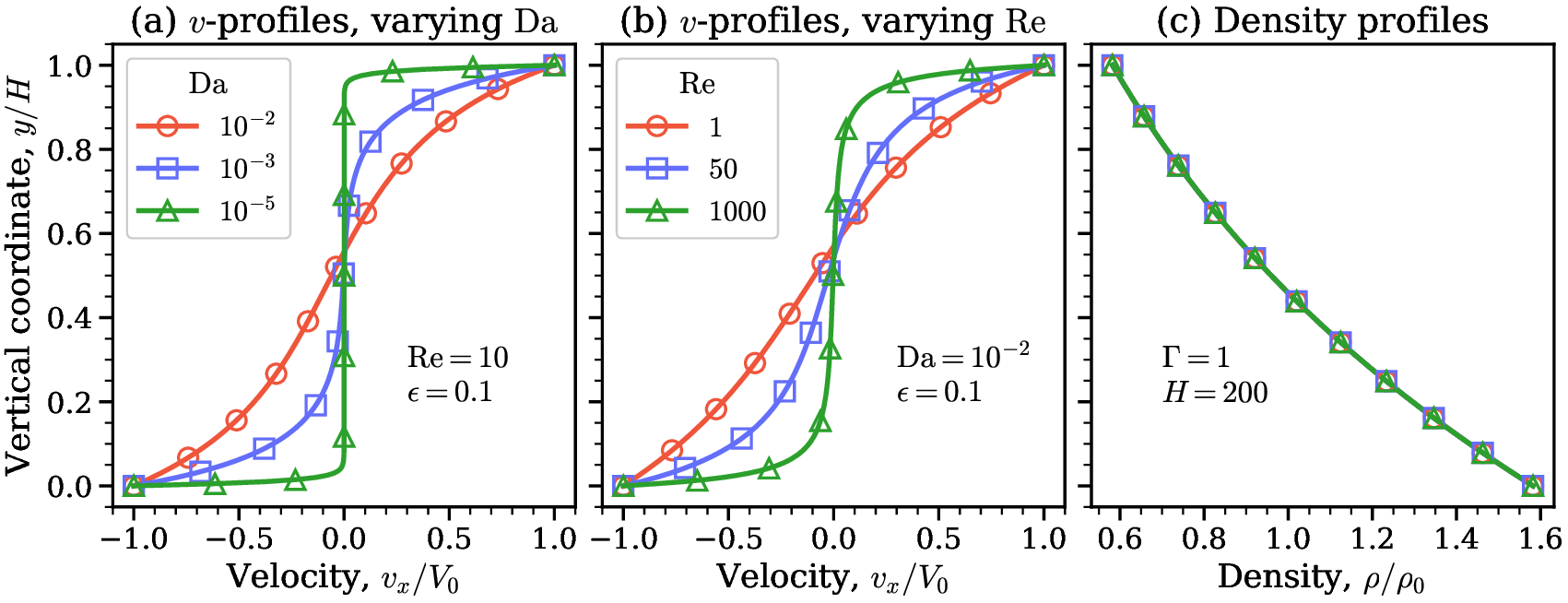}
\caption{Profiles of the superficial flow velocity for Couette flow in a homogeneous porous medium under a gravitational field, shown for different values of (a) the Darcy number and (b) the Reynolds number, together with (c) the corresponding intrinsic fluid density profiles. Solid lines denote the LBM simulation results. Markers indicate the reference solutions: in panels (a,b), the velocity profiles are obtained by numerically solving the BVP (\ref{Couette_v_ODE}, \ref{Couette_BC}), while in panel (c), the density profiles are given by the analytical solution \eqref{Couette_rho_analyt}.}\label{Couette_FlowProfiles}
\end{figure}

To assess convergence to steady state, we use the discrete relative $L_2$ change between two consecutive time levels. The simulation is considered to have reached steady state at time step $t_\mathrm{ss}$ when the superficial velocity field satisfies
\begin{equation} \label{SS_convergence_L2}
    \left[\frac{\sum_{j=0}^{N-1}\sum_{l=0}^{N-1}\left\lvert\bm{v}_{j,l}(t_\mathrm{ss})-\bm{v}_{j,l}(t_\mathrm{ss}-\delta t)\right\rvert^2}{\sum_{j=0}^{N-1}\sum_{l=0}^{N-1}\left\lvert\bm{v}_{j,l}(t_\mathrm{ss}-\delta t)\right\rvert^2}\right]^{1/2} \leq 10^{-10}.
\end{equation}
In the first set of simulations, we consider the two-wall-motion configuration, in which the walls move in opposite directions, and set $\Gamma=1$. In Figure \ref{Couette_FlowProfiles}, the superficial velocity profiles in panels (a,b) and the intrinsic density profiles in panel (c), obtained with the proposed LBM model, are compared against reference solutions for different values of the Darcy number and the Reynolds number. The velocity profiles are benchmarked against the numerical solution of the BVP (\ref{Couette_v_ODE}, \ref{Couette_BC}), while the density profiles are compared with the analytical solution \eqref{Couette_rho_analyt}. Excellent agreement is observed in all cases. Note that, as shown in Figure \ref{Couette_FlowProfiles}(c), the gravitational field induces a pronounced nonlinear density variation across the channel. The close match between the LBM results and the reference solutions confirms the validity of the present model, particularly with respect to the boundary-condition implementation and the consistency of the forcing scheme \eqref{F_RtFM}.

In the second set of simulations, following a study by \cite{Guo_Porous_2002} for a similar system, we examine the nonlinear inertial effect introduced by the Forchheimer drag contribution by comparing the Darcy--Forchheimer system with its Darcy-limit counterpart \eqref{Couette_v_Darcy_analyt}. We consider the one-wall-motion configuration, in which only the top wall moves tangentially, set $\Gamma=0$ to remove the gravitational field, and perform simulations both with $\mu_\mathrm{F}$ prescribed according to \eqref{mu_F_Ergun_perm} and with $\mu_\mathrm{F}=0$.

\begin{figure}[h]
\centering
\includegraphics[width=0.9\textwidth]{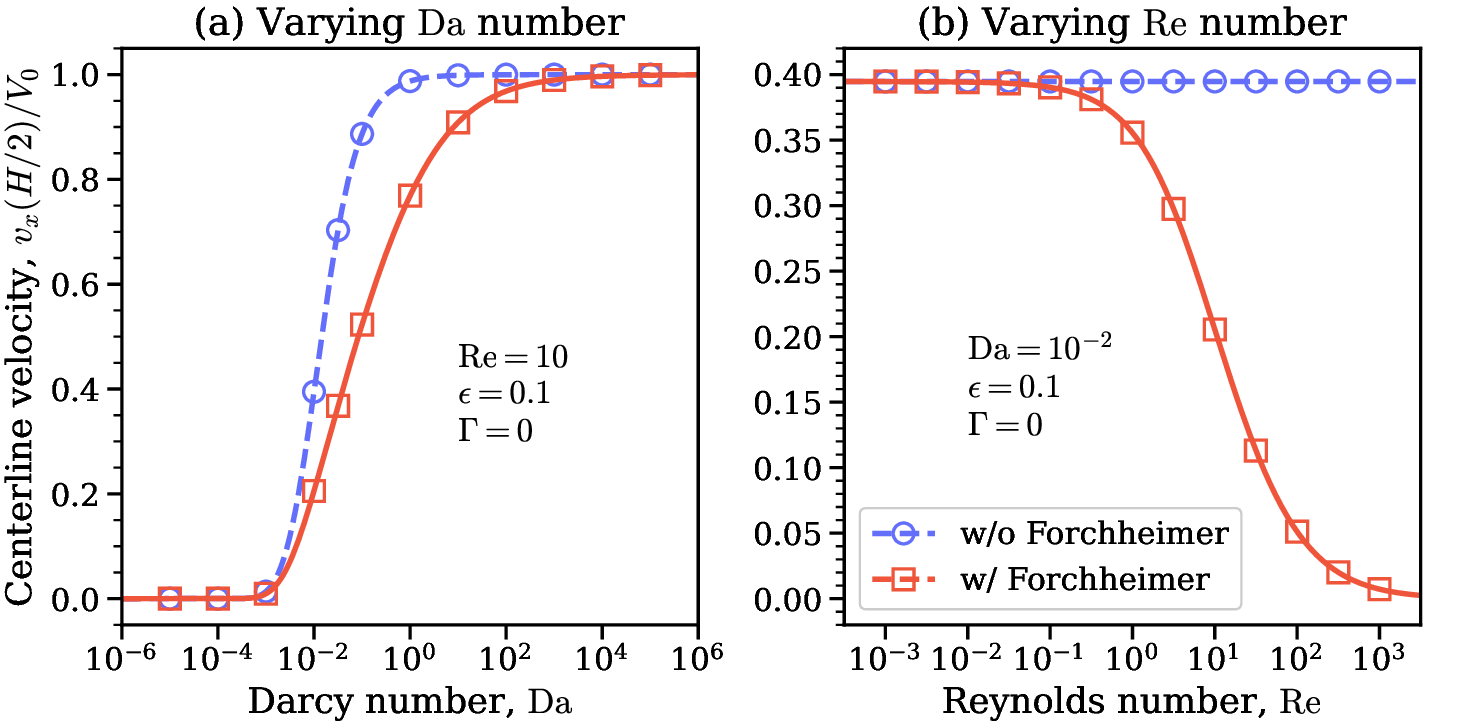}
\caption{Centerline peak superficial flow velocity at the midway ($y=H/2$) of the Couette channel with homogeneous porosity, as a function of (a) Darcy number and (b) Reynolds number, for the Darcy-Forchheimer and the Darcy-limit systems. The red solid lines show the reference numerical solutions of the BVP (\ref{Couette_v_ODE}, \ref{Couette_BC}) with $\mu_\mathrm{F}$ prescribed according to \eqref{mu_F_Ergun_perm}, while the blue dashed lines show the analytical solution \eqref{Couette_v_Darcy_analyt} for the case $\mu_\mathrm{F}=0$. Markers denote the corresponding LBM results: red squares for simulations with $\mu_\mathrm{F}$ set according to \eqref{mu_F_Ergun_perm}, and blue circles for $\mu_\mathrm{F}=0$.}\label{Couette_ReDaRange_Plots}
\end{figure}

Figure \ref{Couette_ReDaRange_Plots} shows the resulting centerline superficial velocity obtained from the LBM simulations and the corresponding reference solutions as a function of (a) the Darcy number and (b) the Reynolds number. Good agreement is observed between the LBM results and the reference solutions.

As shown in Figure \ref{Couette_ReDaRange_Plots}(a), at fixed Reynolds number, the effect of the nonlinear Forchheimer drag is negligible at very small $\mathrm{Da}$, becomes more pronounced as $\mathrm{Da}$ increases, and then decreases again, eventually becoming negligible in the large-$\mathrm{Da}$ limit. In realistic porous media, the Darcy number typically lies in the range $10^{-1}$ to $10^{-8}$, as follows from the permeability correlation \eqref{kappa_Ergun}. For relatively large values of $\mathrm{Da}$, e.g., $10^{-2}$ to $10^{-1}$, corresponding to highly permeable media, the nonlinear inertial resistance becomes important and must therefore be accounted for.

Although $\mathrm{Da}>1$ is not realistic for a bulk porous medium when using correlations such as \eqref{kappa_Ergun}, such values can arise locally within the diffuse transition layer between the porous medium and the free-fluid region in the mesoscopic one-domain continuous model proposed here. For this reason, we also examine the behavior of the system for $\mathrm{Da}>1$, which corresponds to the transition from porous-medium flow to the free-fluid limit. In the limiting case $\mathrm{Da}\rightarrow\infty$, for a fixed domain size, $\kappa\rightarrow\infty$, and the system approaches the free-fluid Couette-flow limit. Consistently, Figure \ref{Couette_ReDaRange_Plots}(a) shows that, as $\mathrm{Da}\rightarrow\infty$, the Darcy--Forchheimer and Darcy-limit solutions overlap. The centerline velocity also approaches one half of the top-wall velocity, as expected for classical free-fluid Couette flow.

The results in Figure \ref{Couette_ReDaRange_Plots}(b) show that, at fixed Darcy number, the effect of the nonlinear drag increases with $\mathrm{Re}$. In particular, for $\mathrm{Da}=10^{-2}$, the nonlinear drag contribution is negligible for $\mathrm{Re}<10^{-1}$. For $\mathrm{Re}>10^{-1}$, however, its influence becomes significant and grows rapidly with increasing $\mathrm{Re}$. This behavior can be understood from the momentum balance \eqref{Couette_v_ODE} and correlations (\ref{mu_D_Ergun_perm}, \ref{mu_F_Ergun_perm}, \ref{F_eps_Ergun}), which imply that the ratio between the nonlinear and linear drag contributions scales as (\cite{Guo_Porous_2002}):
\begin{equation} \label{non_dim_drag_ratio}
    \frac{V_0\mu_\mathrm{F}}{\mu_\mathrm{D}} = \frac{V_0F_\epsilon\sqrt{\kappa}}{\nu_\mathrm{f}} \sim \mathrm{Re}\sqrt{\mathrm{Da}} .
\end{equation}
Thus, when either the Reynolds number or the Darcy number is sufficiently small, the nonlinear drag contribution can be neglected and the Darcy-Forchheimer model reduces to the Darcy-limit system. Conversely, for sufficiently large $\mathrm{Re}$ or $\mathrm{Da}$, nonlinear inertial resistance becomes important and must be considered.

\subsection{Poiseuille flow}\label{numerical_results:Poiseuille}
Poiseuille flow problem is also considered in a two-dimensional channel, similar to the Couette-flow configuration described in Section \ref{numerical_results:Couette}. In this case, however, the flow is driven by a constant and uniform body acceleration $b$ applied in the streamwise direction, i.e., along the $x$-axis. No gravitational field is imposed, and both walls are stationary. We consider two configurations: a homogeneous porous medium in Section \ref{numerical_results:Poiseuille:Homogeneous}, and a heterogeneous porous medium in Section \ref{numerical_results:Poiseuille:Heterogeneous}.

At steady state, the flow is fully developed along the channel, and from the governing hydrodynamic equations (\ref{continuity_target}, \ref{momentum_target}, \ref{visc_stress_target}) it follows that the streamwise superficial velocity satisfies the following boundary-value problem:
\begin{equation}
    \frac{d^2v}{dy^2} + \frac{\epsilon b - \mu_\mathrm{D}v - \mu_\mathrm{F}\lvert v \rvert v}{\nu_\mathrm{eff}} = 0;\qquad v(0)=0,\quad v(H)=0. \label{Poiseuille_v_BVP}
\end{equation}
Unless otherwise noted, all LBM simulations use the same parameters and settings as in the Couette-flow problem in Section \ref{numerical_results:Couette}. This includes the computational domain, discrete lattice, initialization and boundary-condition schemes, and steady-state convergence criteria. As before, validation is performed against reference solutions obtained by solving the one-dimensional BVP \eqref{Poiseuille_v_BVP} with the standard SciPy BVP solver.

\subsubsection{Homogeneous porous channel}\label{numerical_results:Poiseuille:Homogeneous}
Here, we consider a channel filled with a homogeneous porous medium of porosity $\epsilon=0.1$. In the absence of nonlinear Forchheimer drag, i.e., for $\mu_\mathrm{F}=0$, the BVP \eqref{Poiseuille_v_BVP} admits the following closed-form Darcy-limit analytical solution for the superficial velocity profile:
\begin{equation}
    v_\mathrm{D}(y) = \frac{\epsilon b}{\mu_\mathrm{D}}\left\{1-\cosh{\left[r\left(y-\frac{H}{2}\right)\right]}\mathrm{sech}\,{\left(\frac{rH}{2}\right)}\right\}, \quad r=\sqrt{\frac{\mu_\mathrm{D}}{\nu_\mathrm{eff}}}. \label{Poiseuille_v_Darcy_analyt}
\end{equation}
We take the peak superficial velocity of this Darcy-limit solution, attained at the channel centerline, as the characteristic velocity, $V_0=v_\mathrm{D}(H/2)$. The Reynolds number is then defined as $\mathrm{Re}=V_0H/\nu_\mathrm{f}$, and the streamwise acceleration is prescribed accordingly as
\begin{equation}
    b = \frac{\mu_\mathrm{D}V_0}{\epsilon\left[1-\mathrm{sech}\,{\left(\frac{rH}{2}\right)}\right]}.
\end{equation}

\begin{figure}[h]
\centering
\includegraphics[width=0.85\textwidth]{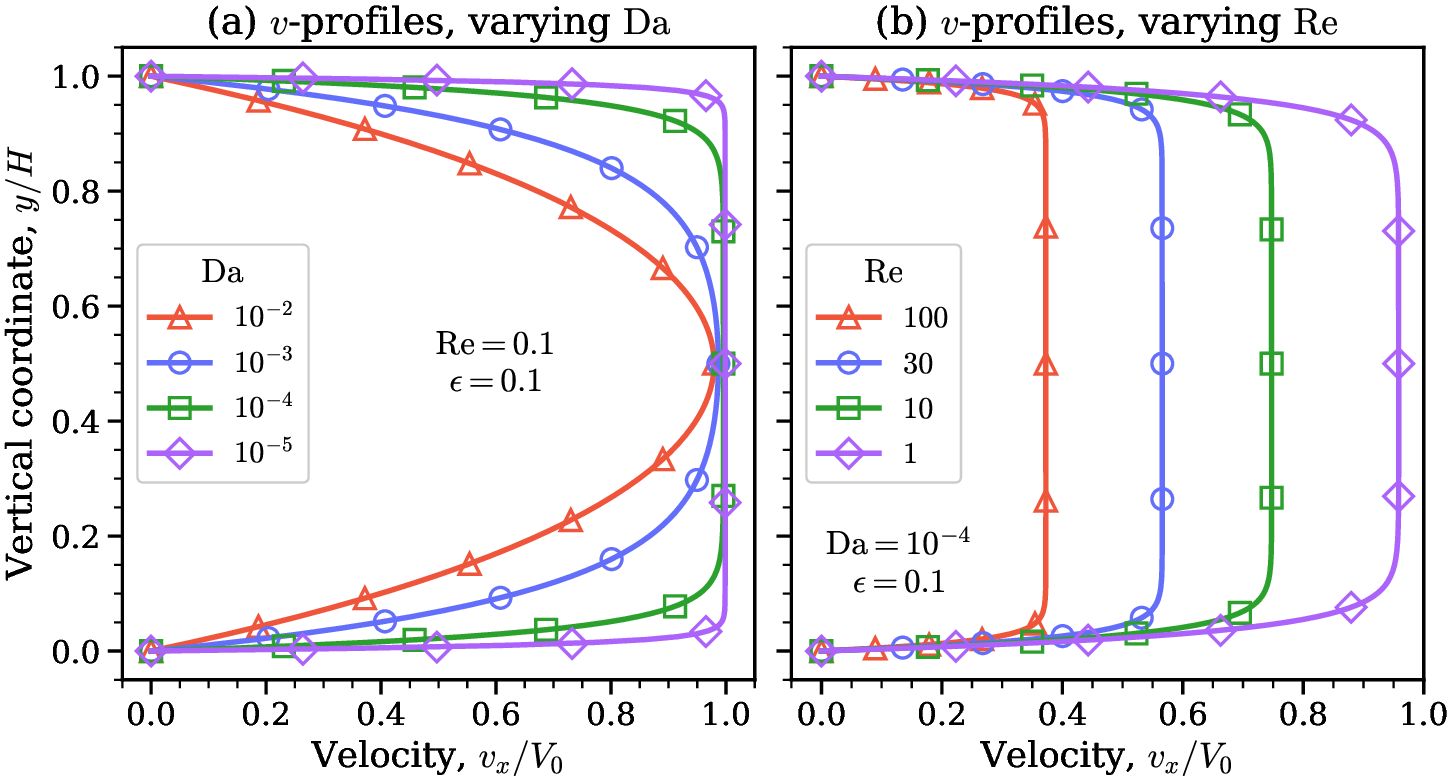}
\caption{Profiles of superficial flow velocity for Poiseuille flow in a homogeneous porous medium for different (a) Darcy numbers and (b) Reynolds numbers. Solid lines denote the LBM simulation results, while markers indicate the reference numerical solutions of the BVP \eqref{Poiseuille_v_BVP}.}\label{Poiseuille_FlowProfiles}
\end{figure}

We first examine the superficial velocity profiles for different values of $\mathrm{Da}$ and $\mathrm{Re}$. Figure \ref{Poiseuille_FlowProfiles} compares the results obtained with the proposed LBM model against the corresponding reference solutions. Excellent agreement is observed across the considered parameter range, supporting the validity of the present LBM model for heterogeneous porous media flows.

To assess the influence of the nonlinear Forchheimer drag, we measure the superficial velocity at the channel centerline for different values of $\mathrm{Re}$ and $\mathrm{Da}$. Two cases are considered: the Darcy--Forchheimer model, with $\mu_\mathrm{F}$ prescribed according to \eqref{mu_F_Ergun_perm}, and the Darcy-limit model, with $\mu_\mathrm{F}=0$.

\begin{figure}[h]
\centering
\includegraphics[width=0.9\textwidth]{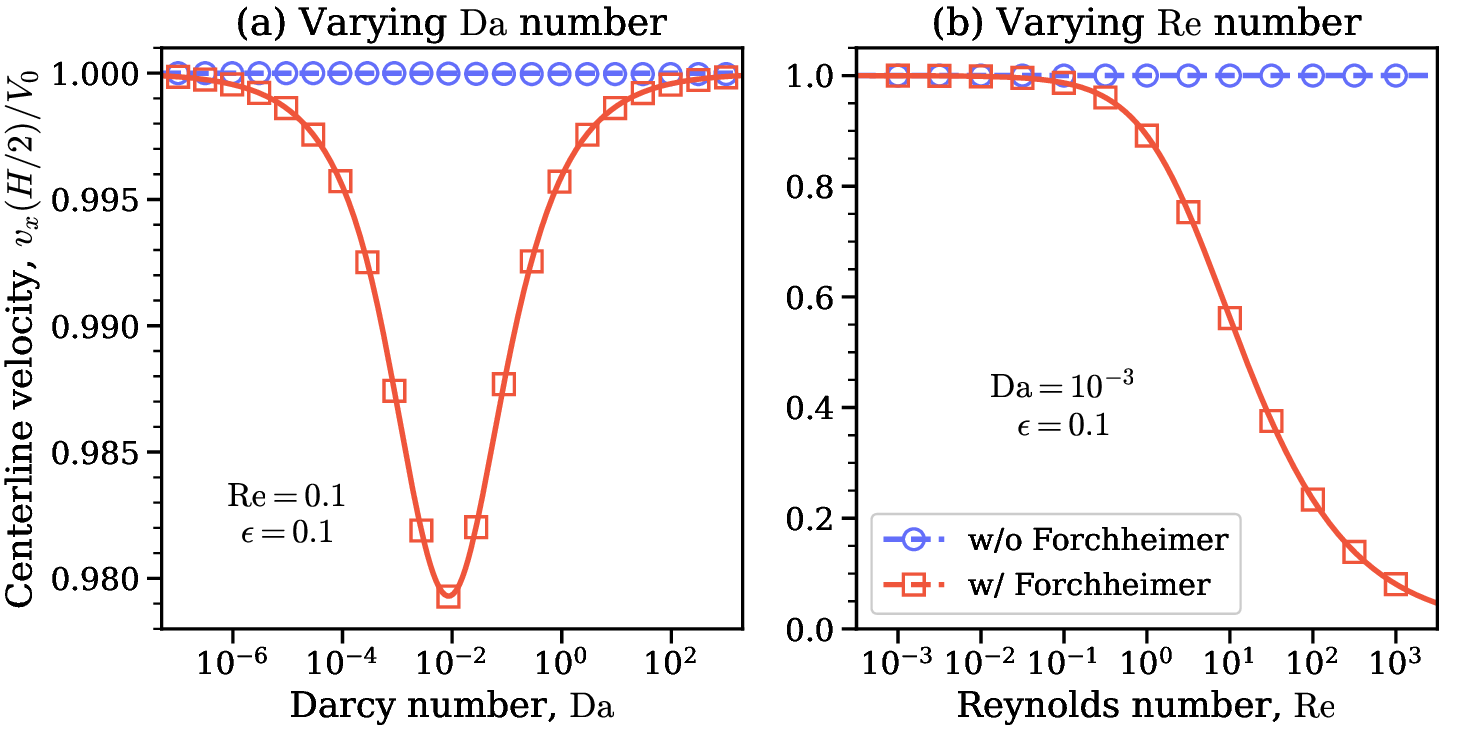}
\caption{Centerline peak superficial flow velocity at the midway ($y=H/2$) of the Poiseuille channel with homogeneous porosity, as a function of (a) Darcy number and (b) Reynolds number, for the Darcy-Forchheimer and the Darcy-limit systems. The red solid lines show the reference numerical solutions of the BVP \eqref{Poiseuille_v_BVP} with $\mu_\mathrm{F}$ prescribed according to \eqref{mu_F_Ergun_perm}, while the blue dashed lines show the analytical solution \eqref{Poiseuille_v_Darcy_analyt} for the case $\mu_\mathrm{F}=0$. Markers denote the corresponding LBM results: red squares for simulations with $\mu_\mathrm{F}$ set according to \eqref{mu_F_Ergun_perm}, and blue circles for $\mu_\mathrm{F}=0$.}\label{Poiseuille_ReDaRange_Plots}
\end{figure}

Figure \ref{Poiseuille_ReDaRange_Plots} shows the centerline velocity obtained from the LBM simulations and the corresponding reference solutions as a function of (a) the Darcy number and (b) the Reynolds number. Excellent agreement is observed between the LBM results and the reference solutions. For small $\mathrm{Re}$ and/or $\mathrm{Da}$, the Darcy--Forchheimer and Darcy-limit systems yield nearly identical results, indicating that the nonlinear drag contribution is negligible. As either $\mathrm{Re}$ or $\mathrm{Da}$ increases, however, the nonlinear drag increasingly suppresses the flow and can no longer be neglected, consistent with the scaling in \eqref{non_dim_drag_ratio}. In addition, for very large Darcy numbers, approaching the free-fluid Poiseuille-flow limit, the Darcy--Forchheimer and Darcy-limit solutions overlap again.

\subsubsection{Heterogeneous porous channel}\label{numerical_results:Poiseuille:Heterogeneous}
\begin{figure}[h]
\centering
\includegraphics[width=0.90\textwidth]{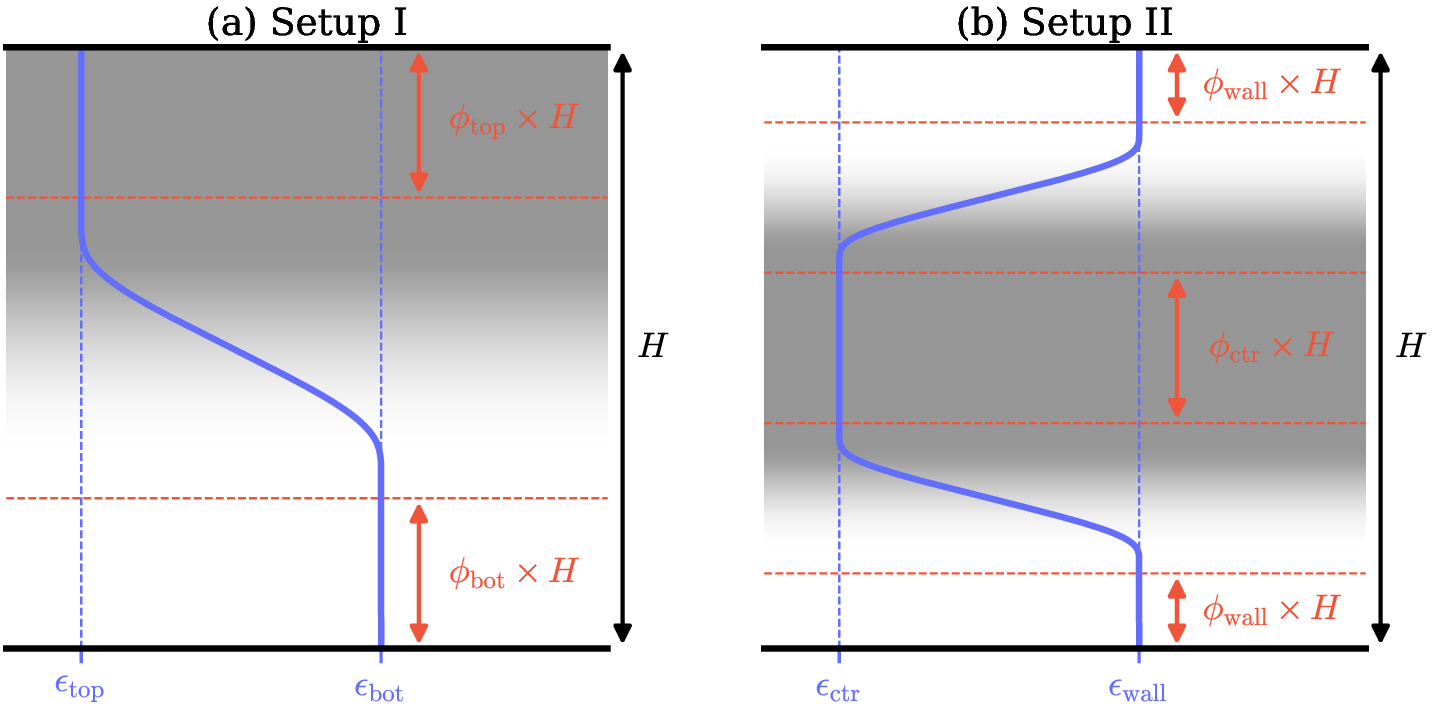}
\caption{Variable-porosity domain setups. (a) Setup I and (b) Setup II porosity profiles defined by \eqref{Hetero_Channel_Setup_I_Profile} and \eqref{Hetero_Channel_Setup_II_Profile}, respectively. The horizontal axis represents the porosity $\epsilon(y)$, while the vertical axis corresponds to the wall-normal coordinate $y$. Blue solid lines, together with the background colormaps, indicates the porosity profiles. Vertical dashed blue lines mark the plateau porosity levels, and horizontal dashed red lines indicate the locations of the matching points of the porosity profiles.}\label{hetero_channel_domain}
\end{figure}

We now consider Poiseuille flow in a channel filled with a heterogeneous porous medium characterized by a smooth spatially varying porosity profile. As discussed in Section \ref{sec:target_macro}, such a smooth porosity field is not merely an arbitrary fitting choice; it can arise naturally from averaging a sharp microscopic pore structure over a finite control volume. We consider two variable-porosity configurations, referred to as Setup I and Setup II, which are visualized in Figure \ref{hetero_channel_domain}.

In Setup I, a wall-normal porosity distribution $\epsilon(y)$ is prescribed as a two-plateau profile connected by a smooth transition layer, as illustrated in Figure \ref{hetero_channel_domain}(a). Specifically, the porosity is set to $\epsilon_\mathrm{bot}$ in the lower part of the channel, $y\le \phi_\mathrm{bot} H$, and to $\epsilon_\mathrm{top}$ in the upper part, $y\ge (1-\phi_\mathrm{top})H$. In the intermediate region, $\phi_\mathrm{bot}<(y/H)<(1-\phi_\mathrm{top})$, the profile is defined using a $C^\infty$ smoothstep function,
\begin{equation} \label{Hetero_Channel_Setup_I_Profile}
    \epsilon(y) = \epsilon_\mathrm{bot} + (\epsilon_\mathrm{top}-\epsilon_\mathrm{bot})\mathcal{S}\left(\frac{(y/H)-\phi_\mathrm{bot}}{1-\phi_\mathrm{top}-\phi_\mathrm{bot}}\right),
\end{equation}
with
\begin{equation} \label{s_cinf}
    \mathcal{S}(\psi)=
    \begin{cases}
        0, & \psi\leq 0, \\
        \frac{\exp{(-1/\psi)}}{\exp{(-1/\psi)}+\exp{[-1/(1-\psi)]}}, & 0<\psi<1, \\
        1, & \psi\geq 1.
    \end{cases}
\end{equation}
This construction yields a monotonic transition between the two constant porosity levels, with all derivatives vanishing at the endpoints of the transition region. %The analytical derivatives of the resulting porosity profile \eqref{Hetero_Channel_Setup_I_Profile} can then be used in simulations to evaluate the variable-porosity correction \eqref{porosity_correction}, as detailed in Appendix \ref{Append:visc_stress_correction}.

In Setup II, the wall-normal porosity distribution $\epsilon(y)$ is prescribed as a centerline-symmetric profile with a central plateau and two near-wall plateau regions, as illustrated in Figure \ref{hetero_channel_domain}(b). Specifically, the porosity is set to $\epsilon_\mathrm{ctr}$ in the central part of the channel, $\lvert (y/H)-1/2\rvert \le \phi_\mathrm{ctr}/2$, and to $\epsilon_\mathrm{wall}$ in the near-wall regions, $\lvert (y/H)-1/2\rvert \ge 1/2-\phi_\mathrm{wall}$. In the intermediate regions, the profile is defined using \eqref{s_cinf} as
\begin{equation} \label{Hetero_Channel_Setup_II_Profile}
    \epsilon(y)=\epsilon_\mathrm{ctr}+(\epsilon_\mathrm{wall}-\epsilon_\mathrm{ctr})\mathcal{S}\left(\frac{\lvert (y/H)-1/2\rvert-\phi_\mathrm{ctr}/2}{1/2-\phi_\mathrm{wall}-\phi_\mathrm{ctr}/2}\right).
\end{equation}

We consider constant effective viscosities and set the effective-viscosity ratios defined in \eqref{ratios_def} to unity. Under steady-state conditions, the flow should therefore still satisfy the target BVP \eqref{Poiseuille_v_BVP}, whose governing ODE is based on the target viscous stress \eqref{visc_stress_target} written in terms of the superficial velocity.

As discussed in Section \ref{sec:lbe}, if the variable-porosity correction \eqref{porosity_correction} is omitted, the macroscopic momentum balance \eqref{momentum_LBM} recovered by the LBM model contains a spurious pressure-gradient contribution in the presence of porosity gradients. In addition, the recovered viscous stress \eqref{visc_stress_LBM} depends on gradients of the intrinsic velocity, in contrast to the target viscous stress \eqref{visc_stress_target}, which depends on gradients of the superficial velocity. Omitting the pressure-related part of the correction \eqref{porosity_correction} therefore does not yield physically meaningful hydrodynamic results. Nevertheless, it is instructive to isolate the effect of the viscous-stress part of the correction by considering the model without it. This allows us to assess its influence on the velocity profiles, validate the implementation algorithm (\ref{v_(0)}--\ref{v_final}), and highlight the importance of including this correction.

From (\ref{continuity_target}, \ref{momentum_target}, \ref{visc_stress_LBM}), it follows that, for fully developed channel flow, the streamwise intrinsic velocity satisfies the following BVP:
\begin{equation} \label{Poiseuille_u_BVP_wrong_visc_stress}
    \frac{d^2 u}{dy^2} + \frac{1}{\epsilon}\left(\frac{d\epsilon}{d y}\right)\frac{du}{dy} + \frac{b - \mu_\mathrm{D}u - \epsilon\mu_\mathrm{F}\lvert u \rvert u}{\nu_\mathrm{eff}} = 0;\qquad u(0)=0,\quad u(H)=0.
\end{equation}
Here, the viscous stress is written in terms of gradients of the intrinsic velocity, consistent with \eqref{visc_stress_LBM}. In the special case of homogeneous porosity, the BVPs \eqref{Poiseuille_v_BVP} and \eqref{Poiseuille_u_BVP_wrong_visc_stress} become essentially equivalent.

For the heterogeneous porosity profiles considered here, however, the drag coefficients $\mu_\mathrm{D}$ and $\mu_\mathrm{F}$ also vary spatially. As a result, analytical solutions of the BVPs \eqref{Poiseuille_v_BVP} and \eqref{Poiseuille_u_BVP_wrong_visc_stress} are difficult to obtain, even in the Darcy limit. Therefore, all reference solutions in this section are obtained numerically using the BVP solver described in Section \ref{numerical_results:Couette}.

To parameterize the flow, we introduce the Hagen number and define a reference velocity as
\begin{equation} \label{Hg_V0_def}
    \mathrm{Hg} = \frac{bH^3}{\nu_\mathrm{f}^2},\quad V_0=\frac{bH^2}{8\nu_\mathrm{f}} = \frac{\nu_\mathrm{f}\mathrm{Hg}}{8H}.
\end{equation}
Here, $V_0$ corresponds to the peak velocity of a free-fluid Poiseuille flow. Consequently, when scaled by $V_0$, the free-fluid velocity profile is independent of $\mathrm{Hg}$.

We first examine how the velocity profile is affected by expressing the viscous stress in terms of either the superficial or intrinsic velocity. To focus specifically on porosity-gradient effects, we consider a synthetic case with spatially varying porosity profile \eqref{Hetero_Channel_Setup_I_Profile}, corresponding to Setup I, but impose a fixed permeability: instead of using the correlation \eqref{kappa_Ergun}, we prescribe a constant Darcy number $\mathrm{Da}=\kappa/H^2$ throughout the domain. In all simulations, we set $\nu_\mathrm{f}=0.1$, $\phi_\mathrm{top}=0.1$, and $\phi_\mathrm{bot}=0.1$.

\begin{figure}[h]
\centering
\includegraphics[width=0.95\textwidth]{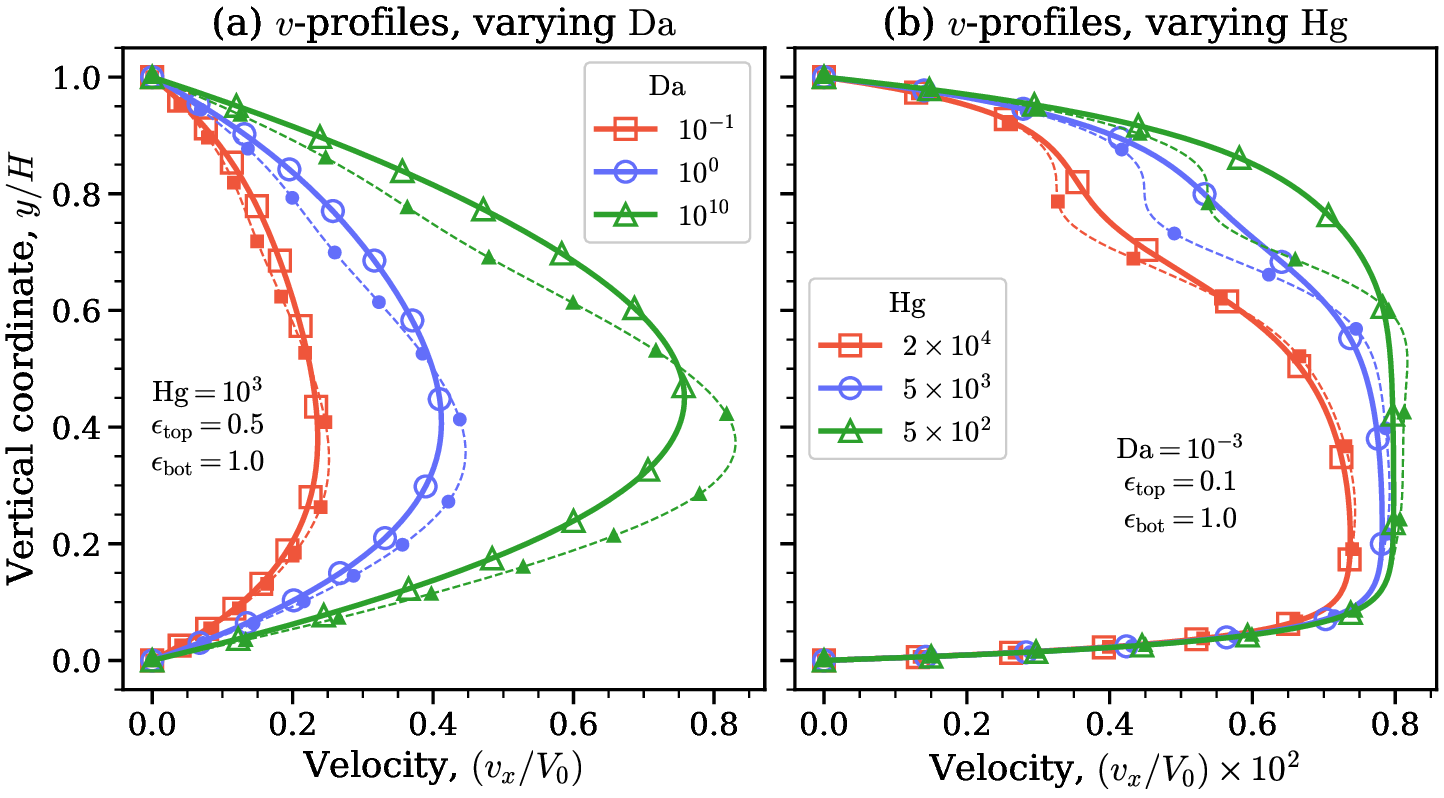}
\caption{Profiles of superficial flow velocity for Poiseuille flow in a heterogeneous porous medium with fixed permeability for different (a) Darcy numbers and (b) Hagen numbers. The porosity profile is given by \eqref{Hetero_Channel_Setup_I_Profile}, corresponding to Setup I. Solid lines show the LBM simulation results obtained with the full variable-porosity correction \eqref{porosity_correction}, while unfilled markers indicate the corresponding reference numerical solutions of the BVP \eqref{Poiseuille_v_BVP}. Dashed lines show the LBM results obtained without the viscous-stress correction part of \eqref{porosity_correction}, while small filled markers indicate the corresponding reference numerical solutions of the BVP \eqref{Poiseuille_u_BVP_wrong_visc_stress}.}\label{Poiseuille_Hetero_Fixed_Drag}
\end{figure}

Figure \ref{Poiseuille_Hetero_Fixed_Drag} compares superficial velocity profiles obtained with the proposed LBM model using the full variable-porosity correction \eqref{porosity_correction} and with the viscous-stress part of this correction omitted. The corresponding reference solutions of \eqref{Poiseuille_v_BVP} and \eqref{Poiseuille_u_BVP_wrong_visc_stress} are also plotted. Figure \ref{Poiseuille_Hetero_Fixed_Drag}(a) shows profiles for different values of $\mathrm{Da}$ at $\mathrm{Hg}=1000$, with $\epsilon_\mathrm{top}=0.5$ and $\epsilon_\mathrm{bot}=1$. The Darcy numbers considered are intentionally large, corresponding to very high permeability. Although such values, together with the assumption of fixed permeability, are not physically realistic, they allow the effect of porosity variability on momentum-flux transfer to be isolated without additional damping variations caused by drag.

The resulting profiles retain the general parabolic character of Poiseuille flow, as for the homogeneous-porosity case shown in Figure \ref{Poiseuille_FlowProfiles}, but the peak velocity is shifted away from the channel centerline toward the high-porosity bottom region. The profiles obtained with the target viscous stress \eqref{visc_stress_target}, written in terms of the superficial velocity, differ clearly from those obtained with the intrinsic-velocity viscous stress \eqref{visc_stress_LBM}. This demonstrates the influence of porosity gradients on viscous momentum transfer.

Figure \ref{Poiseuille_Hetero_Fixed_Drag}(b) further illustrates the effect for different values of $\mathrm{Hg}$ at $\mathrm{Da}=10^{-3}$, with $\epsilon_\mathrm{top}=0.5$ and $\epsilon_\mathrm{bot}=1$. The deviation between the two viscous-stress formulations is most pronounced in the upper, lower-porosity region, highlighting the importance of the correction \eqref{porosity_correction}. In both cases, the LBM results agree well with the corresponding reference solutions across the considered parameter range. This confirms that the base LBM model recovers the viscous stress \eqref{visc_stress_LBM} in terms of gradients of the intrinsic velocity, in accordance with the multiscale analysis in Appendix \ref{Append:CE_analysis}, and therefore requires the correction \eqref{porosity_correction} to recover the target superficial-velocity stress \eqref{visc_stress_target}. The agreement also validates the implicit implementation algorithm (\ref{v_(0)}--\ref{v_final}) used to include this correction.

\begin{figure}[ht!]
\centering
\includegraphics[width=0.85\textwidth]{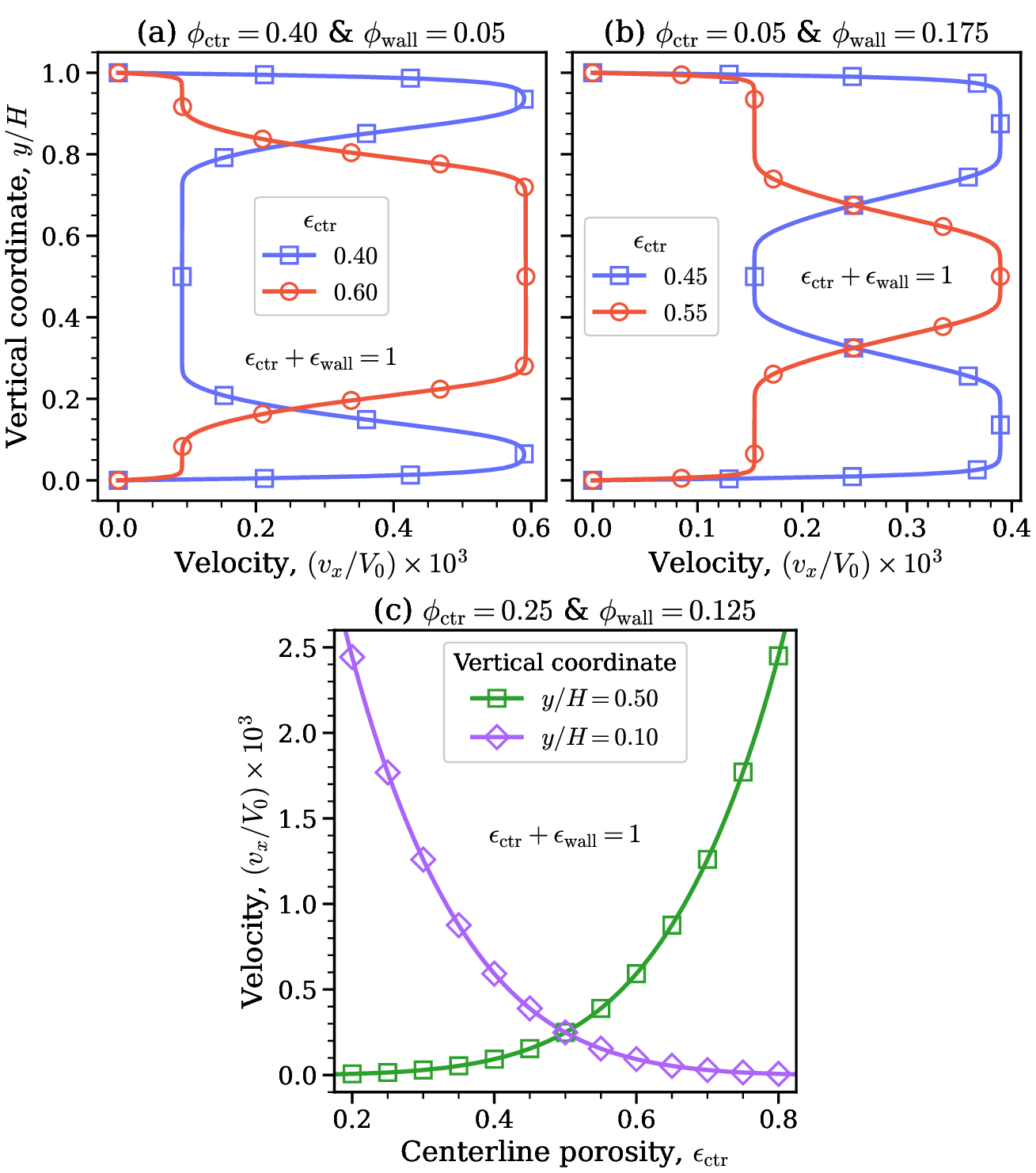}
\caption{Results for Poiseuille flow in a heterogeneous porous medium with the Setup II porosity profile defined by \eqref{Hetero_Channel_Setup_II_Profile}. Panels (a) and (b) show superficial velocity profiles for different values of $\epsilon_\mathrm{ctr}$ and different choices of $\phi_\mathrm{ctr}$ and $\phi_\mathrm{wall}$. Panel (c) shows the superficial velocity at $y=H/10$ and $y=H/2$ as a function of $\phi_\mathrm{ctr}$. In panels (a,b), solid lines denote the LBM results obtained with the full variable-porosity correction \eqref{porosity_correction}, while unfilled markers indicate the corresponding numerical reference solutions of the BVP \eqref{Poiseuille_v_BVP}. In panel (c), solid lines denote the reference solutions, and unfilled markers denote the LBM results.}\label{hetero_channel_FlowPlots}
\end{figure}

Next, we simulate channel flow with a symmetric wall-normal porosity variation, using the profile \eqref{Hetero_Channel_Setup_II_Profile} corresponding to Setup II. Unlike in the previous case, the permeability is now spatially varying and is determined by the local porosity through the correlation \eqref{kappa_Ergun}, with $d_\mathrm{p}=H/10$. The simulations are performed using the full variable-porosity correction \eqref{porosity_correction}, implemented through (\ref{v_(0)}--\ref{v_final}). The LBM simulation results, obtained for different combinations of $\phi_\mathrm{ctr}$ and $\phi_\mathrm{wall}$ and different values of $\epsilon_\mathrm{ctr}$, are compared with the reference solutions in Figure \ref{hetero_channel_FlowPlots}. In all cases, we set $\mathrm{Hg}=10^{6}$, and the porosity levels at the channel center and near the walls satisfy $\epsilon_\mathrm{wall}+\epsilon_\mathrm{ctr}=1$.

Panels (a) and (b) of Figure \ref{hetero_channel_FlowPlots} show the superficial velocity profiles. The results exhibit pronounced flow channeling, a characteristic consequence of porosity heterogeneity, whereby the fluid preferentially passes through regions of higher porosity while bypassing more densely packed regions. Accordingly, the velocity peaks in the channel center when $\epsilon_\mathrm{ctr}>\epsilon_\mathrm{wall}$, and near the walls when $\epsilon_\mathrm{wall}>\epsilon_\mathrm{ctr}$. The profiles also show that increasing $\phi_\mathrm{wall}$ while decreasing $\phi_\mathrm{ctr}$ widens the near-wall velocity-plateaus and narrows the central plateau, as expected. We note, however, that the central velocity-plateau generally remains wider than the wall velocity plateaus for the same porosity-plateau width, due to the larger viscous shear stress near the stationary walls compared with the porosity-transition regions. In addition, a larger difference between $\epsilon_\mathrm{wall}$ and $\epsilon_\mathrm{ctr}$ leads to a stronger velocity contrast between the two plateau regions, and therefore to more pronounced flow channeling.

The relationship between the velocity magnitude and the plateau porosity is very similar in the wall and central regions: for the same plateau porosity, the corresponding velocity values are nearly identical. This is further illustrated in Figure \ref{hetero_channel_FlowPlots}(c), which shows the velocity at $y=H/2$ and $y=H/10$, corresponding respectively to the centers of the central and near-wall velocity-plateaus, as a function of $\phi_\mathrm{ctr}$. The curves display nonlinear, approximately exponential behavior and are nearly symmetric about $\phi_\mathrm{ctr}=0.5$, where they overlap, as expected. These results show that even modest porosity variations can produce substantial velocity non-uniformity across the channel, corresponding to strong flow channeling.

The LBM results are in excellent agreement with the reference solutions, confirming the validity of the proposed model for simulating the volume-averaged dynamics of the fluid phase in heterogeneous porous structures governed by the target macroscopic equations (\ref{continuity_target}, \ref{momentum_target}, \ref{visc_stress_target}). Such nonuniform radial porosity distributions are common in practical industrial systems like packed beds, where they strongly influence equipment performance (\cite{Roblee_Baird_Tierney_1958,White_Tien_1987,duToit_2008}). The present model can therefore be used to support early stage process equipment design by quantifying how porosity variations affect the velocity distribution across the channel.

\section{Numerical studies}\label{sec:Numerical_studies}
To further demonstrate the ability of the proposed LBM model to capture transport between free-fluid and porous-medium regions within a one-domain framework, we consider two illustrative flow problems: confined flow past a permeable cylinder and lid-driven cavity flow with a porous obstacle. In both cases, we examine how the flow dynamics are governed by the obstacle porosity and the Reynolds number, which control the permeability of the porous region and the relative importance of inertial and viscous effects. The resulting flow structures are analyzed qualitatively and compared with available numerical or experimental studies of relevant systems.

\subsection{Confined flow past a permeable cylinder} \label{numerical_results:Poiseuille:Cylinder}
Here we consider two-dimensional channel flow of a free-fluid past a porous cylinder with a circular cross-section. The porous cylinder is represented by a smooth circular, radially symmetric inclusion of reduced porosity embedded in a homogeneous free-fluid background. Let $(x_\mathrm{c},y_\mathrm{c})$ denote the center of the porosity inhomogeneity and introduce a normalized circular radius function
\begin{equation} \label{s_cylinder_circle}
    s_\mathrm{cyl}(x,y) = \left(\frac{2(x-x_\mathrm{c})}{D}\right)^2 + \left(\frac{2(y-y_\mathrm{c})}{D}\right)^2,
\end{equation}
where $D$ is the characteristic diameter of the porous inhomogeneity. The porosity field $\epsilon(x,y)$ is then defined as
\begin{equation} \label{Porous_Cylinder_Profile}
    \epsilon(x,y)=1-(1-\epsilon_\mathrm{core})\mathcal{B}_1\left(s_\mathrm{cyl}\right),
\end{equation}
where $\mathcal{B}$ is a localized generalized bump function defined as
\begin{equation} \label{b_cinf}
    \mathcal{B}_m(s)=
    \begin{cases}
        \exp{\left(1-\frac{1}{1-s^m}\right)}, & s<1,\\
        0, & s\geq 1,
    \end{cases}\quad m\in\mathbb{Z}^+.
\end{equation}
With the construction (\ref{s_cylinder_circle}-\ref{b_cinf}), the porosity equals $\epsilon_\mathrm{core}$ at the center of the inclusion, where $s=0$, and smoothly increases toward the outer value $\epsilon=1$ as $s\to 1$. For $s\ge 1$, the porosity is constant, and the profile joins the exterior region smoothly at $s=1$, with all derivatives vanishing at the interface.

The power $m$ in \eqref{b_cinf} controls the radial distribution inside the inclusion. Larger values of $m$ flatten the profile near the center and shift the main porosity variation closer to the boundary, thereby producing a broader, more plateau-like core. Here we consider only $m=1$, corresponding to a radially symmetric cylinder with continuously varying permeability. Higher values of $m$ may instead be used to approximate a homogeneous porous cylinder surrounded by a diffuse transition layer.

Based on \eqref{Porous_Cylinder_Profile}, we define $D_{50}\approx 0.64D$ as the diameter of the midpoint porosity isocontour, $\epsilon_{50}=(1+\epsilon_\mathrm{core})/2$. This quantity provides a convenient effective diameter of the porous cylinder.

\begin{figure}[h]
\centering
\includegraphics[width=1.0\textwidth]{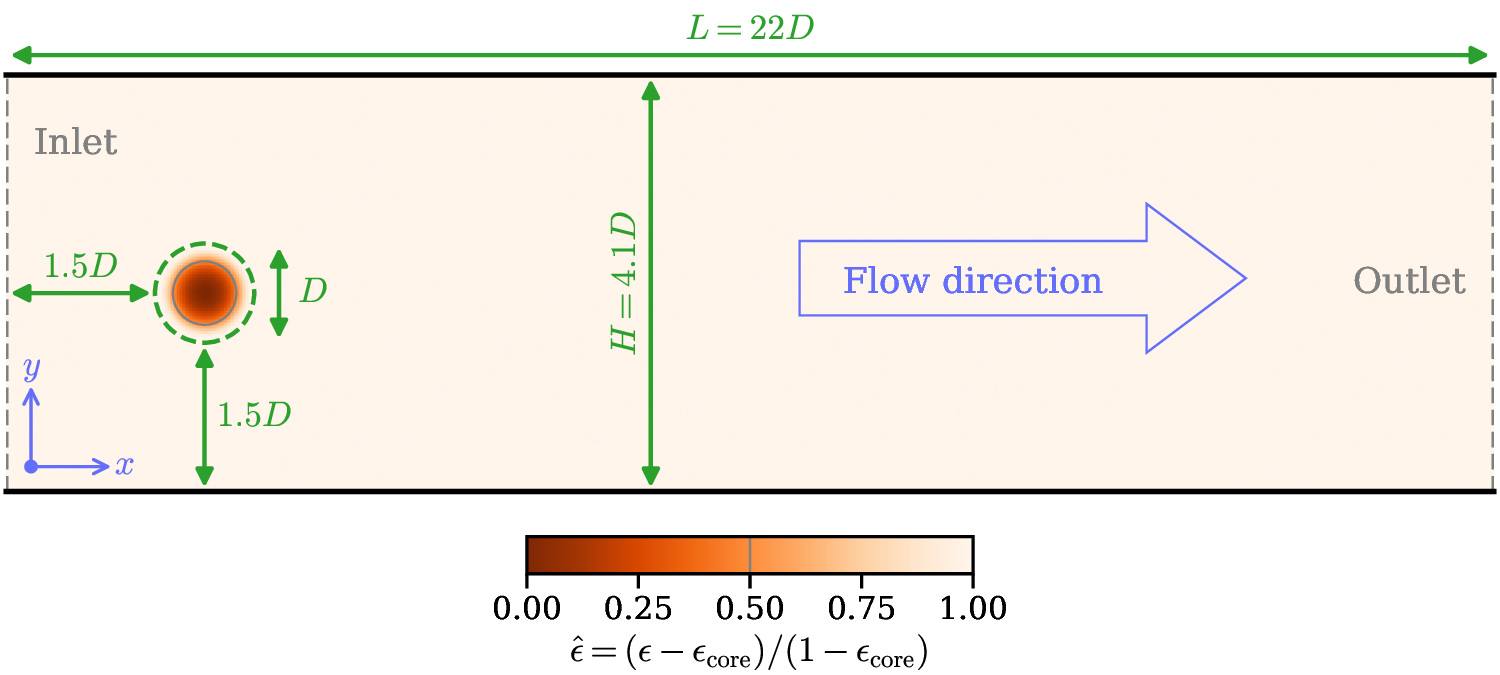}
\caption{Domain geometry for the channel-flow problem with a cylindrical porous obstacle. The black solid horizontal lines denote the top and bottom solid walls of the channel, while the grey dashed lines mark the left and right domain boundaries, corresponding to the inlet and outlet, respectively. The background colormap shows the normalized porosity field $\hat{\epsilon}(x,y)$ defined by \eqref{Porous_Cylinder_Profile}. The green dashed circle of diameter $D$ marks the matching point of the porosity profile, where its derivatives vanish. The grey solid circle of diameter $D_\mathrm{50}\approx0.64D$ denotes the $\hat{\epsilon}=0.5$ porosity isocontour corresponding to the midpoint of the transition.}\label{circular_cylinder_domain}
\end{figure}

The geometry and boundary conditions are chosen analogously to the canonical confined channel-flow configuration with a cylindrical obstacle described in \cite{Schaefer1996}. The domain geometry, together with the porosity profile defined by \eqref{Porous_Cylinder_Profile}, is illustrated in Figure \ref{circular_cylinder_domain}. The configuration consists of a channel of width $H=4.1D$ and length $L=22D$. The porous cylinder is placed near the inlet, slightly offset from the channel centerline, with $(x_\mathrm{c},y_\mathrm{c})=(2D,2D)$.

At the left boundary of the domain (inlet), a velocity boundary condition is imposed using a Poiseuille profile with average flow velocity $V_0$,
\begin{equation}
    v_x(0,y)=\frac{6V_0}{H^2}y(H-y),\quad v_y(0,y)=0, \label{cylinder_inlet_BC}
\end{equation}
and the corresponding Reynolds number is defined as $ \mathrm{Re} = V_0D/\nu_\mathrm{f}$. At the top and bottom stationary walls of the channel, no-slip boundary conditions are prescribed,
\begin{equation}
    \bm{v}(x,0)=(0,0),\quad \bm{v}(x,H)=(0,0). \label{cylinder_wall_BC}
\end{equation}
At both the inlet and wall boundaries, the boundary density is evaluated from the mass continuity using \eqref{bc_density_impermWall}, with the density at the corner nodes obtained by extrapolation from neighboring nodes. The corresponding velocity boundary conditions (\ref{cylinder_inlet_BC}, \ref{cylinder_wall_BC}) are then imposed using the non-equilibrium extrapolation scheme \eqref{NEEM_bc_f_bar}. At the right boundary of the domain (outlet), a zero-gradient outflow boundary condition is applied. Accordingly, the unknown post-advection populations are prescribed as
\begin{equation}
    \bar{f}_i(L,y,t) = \bar{f}_i(L-\delta x,y,t),\quad \forall\ i\in P_- .\label{f_bar_outflow_BC}
\end{equation}
The populations are initialized at equilibrium using \eqref{eq_bc_f_bar}, with the velocity field initially set to zero and the density prescribed uniformly as unity throughout the domain. Each simulation is then advanced until the flow is fully developed. For steady flow modes, convergence is assessed by requiring the instantaneous velocity field to become time independent, using the steady-state criterion \eqref{SS_convergence_L2} based on the relative difference between consecutive time steps. For unsteady flow modes, characterized by wake oscillations or vortex shedding, the flow remains intrinsically time dependent. In this case, the flow is considered fully developed once the wake reaches a periodic or statistically stationary regime.

\begin{figure}[h]
\centering
\includegraphics[width=1.0\textwidth]{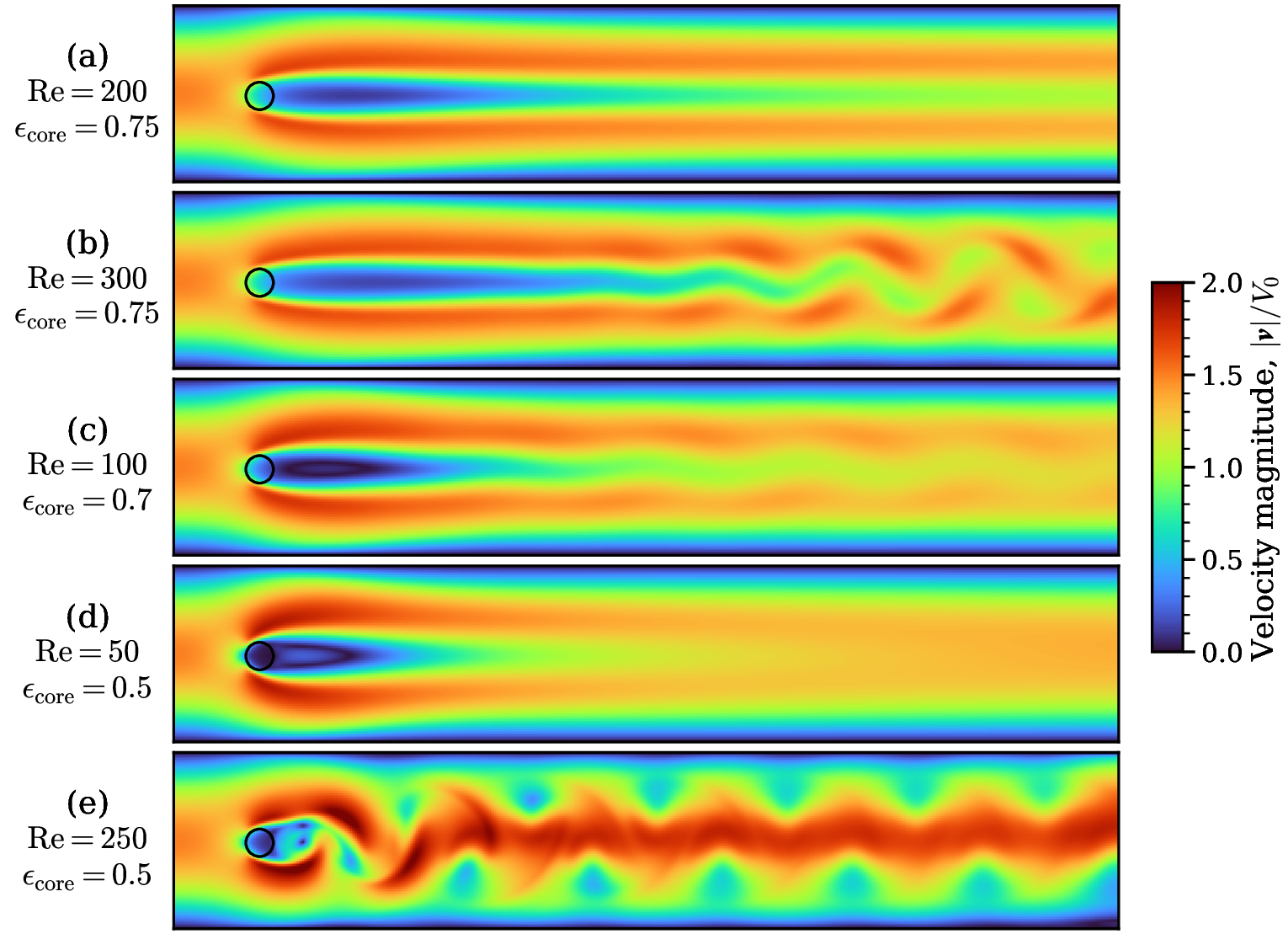}
\caption{Colormap plots of instantaneous superficial velocity magnitude for fully developed channel flow past a porous cylinder, shown for different values of Reynolds number $\mathrm{Re}$ and core porosity $\epsilon_\mathrm{core}$. The horizontal axis spans $x\in[0,L]$, and the vertical axis spans $y\in[0,H]$. The solid black circle of diameter $D_\mathrm{50}\approx0.64D$ denotes the $\epsilon_{50}$ porosity isocontour, corresponding to the midpoint of the porosity transition.}\label{circular_cylinder_vmag}
 \end{figure}

For a porous cylinder, the dominant unsteadiness may occur primarily in the downstream wake, without a clearly defined vortex-shedding signal associated with an alternating force on the cylinder. Therefore, rather than relying on lift or drag signals, as is common for a solid cylinder, we monitor the downstream oscillatory motion of the wake directly. To this end, we define the transverse kinetic energy in the wake as
\begin{equation} \label{ke_wake}
    K(t)=\frac{1}{2}\int_{\Omega_\mathrm{W}}\rho(\bm{x},t)v_y^2(\bm{x},t)\,d\Omega,
\end{equation}
where $\Omega_\mathrm{W}$ denotes the downstream wake region,
\begin{equation} \label{wake_region}
    \Omega_\mathrm{W} = \left\{(x,y):\,6D\leq x\leq 18D,\,0\leq y \leq H\right\}.
\end{equation}
The quantity \eqref{ke_wake} captures transverse wake oscillations independently of whether a classical vortex-shedding pattern is formed. Since the wake signal may contain multiple harmonics or mode interactions, convergence to a fully developed regime is assessed using window-averaged statistics rather than the spacing between individual peaks alone. Specifically, we compute the window average and the root-mean-square (RMS) fluctuation of $K(t)$ over consecutive non-overlapping time windows of length $\Delta T$, as
\begin{gather}
    K_\mathrm{avg}^n = \frac{1}{\Delta T}\int_{n\Delta T}^{(n+1)\Delta T}K(t)\,dt, \label{ke_ma} \\
    K_\mathrm{rms}^n = \left[\frac{1}{\Delta T}\int_{n\Delta T}^{(n+1)\Delta T}\left(K(t)-K_\mathrm{avg}^n\right)^2\,dt\right]^{1/2}. \label{ke_rms}
\end{gather}
Here $\Delta T=12D/V_0$ is the averaging time interval, corresponding to the convective time across the selected wake-monitoring region \eqref{wake_region}. The flow is considered fully developed once both the mean value \eqref{ke_ma} and the RMS fluctuation \eqref{ke_rms} vary by less than the prescribed tolerance between consecutive windows,
\begin{equation}
    \frac{\lvert K_\mathrm{avg}^{n} - K_\mathrm{avg}^{n-1}\rvert}{\lvert K_\mathrm{avg}^{n-1}\rvert} \leq 10^{-5} \quad \text{and} \quad \frac{\lvert K_\mathrm{rms}^{n} - K_\mathrm{rms}^{n-1}\rvert}{K_\mathrm{rms}^{n-1}} \leq 10^{-5}.
\end{equation}
Using numerical simulations with the proposed LBM model, we examine the development of different downstream flow patterns behind the porous cylinder. In all simulations, the temperature and effective-viscosity ratios \eqref{ratios_def} are set to unity, and the average inlet velocity is prescribed as $V_0=0.05$. The geometry is fixed by setting $D=40$, while $d_\mathrm{p}=D/5$ is used in the permeability correlation \eqref{kappa_Ergun}. The core porosity $\epsilon_\mathrm{core}$ and Reynolds number $\mathrm{Re}$ control the permeability and the relative importance of inertial and viscous effects, and therefore strongly influence the resulting wake structure. Figure \ref{circular_cylinder_vmag} shows snapshots of the instantaneous superficial-velocity magnitude in the fully developed regime for different combinations of $\mathrm{Re}$ and $\epsilon_\mathrm{core}$, illustrating the distinct flow modes observed in the simulations.

Based on the downstream flow patterns that emerge behind the porous cylinder, we identify five distinct flow modes. Mode I, shown in Figure \ref{circular_cylinder_vmag}(a), occurs at high cylinder porosity, where the fluid can pass readily through the cylinder. This permeability stabilizes the wake even at moderate $\mathrm{Re}$, producing a steady, unseparated flow regime with no downstream recirculation zones or oscillatory disturbances. Mode II, shown in Figure \ref{circular_cylinder_vmag}(b), also corresponds to a highly permeable cylinder; however, the larger $\mathrm{Re}$ leads to unsteady downstream wake oscillations, although no coherent vortices are formed behind the cylinder.

Mode III, illustrated in Figure \ref{circular_cylinder_vmag}(c), arises when the lower $\epsilon_\mathrm{core}$ restricts the flow through the porous cylinder. In combination with moderate $\mathrm{Re}$, this produces a detached recirculation region behind the cylinder, consisting of an asymmetric pair of counter-rotating vortices. This recirculation structure remains bounded but undergoes stable oscillations, giving rise to an oscillatory downstream wake pattern. Mode IV, corresponding to Figure \ref{circular_cylinder_vmag}(d), is observed at moderate to low $\epsilon_\mathrm{core}$ and low $\mathrm{Re}$. In this regime, a stationary recirculation zone forms behind the cylinder, containing a steady pair of coupled F{\"o}ppl vortices and producing a stable downstream wake. Finally, Mode V, shown in Figure \ref{circular_cylinder_vmag}(e), also occurs for moderate to low $\epsilon_\mathrm{core}$ but at high $\mathrm{Re}$. This regime is characterized by unsteady flow separation and periodic vortex shedding, in which swirling vortices detach from the cylinder and develop into a von K{\'a}rm{\'a}n vortex street.

Modes I-III are specific to permeable bluff bodies and have no direct counterpart in flows past impermeable obstacles. The results show that the cylinder permeability strongly modifies both the wake structure and the associated flow instabilities. Overall, the observed flow features are in good qualitative agreement with those reported in the literature for flows past permeable obstacles.

In regimes with low $\epsilon_\mathrm{core}$, corresponding to low permeability, the wake behavior closely resembles that behind a solid circular cylinder (see e.g., \cite{Nguyen_2023_confined_cylinders_review,Forouzi_2022_review_experimental_bluff_bodies}). At low Reynolds numbers, the flow remains stable but develops a stationary recirculation region. At higher $\mathrm{Re}$, the wake becomes unstable, leading to the formation of a von K{\'a}rm{\'a}n vortex street. For moderate to high $\epsilon_\mathrm{core}$, the increased permeability of the cylinder displaces the recirculation region downstream and progressively reduces its size until it eventually disappears. This behavior is well known and has been demonstrated numerically for flows past permeable circular cylinders (\cite{Bhattacharyya_2006_porous_cylinder,Yu_2011_permeable_circular_cylinder}), as well as for other permeable structures, including rectangular cylinders (\cite{Cummins_2017_permeable_disk,Tang_2021_permeable_disk,Ledda_2018_porous_rectangular_cylinders}), square cylinders (\cite{Yu_2010_porous_square_cylinder,Chen_2008_porous_square,Jue_2004_porous_square}), and spheres (\cite{Ciuti_2021_porous_sphere,Peng_2012_porous_sphere,Nandakumar_1982_porous_sphere}).

At high Reynolds numbers, the flow generally exhibits periodic instabilities, but increasing cylinder permeability has a stabilizing effect, in agreement with the comprehensive numerical stability analysis of \cite{Caruso_2023_permeable_circular_cylinder}. However, at high porosities, an additional flow feature appears: the primary vortex street is suppressed, while oscillations arise farther downstream through a far-wake instability, as observed in Modes II and III. Similar behavior has been reported experimentally by \cite{Castro_1971} and \cite{Cimbala_Nagib_Roshko_1988_far_wakes}, predicted by the global stability analysis of \cite{Ledda_2018_porous_rectangular_cylinders}, and recently investigated using fully nonlinear dynamic simulations by \cite{Chen_PorousCylinder_2025}.

In the study by \cite{Chen_PorousCylinder_2025}, it was suggested that far-wake oscillations may share a common underlying mechanism with the formation of secondary vortex streets (\cite{Williamson_Prasad_1993_wave_resonance,Jiang_2021,Kumar_Mittal_2012}). The present results support the interpretation that these oscillations most likely arise from a mean-flow instability. Figures \ref{circular_cylinder_vmag}(b) and \ref{circular_cylinder_vmag}(c), corresponding to Modes II and III, respectively, show no primary vortex street in the near wake. In Mode II, coherent vortices do not form behind the cylinder, whereas Mode III is characterized by a stationary recirculation region. In both cases, the far-wake oscillations appear only after a relatively quiescent downstream region, indicating that they originate from an instability of the locally averaged mean flow rather than from direct near-wake vortex shedding. The reduced shear and through-flow induced by the porous cylinder evidently suppress near-wake vortex roll-up, thereby delaying the onset of instability until farther downstream, where small perturbations can grow within the mean flow and eventually give rise to the observed far-wake oscillations.

A detailed investigation of the mechanisms governing these unsteady modes, their stability properties, and the transitions between them is beyond the scope of the present work. We note, however, that the marginal stability curves and mode diagram for the present configuration are expected to differ from those reported by \cite{Caruso_2023_permeable_circular_cylinder} and \cite{Chen_PorousCylinder_2025}, because the present study considers the wake dynamics of a confined porous cylinder. For example, Mode III features both a detached stationary recirculation region and far-wake oscillations. This combination of flow features was not observed by \cite{Chen_PorousCylinder_2025}, who considered an unconfined free-flow configuration. In the present confined setting, the bounding walls provide additional stabilization and allow a stationary recirculation region, composed of weakly oscillating vortices, to coexist with far-wake unsteadiness.

\begin{figure}[h]
\centering
\includegraphics[width=1.0\textwidth]{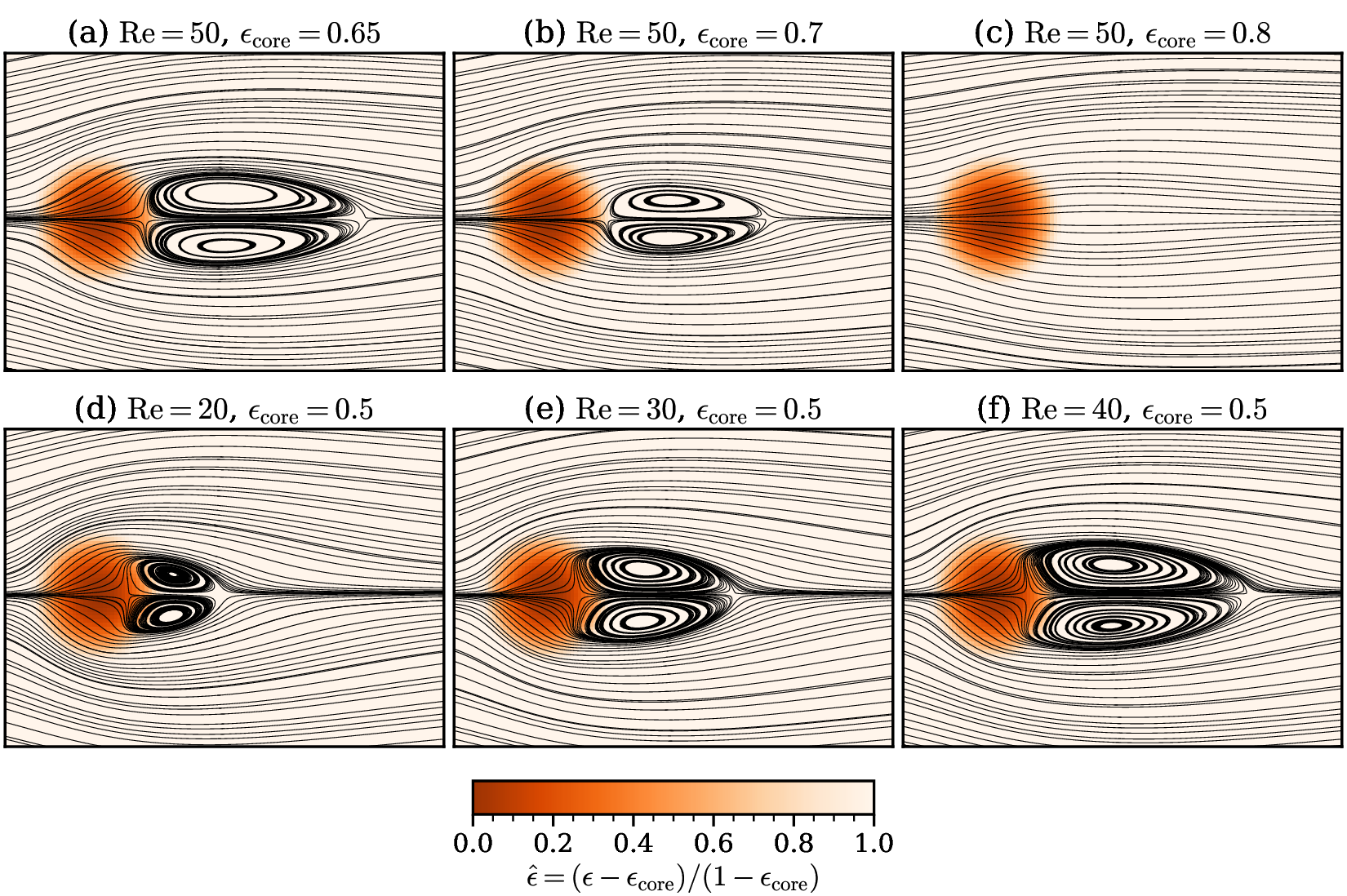}
\caption{Streamlines (black solid lines) of the steady-state superficial velocity field for channel flow past a porous cylinder, shown for different values of Reynolds number $\mathrm{Re}$ and core porosity $\epsilon_\mathrm{core}$. The background colormap represents the normalized porosity field $\hat{\epsilon}(x,y)$ defined in \eqref{Porous_Cylinder_Profile}. The plotted region is restricted to $x\in[1.4,4.3]D$ and $y\in[1,3.1]D$.}\label{circular_cylinder_streamlines}
\end{figure}

We next examine the steady flow regimes, corresponding to Modes I and IV, in more detail. Figure \ref{circular_cylinder_streamlines} shows the streamlines for different Reynolds numbers $\mathrm{Re}$ and core porosities $\epsilon_\mathrm{core}$. The results are qualitatively consistent with previous studies (\cite{Yu_2011_permeable_circular_cylinder,Bhattacharyya_2006_porous_cylinder}), although they are obtained here using a different formulation for the flow through the porous medium.

Unlike flow past a solid cylinder, part of the incoming fluid penetrates the permeable body, thereby modifying the overall flow structure. The porous cylinder therefore only partially diverts the flow: some streamlines bypass the cylinder, while others enter the porous region and pass through part of it. The streamline pattern is not fully symmetric about the vertical centerline of the cylinder because the cylinder is placed slightly off the channel centerline.

The streamlines show that the wake behind the permeable cylinder contains a recirculation region whose position and size depend strongly on permeability. In Figures \ref{circular_cylinder_streamlines}(a-c), increasing $\epsilon_\mathrm{core}$ causes the recirculation region, initially attached to the rear of the cylinder, to detach from the body and eventually disappear from the near wake, corresponding to the transition from Mode IV to Mode I. Figures \ref{circular_cylinder_streamlines}(d-f) show that increasing $\mathrm{Re}$ shifts the recirculation region downstream while increasing its size.

A notable feature in Figures \ref{circular_cylinder_streamlines}(d-f) is that, at low cylinder permeability, the recirculation region penetrates into the rear part of the cylinder. Closer inspection of the front part of the cylinder shows that streamlines entering the porous region do not pass directly through it. Instead, the recirculation region inside the rear part of the body blocks through-flow, forcing these streamlines to turn and bypass the wake. Thus, streamlines entering the porous region tend to diverge, and the fluid exits the cylinder upstream of the point where the streamline bounding the recirculation region begins.

Penetration of the recirculation region is commonly observed for permeable spheres over a wide range of parameters (\cite{Ciuti_2021_porous_sphere,Peng_2012_porous_sphere,Nandakumar_1982_porous_sphere}). Similar penetrating recirculation regions have also been reported, over more limited parameter ranges, by \cite{Yu_2011_permeable_circular_cylinder} for a permeable circular cylinder and by \cite{Tang_2021_permeable_disk} for thick disks. In general, penetration of the recirculation region into the body is weaker for two-dimensional bodies than for spheres. In the present case, however, the cylinder is heterogeneous and contains a broad porosity transition layer. This weakens the disturbance imposed on the flow and reduces the resistance to penetration, particularly near the outer part of the cylinder, where the porosity is higher.

\begin{figure}[h]
\centering
\includegraphics[width=0.9\textwidth]{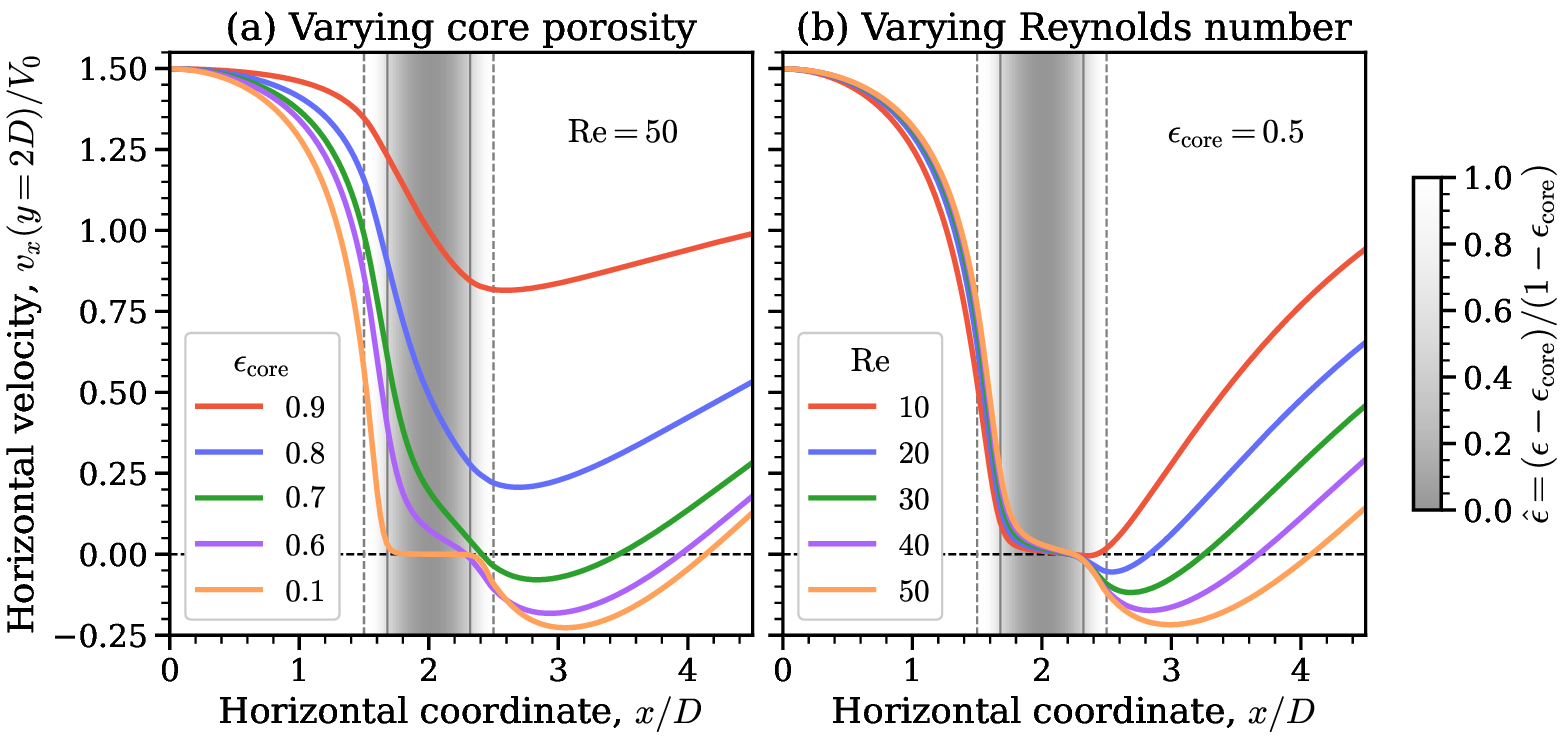}
\caption{Profile of the horizontal component of the steady-state superficial flow velocity along the cylinder centerline ($y=2D$), plotted as a function of the horizontal position for different values of (a) core porosity $\epsilon_\mathrm{core}$ and (b) Reynolds number $\mathrm{Re}$. The dashed black horizontal line indicates zero velocity. The background colormap represents the normalized porosity field $\hat{\epsilon}(x,2D)$ defined in \eqref{Porous_Cylinder_Profile}. The dashed gray vertical lines bound a region of width $D$ and mark the matching points of the porosity profile. The solid gray vertical lines bound a region of width $D_\mathrm{50}\approx0.64D$ and correspond to the location of $\epsilon_{50}$ porosity isolines, marking the midpoints of the porosity transition on either side of the cylinder.}\label{circular_cylinder_horizontal_plots}
\end{figure}

The comparisons of the horizontal superficial velocity profiles for different values of $\mathrm{Re}$ and $\epsilon_\mathrm{core}$ are shown in Figure \ref{circular_cylinder_horizontal_plots}. The results are in good qualitative agreement with those reported in \cite{Yu_2011_permeable_circular_cylinder}. Figure \ref{circular_cylinder_horizontal_plots}(a) shows that, for $\epsilon_\mathrm{core}=0.1$, the velocity distribution is close to that of flow past a solid cylinder: the horizontal velocity is nearly zero within the effective diameter $D_{50}$ due to the large flow resistance inside the porous region. For $\epsilon_\mathrm{core}=0.6$, the velocity increases substantially as the flow enters the porous cylinder, but then decreases rapidly inside it and reaches nearly zero by the downstream edge of the cylinder. For $\epsilon_\mathrm{core}=0.7$, the velocity remains non-zero as the flow exits the effective cylinder, but continues to decrease downstream and eventually becomes negative, indicating the presence of a detached recirculation region.

These results suggest that the exit velocity plays an important role in determining the wake structure behind the porous cylinder. At sufficiently high values of $\epsilon_\mathrm{core}$, the flow exits the cylinder with a relatively large velocity; although the velocity still decreases downstream, this through-flow is strong enough to suppress the formation of a recirculating wake. More generally, the effective deceleration of the flow inside the porous cylinder depends nonlinearly on the entry velocity, due to the combined effects of the porosity-dependent permeability and the Forchheimer drag contribution. Figure \ref{circular_cylinder_horizontal_plots}(b) shows that, for moderate permeability, represented by $\epsilon_\mathrm{core}=0.5$, the velocity upstream of and inside the cylinder is only weakly affected by $\mathrm{Re}$ over the range considered. In contrast, $\mathrm{Re}$ has a strong influence on the wake: larger values of $\mathrm{Re}$ produce a larger recirculation region and a higher backflow velocity downstream of the cylinder.

Overall, although the channel and cylinder configuration considered here differs from those commonly studied in the literature, the present model reproduces the relevant physics and characteristic flow features of porous bluff bodies. A closer correspondence with homogeneous porous cylinders can be obtained by choosing a radial porosity distribution with a broader, flatter core and a narrower transition region, for example by increasing the power $m$ in the bump function \eqref{b_cinf}.

\subsection{Lid-driven cavity with a porous obstacle} \label{sec:results:cavity}
\begin{figure}[h]
\centering
\includegraphics[width=0.6\textwidth]{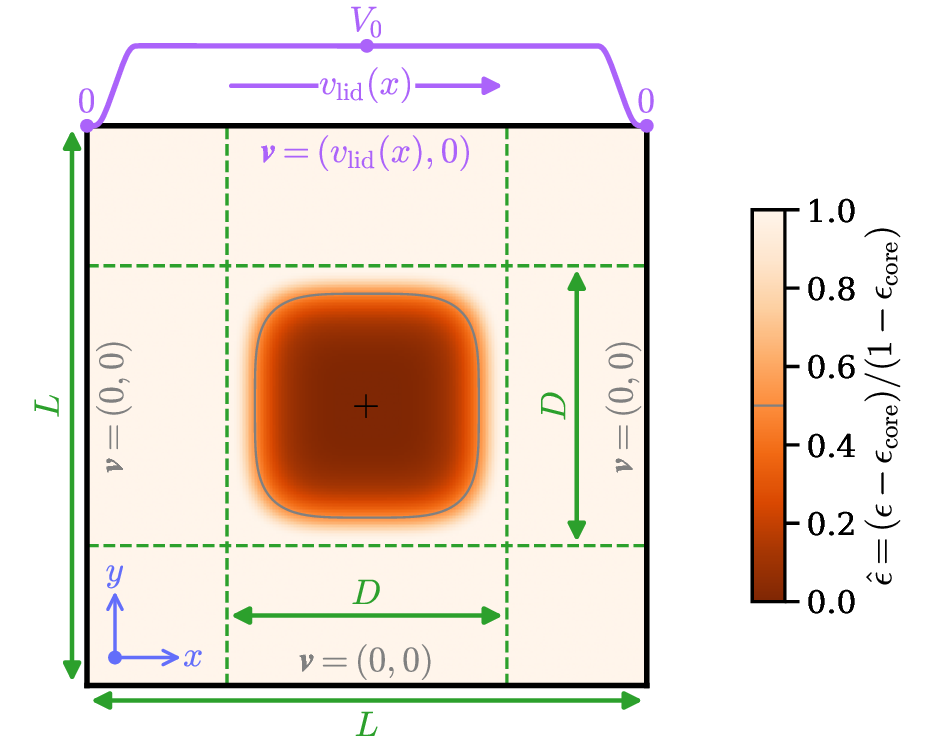}
\caption{Domain geometry and boundary conditions \eqref{cavity_BC} for the lid-driven flow in a cavity with an internal porous obstacle problem. The background colormap shows the normalized porosity field $\hat{\epsilon}(x,y)$ defined by \eqref{Cavity_Porosity_Profile}. The black plus marker denotes the centre of the domain. The grey solid superellipse of diameter $D_{50}\approx0.8D$ is the $\hat{\epsilon}=0.5$ porosity isocontour corresponding to the midpoint of the transition. The purple line above the top wall indicates the prescribed tangential lid velocity profile $v_\mathrm{lid}(x)$ given by \eqref{cavity_velocity_profile}.}\label{cavity_domain}
\end{figure}

Here we consider a two-dimensional lid-driven cavity flow with an internal porous obstacle. The configuration consists of a central porous core surrounded by free-fluid regions near the walls. A smooth transition layer is introduced between these regions, across which both the porosity and permeability vary continuously. The geometry, porosity distribution, and boundary conditions are shown in Figure \ref{cavity_domain}.

The computational domain is a wall-bounded square cavity of size $L\times L$, containing a smooth squircle-like porosity inhomogeneity centered in the domain. The porosity field $\epsilon(x,y)$ is prescribed using a compactly supported bump function. We first define the normalized superelliptic radius, based on a Lam{\'e} curve, as
\begin{equation}
    s_\mathrm{se}(x,y) = \left(\frac{2x-L}{D}\right)^4 + \left(\frac{2y-L}{D}\right)^4,
\end{equation}
where $D$ sets the characteristic diameter of the central inhomogeneity. The porosity field is then given by
\begin{equation} \label{Cavity_Porosity_Profile}
    \epsilon(x,y)=1+(\epsilon_\mathrm{core}-1)\mathcal{B}_1(s_\mathrm{se}),
\end{equation}
where $\mathcal{B}_1$ is the localized bump function defined in \eqref{b_cinf}.

This construction yields a smooth, squircle-like porous core embedded in an otherwise homogeneous free-fluid region. The porosity reaches $\epsilon_\mathrm{core}$ at the center of the cavity and smoothly approaches the outer value $\epsilon=1$ toward the boundary of the superelliptic core, beyond which it remains constant. Based on \eqref{Cavity_Porosity_Profile}, we define the effective diameter $D_{50}\approx0.8D$ as the diameter of the midpoint porosity isocontour, corresponding to $\epsilon_{50}=(1+\epsilon_\mathrm{core})/2$.

No-slip boundary conditions are imposed on all walls. The top wall acts as the moving lid, while the remaining three walls are stationary, so that
\begin{equation} \label{cavity_BC}
    \begin{split}
        \bm{v}(0,y) &= (0,0),\quad \bm{v}(L,y) = (0,0),\\
        \bm{v}(x,0) &= (0,0),\quad \bm{v}(x,L) = \left(v_\mathrm{lid}(x),0\right),
    \end{split}
\end{equation}
where $v_\mathrm{lid}(x)$ is the prescribed tangential lid velocity profile. A uniform lid velocity would introduce a sharp discontinuity at the top corners, where the moving lid meets the stationary side walls. To remove this corner singularity, the lid velocity is regularized so that it vanishes smoothly at both ends of the top wall. Specifically, the profile consists of a constant central plateau and two smooth corner tapers constructed using the smoothstep function $\mathcal{S}$ defined in \eqref{s_cinf},
\begin{equation} \label{cavity_velocity_profile}
    v_\mathrm{lid}(x)=V_0 \mathcal{S}\left(\frac{\min{\left(x,L-x\right)}}{\delta L}\right),
\end{equation}
where $V_0$ is the nominal lid velocity and $\delta$ is the relative width of each transition region. In the present setup, we set $\delta=0.1$, so that the lid moves with the uniform velocity $V_0$ over the central $80\%$ of the wall, while over the leftmost and rightmost $10\%$ the velocity is smoothly reduced to zero. This regularization ensures a smooth matching with the stationary side walls and avoids corner singularities at the top corners. The Reynolds number for this flow configuration is defined as $\mathrm{Re}=V_0L/\nu_\mathrm{f}$.

The no-slip wall boundary conditions \eqref{cavity_BC} are imposed using the non-equilibrium extrapolation scheme \eqref{NEEM_bc_f_bar}, with the boundary density obtained through the bounce-back operation \eqref{bc_density_BounceBack}. The populations are initialized at equilibrium according to \eqref{eq_bc_f_bar}, with the velocity field initially set to zero and the density set uniformly to unity throughout the domain. Each simulation is advanced until the steady-state convergence criterion \eqref{SS_convergence_L2} is satisfied.

\begin{figure}[h]
\centering
\includegraphics[width=1.0\textwidth]{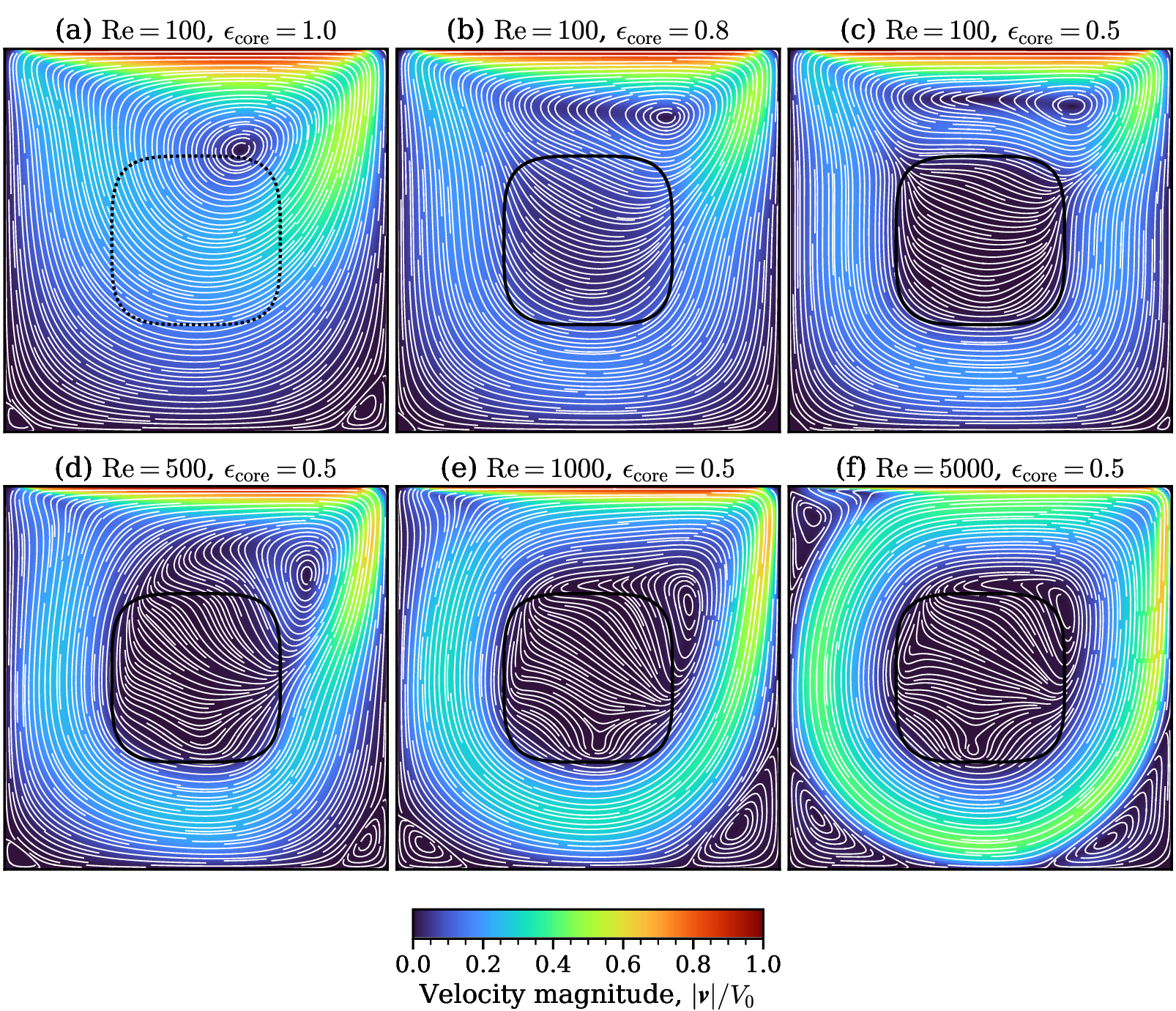}
\caption{Colormap plots of the steady-state superficial-velocity magnitude in the lid-driven cavity with an internal porous obstacle, shown for different Reynolds numbers $\mathrm{Re}$ and core porosities $\epsilon_\mathrm{core}$. White solid lines denote streamlines of the superficial velocity field. Both axes span ${x,y}\in[0,L]$. In panels (b-f), the solid black superellipse, with diameter $D_{50}\approx0.8D$, marks the $\epsilon_{50}$ porosity isocontour, corresponding to the midpoint of the porosity transition. In panel (a), which corresponds to the homogeneous free-fluid case, the dotted centered superellipse of diameter $D_{50}$ is included only as a visual reference, since no porous obstacle is present.}\label{cavity_inner_vmag}
\end{figure}

In all simulations, the temperature and effective-viscosity ratios \eqref{ratios_def} are set to unity. The geometry is fixed at $L=300$ and $D=165$, and $d_\mathrm{p}=D/5$ is used in the permeability correlation \eqref{kappa_Ergun}. The kinematic shear viscosity is prescribed according to
\begin{equation} \label{Cavity_nu}
    \nu_\mathrm{f} = \min{\left(0.1,\ \frac{L \sqrt{RT}}{4\mathrm{Re}}\right)}.
\end{equation}
Figure \ref{cavity_inner_vmag} shows the steady-state distributions of the superficial-velocity magnitude, overlaid with streamlines, for different combinations of core porosity $\epsilon_\mathrm{core}$ and Reynolds number $\mathrm{Re}$. Figure \ref{cavity_inner_vmag}(a) corresponds to $\epsilon_\mathrm{core}=1$, i.e., a homogeneous free-fluid cavity without a porous obstacle, and is included as a reference case. A comparison of the velocity profiles along the centerlines of the corresponding free-fluid cavity with the benchmark data of \cite{Ghia_1982_Cavity_Reference} is provided in Appendix \ref{Append:CavityFreeFluidRef}. Although the present simulations use a regularized lid velocity \eqref{cavity_velocity_profile}, whereas the benchmark data correspond to the classical cavity with a uniformly moving lid, good agreement is observed, with only small differences over the range of Reynolds numbers considered.

As in the pure-fluid cavity, the lid motion in the cases with a porous obstacle, $\epsilon_\mathrm{core}<1$, induces a primary vortex that occupies a large portion of the cavity and both surrounds and partially penetrates the porous region. Figures \ref{cavity_inner_vmag}(a-c), corresponding to $\mathrm{Re}=100$, show that the position of the vortex center differs from that in the pure-fluid case. In particular, as $\epsilon_\mathrm{core}$ decreases, and hence the obstacle permeability is reduced, the flow through the obstacle weakens and the primary-vortex core is displaced toward the upper-right corner, away from the obstacle. This produces steeper velocity gradients near the top wall. The porous obstacle also suppresses the vortical motion: the primary vortex becomes slightly weaker, while the secondary vortices in the lower corners, which are pronounced in the pure-fluid case, are significantly weakened. The flow pattern also becomes more separated. For $\mathrm{Re}=100$ and $\epsilon_\mathrm{core}=0.5$, shown in Figures \ref{cavity_inner_vmag}(c), a pronounced elongated recirculation pattern develops above the obstacle. The fluid accelerated by the moving lid impinges on the right wall and is redirected toward the cavity interior. Upon reaching the porous obstacle, the flow splits into three parts: the dominant portion passes below the obstacle and circulates along the cavity walls, a smaller portion passes over the obstacle, and the smallest portion penetrates through the porous region. These branches then recombine near the upper-left corner of the obstacle and feed into the lid-driven boundary layer.

Figures \ref{cavity_inner_vmag}(c-f) correspond to $\epsilon_\mathrm{core}=0.5$, representing a low-permeability superelliptic obstacle. In this regime, the flow qualitatively resembles lid-driven cavity flows with solid obstacles placed at the cavity center, as reported in \cite{Huang_2020_Cavity_Circular_Obstacles,Rajan_2021_Cavity_Obstacles}. Unlike in the solid-obstacle case, however, the primary vortex does not completely encapsulate the obstacle, since a small amount of fluid can still penetrate the porous region.

Figures~\ref{cavity_inner_vmag}(c-f) also illustrate the effect of increasing $\mathrm{Re}$ at fixed $\epsilon_\mathrm{core}=0.5$. As the Reynolds number increases, the core of the primary vortex moves closer to the porous obstacle, especially toward its upper-right corner. This produces flatter velocity gradients near the moving lid, similar to the behavior observed for cavities with solid internal obstacles. At the same time, secondary vortices gradually appear in the lower corners and grow in size. These vortices initially form very close to the corners, but their centers slowly move toward the cavity center as $\mathrm{Re}$ increases. With a further increase in $\mathrm{Re}$, as shown in Figure \ref{cavity_inner_vmag}(f) for $\mathrm{Re}=5000$, their centers shift along the flow direction, approaching the locations where the primary vortex meets the cavity walls. Figure \ref{cavity_inner_vmag}(f) also shows the emergence of an additional corner vortex near the upper-left corner. At high Reynolds numbers, most of the flow is diverted around the obstacle and follows the cavity boundaries, forming a circulating ring.

Interestingly, Figures \ref{cavity_inner_vmag}(e,f) show that, at high $\mathrm{Re}$ and low $\epsilon_\mathrm{core}$, the strong shear along the lower boundary of the obstacle induces an internal vortical structure inside the porous region, near its bottom side. This vortex forms between the fast-moving primary-vortex flow outside the obstacle and the slow flow passing through the porous medium, despite the strong local dissipation associated with the low permeability.

\begin{figure}[h]
\centering
\includegraphics[width=1.0\textwidth]{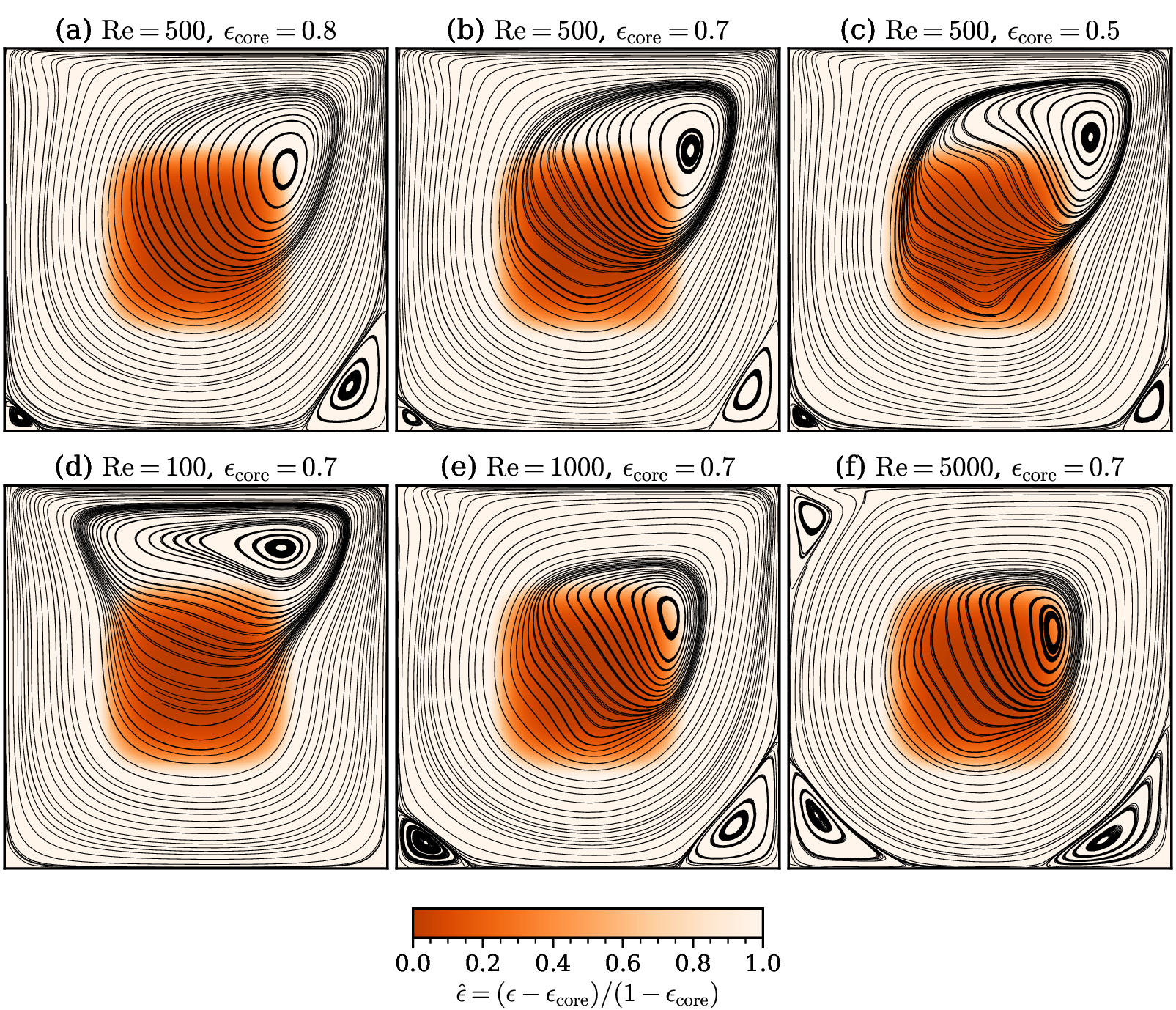}
\caption{Streamline patterns (black solid lines) of the steady-state superficial velocity field in the lid-driven cavity with an internal porous obstacle, shown for different values of Reynolds number $\mathrm{Re}$ and core porosity $\epsilon_\mathrm{core}$. Both axes span $\{x,y\}\in[0,L]$. The background colormap represents the normalized porosity field $\hat{\epsilon}(x,y)$ defined by \eqref{Cavity_Porosity_Profile}.}\label{cavity_inner_streamlines}
\end{figure}

Figure \ref{cavity_inner_streamlines} shows the streamline patterns for cavity-flow configurations with different Reynolds numbers and core porosities. The combination of $\mathrm{Re}$ and $\epsilon_\mathrm{core}$ determines the permeability of the obstacle and the relative importance of inertial and viscous effects, and therefore strongly affects the resulting flow structure. Unlike in cavity flows with solid internal obstacles, part of the primary-vortex flow penetrates the permeable body, modifying the overall streamline pattern. Thus, the porous obstacle only partially diverts the flow: some streamlines bypass it, whereas others enter the porous region and pass through part of it.

Figures \ref{cavity_inner_streamlines}(a-c), corresponding to $\mathrm{Re}=500$, show the effect of progressively decreasing the obstacle permeability. As $\epsilon_\mathrm{core}$ decreases, the center of the primary vortex is pushed away from the obstacle and toward the upper-right corner of the cavity. At the same time, the secondary counter-rotating vortices in the cavity corners are suppressed: they become smaller, and their centers move closer to the corners. This behavior can be attributed to the way in which the porous obstacle modifies the transfer of momentum from the lid to the primary vortex. Decreasing the permeability increases the total dissipation and reduces the total kinetic energy in the system. As a result, the localized corner shear is weakened, and the corner vortical structures are suppressed.

The streamline pattern inside the obstacle also changes markedly as $\epsilon_\mathrm{core}$ is decreased. For the high-permeability case $\epsilon_\mathrm{core}=0.8$, shown in Figure \ref{cavity_inner_streamlines}(a), the streamlines entering the porous region pass almost directly through it. They follow a curved pattern similar to that of the surrounding free-fluid streamlines, indicating that the resistance to entering and traversing the obstacle is low. In this case, the streamlines do not exhibit a clear preferential location for entering or exiting the obstacle. As $\epsilon_\mathrm{core}$ decreases, the streamlines inside the obstacle become increasingly deformed and are deflected away from its center. In particular, for $\epsilon_\mathrm{core}=0.5$, shown in Figure~\ref{cavity_inner_streamlines}(c), the streamlines enter the obstacle preferentially through the central part of its right boundary, diverge within the porous region, and exit near the upper-left corner.

Figures \ref{cavity_inner_streamlines}(d-f) show the effect of increasing $\mathrm{Re}$ at moderately high permeability, $\epsilon_\mathrm{core}=0.7$. In this case, increasing the Reynolds number shifts the center of the primary vortex toward, and eventually into, the porous obstacle. This behavior contrasts with the low-permeability cases shown in Figures \ref{cavity_inner_vmag}(d-f), as well as with cavity flows containing solid obstacles, where the primary-vortex center remains in the vicinity of the obstacle but does not penetrate into it. The difference arises because the larger value of $\epsilon_\mathrm{core}$ weakens the resistance imposed on the flow, particularly near the outer part of the obstacle, where the porosity is higher.

\begin{figure}[ht!]
\centering
\includegraphics[width=1.0\textwidth]{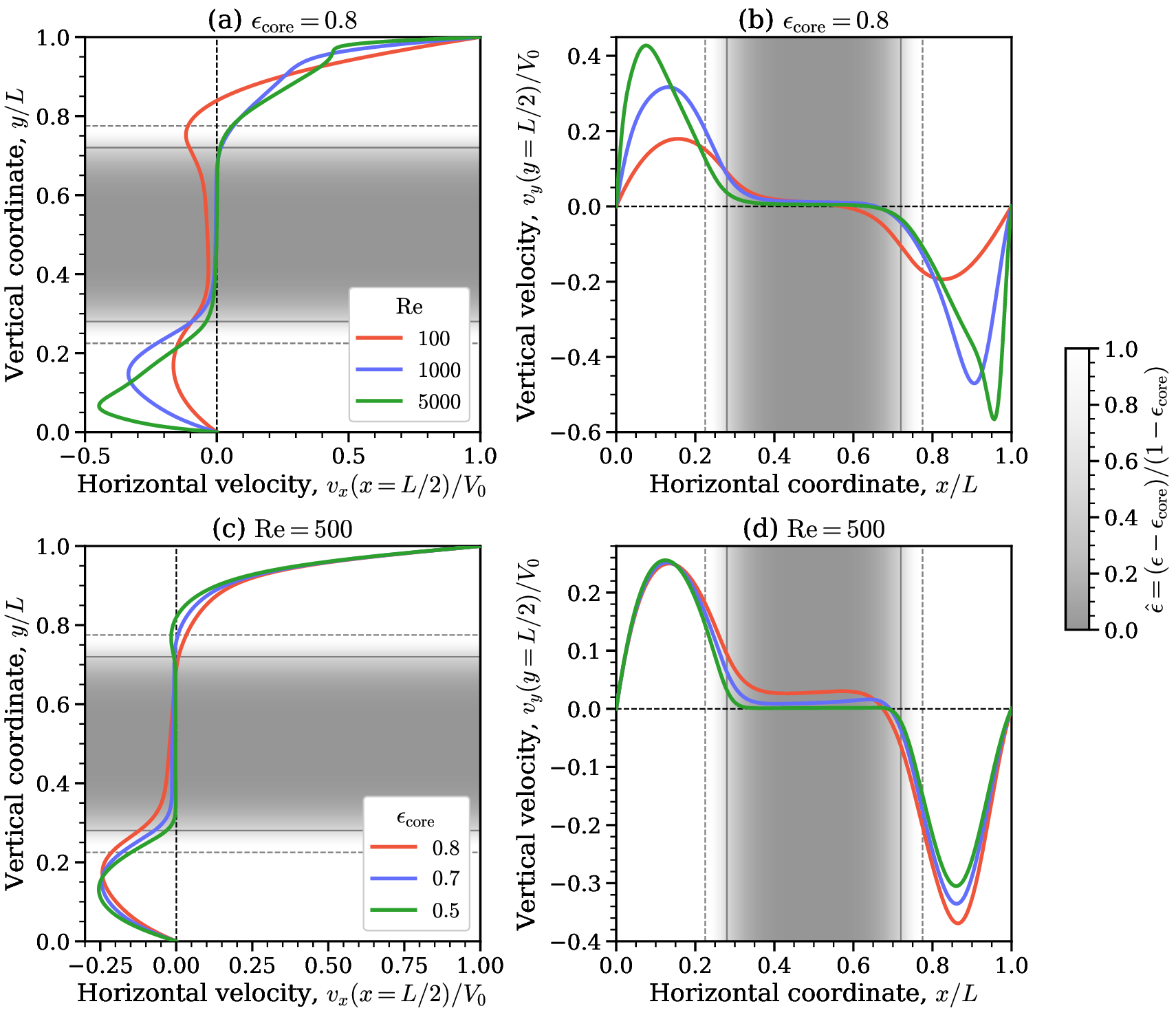}
\caption{Horizontal and vertical profiles of the steady-state superficial velocity in the lid-driven cavity flow with a porous obstacle, shown for different values of (a,b) Reynolds number $\mathrm{Re}$ and (c,d) core porosity $\epsilon_\mathrm{core}$. Panels (a,c) show the horizontal velocity component $v_x$ along the vertical centerline of the cavity, while panels (b,d) show the vertical velocity component $v_y$ along the horizontal centerline. The dashed black lines indicate zero velocity. The background colormap shows the normalized porosity field defined in \eqref{Cavity_Porosity_Profile}. The dashed gray lines bound the region of width $D$ and mark the matching points of the porosity profile. The solid gray lines bound the effective region of width $D_{50}\approx0.8D$, corresponding to the $\epsilon_{50}$ porosity isolines, marking the midpoints of the porosity transition on either side of the porous obstacle.}\label{cavity_hor_vert_plots}
\end{figure}

Figure \ref{cavity_hor_vert_plots} shows the horizontal and vertical velocity profiles along the cavity centerlines. Similarly to the free-fluid cavity, the vertical velocity $v_y$, shown in Figures \ref{cavity_hor_vert_plots}(b,d), exhibits a positive extremum near the left cavity wall and a negative extremum near the right wall. The horizontal velocity $v_x$, shown in Figures \ref{cavity_hor_vert_plots}(a,c), also exhibits a negative extremum close to the bottom wall. As observed in Figures \ref{cavity_hor_vert_plots}(a,b), increasing $\mathrm{Re}$ increases the magnitudes of these velocity extrema, while their locations move closer to the corresponding walls, indicating the thinning of the wall boundary layers. The boundary layer near the moving lid also becomes slightly thinner as $\mathrm{Re}$ increases. In particular, the $v_x$ profile for $\mathrm{Re}=5000$ in Figure \ref{cavity_hor_vert_plots}(a) develops a kink near the top lid, while a similar behavior is observed for the corresponding $v_y$ profile near the right wall in Figure \ref{cavity_hor_vert_plots}(b). Such features are characteristic of high-$\mathrm{Re}$ lid-driven cavity flows.

Similarly to cavity flows with internal solid obstacles, as reported in \cite{Huang_2020_Cavity_Circular_Obstacles,Rajan_2021_Cavity_Obstacles}, Figure \ref{cavity_hor_vert_plots}(a) shows that, for $\mathrm{Re}=100$, the $v_x$ profile exhibits an additional negative extremum near the upper boundary of the obstacle. This feature corresponds to the recirculation pattern formed above the obstacle, as shown in Figures \ref{cavity_inner_vmag}(b,c). As $\mathrm{Re}$ increases, more of the flow is diverted around the obstacle and the center of the primary recirculation region moves downward toward the upper-right corner of the obstacle; consequently, this additional extremum disappears. A similar negative extremum is also observed for $\mathrm{Re}=500$ and $\epsilon_\mathrm{core}=0.5$ in Figure \ref{cavity_hor_vert_plots}(c), although it is much weaker in this case and disappears as $\epsilon_\mathrm{core}$ is increased. Figure \ref{cavity_hor_vert_plots}(d) also shows that the core porosity affects the negative extremum of $v_y$, whose peak magnitude increases with increasing $\epsilon_\mathrm{core}$. Its influence on the other extrema, however, is much weaker.

Overall, the velocity profiles show a strong deceleration of the fluid inside the porous region. For high $\epsilon_\mathrm{core}$ or low $\mathrm{Re}$, corresponding to high permeability or weak inertial drag, this deceleration is less pronounced, and the velocity profiles do not collapse onto the zero line, indicating the absence of a fully stagnant region. In contrast, for low $\epsilon_\mathrm{core}$ or high $\mathrm{Re}$, the velocity profiles inside the porous obstacle lie close to the zero line, indicating an almost uniform quiescent region with very little fluid motion.

\section{Conclusions and Outlook}\label{sec:summary}
This work presented a monolithic lattice Boltzmann formulation for weakly compressible, isothermal flow in heterogeneous isotropic porous media. The model recovers a generalized Darcy-Forchheimer-Brinkman macroscopic description in a continuous one-domain form, allowing free-fluid and porous-medium regions to be treated within a single computational framework. A key feature of the formulation is the consistent porosity scaling of the pressure, convective, and Brinkman viscous terms, with the effective viscous stress expressed in terms of gradients of the superficial velocity. This makes the model suitable for spatially varying porosity fields and for coupled free-fluid/porous-medium configurations without requiring explicit interfacial boundary conditions.

The consistency and accuracy of the method were assessed through dispersion--dissipation analysis and benchmark simulations, including homogeneous and heterogeneous channel flows. The numerical demonstrations further showed that the method can capture the interaction between free flow and permeable structures in more complex configurations, such as confined flow past a permeable cylinder and lid-driven cavity flow with a porous obstacle. These examples indicate that the present model is already useful for obtaining physical insight into steady and unsteady porous-body flows, including wake modification, flow penetration, recirculation dynamics, and the influence of permeability on the surrounding free-fluid motion. In this sense, the formulation provides a practical tool for early-stage engineering design and for exploring complex interactions between free fluids and porous structures.

Although the formulation is multidimensional by construction and is compatible with standard three-dimensional lattices, the present study focused on two-dimensional validation and demonstration cases. A natural next step is therefore the validation and showcase of the model in fully three-dimensional geometries, including realistic porous inserts, permeable bluff bodies, and device-scale configurations with spatially varying porosity.

The most important future extension is the inclusion of an energy balance and heat-transfer model. This would enable the analysis of thermal--hydrodynamic coupling in porous systems, where flow redistribution, pressure losses, heat transport, and effective material properties are strongly interconnected. Such a development is essential for applying the framework to realistic engineering systems, including heat exchangers, catalytic supports, thermal storage devices, fuel-cell components, and other porous structures designed for coupled momentum and heat transport.

Another important direction is the extension to multicomponent mixtures in porous media. This would allow the model to describe species transport, molecular diffusion, cross-diffusion effects, and possibly chemical reactions within heterogeneous porous structures. Coupling the present hydrodynamic formulation with consistent multicomponent kinetic models would therefore broaden its applicability to filtration, separation, catalysis, electrochemical devices, and reactive porous-media flows.

Finally, extending the framework to multi-fluid and multiphase flows in porous media would open the possibility of simulating partially saturated systems, dispersed fluid phases, immiscible fluid displacement, and phase interactions in domains containing both free-fluid and porous regions. Together with thermal and multicomponent extensions, this would move the present formulation toward a more general kinetic framework for multiphysics transport in heterogeneous porous structures.

\section*{Declarations}
\backmatter
\bmhead{Funding}
This work was supported by the Swiss National Science Foundation (SNSF) project 10001232 \emph{"Mesoscale multiphysics modeling for fuel cells"}.

\bmhead{Conflict of interest}
The authors report no conflict of interest.

\bmhead{Materials availability}
The materials that support the findings of this study are available from the corresponding author upon request.

\bmhead{CRediT author statement}
\textbf{Nikita O. Gusev:} Conceptualization, Methodology, Software, Formal analysis, Investigation, Visualization, Writing - Original Draft. \textbf{Ilya V. Karlin:} Conceptualization, Resources, Supervision, Project administration, Funding acquisition, Writing - Review \& Editing.

\begin{appendices}
\section{Method of volume averaging}\label{Append:volume_averaging}
\subsection{Local averaging procedure} \label{Append:Local_averaging_procedure}
In the method of volume averaging (see e.g., \cite{Whitaker_1967,Whitaker_Advances_1969,Whitaker_1986,Whitaker_VA_method_1998}), we associate each point $\bm{x}$ in the space domain with a representative elementary volume $V_{\mathrm{REV}}$. Then consider a conserved field $\phi(\bm{x},t)$. The volume $V_{\mathrm{REV}}$ can be used as an averaging volume, provided that  the scale constraint \eqref{VA_length_scale_assumption} holds.

Suppose that the respective elementary volume $V_{\mathrm{REV}}$ is comprised of multiple phases $k$, and that the volume of each phase $V_{k}(\bm{x},t)$, may be a temporally variable spatially inhomogeneous function, even though the averaging volume is a constant,
\begin{equation}
    V_{\mathrm{REV}} = \sum_{k}{V_{k}(\bm{x},t)} \label{V_REV_general_def}.
\end{equation}
We then define a (superficial) phase average of a field $\phi$ in phase $k$ as:
\begin{equation}
    \langle\phi_{k}\rangle \equiv \frac{1}{V_{\mathrm{REV}}}\int_{V_{k}(\bm{x},t)}{\phi_{k}}\,dV,\label{phase_avg}
\end{equation}
and the intrinsic (also called interstitial) phase average as:
\begin{equation}
    \langle\phi_{k}\rangle^{k} \equiv \frac{1}{V_{k}(\bm{x},t)}\int_{V_{k}(\bm{x},t)}{\phi_{k}}\,dV.\label{intrinsic_phase_avg}
\end{equation}
These two averages can be related as
\begin{equation}
    \langle\phi_{k}\rangle = \epsilon_{k} \langle\phi_{k}\rangle^{k},\label{avg_relation}
\end{equation}
With $\epsilon_{k}$ being the local volume fraction of the $k$-phase, defined by
\begin{equation}
    \epsilon_{k}(\bm{x},t)=\frac{V_{k}(\bm{x},t)}{V_{\mathrm{REV}}}.\label{lov_vol_frac}
\end{equation}
Using the definitions \eqref{phase_avg} and \eqref{intrinsic_phase_avg}, and the linearity of the integral, we can get the following relations:
\begin{align}
    \langle{A\phi_{k}+B\psi_{k}}\rangle &= A\langle{\phi_{k}}\rangle + B\langle{\psi_{k}}\rangle, \\
    \langle{A\phi_{k}+B\psi_{k}}\rangle^{k} &= A\langle{\phi_{k}}\rangle^{k} + B\langle{\psi_{k}}\rangle^{k},
\end{align}
with $A$ and $B$ being some arbitrary constants. Also, as suggested in \cite{Gray_1975,Hsu_Cheng_1990,Ni_Beckermann_1991,Ochoa_Tapia_Whitaker_1998}, we can decompose a microscopic field $\phi_{k}$ into its intrinsic phase average $\langle{\phi_{k}}\rangle^{k}$ and a fluctuating component $\tilde{\phi}_{k}$ as:
\begin{equation}
    \phi_{k} = \langle{\phi_{k}}\rangle^{k} + \tilde{\phi}_{k}. \label{field_decompose}
\end{equation}
Here $\tilde{\phi}_{k}$ is considered to be a deviation of $\phi_{k}$ from the expected mean value $\langle{\phi_{k}}\rangle^{k}$, such that $\langle{\tilde{\phi}_{k}}\rangle^{k}=\langle{\tilde{\phi}_{k}}\rangle=0\,\forall\,V_{k}(\bm{x},t)$. Then, using \eqref{field_decompose}, it can be shown that:
\begin{equation}
    \langle{\phi_{k}\psi_{k}}\rangle = \langle{\phi_{k}}\rangle\langle{\psi_{k}}\rangle^{k} + \langle{\tilde{\phi}_{k}\tilde{\psi}_{k}}\rangle,\label{split_two}
\end{equation}
\begin{equation} \label{split_three}
\begin{split}
\langle{\phi_{k}\psi_{k}\varphi_{k}}\rangle & = \langle{\phi_{k}}\rangle \langle{\psi_{k}}\rangle^{k} \langle{\varphi_{k}}\rangle^{k} + \langle{\phi_{k}}\rangle\langle{\tilde{\psi}_{k} \tilde{\varphi}_{k}}\rangle \\
 & + \langle{\psi_{k}}\rangle\langle{\tilde{\phi}_{k} \tilde{\varphi}_{k}}\rangle + \langle{\varphi_{k}}\rangle\langle{\tilde{\phi}_{k} \tilde{\psi}_{k}}\rangle + \langle{\tilde{\phi}_{k}\tilde{\psi}_{k}\tilde{\varphi}_{k}}\rangle
\end{split}
\end{equation}

\subsection{Averaging theorems}
Consider a system with two phases $\{k,j\}$ occupying the REV. Then, suppose $\phi_{k}$, $\bm{\phi}_{k}$ and $\bm{\Phi}_{k}$ are some continuous scalar, vectorial and tensorial fields, respectively, of the $k$-phase within $V_{\mathrm{REV}}$. As discussed in detail elsewhere (\cite{Gray_Lee_1977,Miller_Gray_2005,Gray_Miller_2013}), the following averaging theorems apply:
\begin{align}
    \langle{\partial_t\phi_{k}}\rangle &= \partial_t\langle{\phi_{k}}\rangle - \frac{1}{V_{\mathrm{REV}}}\int_{A_{kj}(\bm{x},t)}{\phi_{k}(\bm{w}_{kj}\cdot\bm{n}_{kj})}\,dA, \label{theo_dt} \\
    \langle{\bm{\nabla}\phi_{k}}\rangle &= \bm{\nabla}\langle{\phi_{k}}\rangle + \frac{1}{V_{\mathrm{REV}}}\int_{A_{kj}(\bm{x},t)}{\phi_{k}\bm{n}_{kj}}\,dA, \label{theo_grad_scal} \\
    \langle{\bm{\nabla}\bm{\phi}_{k}}\rangle &= \bm{\nabla}\langle{\bm{\phi}_{k}}\rangle + \frac{1}{V_{\mathrm{REV}}}\int_{A_{kj}(\bm{x},t)}{(\bm{\phi}_{k}\otimes\bm{n}_{kj})}\,dA, \label{theo_grad_vect} \\
    \langle{\bm{\nabla}\cdot\bm{\phi}_{k}}\rangle &= \bm{\nabla}\cdot\langle{\bm{\phi}_{k}}\rangle + \frac{1}{V_{\mathrm{REV}}}\int_{A_{kj}(\bm{x},t)}{(\bm{\phi}_{k}\cdot\bm{n}_{kj})}\,dA, \label{theo_div_vect} \\
    \langle{\bm{\nabla}\cdot\bm{\Phi}_{k}}\rangle &= \bm{\nabla}\cdot\langle{\bm{\Phi}_{k}}\rangle + \frac{1}{V_{\mathrm{REV}}}\int_{A_{kj}(\bm{x},t)}{\bm{\Phi}_{k}\bm{n}_{kj}}\,dA, \label{theo_div_tens}
\end{align}
where $A_{kj}(\bm{x},t)$ is the interfacial area between the phases $k$ and $j$ within REV; $\bm{w}_{kj}$ is the velocity of the interface $A_{kj}$; and $\bm{n}_{kj}$ denotes the unit normal vector outwardly directed from $k$-phase to $j$-phase.

Following \cite{Ni_Beckermann_1991,Gray_1975}, using \eqref{theo_grad_scal} with $\phi_{k}=1$ and the fact that $\langle{1}\rangle=\epsilon_k$ we obtain the identity concerning the local volume fraction
\begin{equation}
    \frac{1}{V_{\mathrm{REV}}}\int_{A_{kj}(\bm{x},t)}{\bm{n}_{kj}}\,dA = -\bm{\nabla}\epsilon_{k},
\end{equation}
and thus, since $\langle{\phi_{k}}\rangle^{k}$ is considered constant across the averaging volume, we also have
\begin{equation}
    \frac{1}{V_{\mathrm{REV}}}\int_{A_{kj}(\bm{x},t)}{\langle{\phi_{k}}\rangle^{k}\bm{n}_{kj}}\,dA = -\langle{\phi_{k}}\rangle^{k}\bm{\nabla}\epsilon_{k}, \label{avg_interfacial}
\end{equation}

\section{Derivation of volume-averaged governing equations}\label{Append:derivation_va_equations}
We consider an isothermal two-phase system comprised of a moving fluid phase and a stationary solid phase, $k\in\{f,s\}$, such that $V_{\mathrm{REV}}=V_{f}(\bm{x})+V_{s}(\bm{x})$. The dynamics of the fluid phase are governed by the continuity and momentum equations in weakly-compressible form,
\begin{gather}
    \partial_t{\rho_f} + \bm{\nabla}\cdot(\rho_f\bm{u}_f) = 0, \label{fluid_continuity} \\
    \partial_t{(\rho_f\bm{u}_f)} + \bm{\nabla}\cdot(\rho_f\bm{u}_f\bm{u}_f^{\mathrm{T}}) = \bm{\nabla}\cdot(\bm{\sigma}_f-p_f\mathbf{I})+\rho_f\bm{b}_f, \label{fluid_momentum}
\end{gather}
where $\rho_f$ is the fluid density, $\bm{u}_f$ is the fluid velocity, $p_f$ is the pressure, $\bm{b}_f$ is the acceleration due to some external body force, and $\bm{\sigma}_f$ is the viscous stress tensor given by
\begin{equation}
    \bm{\sigma}_f = \rho_f\nu_f\left[\bm{\nabla}\bm{u}_f + (\bm{\nabla}\bm{u}_f)^{\mathrm{T}} + \left(\frac{\eta_f}{\nu_f} -\frac{2}{D}\right)(\bm{\nabla}\cdot\bm{u}_f)\mathbf{I} \right], \label{fluid_visc_stress}
\end{equation}
with $\nu_f$ and $\eta_f$ being the shear and bulk kinematic viscosities, respectively, and $D$ the number of dimensions.

The following derivation follows from the works of \cite{Whitaker_1973_multiphase_systems,Gray_1975,Gray_ONeill_1976,Sha_Chao_Soo_1984,Hassanizadeh_Gray_1980,Ganesan_Poirier_1990,Ni_Beckermann_1991,Ishii_Hibiki_2005}. Here we will assume that the correlation between the fluctuating components of $\rho_f$ and any fluid field $\phi_f$ is zero, i.e.,
\begin{equation}
    \langle{\tilde{\rho}_f\tilde{\phi}_f}\rangle=0.\label{uncorr_rho_fluc}
\end{equation}
Alternatively, one could define density-weighted variables, however, the resulting form of the equations is virtually identical (\cite{Ni_Beckermann_1991}).

Now, performing phase averaging \eqref{phase_avg} of the continuity equation \eqref{fluid_continuity}, and applying theorems (\ref{theo_dt}, \ref{theo_div_vect}), we obtain:
\begin{equation}
    \partial_{t}\langle{\rho_f}\rangle + \bm{\nabla}\cdot\langle{\rho_f\bm{u}_f}\rangle = \Gamma_{fs}, \label{cont_av_1}
\end{equation}
where $\Gamma_{fs}$ is the interfacial mass transfer from the fluid phase to the solid phase given by
\begin{equation}
    \Gamma_{fs} = -\frac{1}{V_{\mathrm{REV}}}\int_{A_{fs}(\bm{x})}\rho_f(\bm{u}_{f}-\bm{w}_{fs})\cdot\bm{n}_{fs}\,dA. \label{int_mass_trans}
\end{equation}
Using (\ref{avg_relation}, \ref{split_two}, \ref{uncorr_rho_fluc}) in \eqref{cont_av_1} we get
\begin{equation}
    \partial_{t}\bigl(\epsilon_{f}\langle{\rho_f}\rangle^f\bigr) + \bm{\nabla}\cdot\bigl(\langle{\rho_f}\rangle^{f}\langle{\bm{u}_f}\rangle\bigr) = \Gamma_{fs}, \label{cont_av_2}
\end{equation}
Next, we perform the phase averaging operation \eqref{phase_avg} of the momentum equation \eqref{fluid_momentum}, and apply theorems (\ref{theo_dt}, \ref{theo_div_tens}) to get:
\begin{equation}
    \partial_t\langle{\rho_f\bm{u}_f}\rangle + \bm{\nabla}\cdot\langle\rho_f\bm{u}_f\bm{u}_f^{\mathrm{T}}\rangle = -\bm{\nabla}\langle p_f\rangle + \bm{\nabla}\cdot\langle\bm{\sigma}_f\rangle + \bm{M}_{fs}^{\Gamma} + \bm{M}_{fs}^{\sigma} + \langle{\rho_f\bm{b}_f}\rangle, \label{momentum_av_1}
\end{equation}
where $\bm{M}_{fs}^{\Gamma}$ is the momentum change due to mass transfer across the interface and $\bm{M}_{fs}^{\sigma}$ is the total interfacial stress due to fluid-solid interaction. These are given by
\begin{align}
    \bm{M}_{fs}^{\Gamma} &= -\frac{1}{V_{\mathrm{REV}}}\int_{A_{fs}(\bm{x})}\rho_f\bm{u}_{f}\bigl((\bm{u}_{f}-\bm{w}_{fs})\cdot\bm{n}_{fs}\bigr)\,dA, \label{int_mom_transfer} \\
    \bm{M}_{fs}^{\sigma} &= \frac{1}{V_{\mathrm{REV}}}\int_{A_{fs}(\bm{x})}(\bm{\sigma}_f-p_f\mathbf{I})\bm{n}_{fs}\,dA. \label{int_mom_stress}
\end{align}
Then, using (\ref{avg_relation}, \ref{split_two}, \ref{split_three}, \ref{uncorr_rho_fluc}) in \eqref{momentum_av_1}, and considering constant acceleration $\bm{b}_f=\langle{\bm{b}_f}\rangle^f$ across the REV, i.e., $\tilde{\bm{b}}_f=0$, we get:
\begin{equation} \label{momentum_av_2}
    \begin{split}
        \partial_t\bigl(\langle{\rho_f}\rangle^{f}\langle{\bm{u}_f}\rangle\bigr) &+ \bm{\nabla}\cdot\bigl(\langle\rho_f\rangle^f\langle\bm{u}_f\rangle^f\langle\bm{u}_f^{\mathrm{T}}\rangle\bigr) + \bm{\nabla}\bigl(\epsilon_f\langle p_f\rangle^f\bigr) \\
        &= \bm{\nabla}\cdot\bigl(\langle\bm{\sigma}_f\rangle + \bm{\tau}_d\bigr) + \bm{M}_{fs}^{\Gamma} + \bm{M}_{fs}^{\sigma} + \epsilon_f\langle{\rho_f}\rangle^f\langle{\bm{b}_f}\rangle^f, 
    \end{split}
\end{equation}
where $\bm{\tau}_d$ is the dispersive flux given by
\begin{equation}
    \bm{\tau}_d = -\langle{\rho_f}\rangle^f\langle{\tilde{\bm{u}}_f\tilde{\bm{u}}_f^{\mathrm{T}}}\rangle
\end{equation}
Following \cite{Sha_Chao_Soo_1984,Ni_Beckermann_1991,Ishii_Hibiki_2005}, by using \eqref{avg_interfacial}, we separate the total interfacial stress $\bm{M}_{fs}^{\sigma}$, given by \eqref{int_mom_stress}, as:
\begin{equation}
    \bm{M}_{fs}^{\sigma} = \bar{p}_{fi}\bm{\nabla}\epsilon_f - \bm{R}_{fs}, \label{stress_separation}
\end{equation}
where $\bm{R}_{fs}$ is the dissipative part of the interfacial stress and $\bar{p}_{fi}$ is the average interfacial pressure of the fluid (buoyant force). Considering instantaneous microscopic pressure equilibration in the fluid, the average interfacial pressure, and the intrinsic phase average pressure are equal,
\begin{equation}
    \bar{p}_{fi} = \langle p_f\rangle^f. \label{pressure_equality}
\end{equation}
Then, using \eqref{stress_separation} and \eqref{pressure_equality}, the momentum balance \eqref{momentum_av_2} becomes:
\begin{equation} \label{momentum_av_3}
    \begin{split}
        \partial_t\bigl(\langle{\rho_f}\rangle^{f}\langle{\bm{u}_f}\rangle\bigr) &+ \bm{\nabla}\cdot\bigl(\langle\rho_f\rangle^f\langle\bm{u}_f\rangle^f\langle\bm{u}_f^{\mathrm{T}}\rangle\bigr) + \epsilon_f\bm{\nabla}\langle p_f\rangle^f \\
        &= \bm{\nabla}\cdot\bigl(\langle\bm{\sigma}_f\rangle + \bm{\tau}_d\bigr) + \bm{M}_{fs}^{\Gamma} - \bm{R}_{fs} + \epsilon_f\langle{\rho_f}\rangle^f\langle{\bm{b}_f}\rangle^f, 
    \end{split}
\end{equation}
Next, for the macroscopic viscous stress $\langle\bm{\sigma}_f\rangle$, using (\ref{split_two}, \ref{split_three}, \ref{fluid_visc_stress}, \ref{uncorr_rho_fluc}) and assuming negligible viscosity fluctuations within REV, i.e., $\tilde{\nu}_f=0$, and $\tilde{\eta}_f=0$ we get
\begin{equation}
    \langle\bm{\sigma}_f\rangle = \langle{\rho_f}\rangle^{f}\langle{\nu_f}\rangle^{f}\left[\langle{\bm{\nabla}\bm{u}_f}\rangle + \langle{\bm{\nabla}\bm{u}_f}\rangle^{\mathrm{T}} + \left(\frac{\langle{\eta_f}\rangle^{f}}{\langle{\nu_f}\rangle^{f}} -\frac{2}{D}\right)\langle{\bm{\nabla}\cdot\bm{u}_f}\rangle\mathbf{I} \right]. \label{av_visc_stress_1}
\end{equation}
Using theorems (\ref{theo_grad_vect}, \ref{theo_div_vect}) and \eqref{av_visc_stress_1} we obtain
\begin{equation}
    \langle\bm{\sigma}_f\rangle = \langle{\rho_f}\rangle^{f}\langle{\nu_f}\rangle^{f}\left[\bm{\nabla}\langle{\bm{u}_f}\rangle + \bigl(\bm{\nabla}\langle{\bm{u}_f}\rangle\bigr)^{\mathrm{T}} + \left(\frac{\langle{\eta_f}\rangle^{f}}{\langle{\nu_f}\rangle^{f}} -\frac{2}{D}\right)\bigl(\bm{\nabla}\cdot\langle{\bm{u}_f}\rangle\bigr)\mathbf{I} \right] + \bm{S}^{\sigma}_{fs}, \label{av_visc_stress_2}
\end{equation}
with $\bm{S}^{\sigma}_{fs}$ accounting for viscous drag effects due to the motion of the fluid-solid interface, given by:
\begin{multline}
    \bm{S}^{\sigma}_{fs} = \frac{1}{V_{\mathrm{REV}}} \int_{A_{fs}(\bm{x})}\langle{\rho_f}\rangle^f\biggl(\langle{\nu_f}\rangle^f\left[\bm{u}_f\otimes\bm{n}_{fs} + \bm{n}_{fs}\otimes\bm{u}_f -\frac{2}{D}(\bm{u}_f\cdot\bm{n}_{fs})\mathbf{I}\right] \\ + \langle{\eta_f}\rangle^f(\bm{u}_f\cdot\bm{n}_{fs})\mathbf{I}\biggr)\,dA \label{interface_visc}
\end{multline}
Note that the averaged viscous stress tensor $\langle\bm{\sigma}_f\rangle$ in \eqref{av_visc_stress_2} is given in terms of the gradients of the superficial fluid velocity, $\langle{\bm{u}_f}\rangle=\epsilon_f\langle{\bm{u}_f}\rangle^f$. Following (\cite{Brinkman_1949a,Brinkman_1949b,Ni_Beckermann_1991,Ishii_Hibiki_2005}), to model the viscous effects in the unresolved scales and the stress resulting from the macroscopic dispersive flux $\bm{\nabla}\cdot\bm{\tau}_d$, we employ a generalized model that uses effective viscosities $\langle{\nu_f}\rangle^f\rightarrow\nu_{\mathrm{eff}}$ and $\langle{\eta_f}\rangle^f\rightarrow\eta_{\mathrm{eff}}$ inside the viscous stress tensor as:
\begin{equation}
    \langle\bm{\sigma}_f\rangle\bigl(\langle{\nu_f}\rangle^f,\langle{\eta_f}\rangle^f\bigr) + \bm{\tau}_d = \bm{\sigma}_{\mathrm{eff}}(\nu_{\mathrm{eff}}, \eta_{\mathrm{eff}}), \label{eff_visc_model}
\end{equation}
In the system under consideration, there is no mass flux across the interface, the interface is stationary, i.e., $\bm{w}_{fs}=\bm{0}$, and $\bm{u}_f$ and $\bm{n}_{fs}$ are orthogonal, i.e.,  $\bm{u}_f\cdot\bm{n}_{fs}=0$, and the fluid velocity at the interface is $0$. Therefore, from (\ref{int_mass_trans}, \ref{int_mom_transfer}, \ref{interface_visc}) we have $\Gamma_{fs}=0$, $\bm{M}_{fs}^{\Gamma}=\bm{0}$ and $\bm{S}^{\sigma}_{fs}=\bm{0}$. Then, from equations (\ref{cont_av_2}, \ref{momentum_av_3}, \ref{av_visc_stress_2}, \ref{eff_visc_model}) with aforementioned considerations, and renaming variables to ease the notation: $\epsilon_f\rightarrow\epsilon$, $\langle{\rho_f}\rangle^f\rightarrow\rho$, $\langle{p_f}\rangle^f\rightarrow p$, $\langle{\bm{u}_f}\rangle\rightarrow\bm{v}$, $\langle{\bm{u}_f}\rangle^f\rightarrow\bm{u}$, $\langle{\bm{b}_f}\rangle^f\rightarrow \bm{b}$, $\bm{R}_{fs}\rightarrow\bm{R}$, we obtain final governing equations (\ref{continuity_target}, \ref{momentum_target}, \ref{visc_stress_target}) in the main text.

We can also express the viscous stress tensor $\bm{\sigma}_{\mathrm{eff}}$ (in \ref{av_visc_stress_2} or \ref{visc_stress_target}) is in terms of the gradients of the intrinsic fluid velocity, $\bm{u}=\bm{v}/\epsilon$ and porosity $\epsilon$. To that end, we split it as: $\bm{\sigma}_{\mathrm{eff}} = \bm{\sigma}_{\mathrm{eff}}' + \bm{\sigma}_{\mathrm{eff}}''$, with
\begin{gather}
    \bm{\sigma}_{\mathrm{eff}}' = \epsilon\rho\nu_{\mathrm{eff}}\left[\bm{\nabla}\bm{u} + (\bm{\nabla}\bm{u})^{\mathrm{T}} + \left(\frac{\eta_{\mathrm{eff}}}{\nu_{\mathrm{eff}}}-\frac{2}{D}\right)(\bm{\nabla}\cdot\bm{u})\mathbf{I}\right], \label{visc_stress_u} \\
    \bm{\sigma}_{\mathrm{eff}}'' = \rho\nu_{\mathrm{eff}}\left[\bm{u}(\bm{\nabla}\epsilon)^{\mathrm{T}} + (\bm{\nabla}\epsilon)\bm{u}^{\mathrm{T}} + \left(\frac{\eta_{\mathrm{eff}}}{\nu_{\mathrm{eff}}}-\frac{2}{D}\right)(\bm{u}\cdot\bm{\nabla}\epsilon)\mathbf{I}\right]. \label{visc_stress_delta_eps}
\end{gather}
Note that in homogeneous porous media ($\epsilon=\text{const.}$) the porosity gradients will vanish ($\bm{\nabla}\epsilon=\bm{0}$), and so $\bm{\sigma}_{\mathrm{eff}}''=\bm{0}$ and $\bm{\sigma}_{\mathrm{eff}} =  \bm{\sigma}_{\mathrm{eff}}'$.

\section{Discrete equilibria}\label{Append:eq_moments}
We choose the target moments of a discrete equilibrium function to be
\begin{gather}
    \Pi_{0}^{\mathrm{tgt}} = \epsilon\rho, \label{Pi_0_tgt} \\
    \Pi_{\alpha}^{\mathrm{tgt}} = \epsilon\rho\xi_\alpha, \label{Pi_1_tgt} \\
    \Pi_{\alpha\beta}^{\mathrm{tgt}} = \epsilon\rho\left[\xi_\alpha\xi_\beta+ RT\delta_{\alpha\beta}\right], \label{Pi_2_tgt} \\
    \Pi_{\alpha\beta\gamma}^{\mathrm{tgt}} = \epsilon\rho\left[\xi_{\alpha}\xi_{\beta}\xi_{\gamma} + RT(\xi_\alpha\delta_{\beta\gamma}+\xi_\beta\delta_{\alpha\gamma} +\xi_\gamma\delta_{\alpha\beta})\right]. \label{Pi_3_tgt}
\end{gather}
where $\bm{\xi}$ is a velocity variable, set to $\bm{v}/\epsilon$ for the equilibrium and $\bm{v}_{\lambda}^{*}/\epsilon$ for the shifted-equilibrium, as detailed in Section \ref{sec:equilibrium}.

In general, for the product-form equilibrium function $f_{i}^{\mathrm{prod}}(\bm{\xi},\bm{\mathcal{P}})$ defined by \eqref{prod_form} with \eqref{1d_scaled_eq}, the first-, second- and third-order scaled moments, following \cite{Karlin_Factorization_2010}, are given by
\begin{align}
    \frac{\Pi_{\alpha}^{\mathrm{prod}}}{\Pi_{0}^{\mathrm{prod}}} &= \xi_{\alpha} \label{Pi_1_prod_general} \\
    \frac{\Pi_{\alpha\beta}^{\mathrm{prod}}}{\Pi_{0}^{\mathrm{prod}}} &= \begin{cases}
            \xi_{\alpha}\xi_{\beta}, & \alpha\neq\beta \\
            \mathcal{P}_{\alpha\alpha}, & \alpha=\beta
    \end{cases} \label{Pi_2_prod_general} \\
    \frac{\Pi_{\alpha\beta\gamma}^{\mathrm{prod}}}{\Pi_{0}^{\mathrm{prod}}} &= \begin{cases}
            \xi_{\alpha}\xi_{\beta}\xi_{\gamma}, & \alpha\neq\beta\neq\gamma \\
            \mathcal{P}_{\alpha\alpha}\xi_{\beta}, & \alpha=\gamma\neq\beta \\
            \xi_{\alpha}, & \alpha=\gamma=\beta
    \end{cases} \label{Pi_3_prod_general}
\end{align}
Thus, by setting $\Pi_{0}^{\mathrm{prod}} = \epsilon\rho$ and $\mathcal{P}_{\alpha\alpha}=\xi_{\alpha}^2+RT$, the first three moments will match the target ones (\ref{Pi_0_tgt}, \ref{Pi_1_tgt}, \ref{Pi_2_tgt}) exactly. The third-order moment tensor in expanded form reads
\begin{equation}
    \Pi_{\alpha\beta\gamma}^{\mathrm{prod}} = \epsilon\rho\left[\xi_{\alpha}\xi_{\beta}\xi_{\gamma} + RT(\xi_\alpha\delta_{\beta\gamma}+\xi_\beta\delta_{\alpha\gamma} +\xi_\gamma\delta_{\alpha\beta}) +\delta_{\alpha\beta\gamma}\left(1-3RT-\xi_\alpha^2\right)\xi_\alpha\right], \label{Pi_3_prod_generic}
\end{equation}
and by comparison with \eqref{Pi_3_tgt}, it contains an anomalous term in its diagonal.

The product-form equilibrium (\ref{prod_form}, \ref{1d_scaled_eq}) can also be equivalently expressed as
\begin{equation}
    f_{i}^{\mathrm{prod}}=\epsilon\rho\omega_i\prod_{\alpha}{\left[1+\frac{c_{i\alpha}\xi_\alpha}{RT}+\frac{(c_{i\alpha}^2-RT)\xi_\alpha^2}{RT(1-RT)}\right]}, \label{prod_form_v2}
\end{equation}
where $w_{i}$ are the so-called lattice weights, given in a product-form using \eqref{1d_scaled_eq} as
\begin{equation}
     w_{i} = \prod_{\alpha}{\Psi_{i\alpha}(0,RT)}=\prod_{\alpha}{\left[1-c_{i\alpha}^2+\left(\frac{3}{2}c_{i\alpha}^2-1\right)RT\right]}. \label{eq_weights}
\end{equation}
Expanding the product in \eqref{prod_form_v2} and retaining terms up to $\mathcal{O}\left(\lvert\bm{\xi}\rvert^2\right)$, we obtain a simplified second-order equilibrium function
\begin{equation}
     f_i^{\mathrm{so}} = \epsilon\rho w_{i}\left[1 + \frac{\bm{c}_i\cdot\bm{\xi}}{RT} + \frac{\left(\bm{c}_i^{\odot2}-RT\bm{1}\right)\cdot\bm{\xi}^{\odot2}}{RT(1-RT)}+\frac{(\bm{c}_i\cdot\bm{\xi})^{2}-\bm{c}_i^{\odot2}\cdot \bm{\xi}^{\odot2}}{2(RT)^2}\right]+\mathcal{O}\left(\lvert\bm{\xi}\rvert^3\right), \label{so_eq_v1}
\end{equation}
where $(\cdot)^{\odot 2}$ denotes the elementwise (Hadamard) square. For illustrative purposes, it is instructive to rewrite \eqref{so_eq_v1} as
\begin{equation}
     f_i^{\mathrm{so}} = \epsilon\rho w_{i}\left[1 + \frac{\bm{c}_i\cdot\bm{\xi}}{RT} + \frac{(\bm{c}_i\cdot\bm{\xi})^{2}}{2(RT)^2}-\frac{\lvert\bm{\xi}\rvert^{2}}{2RT}+\frac{3RT-1}{2(RT)^2(1-RT)}\left(\bm{c}_i^{\odot2}-RT\bm{1}\right)\cdot\bm{\xi}^{\odot2}\right], \label{so_eq_v2}
\end{equation}
where the first four terms correspond to the second-order equilibrium commonly used in LBM formulations for volume-averaged Navier-Stokes equations (\cite{Wang_Wang_2005,Xiong_Madadi_Lorenzini_2014,Zhang_2014}), while the last term represents an explicit diagonal anisotropic correction, which vanishes for the lattice reference temperature $RT= RT_\mathrm{L} = 1/3$.

For this second-order equilibrium $f_{i}^{\mathrm{so}}(\bm{\xi})$, defined by (\ref{so_eq_v1} or \ref{so_eq_v2}) with weights \eqref{eq_weights}, the first three pertinent moments also match the target moments, and the third-order moment tensor is given by
\begin{equation}
    \Pi_{\alpha\beta\gamma}^{\mathrm{so}} = \epsilon\rho\left[RT(\xi_\alpha\delta_{\beta\gamma}+\xi_\beta\delta_{\alpha\gamma} +\xi_\gamma\delta_{\alpha\beta}) +\delta_{\alpha\beta\gamma}\left(1-3RT\right)\xi_\alpha\right]. \label{Pi_3_so_generic}
\end{equation}
The deviations of the third-order moments for both the product-form and second-order equilibrium from the target form \eqref{Pi_3_tgt} can be highlighted by defining
\begin{gather}
    \tilde{\Pi}^\mathrm{so}_{\alpha\beta\gamma}=\Pi^\mathrm{so}_{\alpha\beta\gamma}-\Pi^{\mathrm{tgt}}_{\alpha\beta\gamma}=\epsilon\rho\left[ \delta_{\alpha\beta\gamma}\left(1-3RT\right)\xi_\alpha - \xi_{\alpha}\xi_{\beta}\xi_{\gamma} \right], \label{so_deviation}\\
    \tilde{\Pi}^\mathrm{prod}_{\alpha\beta\gamma}=\Pi^\mathrm{prod}_{\alpha\beta\gamma}-\Pi^{\mathrm{tgt}}_{\alpha\beta\gamma} = \epsilon\rho\left[\left(1-3RT\right)\xi_\alpha - \xi_\alpha^3\right]\delta_{\alpha\beta\gamma} \label{prod_deviation}.
\end{gather}

\section{Derivation of the LBE}\label{Append:lbe_derivation}
\subsection{Time discretization}\label{Append:time_discretization}
Integrating \eqref{kinetic-eq} along a characteristic $s$ for a time interval $\delta t$:
\begin{equation}
    f_{i}(\bm{x}+\bm{c}_{i}\delta t, t+\delta t)-f_{i}(\bm{x}, t) = \int_{0}^{\delta t}{\left[\Omega_{i}(\bm{x}+\bm{c}_{i}s, t+s) + \mathcal{K}_{i}(\bm{x}+\bm{c}_{i}s, t+s)\right]}\,ds. \label{int_char}
\end{equation}
The integral on the right-hand side of \eqref{int_char} is then approximated by a trapezoidal rule with second-order accuracy, as
\begin{equation}
    \begin{split}
        f_{i}(\bm{x}+\bm{c}_{i}\delta t, t+\delta t)-f_{i}(\bm{x}, t) = &\frac{\delta t}{2} \bigl[\Omega_{i}(\bm{x}, t) + \Omega_{i}(\bm{x}+\bm{c}_{i}\delta t, t+\delta t) \\
        & + \mathcal{K}_{i}(\bm{x}, t) + \mathcal{K}_{i}(\bm{x}+\bm{c}_{i}\delta t, t+\delta t) \bigr] + \mathcal{O}(\delta t^3),\label{implicit_scheme}
    \end{split}
\end{equation}
resulting in a time-implicit scheme. The system can be rendered fully explicit by a change of variables, as introduced in \cite{He_Shan_Doolen_1998}. Defining
transformed populations $\bar{f}_i$ as
\begin{equation}
    \bar{f}_i = f_i - \frac{\delta t}{2}(\Omega_i + \mathcal{K}_i) \label{general_transform}
\end{equation}
Now, by using \eqref{general_transform} in \eqref{implicit_scheme}, we get an explicit scheme
\begin{equation}
    \bar{f}_i(\bm{x}+\bm{c}_{i}\delta t, t+\delta t)-\bar{f}_i(\bm{x}, t) = \delta t [\Omega_i(\bm{x}, t) + \mathcal{K}_i(\bm{x}, t)]. \label{explicit_scheme}
\end{equation}
Finally, using the LBGK model for $\Omega_i$ \eqref{omega_BGK} and the RtFM model for $\mathcal{K}_i$ \eqref{F_RtFM}, we can express \eqref{explicit_scheme} in terms of $\bar{f}_{i}$, $f_{i}^{\mathrm{eq}}$ and $f_{i}^{*}$ as
\begin{equation} \label{lbe_expanded}
    \begin{split}
        \bar{f}_i(\bm{x}+\bm{c}_{i}\delta t, t+\delta t)-\bar{f}_i(\bm{x}, t) & = 2\beta(\bm{x}, t)[f_{i}^{\mathrm{eq}}(\bm{x}, t)-\bar{f}_i(\bm{x}, t)] \\
        & + \delta t \underbrace{\left(\frac{1-\beta(\bm{x}, t)}{\lambda}\right)[f_{i}^{*}(\bm{x}, t)-f_{i}^{\mathrm{eq}}(\bm{x}, t)]}_{=S_i^{\lambda}(\bm{x}, t)}
    \end{split}
\end{equation}
with $\beta(\bm{x}, t)=\delta t/[2\tau(\bm{x}, t) + \delta t]$ as defined by \eqref{beta_def}.

\subsection{Transform of the locally conserved fields}\label{Append:transform_fields}
Using the transform \eqref{general_transform} and the moments (\ref{density_moment}, \ref{momentum_moment}) we get
\begin{align}
    \underbrace{\sum_{i=0}^{Q-1}{f_i}}_{\epsilon\rho} &= \sum_{i=0}^{Q-1}{\bar{f}_i} + \frac{\delta t}{2}\underbrace{\sum_{i=0}^{Q-1}{\Omega_i}}_{=0} + \frac{\delta t}{2}\underbrace{\sum_{i=0}^{Q-1}{\mathcal{K}_i}}_{=0}, \label{transformed_density_moment_2} \\
    \underbrace{\sum_{i=0}^{Q-1}{\bm{c}_{i}f_i}}_{\rho\bm{v}} &= \sum_{i=0}^{Q-1}{\bm{c}_{i}\bar{f}_i} + \frac{\delta t}{2}\underbrace{\sum_{i=0}^{Q-1}{\bm{c}_{i}\Omega_i}}_{=0} + \frac{\delta t}{2}\underbrace{\sum_{i=0}^{Q-1}{\bm{c}_{i}\mathcal{K}_i}}_{=\rho\bm{a}},  \label{transformed_momentum_moment_2}
\end{align}
which gives us the redefined moments (\ref{transformed_density_moment}, \ref{transformed_momentum_moment}). In homogeneous porous media, from \eqref{accel} with $\bm{\Delta}_\nu=\bm{0}\in\Omega$, we have $\bm{a}(\bm{v})=\epsilon \bm{b} - \mu_\mathrm{D} \bm{v} -\mu_\mathrm{F}|\bm{v}|\bm{v}$. Due to the quadratic nature of the equation \eqref{transformed_momentum_moment_2}, the velocity $\bm{v}$ can be solved for explicitly. To that end, we introduce $\bm{v}'$, defined by
\begin{equation}
    \rho\bm{v}' = \sum_{i=0}^{Q-1}{\bm{c}_{i}\bar{f}_i} + \frac{\delta t}{2}\epsilon\rho\bm{b}, \label{v'_def}
\end{equation}
allowing us to express $\bm{v}$ as
\begin{equation}
    \bm{v} = \bm{v}' + \frac{\delta t}{2}(\mu_\mathrm{D}\bm{v} + \mu_\mathrm{F}|\bm{v}|\bm{v}). \label{v_in_v'}
\end{equation}
Equation \eqref{v_in_v'} can then be rearranged as
\begin{equation}
    \underbrace{\left(1+\frac{\delta t}{2}\mu_\mathrm{D}\right)}_{=c_1}\bm{v} + \underbrace{\frac{\delta t}{2}\mu_\mathrm{F}}_{=c_2}|\bm{v}|\bm{v} - \bm{v}' = \bm{0}, \label{v_quadratic}
\end{equation}
with $c_1$ and $c_2$ as defined by \eqref{velocity_coefficients}. Since both $\bm{v}$ and $|\bm{v}|\bm{v}$ are co-linear with $\bm{v}$, then $\bm{v}'$ must also be co-linear with $\bm{v}$, i.e., $\bm{v}\parallel\bm{v}'$ and $\bm{v}=\sigma\bm{v}'$ for some scalar $\sigma\geq0$. Then, taking norms of \eqref{v_quadratic} gives
\begin{equation}
    c_2|\bm{v}|^2 + c_1|\bm{v}| - |\bm{v}'| = 0,
\end{equation}
whose physically relevant (non-negative) root is
\begin{equation}
    |\bm{v}| = \frac{-c_1+\sqrt{c_1^2 + 4c_2 |\bm{v}'|}}{2c_2},
\end{equation}
that can be rearranged as
\begin{equation}
    \frac{|\bm{v}|}{|\bm{v}'|} = \frac{2}{c_1+\sqrt{c_1^2+4c_2|\bm{v}'|}}. \label{root_abs_v}
\end{equation}
Then, since $\bm{v}\parallel\bm{v}'$,
\begin{equation}
    \frac{\bm{v}}{|\bm{v}|} = \frac{\bm{v}'}{|\bm{v}'|}\ \Longrightarrow\ \bm{v}=\frac{|\bm{v}|}{|\bm{v}'|}\bm{v}'. \label{paralell_property}
\end{equation}
Finally, using \eqref{root_abs_v} and \eqref{paralell_property}, we obtain an explicit equation for $\bm{v}$ as
\begin{equation}
    \bm{v} = \frac{2\bm{v}'}{c_1+\sqrt{c_1^2+4c_2|\bm{v}'|}}.
\end{equation}
Note that this matches the steps (\ref{v_(0)}, \ref{v_(1)}) for $\bm{\nabla}\epsilon=0$ in the main text.

\section{Hydrodynamic limit of the LBE}\label{Append:CE_analysis}
Taylor expanding the lattice Boltzmann Equation \eqref{lbe_expanded} around $(\bm{x},t)$ results in:
\begin{equation}
    \delta t\mathcal{D}_i\bar{f}_i + \frac{\delta t ^2}{2}\mathcal{D}_i^2\bar{f}_i + \mathcal{O}(\delta t ^3) = 2\beta(f^{\mathrm{eq}}_{i}-\bar{f}_{i}) + \frac{\delta t}{\lambda}(1-\beta)(f_{i}^{*}-f^{\mathrm{eq}}_{i})
\end{equation}
where $\mathcal{D}_i=\partial_t+c_{i\alpha} \partial_{\alpha}$ is a directional derivative, using index notation and Einstein summation convention. Then, subtracting $(\delta t/2)\mathcal{D}_i$ applied to the equation itself, and neglecting the $\mathcal{O}(\delta t ^3)$ terms we obtain,
\begin{multline} \label{dim_Taylor_LBE}
    \delta t\mathcal{D}_i\bar{f}_i = 2\beta(f^{\mathrm{eq}}_{i}-\bar{f}_{i}) + \frac{\delta t}{\lambda}(1-\beta)(f_{i}^{*}-f^{\mathrm{eq}}_{i}) - \delta t\mathcal{D}_i\left[\beta(f^{\mathrm{eq}}_{i}-\bar{f}_{i})\right] \\ - \delta t\mathcal{D}_i\left[\frac{\delta t}{2\lambda}(1-\beta)(f_{i}^{*}-f^{\mathrm{eq}}_{i})\right].
\end{multline}
Introducing a characteristic time scale $\mathcal{T}$ and length scale $\mathcal{L}$, as well as a reference density $\rho_0$, the following non-dimensional quantities are introduced
\begin{equation}
    t'=\frac{t}{\mathcal{T}}, \quad c_{\alpha}'=c_{\alpha}\frac{\mathcal{T}}{\mathcal{L}}, \quad x_{\alpha}'=\frac{x_{\alpha}}{\mathcal{L}}, \quad f_i'=\frac{f_i}{\rho_0}. \label{non_dim_quants}
\end{equation}
Using \eqref{non_dim_quants} above, equation \eqref{dim_Taylor_LBE} is made non-dimensional as follows:
\begin{multline}
    \delta t'\mathcal{D}_i'\bar{f}_i' = 2\beta(f^{\mathrm{eq}}_{i}{'}-\bar{f}_{i}') + \frac{\delta t}{\lambda}(1-\beta)(f_{i}^{*}{'}-f^{\mathrm{eq}}_{i}{'}) - \delta t'\mathcal{D}_i'\left[\beta(f^{\mathrm{eq}}_{i}{'}-\bar{f}_{i}')\right] \\ - \delta t'\mathcal{D}_i'\left[\frac{\delta t}{2\lambda}(1-\beta)(f_{i}^{*}{'}-f^{\mathrm{eq}}_{i}{'})\right],
\end{multline}
and then introducing a smallness parameter $\varepsilon=\delta t'=\delta t/\mathcal{T}$ and dropping the primes, the equation becomes
\begin{multline} \label{nondim_Taylor_LBE}
    \varepsilon\mathcal{D}_i\bar{f}_i = 2\beta(f^{\mathrm{eq}}_{i}-\bar{f}_{i}) + \frac{\delta t}{\lambda}(1-\beta)(f_{i}^{*}-f^{\mathrm{eq}}_{i}) - \varepsilon\mathcal{D}_i\left[\beta(f^{\mathrm{eq}}_{i}-\bar{f}_{i})\right] \\ - \varepsilon\mathcal{D}_i\left[\frac{\delta t}{2\lambda}(1-\beta)(f_{i}^{*}-f^{\mathrm{eq}}_{i})\right].
\end{multline}
Introducing the following multiscale expansions:
\begin{gather}
    \bar{f}_i = f_i^\mathrm{eq} + \varepsilon\bar{f}_i^{(1)} + \varepsilon^2\bar{f}_i^{(2)}, \label{expansion_f}\\
    f_i^* = f_i^\mathrm{eq} + \varepsilon f_i^{*(1)} + \varepsilon^2 f_i^{*(2)},\label{expansion_f*}\\
    \partial_t = \partial_t^{(1)} + \varepsilon\partial_t^{(2)}, \quad \mathcal{D}_i = \partial_t^{(1)} + c_{i\alpha}\partial_{\alpha}.\label{expansion_der}
\end{gather}
We then substitute expansions (\ref{expansion_f}, \ref{expansion_f*}, \ref{expansion_der}) into \eqref{nondim_Taylor_LBE} and proceed by collecting terms of the same order in $\varepsilon$
\begin{align}
    \mathcal{O}(\varepsilon)\,&:\quad \mathcal{D}_i^{(1)}f_i^\mathrm{eq} = -2\beta \bar{f}_i^{(1)} + \frac{\delta t}{\lambda}(1-\beta)f_i^{*(1)}, \label{O_eps_1} \\
    \mathcal{O}(\varepsilon^2)\,&:\quad \partial_t^{(2)}f_i^\mathrm{eq} + \mathcal{D}_i^{(1)}\left[(1-\beta)\left(\bar{f}_i^{(1)}+\frac{\delta t}{2\lambda}f_i^{*(1)}\right)\right] = -2\beta f_i^{*(2)} + \frac{\delta t}{\lambda}(1-\beta)f_i^{*(2)}. \label{O_eps_2}
\end{align}
Taking zeroth and first moments of \eqref{O_eps_1} gives
\begin{gather}
    \partial_t^{(1)}\Pi_0^{\mathrm{eq}} + \partial_{\alpha}\Pi_{\alpha}^{\mathrm{eq}} = -2\beta \underbrace{\bar{\Pi}_{0}^{(1)}}_{=0} + \frac{\delta t}{\lambda}(1-\beta)\underbrace{\Pi_0^{*(1)}}_{=0}, \label{0th_moment_O_eps_1}\\
    \partial_t^{(1)}\Pi_{\alpha}^{\mathrm{eq}} + \partial_{\beta}\Pi_{\alpha\beta}^{\mathrm{eq}} = \underbrace{\frac{\delta t}{\lambda}\Pi_{\alpha}^{*(1)}}_{=-2\bar{\Pi}_{\alpha}^{(1)}} - 2\beta\underbrace{\left(\frac{\delta t}{2\lambda}\Pi_\alpha^{*(1)}+\bar{\Pi}_{\alpha}^{(1)}\right)}_{=0},\label{1st_moment_O_eps_1}
\end{gather}
where we have used the short-hand notation for the moments $\Pi_\alpha=\sum_ic_{i\alpha}(\cdot)$, and applied the following solvability conditions ensuring mass and momentum conservation,
\begin{gather}
    \bar{\Pi}_0^{(k)}=\Pi_0^{*(k)} = 0,\, \forall \, k \geq 1, \label{solvability_1}\\
    \bar{\Pi}_{\alpha}^{(1)} + \frac{\delta t}{2\lambda}\Pi_{\alpha}^{*(1)} = 0, \label{solvability_2} \\
    \bar{\Pi}_{\alpha}^{(k)} = \Pi_{\alpha}^{*(k)} = 0, \, \forall \, k \geq 2. \label{solvability_3}
\end{gather}
Similarly, taking zeroth and first moments of \eqref{O_eps_2} we get
\begin{gather}
    \begin{split}
        \partial_t^{(2)}\Pi_{0}^{\mathrm{eq}} + \partial_t^{(1)}\left[(1-\beta) \left(\underbrace{\bar{\Pi}_0^{(1)}}_{=0}+\frac{\delta t}{2\lambda}\underbrace{\Pi_0^{*(1)}}_{=0}\right)\right] + \partial_\alpha \left[(1-\beta)\underbrace{\left(\bar{\Pi}_{\alpha}^{(1)}+\frac{\delta t}{2\lambda}\Pi_{\alpha}^{*(1)}\right)}_{=0}\right] \\ = -2\beta\underbrace{\bar{\Pi}_0^{(2)}}_{=0} + \frac{\delta t}{\lambda}(1-\beta)\underbrace{\Pi_0^{*(2)}}_{=0} \label{0th_moment_O_eps_2}
    \end{split} \\
    \begin{split}
        \partial_t^{(2)}\Pi_{\alpha}^{\mathrm{eq}} + \partial_t^{(1)}\left[(1-\beta) \underbrace{\left(\bar{\Pi}_\alpha^{(1)}+\frac{\delta t}{2\lambda}\Pi_\alpha^{*(1)}\right)}_{=0}\right] + \partial_\beta \left[(1-\beta)\left(\bar{\Pi}_{\alpha\beta}^{(1)}+\frac{\delta t}{2\lambda}\Pi_{\alpha\beta}^{*(1)}\right)\right] \\ = -2\beta\underbrace{\bar{\Pi}_\alpha^{(2)}}_{=0} + \frac{\delta t}{\lambda}(1-\beta)\underbrace{\Pi_\alpha^{*(2)}}_{=0} \label{1st_moment_O_eps_2}
    \end{split}
\end{gather}
Summing up the $\mathcal{O}(\varepsilon)$ and $\mathcal{O}(\varepsilon^2)$ component equations in (\ref{0th_moment_O_eps_1}, \ref{1st_moment_O_eps_1}) and (\ref{0th_moment_O_eps_2}, \ref{1st_moment_O_eps_2}), respectively, and reversing the derivative expansions of \eqref{expansion_der}, we obtain
\begin{gather}
    \partial_t\Pi_0^\mathrm{eq} + \partial_\alpha\Pi_\alpha^\mathrm{eq} = 0, \label{mass_1} \\
    \partial_t\Pi_\alpha^\mathrm{eq} + \partial_\beta\Pi_{\alpha\beta}^\mathrm{eq} + \partial_\beta\left[\varepsilon(1-\beta)\left(\bar{\Pi}_{\alpha\beta}^{(1)}+\frac{\delta t}{2\lambda} \Pi_{\alpha\beta}^{*(1)} \right)\right] + 2\bar{\Pi}_\alpha^{(1)} = 0. \label{momentum_1}
\end{gather}
Then, taking the first-order moment of expansion \eqref{expansion_f}
\begin{equation}
    \bar{\Pi}_\alpha = \Pi^\mathrm{eq}_\alpha + \varepsilon\bar{\Pi}_\alpha^{(1)} + \varepsilon^2\underbrace{\bar{\Pi}_\alpha^{(2)}}_{=0}. \label{f_exp_Pi_1}
\end{equation}
Also, from \eqref{transformed_momentum_moment}, in non-dimensional form (lattice units) we have
\begin{equation}
    \rho'v'_\alpha = \sum_{i=0}^{Q-1}c_{i\alpha}'\bar{f}_i'+\frac{\delta t'}{2}\rho'a_\alpha'\quad\Longrightarrow\quad \rho v_\alpha = \bar{\Pi}_\alpha + \varepsilon\frac{\rho a_\alpha}{2} \label{non_dim_mom_transform}
\end{equation}
From zeroth-order equilibrium moment \eqref{Pi_0_eq} and (\ref{f_exp_Pi_1}, \ref{non_dim_mom_transform}) it follows that
\begin{equation}
    \bar{\Pi}_\alpha ^{(1)} = -\frac{\rho a_\alpha}{2}. \label{Pi_1_fbar_(1)}
\end{equation}
Next, using the first three equilibrium moments (\ref{Pi_0_eq}, \ref{Pi_1_eq}, \ref{Pi_2_eq}) and relation \eqref{Pi_1_fbar_(1)}, the conservation equations (\ref{mass_1} and \ref{momentum_1}) can be expressed as
\begin{gather}
    \partial_t(\epsilon\rho) + \partial_\alpha(\rho v_\alpha) = 0, \label{mass_2}\\
    \partial_t(\rho v_\alpha) + \partial_\beta\left(\frac{\rho v_\alpha v_\beta}{\epsilon}\right) = -\partial_\alpha(\epsilon\rho RT) + \partial_\beta \sigma_{\alpha\beta} + \rho a_\alpha, \label{momentum_2}
\end{gather}
which are the continuity and momentum equations with a yet unknown viscous stress tensor $\sigma_{\alpha\beta}$ given by
\begin{equation}
    \sigma_{\alpha \beta} = \delta t (\beta - 1)\left(\bar{\Pi}_{\alpha\beta}^{(1)} + \frac{\delta t}{2\lambda}\Pi_{\alpha\beta}^{*(1)}\right) \label{visc_1}
\end{equation}
by taking the second moment of $\mathcal{O}(\varepsilon)$ level equation \eqref{O_eps_1} and plugging it into \eqref{visc_1} we find
\begin{equation}
    \sigma_{\alpha\beta} = \delta t\left(\frac{1}{2}-\frac{1}{2\beta}\right)\left(\frac{\delta t}{\lambda}\Pi_{\alpha\beta}^{*(1)}-\partial_t^{(1)}\Pi_{\alpha\beta}^{\mathrm{eq}}-\partial_\gamma\Pi^\mathrm{eq}_{\alpha\beta\gamma} \right) \label{visc_2}
\end{equation}
Using \eqref{Pi_2_prod_general}, with $\xi_\alpha^*=v_\alpha/\epsilon+(\lambda/\epsilon)a_\alpha$ and $\mathcal{P}_{\alpha\alpha}^*=(\xi_\alpha^*)^2+RT+(\lambda/\epsilon\rho)\Phi_{\alpha\alpha}$, and the second-order moment of expansion \eqref{expansion_f*}, we get
\begin{gather}
    \Pi^*_{\alpha\beta} = \underbrace{\rho\left(\frac{v_\alpha v_\beta}{\epsilon}+\epsilon RT \delta_{\alpha\beta}\right)}_{=\Pi^\mathrm{eq}_{\alpha\beta}} + \underbrace{\lambda\left(\frac{\rho v_\alpha a_\beta}{\epsilon} + \frac{\rho a_\alpha v_\beta}{\epsilon} + \Phi_{\alpha\alpha}\delta_{\alpha\beta}\right)}_{=\delta t \Pi _{\alpha\beta}^{*(1)}} + \underbrace{\lambda^2\left(\frac{\rho a_\alpha a_\beta}{\epsilon}\right)}_{=\delta t^2\Pi_{\alpha\beta}^{*(2)}}, \label{Pi_2_*} \\
    \Longrightarrow\quad \frac{\delta t}{\lambda} \Pi_{\alpha \beta}^{*(1)} = \frac{\rho v_\alpha a_\beta}{\epsilon} + \frac{\rho a_\alpha v_\beta}{\epsilon} + \Phi_{\alpha\alpha}\delta_{\alpha\beta}. \label{Pi_2_*(1)}
\end{gather}
Note that the term $\Pi_{\alpha\beta}^{*(2)}\propto a_\alpha a_\beta$ in \eqref{Pi_2_*} is of order $\mathcal{O}(\varepsilon^2)$, and will not appear at the Navier-Stokes level hydrodynamics. Next, using the second-order equilibrium moment \eqref{Pi_2_eq} and product rule we can express its partial time derivative as
\begin{equation}
    \partial_t^{(1)}\Pi_{\alpha\beta}^\mathrm{eq} = \frac{v_\alpha}{\epsilon}\partial_t^{(1)}(\rho v_\beta) + \frac{v_\beta}{\epsilon}\partial_t^{(1)}(\rho v_\alpha) - \frac{v_\alpha v_\beta}{\epsilon^2}\partial_t^{(1)}(\epsilon\rho) + RT\delta_{\alpha\beta}\partial_t^{(1)}(\epsilon\rho). \label{dt_Pi_2_eq_1}
\end{equation}
Using $\mathcal{O}(\varepsilon)$ moment equations (\ref{0th_moment_O_eps_1}, \ref{1st_moment_O_eps_1}) with equilibrium moments (\ref{Pi_0_eq}, \ref{Pi_1_eq}, \ref{Pi_2_eq}) and \eqref{Pi_1_fbar_(1)} we obtain Euler-level conservation equations
\begin{gather}
    \partial_t^{(1)}(\epsilon\rho) = -\partial_\alpha (\rho v_\alpha), \label{euler_mass} \\
    \partial_t^{(1)}(\rho v_\alpha) + \partial_\beta\left(\frac{\rho v_\alpha v_\beta}{\epsilon}\right) + RT\partial_\alpha (\epsilon\rho) = \rho a _\alpha. \label{euler_momentum}
\end{gather}
We can then use the equations (\ref{euler_mass}, \ref{euler_momentum}) above to replace the time derivatives in \eqref{dt_Pi_2_eq_1} and subsequently rearrange to obtain
\begin{multline}
    \partial_t^{(1)}\Pi_{\alpha\beta}^\mathrm{eq} = -\partial_\gamma \left(\frac{\rho v_\alpha v_\beta v_\gamma}{\epsilon^2}\right) + \frac{\rho v_\alpha a_\beta}{\epsilon} + \frac{\rho a_\alpha v_\beta}{\epsilon} \\ - RT \left[\frac{v_\alpha}{\epsilon}\partial_\beta(\epsilon\rho) + \frac{v_\beta}{\epsilon}\partial_\alpha (\epsilon\rho) + \delta_{\alpha\beta}\partial_\gamma(\rho v_\gamma)\right]. \label{dt_Pi_2_eq_2}
\end{multline}
Next, using \eqref{Pi_3_prod_general} with $\xi_\alpha^{\mathrm{eq}}=v_\alpha/\epsilon$ and $\mathcal{P}_{\alpha\alpha}^\mathrm{eq}=(\xi_\alpha^{\mathrm{eq}})^2+RT$, we get the third-order equilibrium moment
\begin{equation}
    \Pi^\mathrm{eq}_{\alpha\beta\gamma} = \frac{\rho v_\alpha v_\beta v_\gamma}{\epsilon^2} + \rho RT (v_\alpha\delta_{\beta\gamma} + v_\beta\delta_{\alpha\gamma} + v_\gamma\delta_{\alpha\beta}) - \delta_{\alpha\beta\gamma} \left[\frac{\rho v_\alpha^3}{\epsilon^2} + \rho v_\alpha(3RT-1)\right], \label{Pi_3_eq}
\end{equation}
and find its partial space derivative to be
\begin{multline}
    \partial_\gamma \Pi^\mathrm{eq}_{\alpha\beta\gamma} = \partial_\gamma\left(\frac{\rho v_\alpha v_\beta v_\gamma}{\epsilon^2}\right) + RT\left[\partial_\beta(\rho v_\alpha) + \partial_\alpha(\rho v_\beta) + \delta_{\alpha\beta}\partial_\gamma(\rho v_\gamma) \right] \\ - \delta_{\alpha\beta\gamma}\partial_\gamma \left[\frac{\rho v_\alpha^3}{\epsilon^2} + \rho v_\alpha(3RT-1)\right]. \label{d_Pi_3_eq}
\end{multline}
Then, by plugging (\ref{Pi_2_*(1)}, \ref{dt_Pi_2_eq_2}, \ref{d_Pi_3_eq}) into \eqref{visc_2} and rearranging we obtain
\begin{multline}
    \sigma_{\alpha\beta} = \delta t \left(\frac{1}{2\beta} - \frac{1}{2}\right) \biggl(\epsilon\rho RT\left[\partial_\alpha\left(\frac{v_\beta}{\epsilon}\right) + \partial_\beta\left(\frac{v_\alpha}{\epsilon}\right)\right] \\ - \delta_{\alpha\beta}\biggl\{\Phi_{\alpha\alpha} + \partial_\beta\left[\frac{\rho v_\alpha^3}{\epsilon^2} + \rho v_\alpha (3RT-1)\right]\biggr\}\biggr). \label{visc_3}
\end{multline}
By setting
\begin{gather}
    \nu_{\mathrm{eff}} = \delta t \left(\frac{1}{2\beta}-\frac{1}{2}\right)RT \\
    \Phi_{\alpha\alpha} = \underbrace{\partial_\alpha\left[\rho v_\alpha (1-3RT)-\frac{\rho v_\alpha ^3}{\epsilon^2}\right]}_{=\Phi_{\alpha\alpha}^\mathrm{ex}} + \underbrace{\rho RT\left(\frac{2}{D}-\frac{\eta_\mathrm{eff}}{\nu_\mathrm{eff}}\right)(\partial_\gamma v_\gamma)}_{=\Phi^\eta}, \label{corr_full}
\end{gather}
the viscous stress in \eqref{visc_3} takes the form
\begin{equation}
    \sigma_{\alpha\beta} = \rho \nu_\mathrm{eff}\left[\epsilon\left(\partial_\alpha u_\beta + \partial_\beta u_\alpha\right) + \delta_{\alpha\beta}\left(\frac{\eta_\mathrm{eff}}{\nu_\mathrm{eff}}-\frac{2}{D}\right)(\partial_\gamma v_\gamma)\right], \label{visc_4}
\end{equation}
with $u_\alpha=v_\alpha /\epsilon$ being the intrinsic fluid velocity. Important to note that by setting the correction term $\Phi_{\alpha\alpha}$ as in \eqref{corr_full}, we:
\begin{itemize}
    \item Lift the temperature restriction $RT=RT_\mathrm{L}=1/3$.
    \item Avoid an error term $\mathcal{O}(v^3)$ in the hydrodynamic limit.
    \item Independently set $\eta_\mathrm{eff}$ as a free tunable parameter. 
\end{itemize}

\section{Second-order-in-time velocity evaluation}\label{Append:Picard_iter}
We want to solve the implicit equation
\begin{equation}
    c_1\bm{v} + c_2|\bm{v}|\bm{v} = \frac{\bar{\bm{\Pi}}_1}{\rho} + \frac{\delta t}{2}\left(\epsilon\bm{b}+RT\bm{\nabla}\epsilon\right) + \frac{\delta t}{2\rho}\bm{\nabla}\cdot\left(\rho \nu_\mathrm{eff}\left[\left(\frac{\bm{v}}{\epsilon}\right)\otimes\bm{\nabla}\epsilon +\bm{\nabla}\epsilon\otimes\left(\frac{\bm{v}}{\epsilon}\right)\right]\right),
\end{equation}
with $c_1 = 1+(\delta t/2)\mu_\mathrm{D}$ and $c_2 = (\delta t/2)\mu_\mathrm{F}$. To that end, we can use a Picard (fixed-point) iteration scheme, with the following the algorithm:
\begin{enumerate}
    \item Precompute the base predictor:
    \begin{equation}
        \bm{v}_{0} = \frac{\bar{\bm{\Pi}}_1}{\rho} + \frac{\delta t}{2}\left(\epsilon\bm{b}+RT\bm{\nabla}\epsilon\right).
    \end{equation}
    \item Initialize $\bm{v}^{(0)}$, using either
    \begin{itemize}
        \item previous time step velocity, or
        \item the one-shot scheme (\ref{v_(0)}-\ref{v_final}) results.
    \end{itemize}
    \item Iterate for $k=0,1,2,\dots$
    \begin{enumerate}
        \item Compute the updated RHS:
        \begin{equation}
            \bm{s}^{(k)} = \bm{v}_{0} + \frac{\delta t}{2\rho}\bm{\nabla}\cdot\left(\rho \nu_\mathrm{eff}\left[\left(\frac{\bm{v}^{(k)}}{\epsilon}\right)\otimes\bm{\nabla}\epsilon +\bm{\nabla}\epsilon\otimes\left(\frac{\bm{v}^{(k)}}{\epsilon}\right)\right]\right).
        \end{equation}
        \item Solve for the updated velocity using the drag relation:
        \begin{equation}
            \bm{v}^{(k+1)} = \frac{2\bm{s}^{(k)}}{c_1 + \sqrt{c_1^2 + 4c_2|\bm{s}^{(k)}|}}.
        \end{equation}
        \item (Optional but sometimes helpful) under-relaxation:
        \begin{equation}
            \bm{v}^{(k+1)} \leftarrow(1-\omega)\bm{v}^{(k)} + \omega\bm{v}^{(k+1)},\quad\omega\in(0,1].
        \end{equation}
        \item Stop when $\lvert\lvert\bm{v}^{(k+1)}-\bm{v}^{(k)}\rvert\rvert/\lvert\lvert\bm{v}^{(k+1)}\rvert\rvert$ is below a chosen tolerance $0<\varepsilon_\mathrm{tol}\ll 1$.
    \end{enumerate}
\end{enumerate}

\section{Temporal Fourier analysis} \label{Append:Spectral_Analysis}
\subsection{General solution} \label{Append:Spectral_Analysis:General}
Here we perform a temporal Fourier spectral analysis using a normal mode method. Consider an isothermal fluid flow in homogeneous isotropic stationary porous media with constant effective viscosities, i.e., $\nu_\mathrm{eff}$, $\eta_\mathrm{eff}$, $\epsilon$, $\mu_\mathrm{D}$, $\mu_\mathrm{F}$ and $RT$ are constants. Recall that the governing hydrodynamic equations are the continuity equation \eqref{continuity_target} and the momentum equation \eqref{momentum_target} with a viscous stress tensor \eqref{visc_stress_target}. For convenience in derivation, we introduce a strain operator $\bm{\pi}$ acting on the superficial velocity field $\bm{v}$,
\begin{equation}
    \bm{\pi}(\bm{v})=\frac{\bm{\sigma}_{\mathrm{eff}}}{\rho\nu_{\mathrm{eff}}} = \bm{\nabla}\bm{v} + (\bm{\nabla}\bm{v})^{\mathrm{T}} + \left(\frac{\eta_{\mathrm{eff}}}{\nu_{\mathrm{eff}}}-\frac{2}{D}\right)(\bm{\nabla}\cdot\bm{v})\mathbf{I}. \label{pi_v_strain}
\end{equation}
Now, consider some base flow solution $(\rho_0,\bm{v}_0)$ that satisfies the governing equations (\ref{continuity_target}, \ref{momentum_target}, \ref{visc_stress_target}) with $p=\rho RT$ and the aforementioned considerations,
\begin{gather}
    \epsilon\partial_t \rho_0 + \bm{\nabla}\cdot (\rho_0\bm{v}_0) = 0, \label{base_flow_continuity} \\
    \begin{split}
        \partial_t(\rho_0\bm{v}_0) &+ \frac{1}{\epsilon}\bm{\nabla}\cdot\left(\rho_0\bm{v}_0\bm{v}_0^{\mathrm{T}}\right) + \epsilon RT\bm{\nabla}\rho_0 \\
        &= \nu_\mathrm{eff} \bm{\nabla}\cdot\left[\rho_0\bm{\pi}(\bm{v}_0)\right] - \mu_\mathrm{D}\rho_0\bm{v}_0 - \mu_\mathrm{F}\rho_0\lvert\bm{v}_0\rvert\bm{v}_0+\epsilon\rho_0\bm{b}.
    \end{split} \label{base_flow_momentum}
\end{gather}
Suppose now that this base flow is perturbed by a disturbance flow field $(\tilde{\rho},\tilde{\bm{v}})$, so that the perturbed total flow field reads
\begin{equation}
    \begin{cases}
        \rho = \rho_0 + \tilde{\rho}\\
        \bm{v} = \bm{v}_0 + \tilde{\bm{v}}
    \end{cases}. \label{perturbed_flow}
\end{equation}
The perturbed flow \eqref{perturbed_flow} must also satisfy the governing hydrodynamic equations,
\begin{gather}
    \epsilon\partial_t(\rho_0+\tilde{\rho}) + \bm{\nabla}\cdot\left[(\rho_0+\tilde{\rho})(\bm{v}_0+\tilde{\bm{v}})\right] = 0, \label{perturbed_flow_continuity} \\
    \begin{split}
        \partial_t\left[(\rho_0+\tilde{\rho})(\bm{v}_0+\tilde{\bm{v}})\right] &+ \frac{1}{\epsilon}\bm{\nabla}\cdot\left[(\rho_0+\tilde{\rho})(\bm{v}_0+\tilde{\bm{v}})(\bm{v}_0+\tilde{\bm{v}})^\mathrm{T}\right] + \epsilon RT\bm{\nabla}(\rho_0+\tilde{\rho}) \\
        &= \nu_\mathrm{eff} \bm{\nabla}\cdot\left[(\rho_0+\tilde{\rho})\bm{\pi}(\bm{v}_0+\tilde{\bm{v}})\right] - \mu_\mathrm{D}(\rho_0+\tilde{\rho})(\bm{v}_0+\tilde{\bm{v}}) \\
        & - \mu_\mathrm{F}(\rho_0+\tilde{\rho})(\bm{v}_0+\tilde{\bm{v}})\left(\lvert\bm{v}_0+\tilde{\bm{v}}\rvert\right)+\epsilon(\rho_0+\tilde{\rho})\bm{b}.
    \end{split}\label{perturbed_flow_momentum}
\end{gather}
The general disturbance equations are then obtained by subtracting the governing equations for the base flow (\ref{base_flow_continuity}, \ref{base_flow_momentum}) from those for the perturbed flow (\ref{perturbed_flow_continuity}, \ref{perturbed_flow_momentum}) and read
\begin{gather}
    \epsilon\partial_t\tilde{\rho} + \bm{\nabla}\cdot\left(\rho_0\tilde{\bm{v}}+\tilde{\rho}\bm{v}_0+\tilde{\rho}\tilde{\bm{v}}\right) = 0, \label{general_dist_continuity} \\
    \begin{split}
        \partial_t(\rho_0\tilde{\bm{v}} &+ \tilde{\rho}\bm{v}_0 + \tilde{\rho}\tilde{\bm{v}}) \\ &+ \frac{1}{\epsilon}\bm{\nabla}\cdot\left[\rho_0\left(\bm{v}_0\tilde{\bm{v}}^\mathrm{T} + \tilde{\bm{v}}\bm{v}_0^\mathrm{T} + \tilde{\bm{v}}\tilde{\bm{v}}^\mathrm{T}\right) + \tilde{\rho}\left(\bm{v}_0\bm{v}_0^\mathrm{T} + \bm{v}_0\tilde{\bm{v}}^\mathrm{T} + \tilde{\bm{v}}\bm{v}_0^\mathrm{T} + \tilde{\bm{v}}\tilde{\bm{v}}^\mathrm{T}\right) \right] \\
        &= -\epsilon RT\bm{\nabla}\tilde{\rho} + \nu_\mathrm{eff} \bm{\nabla}\cdot\left[\rho_0\bm{\pi}(\tilde{\bm{v}}) + \tilde{\rho}\bm{\pi}(\bm{v}_0) + \tilde{\rho}\bm{\pi}(\tilde{\bm{v}}) \right] - \mu_\mathrm{D}(\rho_0\tilde{\bm{v}} + \tilde{\rho}\bm{v}_0 + \tilde{\rho}\tilde{\bm{v}}) \\
        & - \mu_\mathrm{F}\left[(\rho_0+\tilde{\rho})(\bm{v}_0+\tilde{\bm{v}})\left(\lvert\bm{v}_0+\tilde{\bm{v}}\rvert\right)-\rho_0\lvert\bm{v}_0\rvert\bm{v}_0\right]+\epsilon\tilde{\rho}\bm{b}.
    \end{split}\label{general_dist_momentum}
\end{gather}
The nonlinear partial differential equations (\ref{general_dist_continuity}, \ref{general_dist_momentum}) are the general disturbance equations. These generally cannot be solved analytically. We now proceed by making several simplifications to approximate an analytical solution for some specific cases of interest.

We assume that the disturbance amplitude is small, such that $\lvert\lvert\tilde{\rho}\rvert\rvert\ll\lvert\lvert\rho_0\rvert\rvert$ and $\lvert\lvert\tilde{\bm{v}}\rvert\rvert\ll\lvert\lvert\bm{v}_0\rvert\rvert$. Under this assumption, and by approximating the norm in \eqref{general_dist_momentum} by expanding it to the first-order in $\tilde{\bm{v}}$,
\begin{equation}
    \lvert\bm{v}_0+\tilde{\bm{v}}\rvert \approx \lvert\bm{v}_0\rvert + \frac{\bm{v}_0\cdot\tilde{\bm{v}}}{\lvert\bm{v}_0\rvert} + \mathcal{O}(\tilde{v}^2),
\end{equation}
we can linearize equations (\ref{general_dist_continuity}, \ref{general_dist_momentum}) by dropping all non-linear disturbance terms to obtain
\begin{gather}
    \epsilon\partial_t\tilde{\rho} + \bm{\nabla}\cdot\left(\rho_0\tilde{\bm{v}}+\tilde{\rho}\bm{v}_0\right) = 0, \label{gen_lin_dist_continuity} \\
    \begin{split}
        \partial_t(\rho_0\tilde{\bm{v}} + \tilde{\rho}\bm{v}_0) &+ \frac{1}{\epsilon}\bm{\nabla}\cdot\left[\rho_0\left(\bm{v}_0\tilde{\bm{v}}^\mathrm{T} + \tilde{\bm{v}}\bm{v}_0^\mathrm{T}\right) + \tilde{\rho}\bm{v}_0\bm{v}_0^\mathrm{T}\right] + \epsilon RT\bm{\nabla}\tilde{\rho} \\
        &= \nu_\mathrm{eff} \bm{\nabla}\cdot\left[\rho_0\bm{\pi}(\tilde{\bm{v}}) + \tilde{\rho}\bm{\pi}(\bm{v}_0)\right] - \mu_\mathrm{D}(\rho_0\tilde{\bm{v}} + \tilde{\rho}\bm{v}_0) \\
        & - \mu_\mathrm{F}\left[\rho_0 \left(\lvert\bm{v}_0\rvert\tilde{\bm{v}} + \frac{\bm{v}_0\cdot\tilde{\bm{v}}}{\lvert\bm{v}_0\rvert}\bm{v}_0\right) + \tilde{\rho}\lvert\bm{v}_0\rvert\bm{v}_0\right]+\epsilon\tilde{\rho}\bm{b}.
    \end{split}\label{gen_lin_dist_momentum}
\end{gather}
To simplify the analysis, we impose a homogeneous and steady base flow such that
\begin{align}
    \partial_t\rho_0&=0, &   \bm{\nabla}\rho_0&=\bm{0},   &   \partial_t\bm{v}_0&=\bm{0}, \nonumber \\
    \bm{\nabla}\cdot\bm{v}_0&=0,  &   \bm{\nabla}\bm{v}_0&=\bm{0},    &   \bm{\pi}(\bm{v}_0)&=\bm{0}. \label{steady_base_flow}
\end{align}
From the base flow momentum equation \eqref{base_flow_momentum}, the condition \eqref{steady_base_flow} implies that there must be a body force that counteracts the drag due to the porous medium, its acceleration being
\begin{equation}
    \bm{b} = \left(\mu_\mathrm{D} + \mu_\mathrm{F} |\bm{v}_0|\right)\frac{\bm{v}_0}{\epsilon}.\label{base_flow_force}
\end{equation}
Using (\ref{steady_base_flow}, \ref{base_flow_force}) and expanding $\bm{\pi}(\tilde{\bm{v}})$ using \eqref{pi_v_strain}, the linearized disturbance equations (\ref{gen_lin_dist_continuity}, \ref{gen_lin_dist_momentum}) simplify to
\begin{gather}
    \epsilon\partial_t\tilde{\rho} + \rho_0(\bm{\nabla}\cdot\tilde{\bm{v}}) + \bm{v}_0\cdot\bm{\nabla}\tilde{\rho} = 0, \label{simple_lin_dist_continuity} \\
    \begin{split}
        \rho_0\partial_t\tilde{\bm{v}} + \frac{\rho_0}{\epsilon}(\bm{v}_0&\cdot\bm{\nabla})\tilde{\bm{v}} + \epsilon RT\bm{\nabla}\tilde{\rho} \\
        &= \rho_0\nu_\mathrm{eff}\bm{\nabla}^2\tilde{\bm{v}} + \rho_0\left[\eta_\mathrm{eff} + \left(1-\frac{2}{D}\right)\nu_\mathrm{eff}\right]\bm{\nabla}(\bm{\nabla}\cdot\tilde{\bm{v}}) \\
        &- \mu_\mathrm{D}\rho_0\tilde{\bm{v}} - \mu_\mathrm{F}\rho_0\left(\lvert\bm{v}_0\rvert\tilde{\bm{v}} + \frac{\bm{v}_0\cdot\tilde{\bm{v}}}{\lvert\bm{v}_0\rvert}\bm{v}_0\right).
    \end{split} \label{simple_lin_dist_momentum}
\end{gather}
Following the normal mode method, we want to decompose the disturbance $(\tilde{\rho},\tilde{\bm{v}})$ into its Fourier modes. To that end, we introduce a normal mode ansatz for plane-wave disturbances of the form
\begin{equation}
   \{\tilde{\rho},\tilde{\bm{v}}\} = \{\hat{\rho},\hat{\bm{v}}\}\exp{\left[\ri(\bm{k}\cdot\bm{x}-\omega t)\right]}, \label{norm_mode_ansatz} 
\end{equation}
where $(\hat{\rho},\hat{\bm{v}})$ are complex amplitudes, $\bm{k}$ is the wave-vector, and $\omega$ is the angular frequency. Note that the complex conjugate term is omitted for the sake of clarity.

For the analysis using simple plane waves we consider constant amplitudes, i.e., $(\hat{\rho},\hat{\bm{v}})$ do not vary in space or time. In addition, the linearized system is time-invariant and, within a single normal mode, all variables $(\tilde{\rho},\tilde{\bm{v}})$ carry the same $(\bm{k},\omega)$. Also, as we are performing a temporal analysis, we consider real $\bm{k}\in\mathbb{R}^2$ and complex $\omega\in\mathbb{C}$. We can then reformulate \eqref{norm_mode_ansatz} as
\begin{equation}
   \{\tilde{\rho},\tilde{\bm{v}}\} = \{\hat{\rho},\hat{\bm{v}}\}\exp{\left[\Im{(\omega)}t\right]}\exp{\left[\ri(\bm{k}\cdot\bm{x}-\Re(\omega) t)\right]}, \label{norm_mode_ansatz_v2} 
\end{equation}
From equation \eqref{norm_mode_ansatz_v2}, we see that the imaginary part of the frequency $\Im{(\omega)}$ is the temporal growth rate while the real part $\Re(\omega)$ corresponds to the propagation of the perturbation at phase speed $c=\Re(\omega)/\lvert\bm{k}\rvert$.

Plugging the normal mode ansatz \eqref{norm_mode_ansatz} into (\ref{simple_lin_dist_continuity}, \ref{simple_lin_dist_momentum}), we obtain
\begin{gather}
    \omega\epsilon\hat{\rho} = (\bm{v}_0\cdot\bm{k})\hat{\rho} + (\bm{k}\cdot\hat{\bm{v}})\rho_0, \label{continuity_ansatz} \\
    \begin{split}
        \ri\omega\rho_0\hat{\bm{v}} &= \ri\frac{\rho_0}{\epsilon}(\bm{v}_0\cdot\bm{k})\hat{\bm{v}} + \ri\epsilon RT \hat{\rho}\bm{k} \\
        &+ \rho_0\nu_\mathrm{eff}(\bm{k}\cdot\bm{k})\hat{\bm{v}} + \rho_0\left[\eta_\mathrm{eff} + \left(1-\frac{2}{D}\right)\nu_\mathrm{eff}\right](\bm{k}\cdot\hat{\bm{v}})\bm{k} \\
        &+ \mu_\mathrm{D}\rho_0\hat{\bm{v}} + \mu_\mathrm{F}\rho_0\left(\lvert\bm{v}_0\rvert\hat{\bm{v}}+\frac{\bm{v}_0\cdot\hat{\bm{v}}}{\lvert\bm{v}_0\rvert}\bm{v}_0\right).
    \end{split} \label{momentum_ansatz}
\end{gather}
From \eqref{momentum_ansatz} we note that the base flow $\bm{v}_0$ not only enters through the scalar shift $\bm{v}_0\cdot\bm{k}$, but it also appears inside the Forchheimer drag term: $\left[\left(\bm{v}_0\cdot\hat{\bm{v}}\right)/\lvert\bm{v}_0\rvert\right]\bm{v}_0\propto\bm{v}_0\bm{v}_0^\mathrm{T}$. This term is anisotropic, as $\bm{v}_0\bm{v}_0^\mathrm{T}$ projects along $\bm{v}_0$ and thus couples the transverse velocity to the longitudinal part when $\bm{v}_0$ is not parallel to $\bm{k}$. The significance of this anisotropy will be addressed later.

In two-dimensional Cartesian space $\bm{x}=(x,y)$, the system of equations (\ref{continuity_ansatz}, \ref{momentum_ansatz}) can be expressed as a standard $3\times3$ eigenproblem: $\omega\bm{q}=\bm{M}\bm{q}$, with $\bm{q}=(\hat{\rho}, \hat{v}_x, \hat{v}_y)$ and
\begin{equation}
    \bm{M} =
    \begin{bmatrix}
        \frac{\bm{v}_0\cdot\bm{k}}{\epsilon} & \frac{\rho_0 k_x}{\epsilon} & \frac{\rho_0 k_y}{\epsilon}\\
        \frac{\epsilon RT k_x}{\rho_0} & \frac{\bm{v}_0\cdot\bm{k}}{\epsilon} - \ri Z_x & -\ri\left(\eta_\mathrm{eff}k_x k_y + \mu_\mathrm{F}\frac{v_{0,x}v_{0,y}}{\lvert\bm{v}_0\rvert}\right)\\
        \frac{\epsilon RT k_y}{\rho_0} & -\ri\left(\eta_\mathrm{eff}k_x k_y + \mu_\mathrm{F}\frac{v_{0,x}v_{0,y}}{\lvert\bm{v}_0\rvert}\right) & \frac{\bm{v}_0\cdot\bm{k}}{\epsilon} - \ri Z_y
    \end{bmatrix}, \label{matrix_M}
\end{equation}
where $Z_\alpha$, $\alpha\in\{x,y\}$, is given by:
\begin{equation}
    Z_\alpha = \nu_\mathrm{eff}(\bm{k}\cdot\bm{k}) + \eta_\mathrm{eff}k_\alpha^2+\mu_\mathrm{D}+\mu_\mathrm{F}\left(\lvert\bm{v}_0\rvert+\frac{v_{0,\alpha}^2}{\lvert\bm{v}_0\rvert}\right). \label{Z_alpha}
\end{equation}
The necessary and sufficient condition for non-trivial solutions of the eigenproblem is given by the cubic equation: $\det{\left(\bm{M}-\omega\mathbf{I}\right)}=0$. Finding the roots of this characteristic equation, i.e., the eigenvalues, through factorization is not trivial, yet it will help us gain some physical understanding about the nature of the roots and the temporal behavior of the system. To factorize the characteristic equation, we proceed by splitting velocity vectors into components parallel and perpendicular to the wave-vector $\bm{k}$. To that end, let us introduce the following notation:
\begin{equation}
    \hat{\bm{k}}_\parallel=\frac{\bm{k}}{\lvert\bm{k}\rvert}=\begin{pmatrix}\hat{k}_x\\ \hat{k}_y \end{pmatrix},\quad \hat{\bm{k}}_\perp=\begin{pmatrix}-\hat{k}_y\\ \hat{k}_x \end{pmatrix},\quad \bm{R} = \begin{bmatrix}\hat{k}_x & \hat{k}_y\\ -\hat{k}_y & \hat{k}_x \end{bmatrix}. \label{change_basis_notation}
\end{equation}
Using the rotation matrix $\bm{R}$, we can then change the basis of the base flow and disturbance velocity as follows,
\begin{gather}
    \begin{pmatrix}v_{0,\parallel}\\ v_{0,\perp} \end{pmatrix} = \bm{R}\begin{pmatrix}v_{0,x}\\ v_{0,y} \end{pmatrix} = \begin{pmatrix}\hat{\bm{k}}_\parallel\cdot\bm{v}_0\\ \hat{\bm{k}}_\perp\cdot\bm{v}_0 \end{pmatrix},\\
    \begin{pmatrix}\hat{v}_{\parallel}\\ \hat{v}_{\perp} \end{pmatrix} = \bm{R}\begin{pmatrix}\hat{v}_{x}\\ \hat{v}_{y} \end{pmatrix} = \begin{pmatrix}\hat{\bm{k}}_\parallel\cdot\hat{\bm{v}}\\ \hat{\bm{k}}_\perp\cdot\hat{\bm{v}} \end{pmatrix}.
\end{gather}
The eigenproblem can then be reformulated as: $\omega\bm{q}'=\bm{M}'\bm{q}'$, with $\bm{q}'=\diag{(1,\bm{R})}\bm{q}=(\hat{\rho}, \hat{v}_\parallel, \hat{v}_\perp)$ and the transformed matrix $\bm{M}'=\diag{(1,\bm{R})}\,\bm{M}\,\diag{(1,\bm{R}^\mathrm{T})}$, expressed explicitly using (\ref{matrix_M}, \ref{Z_alpha}, \ref{change_basis_notation}) as
\begin{equation}
    \bm{M}'=
    \begin{bmatrix}
        \frac{V_0 k\cos{\theta}}{\epsilon} & \frac{\rho_0 k}{\epsilon} & 0\\
        \frac{\epsilon RT k}{\rho_0} & \frac{V_0 k\cos{\theta}}{\epsilon}-\ri Z_\parallel & -\ri\mu_\mathrm{F} V_0 \cos{\theta}\sin{\theta}\\
        0 & -\ri\mu_\mathrm{F} V_0 \cos{\theta}\sin{\theta} & \frac{V_0 k\cos{\theta}}{\epsilon} - \ri Z_\perp
    \end{bmatrix},\label{matrix_M'}
\end{equation}
where $Z_\parallel$ and $Z_\perp$ are given by:
\begin{gather}
   Z_\parallel = (\nu_\mathrm{eff}+\eta_\mathrm{eff})k^2+\mu_\mathrm{D}+\mu_\mathrm{F} V_0(1+\cos^2{\theta}), \label{Z_paral} \\
   Z_\perp = \nu_\mathrm{eff}k^2+\mu_\mathrm{D}+\mu_\mathrm{F} V_0(1+\sin^2{\theta}). \label{Z_perp}
\end{gather}
Note that here we have further simplified some notation: $k=\lvert\bm{k}\rvert$, $V_0=\lvert\bm{v}_0\rvert$, and used $v_{0,\parallel}=V_0\cos{\theta}$ and $v_{0,\perp}=V_0\sin{\theta}$, with $\theta$ being the angle between $\bm{k}$ and $\bm{v}_0$ (taken positive in the anticlockwise direction).

We now have a modified eigenproblem characteristic equation: $\det{(\bm{M}'-\omega\mathbf{I})}=0$. A closer look at the matrix $\bm{M}'$ in \eqref{matrix_M'} with (\ref{Z_paral}, \ref{Z_perp}) reveals that the resulting $3\times3$ polynomial will still not easily factor in the general case. This is due to the anisotropy mentioned before. For $\bm{M}'$ to gain a block-diagonal structure (i.e., $\bm{M}'_{12}=\bm{M}'_{21}=0$) and factorize into transverse and longitudinal modes, there are three possible conditions:
\begin{enumerate}
    \item The base flow is aligned with the wave-vector: $\bm{v}_0\parallel\bm{k}$ (e.g., $\theta=0$).
    \item The base flow is perpendicular to the wave-vector: $\bm{v}_0\perp\bm{k}$ (e.g., $\theta=\pi/2$).
	\item The nonlinear Forchheimer drag is neglected: $\mu_\mathrm{F}=0$.
\end{enumerate}
For the sake of generality, we shall keep the Forchheimer drag and consider the first two conditions. Solving the eigenproblem gives the following shear (transverse) mode and two acoustic (longitudinal) modes,
\begin{gather}
    \omega_\theta^\mathrm{sh} = \frac{V_0 k\cos{\theta}}{\epsilon} - \ri\left[\nu_\mathrm{eff} k^2+\mu_\mathrm{D} + (1+\gamma_\theta)\mu_\mathrm{F} V_0\right], \label{w_sh_general} \\
    \begin{split}
        \omega_\theta^\mathrm{ac\pm} = \frac{V_0 k\cos{\theta}}{\epsilon} &- \frac{\ri}{2}\left[(\nu_\mathrm{eff}+\eta_\mathrm{eff})k^2 + \mu_\mathrm{D} + (2-\gamma_\theta)\mu_\mathrm{F} V_0\right] \\
        &\pm \sqrt{RTk^2 -\frac{1}{4}\left[(\nu_\mathrm{eff}+\eta_\mathrm{eff})k^2 + \mu_\mathrm{D} + (2-\gamma_\theta)\mu_\mathrm{F} V_0\right]^2}. \label{w_ac_general}
    \end{split}
\end{gather}
Here $\gamma_\theta$ is a factor depending on the flow arrangement, i.e., on $\theta$. When the base flow is parallel to the wavevector $\bm{v}_0\parallel\bm{k}$ (i.e., $v_{0,\perp}=0$) it will be $\gamma_\parallel=0$. On the other hand, when the base flow is perpendicular to the wavevector $\bm{v}_0\perp\bm{k}$ (i.e., $v_{0,\parallel}=0$), one has $\gamma_\perp=1$.

We are mostly interested in the imaginary parts of the eigenvalues $(\omega_\theta^\mathrm{sh}, \omega_\theta^\mathrm{ac\pm})$ given by (\ref{w_sh_general}, \ref{w_ac_general}), as these give us the temporal growth/decay rate of the disturbance. The general equation \eqref{w_ac_general} for the acoustic modes $\omega_\theta^\mathrm{ac\pm}$ could use some further simplification of the square root to gain an understanding of whether it will contribute to the real or the imaginary part of the angular frequency. To that end, we consider a flow domain that is space periodic in $\bm{x}$ with a period $\bm{L}$. Then, from the normal mode ansatz \eqref{norm_mode_ansatz}, we must have a wave-vector of the form
\begin{equation}
    \bm{k} = \begin{pmatrix}
        \frac{2\pi n_x}{L_x} \\
        \frac{2\pi n_y}{L_y}
    \end{pmatrix}
\end{equation}
Suppose the flow domain is a square with $L_x=L_y=L$, and the base flow is aligned with one of the coordinate axes. For the two configurations considered ($\bm{v}_0\parallel\bm{k}$ or $\bm{v}_0\perp\bm{k}$), we select the Fourier mode indices as either $(n_x,n_y)=(1,0)$ or $(0,1)$. With this choice, the wave-vector magnitude is the same in all cases, namely $k=2\pi/L$.

We then take $k$ as a small parameter as $k\propto L^{-1}\propto\mathrm{Kn}$, and in a well resolved LBM simulation we will have $k\ll 1$. Then, in the long-wavelength limit $k\rightarrow 0$ ($L\rightarrow\infty$), we can approximate a square root of a generic polynomial in $k$ as
\begin{equation}
    \sqrt{c_1 + c_2 k + c_3 k^2} \approx \sqrt{c_1} + \mathcal{O}(k), \label{sqrt_approx_O_k}
\end{equation}
where $c_i$, $i\in\{1,2,3\}$, are arbitrary constants and $c_1\neq 0$.

With the considerations above and using \eqref{sqrt_approx_O_k}, the root in \eqref{w_ac_general} can be approximated by asymptotic scaling (in the limit $k\rightarrow 0$). Two different approaches, differing in the choice of drag parameters $\mu_\mathrm{D}$ and $\mu_\mathrm{F}$, will be explored in the following two sections.

\subsection{Balanced-damping scaling} \label{Append:Spectral_Analysis:VDB}
To simplify the linear analysis and minimize the influence of higher-order (non-hydrodynamic) moments introduced by the forcing scheme \eqref{F_RtFM}, we can perform matched-asymptotic scaling with a viscous-drag balance, where we choose the parameters so that the damping contributions remain comparable in magnitude, i.e., the viscous damping is of the same order as the drag-induced damping. To that end, we set
\begin{equation}
    \mu_\mathrm{D} = \nu_\mathrm{f} k^2,\quad \mu_\mathrm{F} = \frac{\nu_\mathrm{f} k^2}{\epsilon\sqrt{RT}}, \label{balanced_damping}
\end{equation}
and introduce an advection Mach number for the base flow defined as: $\mathrm{Ma}_0=V_0/(\epsilon\sqrt{RT})$. Using this with the temperature and viscosity ratios introduced in the main text \eqref{ratios_def}, equations (\ref{w_sh_general}, \ref{w_ac_general}) can be rewritten as
\begin{gather}
    \omega^\mathrm{sh}_\theta = \mathrm{Ma}_0\sqrt{J_T RT_\mathrm{L}}\cos{\theta}\,k - \ri\left[J_{\nu_\mathrm{eff}} + 1 + (1+\gamma_\theta)\mathrm{Ma}_0\right]\nu_\mathrm{f} k^2,\label{w_sh_balanced} \\
    \begin{split}
        \omega^\mathrm{ac\pm}_\theta = \mathrm{Ma}_0 & \sqrt{J_T RT_\mathrm{L}}\cos{\theta}\,k - \frac{\ri}{2}\left[J_{\nu_\mathrm{eff}} + J_{\eta_\mathrm{eff}} + 1 + (2-\gamma_\theta)\mathrm{Ma}_0\right]\nu_\mathrm{f} k^2 \\
        &\pm k\sqrt{J_T RT_\mathrm{L} - \frac{1}{4}\left[J_{\nu_\mathrm{eff}} + J_{\eta_\mathrm{eff}} + 1 + (2-\gamma_\theta)\mathrm{Ma}_0\right]^2 \nu_\mathrm{f}^2 k^2}.
    \end{split} \label{w_ac_balanced_sqrt}
\end{gather}
Using \eqref{sqrt_approx_O_k}, $\omega^\mathrm{ac\pm}_\theta$ in \eqref{w_ac_balanced_sqrt} can be further approximated up to $\mathcal{O}(k^2)$ as
\begin{equation}
    \omega^\mathrm{ac\pm}_\theta \approx k\sqrt{J_T RT_\mathrm{L}}\left(\mathrm{Ma}_0\cos{\theta}\pm 1\right) - \frac{\ri}{2}\left[J_{\nu_\mathrm{eff}} + J_{\eta_\mathrm{eff}} + 1 + (2-\gamma_\theta)\mathrm{Ma}_0\right]\nu_\mathrm{f} k^2 + \mathcal{O}(k^2). \label{w_ac_balanced_approx}
\end{equation}

\subsection{Constant Darcy number scaling} \label{Append:Spectral_Analysis:constDa}
Introducing the Darcy number $\mathrm{Da}=\kappa/L^2$ that relates the permeability $\kappa$ to the domain size $L$. Then, with $k=2\pi/L$, the relations for the drag coefficients in (\ref{mu_D_Ergun_perm}, \ref{mu_F_Ergun_perm}) can be expressed as
\begin{equation}
    \mu_\mathrm{D} = \frac{\epsilon\nu_\mathrm{f}}{4\pi^2\mathrm{Da}}k^2,\quad \mu_\mathrm{F} = \frac{\epsilon F_\epsilon}{2\pi\sqrt{\mathrm{Da}}}k. \label{const_Da}
\end{equation}
That is, keeping $\mathrm{Da}$ constant, the drag coefficients scale like $\mu_\mathrm{D}\propto k^2$ and $\mu_\mathrm{F}\propto k$. Then, plugging in \eqref{const_Da} into (\ref{w_sh_general}, \ref{w_ac_general}) and approximating the square root using \eqref{sqrt_approx_O_k}, the eigenvalues can be expressed, with the help of the dimensionless parameters, as follows
\begin{gather}
    \begin{split}
        \omega_\theta^\mathrm{sh} = \left(\frac{2\pi}{L}\right)&\mathrm{Ma}_0\sqrt{J_T RT_\mathrm{L}}\cos{\theta} \\
        &- \ri \left\{ \nu_\mathrm{f}\left[J_{\nu_\mathrm{eff}}\left(\frac{2\pi}{L}\right)^2+\frac{\epsilon}{L^2 \mathrm{Da}}\right] + (1+\gamma_\theta)\frac{\epsilon^2F_\epsilon\mathrm{Ma}_0}{L}\sqrt{\frac{J_T RT_\mathrm{L}}{\mathrm{Da}}} \right\},\label{w_sh_constDa_long}
    \end{split} \\
    \begin{split}
        \omega_\theta^\mathrm{ac\pm} \approx &\left(\frac{2\pi}{L}\right)\sqrt{J_T RT_\mathrm{L}}\left[ \mathrm{Ma}_0\cos{\theta}\pm\sqrt{1-(2-\gamma_\theta)^2\frac{\epsilon^4 F_\epsilon^2 \mathrm{Ma}_0^2}{16\pi^2 \mathrm{Da}}} \right] + \mathcal{O}(k^2)\\
        &- \frac{\ri}{2}\left\{\nu_\mathrm{f}\left[(J_{\nu_\mathrm{eff}}+J_{\eta_\mathrm{eff}})\left(\frac{2\pi}{L}\right)^2 + \frac{\epsilon}{L^2\mathrm{Da}}\right]+ (2-\gamma_\theta)\frac{\epsilon^2F_\epsilon \mathrm{Ma}_0}{L}\sqrt{\frac{J_T RT_\mathrm{L}}{\mathrm{Da}}} \right\}.
    \end{split} \label{w_ac_constDa_approx_long}
\end{gather}
Note that if
\begin{equation}
    \mathrm{Da} > (2-\gamma_\theta)^2\frac{\epsilon^4F_\epsilon^2\mathrm{Ma}_0^2}{16\pi^2},
\end{equation}
and $L$ is sufficiently large, the imaginary parts of the acoustic eigenvalues \eqref{w_ac_constDa_approx_long} remain the same.

Equivalently, we can also express the acoustic eigenvalue \eqref{w_ac_constDa_approx_long} as
\begin{equation}
    \begin{split}
        \omega_\theta^\mathrm{ac\pm} \approx \frac{V_0 k\cos{\theta}}{\epsilon} &\pm \sqrt{RT k^2-\left(\frac{(2-\gamma_\theta)\mu_\mathrm{F} V_0}{2}\right)^2} + \mathcal{O}(k^2) \\
        &- \frac{\ri}{2}\left[(\nu_\mathrm{eff}+\eta_\mathrm{eff})k^2 +\mu_\mathrm{D} + (2-\gamma_\theta)\mu_\mathrm{F} V_0\right].
    \end{split} \label{w_ac_constDa_approx_short}
\end{equation}
We note from the approximated expressions for $\omega_\theta^\mathrm{ac\pm}$ in (\ref{w_ac_constDa_approx_long}, \ref{w_ac_constDa_approx_short}) that in the long-wavelength limit under a constant Darcy number ($\mathrm{Da}=\text{const.}$), the acoustic phase speed is influenced by the inertial Forchheimer drag,
\begin{equation}
    \begin{split}
        c_\mathrm{ac}^\mathrm{approx} &\approx \frac{1}{k}\sqrt{RT k^2-\left(\frac{(2-\gamma_\theta)\mu_\mathrm{F} V_0}{2}\right)^2} + \mathcal{O}(k) \\ &= \underbrace{\sqrt{J_T RT_\mathrm{L}\left(1-(2-\gamma_\theta)^2\frac{\epsilon^4 F_\epsilon^2 \mathrm{Ma}_0^2}{16\pi^2 \mathrm{Da}}\right)}}_{c_\mathrm{s,eff}} + \mathcal{O}(k),
    \end{split} \label{c_ac_approx_append}
\end{equation}
that is, it is a function of base flow velocity, medium porosity and the Darcy number. Note that this approximation to leading order gives the effective (or apparent) speed of sound $c_\mathrm{s,eff}$.

\section{Lid-driven free-fluid cavity}\label{Append:CavityFreeFluidRef}
As discussed in Section \ref{sec:target_macro}, the macroscopic hydrodynamic equations recovered by the present model, namely the volume-averaged Navier-Stokes equations, reduce to the standard Navier-Stokes equations in the limit $\epsilon \to 1$. The model can therefore also be applied to free-fluid flows. To support this validation, we consider the regularized lid-driven flow in a free-fluid cavity, using the domain configuration shown in Figure \ref{cavity_domain} and the same simulation setup as described in Section \ref{sec:results:cavity}. Here, a uniform porosity profile $\epsilon=1$ is prescribed, corresponding to a homogeneous free-fluid domain.

\begin{figure}[h]
\centering
\includegraphics[width=0.95\textwidth]{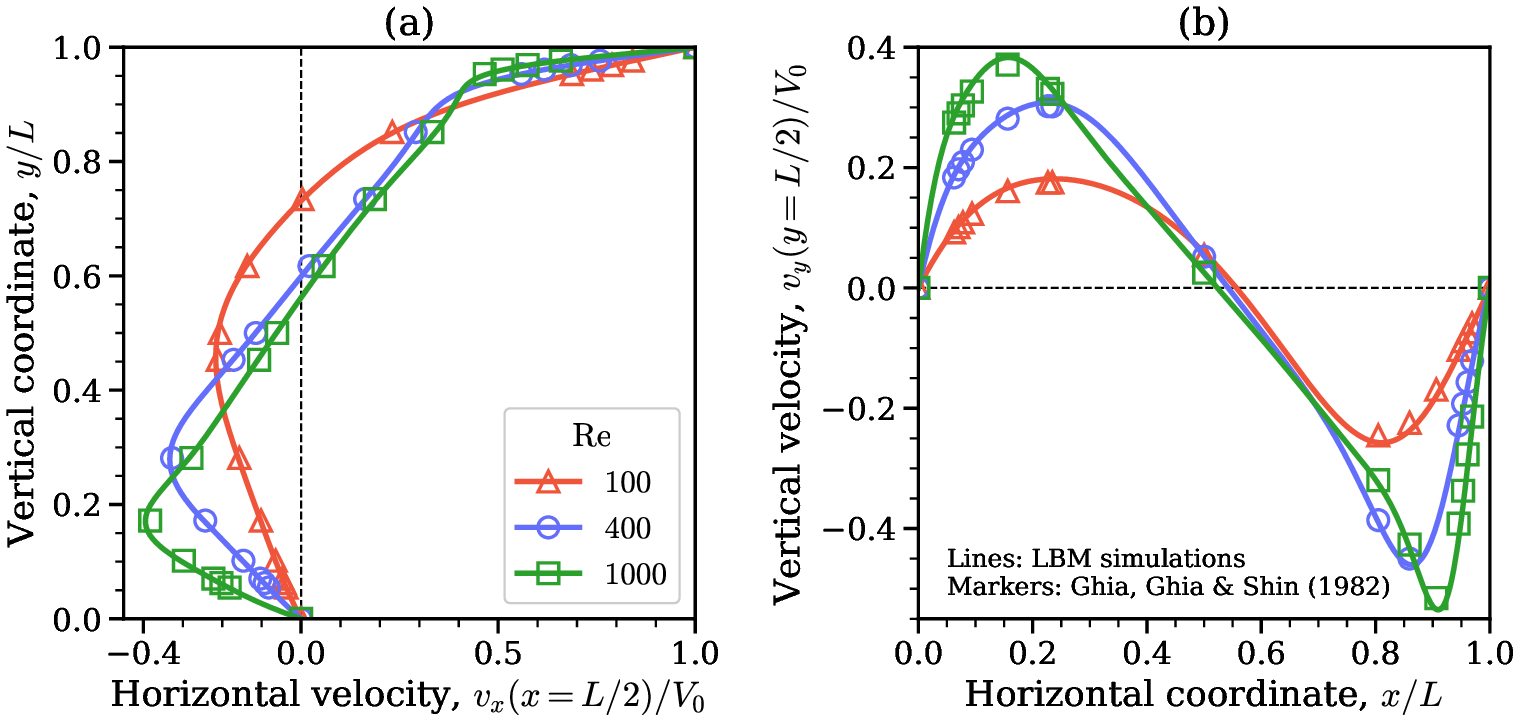}
\caption{Horizontal and vertical velocity profiles for the steady regularized lid-driven flow in a free-fluid cavity at different Reynolds numbers $\mathrm{Re}$. Panel (a) shows the horizontal velocity component $v_x$ along the vertical centerline of the cavity, while panel (b) shows the vertical velocity component $v_y$ along the horizontal centerline. Solid lines denote the present LBM results obtained on a $301\times301$ lattice using the configuration described in Section \ref{sec:results:cavity}, with $\epsilon_\mathrm{core}=1$, corresponding to a homogeneous free-fluid domain. Markers indicate the classical benchmark solutions of \cite{Ghia_1982_Cavity_Reference} for the standard lid-driven cavity. The dashed black lines indicate zero velocity.}\label{cavity_purefluid_ref}
\end{figure}

Figure \ref{cavity_purefluid_ref} compares the steady-state horizontal and vertical velocity profiles along the cavity centerlines, obtained with the present LBM model for different Reynolds numbers $\mathrm{Re}$, with the classical benchmark data of \cite{Ghia_1982_Cavity_Reference}. Overall, the present results show good agreement with the benchmark solutions for the cases considered. Small deviations are observed near the extrema of the velocity profiles, where the regularized-lid simulations predict slightly larger absolute values of both velocity components. These differences are attributed to the fact that the present configuration uses the regularized lid-velocity profile defined by \eqref{cavity_velocity_profile}, for which the velocity smoothly vanishes at the top corners where the moving lid meets the stationary side walls. In contrast, the benchmark data correspond to the classical lid-driven cavity with a uniformly moving lid and discontinuous velocity at the top corners.

The appearance of slightly larger extrema in the regularized case may seem counterintuitive, since the regularized lid imposes a lower average wall velocity than a fully uniform lid. However, the two configurations differ not only in the total imposed wall motion, but also in the way vorticity is generated near the top corners. In the classical cavity, the discontinuity between the moving lid and the stationary side walls produces strong localized corner shear and pronounced corner vortical structures, which may enhance local dissipation. By smoothing the lid velocity near the corners, the present regularization weakens these localized corner effects and modifies the momentum transfer from the lid to the primary vortex. As a result, the bulk circulation, and hence the centerline velocity extrema, can be slightly larger despite the lower average lid velocity.

\end{appendices}

\clearpage

\bibliography{sn-bibliography}

@article{Hosseini_Karlin_Shallow_Water_2025,
    author = {Hosseini, S. A. and Karlin, I. V.},
    title = {{Consistent lattice Boltzmann model for shallow water equations}},
    journal = {Physics of Fluids},
    volume = {37},
    number = {10},
    pages = {102114},
    year = {2025},
    month = {10},
    issn = {1070-6631},
    doi = {10.1063/5.0295340},
}

@book{Kruger_LBM_2016,
    title={The Lattice Boltzmann Method: Principles and Practice},
    author={Kr{\"u}ger, Timm and Kusumaatmaja, Halim and Kuzmin, Alexandr and Shardt, Orest and Silva, Goncalo and Viggen, Erlend Magnus},
    year={2016},
    address = {Cham},
    publisher={Springer}
}

@article{Whitaker_1973_multiphase_systems,
    title = {The transport equations for multi-phase systems},
    journal = {Chemical Engineering Science},
    volume = {28},
    number = {1},
    pages = {139-147},
    year = {1973},
    issn = {0009-2509},
    doi = {10.1016/0009-2509(73)85094-8},
    author = {Stephen Whitaker}
}

@Inbook{Schaefer1996,
    author="Sch{\"a}fer, M. and Turek, S. and Durst, F. and Krause, E. and Rannacher, R.",
    editor="Hirschel, Ernst Heinrich",
    title="Benchmark Computations of Laminar Flow Around a Cylinder",
    bookTitle="Flow Simulation with High-Performance Computers II: DFG Priority Research Programme Results 1993-1995",
    year="1996",
    publisher="Vieweg+Teubner Verlag",
    address="Wiesbaden",
    pages="547-566",
    isbn="978-3-322-89849-4",
    doi="10.1007/978-3-322-89849-4_39"
}

@article{Chen_PorousCylinder_2025,
    author = {Chen, Zepeng and Liu, Yingzheng and Sung, Hyung Jin},
    title = {{Influence of Darcy and Reynolds numbers on flow past a porous circular cylinder}},
    journal = {Physics of Fluids},
    volume = {37},
    number = {8},
    pages = {083635},
    year = {2025},
    month = {08},
    issn = {1070-6631},
    doi = {10.1063/5.0284671}
    }

@article{Nguyen_2023_confined_cylinders_review,
    author = {Nguyen, Quang Duy and Lu, Wilson and Chan, Leon and Ooi, Andrew and Lei, Chengwang},
    title = {A state-of-the-art review of flows past confined circular cylinders},
    journal = {Physics of Fluids},
    volume = {35},
    number = {7},
    pages = {071301},
    year = {2023},
    month = {07},
    issn = {1070-6631},
    doi = {10.1063/5.0157470}
    }

@article{Forouzi_2022_review_experimental_bluff_bodies,
    author = {Forouzi Feshalami, Behzad and He, Shuisheng and Scarano, Fulvio and Gan, Lian and Morton, Chris},
    title = {A review of experiments on stationary bluff body wakes},
    journal = {Physics of Fluids},
    volume = {34},
    number = {1},
    pages = {011301},
    year = {2022},
    month = {01},
    issn = {1070-6631},
    doi = {10.1063/5.0077323}
}

@article{Cummins_2017_permeable_disk,
    author = {Cummins, Cathal and Viola, Ignazio Maria and Mastropaolo, Enrico and Nakayama, Naomi},
    title = {The effect of permeability on the flow past permeable disks at low Reynolds numbers},
    journal = {Physics of Fluids},
    volume = {29},
    number = {9},
    pages = {097103},
    year = {2017},
    month = {09},
    issn = {1070-6631},
    doi = {10.1063/1.5001342}
}

@article{Tang_2021_permeable_disk,
  title = {Effect of aspect ratio on flow through and around a porous disk},
  author = {Tang, Tingting and Xie, Jin and Yu, Shimin and Li, Jianhui and Yu, Peng},
  journal = {Phys. Rev. Fluids},
  volume = {6},
  issue = {7},
  pages = {074101},
  numpages = {25},
  year = {2021},
  month = {Jul},
  publisher = {American Physical Society},
  doi = {10.1103/PhysRevFluids.6.074101}
}

@article{Peng_2012_porous_sphere,
    title = {Numerical simulation on steady flow around and through a porous sphere},
    journal = {International Journal of Heat and Fluid Flow},
    volume = {36},
    pages = {142-152},
    year = {2012},
    issn = {0142-727X},
    doi = {10.1016/j.ijheatfluidflow.2012.03.002},
    author = {Peng Yu and Yan Zeng and Thong See Lee and Xiao Bing Chen and Hong Tong Low}
}

@article{Ciuti_2021_porous_sphere,
    author = {Ciuti, M. and Zampogna, G. A. and Gallaire, F. and Camarri, S. and Ledda, P. G.},
    title = {On the effect of a penetrating recirculation region on the bifurcations of the flow past a permeable sphere},
    journal = {Physics of Fluids},
    volume = {33},
    number = {12},
    pages = {124103},
    year = {2021},
    month = {12},
    issn = {1070-6631},
    doi = {10.1063/5.0075244}
}

@article{Seol_Kim_Kim_2024,
    title={The effect of permeability on the flow structure of porous square cylinders},
    volume={985},
    DOI={10.1017/jfm.2024.311},
    journal={Journal of Fluid Mechanics},
    author={Seol, Chansoo and Kim, Taewoo and Kim, Taehoon},
    year={2024},
    pages={A29}
}

@article{Bhattacharyya_2006_porous_cylinder,
    title = {Fluid motion around and through a porous cylinder},
    journal = {Chemical Engineering Science},
    volume = {61},
    number = {13},
    pages = {4451-4461},
    year = {2006},
    note = {The John Bridgwater Symposium: "Shaping the Future of Chemical Engineering"},
    issn = {0009-2509},
    doi = {10.1016/j.ces.2006.02.012},
    author = {S. Bhattacharyya and S. Dhinakaran and A. Khalili}
}

@article{Yu_2010_porous_square_cylinder,
    title = {Wake structure for flow past and through a porous square cylinder},
    journal = {International Journal of Heat and Fluid Flow},
    volume = {31},
    number = {2},
    pages = {141-153},
    year = {2010},
    issn = {0142-727X},
    doi = {10.1016/j.ijheatfluidflow.2009.12.009},
    author = {P. Yu and Y. Zeng and T.S. Lee and H.X. Bai and H.T. Low}
}

@article{Yu_2011_permeable_circular_cylinder,
    title = {Steady flow around and through a permeable circular cylinder},
    journal = {Computers \& Fluids},
    volume = {42},
    number = {1},
    pages = {1-12},
    year = {2011},
    issn = {0045-7930},
    doi = {10.1016/j.compfluid.2010.09.040},
    author = {Peng Yu and Yan Zeng and Thong See Lee and Xiao Bing Chen and Hong Tong Low}
}

@article{Ledda_2018_porous_rectangular_cylinders,
    title = {Suppression of von K\'arm\'an vortex streets past porous rectangular cylinders},
    author = {Ledda, P. G. and Siconolfi, L. and Viola, F. and Gallaire, F. and Camarri, S.},
    journal = {Phys. Rev. Fluids},
    volume = {3},
    issue = {10},
    pages = {103901},
    numpages = {22},
    year = {2018},
    month = {Oct},
    publisher = {American Physical Society},
    doi = {10.1103/PhysRevFluids.3.103901}
}

@article{Caruso_2023_permeable_circular_cylinder,
    title = {Von K\'arm\'an vortex street past a permeable circular cylinder: Two-dimensional flow and dynamic-mode-decomposition-based secondary stability analysis},
    author = {Caruso Lombardi, F. and Bongarzone, A. and Zampogna, G. A. and Gallaire, F. and Camarri, S. and Ledda, P. G.},
    journal = {Phys. Rev. Fluids},
    volume = {8},
    issue = {8},
    pages = {083901},
    numpages = {29},
    year = {2023},
    month = {Aug},
    publisher = {American Physical Society},
    doi = {10.1103/PhysRevFluids.8.083901}
}

@article{Williamson_Prasad_1993_wave_resonance,
    title={A new mechanism for oblique wave resonance in the 'natural' far wake},
    volume={256},
    doi={10.1017/S0022112093002794},
    journal={Journal of Fluid Mechanics},
    author={Williamson, C. H. K. and Prasad, A.},
    year={1993},
    pages={269-313}
}

@article{Cimbala_Nagib_Roshko_1988_far_wakes,
    title={Large structure in the far wakes of two-dimensional bluff bodies},
    volume={190},
    doi={10.1017/S0022112088001314},
    journal={Journal of Fluid Mechanics},
    author={Cimbala, John M. and Nagib, Hassan M. and Roshko, Anatol},
    year={1988},
    pages={265-298}
}

@article{Huang_2020_Cavity_Circular_Obstacles,
    title = {Simulation of Lid-Driven Cavity Flow with Internal Circular Obstacles},
    journal =  {Applied Sciences},
    volume = {10},
    number = {13},
    pages = {4583},
    year = {2020},
    issn = {2076-3417},
    doi = {10.3390/app10134583},
    author = {Huang, Tingting and Lim, Hee-Chang}
}

@article{Rajan_2021_Cavity_Obstacles,
   title={Flow Dynamics of Lid-Driven Cavities with Obstacles of Various Shapes and Configurations Using the Lattice Boltzmann Method},
   journal={Journal of Thermal Engineering},
   volume={7},
   pages={83-102},
   year={2021},
   doi={10.18186/thermal.869135},
   author={Rajan, Isac and Perumal, D. Arumuga},
   number={2}
}

@article{Ghia_1982_Cavity_Reference,
    title = {High-Re solutions for incompressible flow using the Navier-Stokes equations and a multigrid method},
    journal = {Journal of Computational Physics},
    volume = {48},
    number = {3},
    pages = {387-411},
    year = {1982},
    issn = {0021-9991},
    doi = {10.1016/0021-9991(82)90058-4},
    author = {U Ghia and K.N Ghia and C.T Shin}
}

@article{Castro_1971,
    title={Wake characteristics of two-dimensional perforated plates normal to an air-stream},
    volume={46},
    doi={10.1017/S0022112071000727},
    number={3},
    journal={Journal of Fluid Mechanics},
    author={Castro, I. P.},
    year={1971},
    pages={599-609}
}

@article{Kumar_Mittal_2012,
    title={On the origin of the secondary vortex street},
    volume={711},
    doi={10.1017/jfm.2012.421},
    journal={Journal of Fluid Mechanics},
    author={Kumar, Bhaskar and Mittal, Sanjay},
    year={2012},
    pages={641-666}
}

@article{Jiang_2021,
    title={Formation mechanism of a secondary vortex street in a cylinder wake},
    volume={915},
    doi={10.1017/jfm.2021.195},
    journal={Journal of Fluid Mechanics},
    author={Jiang, Hongyi},
    year={2021},
    pages={A127}
}

@article{Chen_2008_porous_square,
    author = {Chen, Xiaobing and Yu, Peng and Winoto, S.H. and Low, Hong‐Tong},
    title = {Numerical analysis for the flow past a porous square cylinder based on the stress‐jump interfacial‐conditions},
    journal = {International Journal of Numerical Methods for Heat \& Fluid Flow},
    volume = {18},
    number = {5},
    pages = {635-655},
    year = {2008},
    month = {06},
    issn = {0961-5539},
    doi = {10.1108/09615530810879756},
}

@article{Jue_2004_porous_square,
    author = {Jue, Tswen‐Chyuan},
    title = {Numerical analysis of vortex shedding behind a porous square cylinder},
    journal = {International Journal of Numerical Methods for Heat \& Fluid Flow},
    volume = {14},
    number = {5},
    pages = {649-663},
    year = {2004},
    month = {07},
    issn = {0961-5539},
    doi = {10.1108/09615530410539964}
}

@article{Nandakumar_1982_porous_sphere,
    author = {Nandakumar, K. and Masliyah, Jacob H.},
    title = {Laminar flow past a permeable sphere},
    journal = {The Canadian Journal of Chemical Engineering},
    volume = {60},
    number = {2},
    pages = {202-211},
    doi = {10.1002/cjce.5450600202},
    year = {1982}
}

@book{succi_LBM_2018,
  title={The Lattice Boltzmann Equation: For Complex States of Flowing Matter},
  author={Succi, Sauro},
  year={2018},
  address = {Oxford},
  publisher={Oxford University Press}
}

@book{Cercignani_Boltzmann_1988,
  author={Cercignani, Carlo},
  title={The Boltzmann equation and its applications},
  year={1987},
  publisher={Springer},
  address = {New York}
}

@book{Darcy_1856,
    author    = {Darcy, Henry},
    title     = {Les fontaines publiques de la ville de Dijon. Exposition et application des principes {\`a} suivre et des formules {\`a} employer dans les questions de distribution d'eau},
    year      = {1856},
    publisher = {Victor Dalmont},
    address   = {Paris}
}

@article{Sawant_Dorschner_Karlin_2022_Reactive,
    title={Consistent lattice Boltzmann model for reactive mixtures},
    volume={941},
    doi={10.1017/jfm.2022.345},
    journal={Journal of Fluid Mechanics},
    author={Sawant, N. and Dorschner, B. and Karlin, I.V.},
    year={2022},
    pages={A62}
}

@article{He_1997_NonSlipBC,
    author={He, Xiaoyi and Zou, Qisu and Luo, Li Shi and Dembo, Micah},
    title={Analytic solutions of simple flows and analysis of nonslip boundary conditions for the lattice Boltzmann BGK model},
    journal={Journal of Statistical Physics},
    year={1997},
    month={Apr},
    day={01},
    volume={87},
    number={1},
    pages={115-136},
    issn={1572-9613},
    doi={10.1007/BF02181482}
}

@article{Mohamad_2009_EqBC,
    author={Mohamad, A. A. and Succi, S.},
    title={A note on equilibrium boundary conditions in lattice Boltzmann fluid dynamic simulations},
    journal={The European Physical Journal Special Topics},
    year={2009},
    month={Apr},
    day={01},
    volume={171},
    number={1},
    pages={213-221},
    issn={1951-6401},
    doi={10.1140/epjst/e2009-01031-9}
}

@article{Extended_LBM_Saadat_2021,
  author = {Saadat, Mohammad Hossein and Dorschner, Benedikt and Karlin, Ilya},
  title = {Extended Lattice Boltzmann Model},
  journal = {Entropy},
  volume = {23},
  year = {2021},
  number = {4},
  article-number = {475},
  issn = {1099-4300},
  doi = {10.3390/e23040475}
}

@article{Brinkman_1949a,
    author={Brinkman, Henri Coenraad},
    title={A calculation of the viscous force exerted by a flowing fluid on a dense swarm of particles},
    journal={Flow, Turbulence and Combustion},
    year={1949a},
    volume={1},
    number={1},
    pages={27-34},
    issn={1573-1987},
    doi={10.1007/BF02120313}
}

@article{Miller_Gray_2005,
    title = {Thermodynamically constrained averaging theory approach for modeling flow and transport phenomena in porous medium systems: 2. Foundation},
    journal = {Advances in Water Resources},
    volume = {28},
    number = {2},
    pages = {181-202},
    year = {2005},
    issn = {0309-1708},
    doi = {10.1016/j.advwatres.2004.09.006},
    author = {Cass T. Miller and William G. Gray}
}

@article{Gray_Miller_2013,
    title = {A generalization of averaging theorems for porous medium analysis},
    journal = {Advances in Water Resources},
    volume = {62},
    pages = {227-237},
    year = {2013},
    note = {A tribute to Stephen Whitaker},
    issn = {0309-1708},
    doi = {10.1016/j.advwatres.2013.06.006},
    author = {William G. Gray and Cass T. Miller}
}

@article{Hsu_Cheng_1990,
    title = {Thermal dispersion in a porous medium},
    journal = {International Journal of Heat and Mass Transfer},
    volume = {33},
    number = {8},
    pages = {1587-1597},
    year = {1990},
    issn = {0017-9310},
    doi = {10.1016/0017-9310(90)90015-M},
    author = {C.T. Hsu and P. Cheng}
}

@article{Gray_1975,
title = {A derivation of the equations for multi-phase transport},
journal = {Chemical Engineering Science},
volume = {30},
number = {2},
pages = {229-233},
year = {1975},
issn = {0009-2509},
doi = {10.1016/0009-2509(75)80010-8},
author = {William G. Gray}
}

@article{Ni_Beckermann_1991,
    title={A volume-averaged two-phase model for Transport Phenomena during solidification},
    volume={22},
    DOI={10.1007/bf02651234},
    number={3},
    journal={Metallurgical Transactions B},
    author={Ni, J. and Beckermann, C.},
    year={1991},
    month={Jun},
    pages={349-361}
}

@article{Gray_Lee_1977,
    title = {On the theorems for local volume averaging of multiphase systems},
    journal = {International Journal of Multiphase Flow},
    volume = {3},
    number = {4},
    pages = {333-340},
    year = {1977},
    issn = {0301-9322},
    doi = {10.1016/0301-9322(77)90013-1},
    author = {W.G. Gray and P.C.Y. Lee}
}

@article{Brinkman_1949b,
    title={On the permeability of media consisting of closely packed porous particles}, 
    volume={1},
    DOI={10.1007/bf02120318},
    number={1},
    journal={Flow, Turbulence and Combustion},
    author={Brinkman, H. C.},
    year={1949b},
    month={Dec}
}

@article{2020_SciPy,
    author  = {Virtanen, Pauli and Gommers, Ralf and Oliphant, Travis E. and
              Haberland, Matt and Reddy, Tyler and Cournapeau, David and
              Burovski, Evgeni and Peterson, Pearu and Weckesser, Warren and
              Bright, Jonathan and {van der Walt}, St{\'e}fan J. and
              Brett, Matthew and Wilson, Joshua and Millman, K. Jarrod and
              Mayorov, Nikolay and Nelson, Andrew R. J. and Jones, Eric and
              Kern, Robert and Larson, Eric and Carey, C J and
              Polat, {\.I}lhan and Feng, Yu and Moore, Eric W. and
              {VanderPlas}, Jake and Laxalde, Denis and Perktold, Josef and
              Cimrman, Robert and Henriksen, Ian and Quintero, E. A. and
              Harris, Charles R. and Archibald, Anne M. and
              Ribeiro, Ant{\^o}nio H. and Pedregosa, Fabian and
              {van Mulbregt}, Paul and {SciPy 1.0 Contributors}},
    title   = {{{SciPy} 1.0: Fundamental Algorithms for Scientific
              Computing in Python}},
    journal = {Nature Methods},
    year    = {2020},
    volume  = {17},
    pages   = {261-272},
    adsurl  = {https://rdcu.be/b08Wh},
    doi     = {10.1038/s41592-019-0686-2},
}

@article{Zhang_2014,
    doi = {10.1209/0295-5075/107/20001},
    year = {2014},
    month = {Jul},
    publisher = {EDP Sciences, IOP Publishing and Societ{\`a} Italiana di Fisica},
    volume = {107},
    number = {2},
    pages = {20001},
    author = {Zhang, Jinfeng and Wang, Limin and Ouyang, Jie},
    title = {Lattice Boltzmann model for the volume-averaged Navier-Stokes equations},
    journal = {Europhysics Letters}
}

@article{Rong_Guo_Lu_Shi_2011,
author = {Rong, F. M. and Guo, Z. L. and Lu, J. H. and Shi, B. C.},
title = {Numerical simulation of the flow around a porous covering square cylinder in a channel via lattice Boltzmann method},
journal = {International Journal for Numerical Methods in Fluids},
volume = {65},
number = {10},
pages = {1217-1230},
doi = {10.1002/fld.2237},
year = {2011}
}

@article{Pepona_Favier_2016,
    title = {A coupled Immersed Boundary-Lattice Boltzmann method for incompressible flows through moving porous media},
    journal = {Journal of Computational Physics},
    volume = {321},
    pages = {1170-1184},
    year = {2016},
    issn = {0021-9991},
    doi = {10.1016/j.jcp.2016.06.026},
    author = {Marianna Pepona and Julien Favier}
}

@article{Wang_Wang_Guo_Mi_2015,
    title = {Volume-averaged macroscopic equation for fluid flow in moving porous media},
    journal = {International Journal of Heat and Mass Transfer},
    volume = {82},
    pages = {357-368},
    year = {2015},
    issn = {0017-9310},
    doi = {10.1016/j.ijheatmasstransfer.2014.11.056},
    author = {Liang Wang and Lian-Ping Wang and Zhaoli Guo and Jianchun Mi}
}

@article{Bai_Yu_Winoto_Low_2009,
    author = {Bai, Huixing and Yu, P. and Winoto, S. H. and Low, H. T.},
    title = {Lattice Boltzmann method for flows in porous and homogenous fluid domains coupled at the interface by stress jump},
    journal = {International Journal for Numerical Methods in Fluids},
    volume = {60},
    number = {6},
    pages = {691-708},
    doi = {10.1002/fld.1913},
    year = {2009}
}

@article{Zhao_Liu_Qin_Fei_2023,
    title={Pore-scale fluid flow simulation coupling lattice Boltzmann method and pore network model},
    volume={7},
    DOI={10.46690/capi.2023.06.01},
    number={3},
    journal={Capillarity},
    author={Zhao, Jianlin and Liu, Yang and Qin, Feifei and Fei, Linlin},
    year={2023},
    month={May},
    pages={41-46}
}

@article{Fattahi_Waluga_etal_2016,
    title = {Lattice Boltzmann methods in porous media simulations: From laminar to turbulent flow},
    journal = {Computers \& Fluids},
    volume = {140},
    pages = {247-259},
    year = {2016},
    issn = {0045-7930},
    doi = {10.1016/j.compfluid.2016.10.007},
    author = {Ehsan Fattahi and Christian Waluga and Barbara Wohlmuth and Ulrich R{\"u}de and Michael Manhart and Rainer Helmig}
}

@article{Li_Brown_2017,
    author = {Li, Jun and Brown, Donald},
    title = {Upscaled Lattice Boltzmann Method for Simulations of Flows in Heterogeneous Porous Media},
    journal = {Geofluids},
    volume = {2017},
    number = {1},
    pages = {1740693},
    doi = {10.1155/2017/1740693},
    year = {2017}
}

@article{Guo_Porous_2002,
    title = {Lattice Boltzmann model for incompressible flows through porous media},
    author = {Guo, Zhaoli and Zhao, T. S.},
    journal = {Phys. Rev. E},
    volume = {66},
    issue = {3},
    pages = {036304},
    numpages = {9},
    year = {2002},
    month = {Sep},
    publisher = {American Physical Society},
    doi = {10.1103/PhysRevE.66.036304},
}

@article{Whitaker_Forchheimer_1996,
    title={The Forchheimer equation: a theoretical development},
    author={Whitaker, Stephen},
    journal={Transport in Porous media},
    volume={25},
    number={1},
    pages={27-61},
    year={1996},
    publisher={Springer},
    doi = {10.1007/BF00141261},
}

@book{Whitaker_VA_method_1998,
    title={The Method of Volume Averaging},
    author={Whitaker, Stephen},
    year={1998},
    address = {Dordrecht},
    publisher={Springer}
}

@article{Whitaker_Advances_1969,
author = {Whitaker, Stephen},
title = {Advances in theory of fluid motion in porous media},
journal = {Industrial \& Engineering Chemistry},
volume = {61},
number = {12},
pages = {14-28},
year = {1969},
doi = {10.1021/ie50720a004},
}

@article{Ergun_1952,
    title={Fluid flow through packed columns},
    author={Ergun, Sabri},
    journal={Chemical engineering progress},
    volume={48},
    number={2},
    pages={89},
    year={1952}
}

@article{Karlin_Factorization_2010,
    title = {Factorization symmetry in the lattice Boltzmann method},
    journal = {Physica A: Statistical Mechanics and its Applications},
    volume = {389},
    number = {8},
    pages = {1530-1548},
    year = {2010},
    issn = {0378-4371},
    doi = {10.1016/j.physa.2009.12.032},
    author = {Ilya Karlin and Pietro Asinari}
}

@article{Jarauta_Compressible_Porous_2020,
    title={A compressible fluid flow model coupling channel and porous media flows and its application to fuel cell materials},
    volume={134},
    DOI={10.1007/s11242-020-01449-2},
    number={2},
    journal={Transport in Porous Media},
    author={Jarauta, Alex and Zingan, Valentin and Minev, Peter and Secanell, Marc},
    year={2020},
    month={Jul},
    pages={351-386}
}

@article{Sha_Chao_Soo_1984,
    title = {Porous-media formulation for multiphase flow with heat transfer},
    journal = {Nuclear Engineering and Design},
    volume = {82},
    number = {2},
    pages = {93-106},
    year = {1984},
    issn = {0029-5493},
    doi = {10.1016/0029-5493(84)90206-1},
    author = {William T. Sha and B.T. Chao and S.L. Soo}
}

@article{Wang_Wang_2005,
  title = {Two-fluid model based on the lattice Boltzmann equation},
  author = {Wang, Tiefeng and Wang, Jinfu},
  journal = {Phys. Rev. E},
  volume = {71},
  issue = {4},
  pages = {045301(R)},
  numpages = {4},
  year = {2005},
  month = {Apr},
  publisher = {American Physical Society},
  doi = {10.1103/PhysRevE.71.045301}
}

@article{Song_Wang_Li_2013,
    title = {A lattice Boltzmann method for particle-fluid two-phase flow},
    journal = {Chemical Engineering Science},
    volume = {102},
    pages = {442-450},
    year = {2013},
    issn = {0009-2509},
    doi = {10.1016/j.ces.2013.08.037},
    author = {Feifei Song and Wei Wang and Jinghai Li}
}

@article{Xiong_Madadi_Lorenzini_2014,
    title={A LBM-dem solver for fast discrete particle simulation of particle-fluid flows},
    volume={26},
    DOI={10.1007/s00161-014-0351-z},
    number={6},
    journal={Continuum Mechanics and Thermodynamics},
    author={Xiong, Qingang and Madadi-Kandjani, Ehsan and Lorenzini, Giulio},
    year={2014},
    month={Apr},
    pages={907-917}
}

@article{Givler_Altobelli_1994,
    title={A determination of the effective viscosity for the Brinkman-Forchheimer flow model},
    volume={258},
    DOI={10.1017/S0022112094003368},
    journal={Journal of Fluid Mechanics},
    author={Givler, R. C. and Altobelli, S. A.},
    year={1994},
    pages={355-370}
}

@article{Valdes_Ochoa_Tapia_2007,
    title = {On the effective viscosity for the Darcy–Brinkman equation},
    journal = {Physica A: Statistical Mechanics and its Applications},
    volume = {385},
    number = {1},
    pages = {69-79},
    year = {2007},
    issn = {0378-4371},
    doi = {10.1016/j.physa.2007.06.012},
    author = {Francisco J. Valdes-Parada and J. {Alberto Ochoa-Tapia} and Jose Alvarez-Ramirez}
}

@article{Hassanizadeh_Gray_1980,
    title = {General conservation equations for multi-phase systems: 3. Constitutive theory for porous media flow},
    journal = {Advances in Water Resources},
    volume = {3},
    number = {1},
    pages = {25-40},
    year = {1980},
    issn = {0309-1708},
    doi = {10.1016/0309-1708(80)90016-0},
    author = {Majid Hassanizadeh and William G. Gray}
}

@article{Zachariah_etal_2026,
  title = {Lattice Boltzmann approaches to the Euler-Euler equations for two-phase flows},
  author = {Zachariah, Githin Tom and Van den Akker, Harry E. A.},
  journal = {Phys. Rev. Fluids},
  volume = {11},
  issue = {4},
  pages = {044904},
  numpages = {26},
  year = {2026},
  month = {Apr},
  publisher = {American Physical Society},
  doi = {10.1103/b4mq-2hm6}
}

@misc{Rinehart_2021,
      title={The Brinkman viscosity for porous media exposed to a free flow}, 
      author={Aidan Rinehart and U{\v{g}}is L{\=a}cis and Shervin Bagheri},
      year={2021},
      eprint={2106.01879},
      archivePrefix={arXiv},
      primaryClass={physics.flu-dyn},
      url={https://arxiv.org/abs/2106.01879}, 
}

@article{Ganesan_Poirier_1990,
    title={Conservation of mass and momentum for the flow of interdendritic liquid during solidification},
    volume={21},
    DOI={10.1007/bf02658128},
    number={1},
    journal={Metallurgical Transactions B},
    author={Ganesan, S. and Poirier, D. R.},
    year={1990},
    month={Feb},
    pages={173-181}
}

@article{Gray_ONeill_1976,
    author = {Gray, William G. and O'Neill, Kevin},
    title = {On the general equations for flow in porous media and their reduction to Darcy's Law},
    journal = {Water Resources Research},
    volume = {12},
    number = {2},
    pages = {148-154},
    doi = {10.1029/WR012i002p00148},
    year = {1976}
}

@book{Ishii_Hibiki_2005,
    address={New York},
    edition={1},
    title={Thermo-fluid Dynamics of Two-Phase Flow},
    publisher={Springer},
    author={Ishii, Mamoru and Hibiki, Takashi},
    year={2005}
}

@article{Sawant_Karlin_2025,
    title={Mean field lattice Boltzmann model for reactive mixtures in porous media},
    volume={1011},
    doi={10.1017/jfm.2025.359},
    journal={Journal of Fluid Mechanics},
    author={Sawant, N. and Karlin, I.V.},
    year={2025},
    pages={A9}
}

@article{Ginzburg_Silva_Talon_2015,
    title = {Analysis and improvement of Brinkman lattice Boltzmann schemes: Bulk, boundary, interface. Similarity and distinctness with finite elements in heterogeneous porous media},
    author = {Ginzburg, Irina and Silva, Goncalo and Talon, Laurent},
    journal = {Phys. Rev. E},
    volume = {91},
    issue = {2},
    pages = {023307},
    numpages = {32},
    year = {2015},
    month = {Feb},
    publisher = {American Physical Society},
    doi = {10.1103/PhysRevE.91.023307}
}

@article{He_Liu_Li_Tao_2019,
    title = {Lattice Boltzmann methods for single-phase and solid-liquid phase-change heat transfer in porous media: A review},
    journal = {International Journal of Heat and Mass Transfer},
    volume = {129},
    pages = {160-197},
    year = {2019},
    issn = {0017-9310},
    doi = {10.1016/j.ijheatmasstransfer.2018.08.135},
    author = {Ya-Ling He and Qing Liu and Qing Li and Wen-Quan Tao}
}

@article{Dardis_McCloskey_1998,
  title = {Lattice Boltzmann scheme with real numbered solid density for the simulation of flow in porous media},
  author = {Dardis, Orla and McCloskey, John},
  journal = {Phys. Rev. E},
  volume = {57},
  issue = {4},
  pages = {4834-4837},
  numpages = {0},
  year = {1998},
  month = {Apr},
  publisher = {American Physical Society},
  doi = {10.1103/PhysRevE.57.4834}
}

@article{Spaid_Phelan_1997,
    author = {Spaid, Michael A. A. and Phelan, Frederick R., Jr.},
    title = {Lattice Boltzmann methods for modeling microscale flow in fibrous porous media},
    journal = {Physics of Fluids},
    volume = {9},
    number = {9},
    pages = {2468-2474},
    year = {1997},
    month = {09},
    issn = {1070-6631},
    doi = {10.1063/1.869392}
}

@article{Whitaker_1967,
    author = {Whitaker, Stephen},
    title = {Diffusion and dispersion in porous media},
    journal = {AIChE Journal},
    volume = {13},
    number = {3},
    pages = {420-427},
    doi = {10.1002/aic.690130308},
    year = {1967}
}

@article{Whitaker_1986,
    title={Flow in porous media I: A theoretical derivation of Darcy’s Law},
    volume={1},
    DOI={10.1007/bf01036523},
    number={1},
    journal={Transport in Porous Media},
    author={Whitaker, Stephen},
    year={1986},
    pages={3-25}
}

@article{Renard_2021,
    title = {Improved compressible hybrid lattice Boltzmann method on standard lattice for subsonic and supersonic flows},
    journal = {Computers \& Fluids},
    volume = {219},
    pages = {104867},
    year = {2021},
    issn = {0045-7930},
    doi = {10.1016/j.compfluid.2021.104867},
    author = {Florian Renard and Yongliang Feng and Jean-François Boussuge and Pierre Sagaut}
}

@article{He_Shan_Doolen_1998,
    title = {Discrete Boltzmann equation model for nonideal gases},
    author = {He, Xiaoyi and Shan, Xiaowen and Doolen, Gary D.},
    journal = {Phys. Rev. E},
    volume = {57},
    issue = {1},
    pages = {R13-R16},
    numpages = {0},
    year = {1998},
    month = {Jan},
    publisher = {American Physical Society},
    doi = {10.1103/PhysRevE.57.R13}
}

@article{Chikatamarla_Karlin_2013_TMSBC,
    title = {Entropic lattice Boltzmann method for turbulent flow simulations: Boundary conditions},
    journal = {Physica A: Statistical Mechanics and its Applications},
    volume = {392},
    number = {9},
    pages = {1925-1930},
    year = {2013},
    issn = {0378-4371},
    doi = {10.1016/j.physa.2012.12.034},
    author = {S.S. Chikatamarla and I.V. Karlin}
}

@article{Zou_He_1997_BC,
    author = {Zou, Qisu and He, Xiaoyi},
    title = {On pressure and velocity boundary conditions for the lattice Boltzmann BGK model},
    journal = {Physics of Fluids},
    volume = {9},
    number = {6},
    pages = {1591-1598},
    year = {1997},
    month = {06},
    issn = {1070-6631},
    doi = {10.1063/1.869307}
}

@article{Latt_etal_2008_DirichletBC,
    title = {Straight velocity boundaries in the lattice Boltzmann method},
    author = {Latt, Jonas and Chopard, Bastien and Malaspinas, Orestis and Deville, Michel and Michler, Andreas},
    journal = {Phys. Rev. E},
    volume = {77},
    issue = {5},
    pages = {056703},
    numpages = {16},
    year = {2008},
    month = {May},
    publisher = {American Physical Society},
    doi = {10.1103/PhysRevE.77.056703}
}

@article{Karlin_Hosseini_PracticalModels_2025,
    title = {Practical Kinetic Models for Dense Fluids},
    author = {Karlin, Ilya and Hosseini, Seyed Ali},
    journal = {Phys. Rev. Lett.},
    volume = {136},
    issue = {10},
    pages = {104002},
    numpages = {6},
    year = {2026},
    month = {Mar},
    publisher = {American Physical Society},
    doi = {10.1103/fd39-6hmq},
    url = {https://link.aps.org/doi/10.1103/fd39-6hmq}
}

@article{Hosseini_Feinberg_Karlin_2026,
    title={Lattice Boltzmann model for non-ideal compressible fluid dynamics},
    volume={1037},
    DOI={10.1017/jfm.2026.11636},
    journal={Journal of Fluid Mechanics},
    author={Hosseini, Seyed Ali and Feinberg, Milo and Karlin, Ilya},
    year={2026},
    pages={A12}
}

@article{Guo_NEEM_BC_1,
    doi = {10.1088/1009-1963/11/4/310},
    year = {2002},
    month = {apr},
    publisher = {},
    volume = {11},
    number = {4},
    pages = {366},
    author = {Guo, Zhaoli and Zheng, Chuguang and Shi, Baochang},
    title = {Non-equilibrium extrapolation method for velocity and pressure boundary conditions in the lattice Boltzmann method},
    journal = {Chinese Physics}
}

@article{Irmay_1958,
author = {Irmay, S.},
title = {On the theoretical derivation of Darcy and Forchheimer formulas},
journal = {Eos, Transactions American Geophysical Union},
volume = {39},
number = {4},
pages = {702-707},
doi = {10.1029/TR039i004p00702},
year = {1958}
}

@article{Guo_NEEM_BC_2,
    author = {Guo, Zhaoli and Zheng, Chuguang and Shi, Baochang},
    title = {An extrapolation method for boundary conditions in lattice Boltzmann method},
    journal = {Physics of Fluids},
    volume = {14},
    number = {6},
    pages = {2007-2010},
    year = {2002},
    month = {06},
    issn = {1070-6631},
    doi = {10.1063/1.1471914}
}

@article{Mott_Smith_1951,
    title = {The Solution of the Boltzmann Equation for a Shock Wave},
    author = {Mott-Smith, H. M.},
    journal = {Phys. Rev.},
    volume = {82},
    issue = {6},
    pages = {885-892},
    numpages = {0},
    year = {1951},
    month = {Jun},
    publisher = {American Physical Society},
    doi = {10.1103/PhysRev.82.885}
}

@article{Banerjee_2021_PMC_review,
    title = {Developments and applications of porous medium combustion: A recent review},
    journal = {Energy},
    volume = {221},
    pages = {119868},
    year = {2021},
    issn = {0360-5442},
    doi = {10.1016/j.energy.2021.119868},
    author = {Abhisek Banerjee and Diplina Paul}
}

@article{Ladd_Szymczak_2021,
    title={Reactive flows in porous media: Challenges in theoretical and Numerical Methods},
    volume={12},
    doi={10.1146/annurev-chembioeng-092920-102703},
    number={1},
    journal={Annual Review of Chemical and Biomolecular Engineering},
    author={Ladd, Anthony J.C. and Szymczak, Piotr},
    year={2021},
    month={Jun},
    pages={543-571}
}

@incollection{Ahmed_2019_reservoir_rock,
    title = {Fundamentals of Rock Properties},
    booktitle = {Reservoir Engineering Handbook},
    publisher = {Gulf Professional Publishing},
    edition = {Fifth Edition},
    pages = {167-281},
    year = {2019},
    isbn = {978-0-12-813649-2},
    doi = {10.1016/B978-0-12-813649-2.00004-9},
    author = {Tarek Ahmed}
}

@incollection{Ahmed_2019_Fluid_Flow,
    title = {Fundamentals of Reservoir Fluid Flow},
    booktitle = {Reservoir Engineering Handbook},
    publisher = {Gulf Professional Publishing},
    edition = {Fifth Edition},
    pages = {331-456},
    year = {2019},
    isbn = {978-0-12-813649-2},
    doi = {10.1016/B978-0-12-813649-2.00006-2},
    author = {Tarek Ahmed}
}

@article{Khoso_2026_HEX,
    title = {Unveiling thermal-hydraulic effects of porous structure in a flow channel: A comparative experimental study},
    journal = {Applied Thermal Engineering},
    volume = {287},
    pages = {129388},
    year = {2026},
    issn = {1359-4311},
    doi = {10.1016/j.applthermaleng.2025.129388},
    author = {Abdul Qadeer Khoso and Tianxiao Xie and Bernardo Buonomo and Sergio Nardini and Marko Kleissl and Heinz Peter Berg and Oronzio Manca}
}

@article{Rashidi_2019,
    title = {Potentials of porous materials for energy management in heat exchangers -- A comprehensive review},
    journal = {Applied Energy},
    volume = {243},
    pages = {206-232},
    year = {2019},
    issn = {0306-2619},
    doi = {10.1016/j.apenergy.2019.03.200},
    author = {Saman Rashidi and Mohammad Hossein Kashefi and Kyung Chun Kim and Omid Samimi-Abianeh}
}

@article{Gharehghani_2021_PMCT,
    title = {Applications of porous materials in combustion systems: A comprehensive and state-of-the-art review},
    journal = {Fuel},
    volume = {304},
    pages = {121411},
    year = {2021},
    issn = {0016-2361},
    doi = {10.1016/j.fuel.2021.121411},
    author = {Ayat Gharehghani and Kasra Ghasemi and Majid Siavashi and Sadegh Mehranfar}
}

@article{Zhang_2021_FC,
    title = {Application of porous materials for the flow field in polymer electrolyte membrane fuel cells},
    journal = {Journal of Power Sources},
    volume = {492},
    pages = {229664},
    year = {2021},
    issn = {0378-7753},
    doi = {10.1016/j.jpowsour.2021.229664},
    author = {Yinghui Zhang and Youkun Tao and Jing Shao}
}

@article{Kumar_2024_flowdist,
    title = {Performance assessment of a porous flow distributor in arresting the effect of temperature fluctuation in a thermocline system},
    journal = {Thermal Science and Engineering Progress},
    volume = {55},
    pages = {102942},
    year = {2024},
    issn = {2451-9049},
    doi = {10.1016/j.tsep.2024.102942},
    author = {Kapil Kumar and Varun Joshi and Shireesh B. Kedare and Manaswita Bose}
}

@incollection{Flovenz_2012_Geothermal_Energy,
    title = {7.03 - Geothermal Energy Exploration Techniques},
    editor = {Ali Sayigh},
    booktitle = {Comprehensive Renewable Energy},
    publisher = {Elsevier},
    address = {Oxford},
    pages = {51-95},
    year = {2012},
    isbn = {978-0-08-087873-7},
    doi = {10.1016/B978-0-08-087872-0.00705-8},
    author = {Fl{\'o}venz, {\'O}.G. and Hersir, G.P. and S{\ae}mundsson, K. and {\'A}rmannsson, H. and Fri{\dh}riksson, {\TH}.}
}

@incollection{Cumming_2016_Geophysics,
    title = {3 - Geophysics and resource conceptual models in geothermal exploration and development},
    editor = {Ronald DiPippo},
    booktitle = {Geothermal Power Generation},
    publisher = {Woodhead Publishing},
    pages = {33-75},
    year = {2016},
    isbn = {978-0-08-100337-4},
    doi = {10.1016/B978-0-08-100337-4.00003-6},
    author = {W. Cumming}
}

@book{Mohamed_2018_GeoEngineering,
    title = {Fundamentals of Geoenvironmental Engineering},
    author = {Abdel-Mohsen Onsy Mohamed and Evan K. Paleologos},
    year = {2018},
    address = {Oxford},
    publisher = {Butterworth-Heinemann},
    isbn = {978-0-12-804830-6},
    doi = {10.1016/C2015-0-01982-7}
}

@book{Feder_2022_Porous_Media,
    address={Cambridge},
    title={Physics of Flow in Porous Media},
    publisher={Cambridge University Press},
    author={Feder, Jens and Flekkøy, Eirik Grude and Hansen, Alex},
    year={2022},
    isbn = {978-1-108-83911-2},
    doi = {10.1017/9781009100717}
}

@article{Wood_He_Apte_2020,
    title={Modeling Turbulent flows in porous media},
    volume={52},
    doi={10.1146/annurev-fluid-010719-060317},
    number={1},
    journal={Annual Review of Fluid Mechanics},
    author={Wood, Brian D. and He, Xiaoliang and Apte, Sourabh V.},
    year={2020},
    month={Jan},
    pages={171-203}
}

@article{Valverde_Griffiths_2024,
    title={The role of adsorbent microstructure and its packing arrangement in optimising the performance of an adsorption column},
    volume={4},
    doi={10.1007/s43938-024-00064-7},
    number={1},
    journal={Discover Chemical Engineering},
    author={Valverde, Abel and Griffiths, Ian M.},
    year={2024},
    month={Sep}
}

@article{Banhart_2001_metal_foams,
    title = {Manufacture, characterisation and application of cellular metals and metal foams},
    journal = {Progress in Materials Science},
    volume = {46},
    number = {6},
    pages = {559-632},
    year = {2001},
    issn = {0079-6425},
    doi = {10.1016/S0079-6425(00)00002-5},
    author = {John Banhart}
}

@inbook{Mathias_2010,
    author = {Mathias, M. F. and Roth, J. and Fleming, J. and Lehnert, W.},
    publisher = {John Wiley \& Sons, Ltd},
    isbn = {9780470974001},
    title = {Diffusion media materials and characterisation},
    booktitle = {Handbook of Fuel Cells},
    doi = {10.1002/9780470974001.f303046},
    year = {2010}
}

@inbook{Divisek_2010,
    author = {Divisek, J. and Emonts, B.},
    publisher = {John Wiley \& Sons, Ltd},
    isbn = {9780470974001},
    title = {Energy storage via electrolysis/fuel cells},
    booktitle = {Handbook of Fuel Cells},
    pages = {--},
    doi = {10.1002/9780470974001.f104023},
    year = {2010}
}

@incollection{Rabiee_2019,
    title = {Energy-Water Nexus: Renewable-Integrated Hybridized Desalination Systems},
    editor = {Kaveh Rajab Khalilpour},
    booktitle = {Polygeneration with Polystorage for Chemical and Energy Hubs},
    publisher = {Academic Press},
    pages = {409-458},
    year = {2019},
    isbn = {978-0-12-813306-4},
    doi = {10.1016/B978-0-12-813306-4.00013-6},
    author = {Hesamoddin Rabiee and Kaveh Rajab Khalilpour and John M. Betts and Nigel Tapper}
}

@article{Michaud_2023,
    title = {Unprecedented continuous elastic foam-bed reactor for CO2 capture},
    journal = {Chemical Engineering Journal},
    volume = {452},
    pages = {138604},
    year = {2023},
    issn = {1385-8947},
    doi = {10.1016/j.cej.2022.138604},
    url = {https://www.sciencedirect.com/science/article/pii/S1385894722040852},
    author = {Michaud, Ma{\"i}t{\'e} and Bornette, Fr{\'e}deric and Rautu, Eduard and More, Shahaji H. and {Leonardo Martinez Mendez}, Miguel and Jierry, Lo{\"i}c and Edouard, David}
}

@Article{Zhiqiang_2025,
    AUTHOR = {Xu, Zhiqiang and Liu, Wenming and Yu, Zhengyong and Liu, Xuedong},
    TITLE = {Advances and Challenges in Catalyst Dense-Phase Packing Technology: A Review},
    JOURNAL = {Catalysts},
    VOLUME = {15},
    YEAR = {2025},
    NUMBER = {3},
    ARTICLE-NUMBER = {222},
    ISSN = {2073-4344},
    DOI = {10.3390/catal15030222}
}

@article{Shannon_2008,
    title={Science and Technology for water purification in the coming decades},
    volume={452},
    DOI={10.1038/nature06599},
    number={7185},
    journal={Nature},
    author = {Shannon, Mark A. and Bohn, Paul W. and Elimelech, Menachem and Georgiadis, John G. and Mari{\~n}as, Benito J. and Mayes, Anne M.},
    year={2008},
    month={Mar},
    pages={301-310}
}

@Article{Labbe_2019,
    author = {Labb{\'e}, R. and Duprat, C.},
    title  ="Capturing aerosol droplets with fibers",
    journal  ="Soft Matter",
    year  ="2019",
    volume  ="15",
    issue  ="35",
    pages  ="6946-6951",
    publisher  ="The Royal Society of Chemistry",
    doi  ="10.1039/C9SM01205B"
}

@article{Wood_2007_Inertial,
    author = {Wood, Brian D.},
    title = {Inertial effects in dispersion in porous media},
    journal = {Water Resources Research},
    volume = {43},
    number = {12},
    pages = {},
    doi = {10.1029/2006WR005790},
    year = {2007}
}

@article{Ghosh_2023,
    title={Photocatalytically reactive surfaces for simultaneous water harvesting and treatment},
    volume={6},
    DOI={10.1038/s41893-023-01159-9},
    number={12},
    journal={Nature Sustainability},
    author = {Ghosh, Ritwick and Baut, Adrien and Belleri, Giorgio and Kappl, Michael and Butt, Hans-J{\"u}rgen and Schutzius, Thomas M.},
    year={2023},
    month={Aug},
    pages={1663-1672}
}

@article{Foggi_Rota_Monti_Olivieri_Rosti_2024,
    title={Dynamics and fluid-structure interaction in turbulent flows within and above flexible canopies},
    volume={989},
    DOI={10.1017/jfm.2024.481},
    journal={Journal of Fluid Mechanics},
    author={Foggi Rota, Giulio and Monti, Alessandro and Olivieri, Stefano and Rosti, Marco Edoardo},
    year={2024},
    pages={A11}
}

@article{Zampogna_etal_2016,
    author = {Zampogna, Giuseppe A. and Pluvinage, Franck and Kourta, Azeddine and Bottaro, Alessandro},
    title = {Instability of canopy flows},
    journal = {Water Resources Research},
    volume = {52},
    number = {7},
    pages = {5421-5432},
    doi = {10.1002/2016WR018915},
    year = {2016}
}

@article{Battiato_Rubol_2014,
    author = {Battiato, Ilenia and Rubol, Simonetta},
    title = {Single-parameter model of vegetated aquatic flows},
    journal = {Water Resources Research},
    volume = {50},
    number = {8},
    pages = {6358-6369},
    doi = {10.1002/2013WR015065},
    year = {2014}
}

@article{Papke_2013,
    author = {Papke, A. and Battiato, I.},
    title = {A reduced complexity model for dynamic similarity in obstructed shear flows},
    journal = {Geophysical Research Letters},
    volume = {40},
    number = {15},
    pages = {3888-3892},
    doi = {10.1002/grl.50759},
    year = {2013}
}

@book{Nield_Bejan_2017,
    address={Cham},
    title={Convection in Porous Media},
    publisher={Springer},
    edition={4},
    author={Nield, Donald A. and Bejan, Adrian},
    year={2017}
}

@article{Battiato_2012,
    title={Self-similarity in coupled Brinkman/Navier-Stokes flows},
    volume={699},
    DOI={10.1017/jfm.2012.85},
    journal={Journal of Fluid Mechanics},
    author={Battiato, Ilenia},
    year={2012},
    pages={94-114}
}

@article{duToit_2008,
    title = {Radial variation in porosity in annular packed beds},
    journal = {Nuclear Engineering and Design},
    volume = {238},
    number = {11},
    pages = {3073-3079},
    year = {2008},
    note = {HTR-2006: 3rd International Topical Meeting on High Temperature Reactor Technology},
    issn = {0029-5493},
    doi = {10.1016/j.nucengdes.2007.12.018},
    author = {C.G. {du Toit}}
}

@article{Roblee_Baird_Tierney_1958,
    author = {Roblee, L. H. S. and Baird, R. M. and Tierney, J. W.},
    title = {Radial porosity variations in packed beds},
    journal = {AIChE Journal},
    volume = {4},
    number = {4},
    pages = {460-464},
    doi = {10.1002/aic.690040415},
    year = {1958}
}

@article{White_Tien_1987,
    title={Analysis of flow channeling near the wall in packed beds},
    volume={21},
    DOI={10.1007/bf01009290},
    number={5},
    journal={Wärme- und Stoffübertragung},
    author={White, S. M. and Tien, C. L.},
    year={1987},
    month={Sep},
    pages={291-296}
}

@article{He_Liu_Shen_2022,
    title={Simulation-based study of turbulent aquatic canopy flows with flexible stems},
    volume={947},
    DOI={10.1017/jfm.2022.655},
    journal={Journal of Fluid Mechanics},
    author={He, Sida and Liu, Han and Shen, Lian},
    year={2022},
    pages={A33}
}

@article{Jin_Kuznetsov_2024,
    author = {Jin, Yan and Kuznetsov, Andrey V.},
    title = {Multiscale modeling and simulation of turbulent flows in porous media},
    journal = {International Journal of Fluid Engineering},
    volume = {1},
    number = {1},
    pages = {010601},
    year = {2024},
    month = {03},
    issn = {2994-9009},
    doi = {10.1063/5.0190279}
}

@article{Sueki_2009,
    title={Aerodynamic Noise Reduction using Porous Materials and their Application to High-speed Pantographs},
    author={Takeshi Sueki and Mitsuru Ikeda and Takehisa Takaishi},
    journal={Quarterly Report of RTRI},
    volume={50},
    number={1},
    pages={26-31},
    year={2009},
    doi={10.2219/rtriqr.50.26}
}

@article{Sato_Hattori_2021,
    title={Mechanism of reduction of aeroacoustic sound by porous material: comparative study of microscopic and macroscopic models},
    volume={929},
    DOI={10.1017/jfm.2021.884},
    journal={Journal of Fluid Mechanics},
    author={Sato, Yasunori and Hattori, Yuji},
    year={2021},
    pages={A34}
}

@article{Yuan_etal_2021,
    title = {Influence of porous media coatings on flow characteristics and vortex-induced vibration of circular cylinders},
    journal = {Journal of Fluids and Structures},
    volume = {106},
    pages = {103365},
    year = {2021},
    issn = {0889-9746},
    doi = {10.1016/j.jfluidstructs.2021.103365},
    author = {Wenyong Yuan and Shujin Laima and Donglai Gao and Wen-Li Chen and Hui Li}
}

@article{Geyer_2020,
    title={Experimental evaluation of cylinder vortex shedding noise reduction using porous material},
    volume={61},
    DOI={10.1007/s00348-020-02972-0},
    number={7},
    journal={Experiments in Fluids},
    author={Geyer, Thomas F.},
    year={2020},
    month={Jun}
}

@article{Klausmann_Ruck_2017,
    title={Drag reduction of circular cylinders by porous coating on the leeward side},
    volume={813},
    DOI={10.1017/jfm.2016.757},
    journal={Journal of Fluid Mechanics},
    author={Klausmann, Katharina and Ruck, Bodo},
    year={2017},
    pages={382-411}
}

@article{Du_2022a,
    author = {Du, Hai and Li, Qixuan and Zhang, Qinlin and Zhang, Wenxiao and Yang, Lejie},
    title = {Experimental study on drag reduction of the turbulent boundary layer via porous media under nonzero pressure gradient},
    journal = {Physics of Fluids},
    volume = {34},
    number = {2},
    pages = {025110},
    year = {2022a},
    month = {02},
    issn = {1070-6631},
    doi = {10.1063/5.0083143}
}

@article{Du_2022b,
    author = {Du, Hai and Zhang, Qinlin and Li, Qixuan and Kong, Wenjie and Yang, Lejie},
    title = {Drag reduction in cylindrical wake flow using porous material},
    journal = {Physics of Fluids},
    volume = {34},
    number = {4},
    pages = {045102},
    year = {2022b},
    month = {04},
    issn = {1070-6631},
    doi = {10.1063/5.0085990}
}

@article{Xu_2022,
    title = {Structured porous surface for drag reduction and wake attenuation of cylinder flow},
    journal = {Ocean Engineering},
    volume = {247},
    pages = {110444},
    year = {2022},
    issn = {0029-8018},
    doi = {10.1016/j.oceaneng.2021.110444},
    author = {Zhihan Xu and Xu Chang and Haiyang Yu and Wen-Li Chen and Donglai Gao}
}

@article{Icardi_2014_pore_scale,
  title = {Pore-scale simulation of fluid flow and solute dispersion in three-dimensional porous media},
  author = {Icardi, Matteo and Boccardo, Gianluca and Marchisio, Daniele L. and Tosco, Tiziana and Sethi, Rajandrea},
  journal = {Phys. Rev. E},
  volume = {90},
  issue = {1},
  pages = {013032},
  numpages = {13},
  year = {2014},
  month = {Jul},
  publisher = {American Physical Society},
  doi = {10.1103/PhysRevE.90.013032}
}

@article{Feriadi_Arbie_Fauzi_Fariduzzaman_2024,
    title={Pore-scale simulation of flow in porous rocks for wall shear stress analysis},
    volume={10},
    DOI={10.1007/s40808-024-02036-w},
    number={4},
    journal={Modeling Earth Systems and Environment},
    author={Feriadi, Yusron and Arbie, Muhammad Rizqie and Fauzi, Umar and Fariduzzaman},
    year={2024},
    month={May},
    pages={4877-4897}
}

@article{Chen_2022_Pore_scale,
    title = {Pore-scale modeling of complex transport phenomena in porous media},
    journal = {Progress in Energy and Combustion Science},
    volume = {88},
    pages = {100968},
    year = {2022},
    issn = {0360-1285},
    doi = {0.1016/j.pecs.2021.100968},
    author = {Li Chen and An He and Jianlin Zhao and Qinjun Kang and Zeng-Yao Li and Jan Carmeliet and Naoki Shikazono and Wen-Quan Tao}
}

@book{Hornung_1997,
    address={New York, NY},
    series={Interdisciplinary Applied Mathematics},
    title={Homogenization and Porous Media},
    volume={6},
    publisher={Springer},
    author={Hornung, Ulrich},
    editor={John, F. and Wiggins, S. and Sirovich, L. and Marsden, J. E. and Kadanoff, L.},
    year={1997},
    collection={Interdisciplinary Applied Mathematics}
}

@article{Joseph_1982_Nonlinear,
    author = {Joseph, D. D. and Nield, D. A. and Papanicolaou, G.},
    title = {Nonlinear equation governing flow in a saturated porous medium},
    journal = {Water Resources Research},
    volume = {18},
    number = {4},
    pages = {1049-1052},
    doi = {10.1029/WR018i004p01049},
    year = {1982}
}

@article{Forchheimer_1901,
    title={ Wasserbewegung durch Boden},
    author={Forchheimer, Philipp},
    journal={Zeitschrift des Vereins Deutscher Ingenieure},
    volume={45},
    pages={1782-1788},
    year={1901}
}

@article{Beavers_Joseph_1967,
    title={Boundary conditions at a naturally permeable wall},
    volume={30},
    DOI={10.1017/S0022112067001375},
    number={1},
    journal={Journal of Fluid Mechanics},
    author={Beavers, Gordon S. and Joseph, Daniel D.},
    year={1967},
    pages={197-207}
}

@article{Taylor_1971,
    title={A model for the boundary condition of a porous material. Part 1},
    volume={49},
    DOI={10.1017/S0022112071002088},
    number={2},
    journal={Journal of Fluid Mechanics},
    author={Taylor, G. I.},
    year={1971},
    pages={319-326}
}

@article{Richardson_1971,
    title={A model for the boundary condition of a porous material. Part 2},
    volume={49}, DOI={10.1017/S002211207100209X},
    number={2},
    journal={Journal of Fluid Mechanics},
    author={Richardson, S.},
    year={1971},
    pages={327-336}
}

@article{Saffman_1971,
    author = {Saffman, P. G.},
    title = {On the Boundary Condition at the Surface of a Porous Medium},
    journal = {Studies in Applied Mathematics},
    volume = {50},
    number = {2},
    pages = {93-101},
    doi = {10.1002/sapm197150293},
    year = {1971}
}

@article{Ene_Sanchez_1975,
    author    = {Ene, Horia I. and Sanchez-Palencia, Enrique},
    title     = {{\'{E}}quations et ph{\'{e}}nom{\`e}nes de surface pour l'{\'{e}}coulement dans un mod{\`e}le de milieu poreux},
    journal   = {Journal de M{\'{e}}canique},
    volume    = {14},
    pages     = {73-108},
    year      = {1975}
}

@article{Levy_Sanchez_1975,
    title = {On boundary conditions for fluid flow in porous media},
    journal = {International Journal of Engineering Science},
    volume = {13},
    number = {11},
    pages = {923-940},
    year = {1975},
    issn = {0020-7225},
    doi = {10.1016/0020-7225(75)90054-3},
    author = {Th{\'{e}}r{\`e}se Levy and Enrique Sanchez-Palencia}
}

@article{Ochoa_Tapia_Whitaker_1995a,
    title = {Momentum transfer at the boundary between a porous medium and a homogeneous fluid -- I. Theoretical development},
    journal = {International Journal of Heat and Mass Transfer},
    volume = {38},
    number = {14},
    pages = {2635-2646},
    year = {1995},
    issn = {0017-9310},
    doi = {10.1016/0017-9310(94)00346-W},
    author = {J.Alberto Ochoa-Tapia and Stephen Whitaker}
}

@article{Ochoa_Tapia_Whitaker_1995b,
    title = {Momentum transfer at the boundary between a porous medium and a homogeneous fluid -- II. Comparison with experiment},
    journal = {International Journal of Heat and Mass Transfer},
    volume = {38},
    number = {14},
    pages = {2647-2655},
    year = {1995},
    issn = {0017-9310},
    doi = {10.1016/0017-9310(94)00347-X},
    author = {J. Alberto Ochoa-Tapia and Stephen Whitaker}
    }

@article{Ochoa_Tapia_Whitaker_1998,
    title={Momentum Jump Condition at the Boundary Between a Porous Medium and a Homogeneous Fluid: Inertial Effects},
    author={J. Alberto Ochoa-Tapia and Stephen Whitaker},
    journal={Journal of Porous Media},
    year={1998},
    volume={1},
    number={3},
    pages={201-217}
}

@article{Cieszko_Kubik_1999,
    title={Derivation of matching conditions at the contact surface between fluid-saturated porous solid and bulk fluid},
    volume={34},
    DOI={10.1023/a:1006590215455},
    journal={Transport in Porous Media},
    author={Cieszko, Mieczys\l{}aw and Kubik, J{\'o}zef},
    year={1999},
    month={Mar},
    pages={319-336}
}

@article{Alazmi_Vafai_2001,
    title = {Analysis of fluid flow and heat transfer interfacial conditions between a porous medium and a fluid layer},
    journal = {International Journal of Heat and Mass Transfer},
    volume = {44},
    number = {9},
    pages = {1735-1749},
    year = {2001},
    issn = {0017-9310},
    doi = {10.1016/S0017-9310(00)00217-9},
    author = {B. Alazmi and K. Vafai}
}

@article{BARS_WORSTER_2006,
    title={Interfacial conditions between a pure fluid and a porous medium: implications for binary alloy solidification},
    volume={550},
    DOI={10.1017/S0022112005007998},
    journal={Journal of Fluid Mechanics},
    author={Le Bars, Michael and Worster, M. Grae},
    year={2006},
    pages={149-173}
}

@article{Jamet_Chandesris_Goyeau_2008,
    title={On the equivalence of the discontinuous one- and two-domain approaches for the modeling of transport phenomena at a fluid/porous interface},
    volume={78},
    DOI={10.1007/s11242-008-9314-9},
    number={3},
    journal={Transport in Porous Media},
    author={Jamet, D. and Chandesris, M. and Goyeau, B.},
    year={2008},
    month={Dec},
    pages={403-418}
}

@article{Goyeau_Lhuillier_2003,
    title = {Momentum transport at a fluid-porous interface},
    journal = {International Journal of Heat and Mass Transfer},
    volume = {46},
    number = {21},
    pages = {4071-4081},
    year = {2003},
    issn = {0017-9310},
    doi = {10.1016/S0017-9310(03)00241-2},
    author = {B. Goyeau and D. Lhuillier and D. Gobin and M.G. Velarde}
}

@article{Hirata_Goyeau_2009,
    title = {Stability of natural convection in superposed fluid and porous layers: Equivalence of the one- and two-domain approaches},
    journal = {International Journal of Heat and Mass Transfer},
    volume = {52},
    number = {1},
    pages = {533-536},
    year = {2009},
    issn = {0017-9310},
    doi = {10.1016/j.ijheatmasstransfer.2008.07.045},
    author = {S.C. Hirata and B. Goyeau and D. Gobin and M. Chandesris and D. Jamet}
}

@article{Chandesris_Jamet_2006,
    title = {Boundary conditions at a planar fluid-porous interface for a Poiseuille flow},
    journal = {International Journal of Heat and Mass Transfer},
    volume = {49},
    number = {13},
    pages = {2137-2150},
    year = {2006},
    issn = {0017-9310},
    doi = {10.1016/j.ijheatmasstransfer.2005.12.010},
    author = {M. Chandesris and D. Jamet}
}

@article{Chikh_Boumedien_Bouhadef_Lauriat_1998,
    title={Analysis of fluid flow and heat transfer in a channel with intermittent heated porous blocks},
    volume={33},
    DOI={10.1007/s002310050208},
    journal={Heat and Mass Transfer},
    author={Chikh, S. and Boumedien, A. and Bouhadef, K. and Lauriat, G.},
    year={1998},
    month={Apr},
    pages={405-413}
}

@article{Neale_Nader_1974,
    author = {Neale, Graham and Nader, Walter},
    title = {Practical significance of brinkman's extension of darcy's law: Coupled parallel flows within a channel and a bounding porous medium},
    journal = {The Canadian Journal of Chemical Engineering},
    volume = {52},
    number = {4},
    pages = {475-478},
    doi = {10.1002/cjce.5450520407},
    year = {1974}
}

@article{Vafai_Kim_1990,
    title = {Fluid mechanics of the interface region between a porous medium and a fluid layer—an exact solution},
    journal = {International Journal of Heat and Fluid Flow},
    volume = {11},
    number = {3},
    pages = {254-256},
    year = {1990},
    issn = {0142-727X},
    doi = {10.1016/0142-727X(90)90045-D},
    author = {K. Vafai and S.J. Kim}
}

@article{Ruan_Rybak_2026,
    title = {Stokes-Brinkman-Darcy models for fluid-porous systems: derivation, analysis and validation},
    journal = {Applied Mathematics and Computation},
    volume = {510},
    pages = {129687},
    year = {2026},
    issn = {0096-3003},
    doi = {10.1016/j.amc.2025.129687},
    author = {Linheng Ruan and Iryna Rybak}
}

@article{Ruan_Rybak_2025,
    title={A hybrid-dimensional Stokes-Brinkman-Darcy model for arbitrary flows to the fluid-porous interface},
    volume={152},
    DOI={10.1007/s11242-025-02220-1},
    number={10},
    journal={Transport in Porous Media},
    author={Ruan, Linheng and Rybak, Iryna},
    year={2025},
    month={Aug}
}

@article{Basu_Khalili_1999,
    author = {Basu, A. J. and Khalili, A.},
    title = {Computation of flow through a fluid-sediment interface in a benthic chamber},
    journal = {Physics of Fluids},
    volume = {11},
    number = {6},
    pages = {1395-1405},
    year = {1999},
    month = {06},
    issn = {1070-6631},
    doi = {10.1063/1.870004}
}

@article{Verma_Tomar_2023,
    title = {A continuous one-domain framework for fluid flow in superposed clear and porous media},
    journal = {Journal of Computational Physics},
    volume = {495},
    pages = {112554},
    year = {2023},
    issn = {0021-9991},
    doi = {10.1016/j.jcp.2023.112554},
    author = {Abhijit Verma and Gaurav Tomar}
}

@article{Chen_Wang_2014,
    title = {A one-domain approach for modeling and simulation of free fluid over a porous medium},
    journal = {Journal of Computational Physics},
    volume = {259},
    pages = {650-671},
    year = {2014},
    issn = {0021-9991},
    doi = {10.1016/j.jcp.2013.12.008},
    author = {Huangxin Chen and Xiao-Ping Wang}
}

@article{Arico_Helmig_Puleo_Schneider_2024,
    title = {A new numerical mesoscopic scale one-domain approach solver for free fluid/porous medium interaction},
    journal = {Computer Methods in Applied Mechanics and Engineering},
    volume = {419},
    pages = {116655},
    year = {2024},
    issn = {0045-7825},
    doi = {10.1016/j.cma.2023.116655},
    author = {Costanza Aric{\`o} and Rainer Helmig and Daniele Puleo and Martin Schneider}
}

@article{Silva_Lemos_2003,
    author = {Renato A. Silva and Marcelo J. S. de Lemos},
    title = {Numerical Analysis of the Stress Jump Interface Condition for Laminar Flow Over a Porous Layer},
    journal = {Numerical Heat Transfer, Part A: Applications},
    volume = {43},
    number = {6},
    pages = {603-617},
    year = {2003},
    publisher = {Taylor \& Francis},
    doi = {10.1080/10407780307351}
}

@article{Chandesris_Jamet_2007,
    title = {Boundary conditions at a fluid-porous interface: An a priori estimation of the stress jump coefficients},
    journal = {International Journal of Heat and Mass Transfer},
    volume = {50},
    number = {17},
    pages = {3422-3436},
    year = {2007},
    issn = {0017-9310},
    doi = {10.1016/j.ijheatmasstransfer.2007.01.053},
    author = {M. Chandesris and D. Jamet}
}

\end{document}